\documentclass[a4paper,11pt]{article}
\pdfoutput=1

\usepackage[no-natbib-sort]{jheppub}

\usepackage[T1]{fontenc}

\usepackage{mathtools,amsfonts,amssymb}
\usepackage[utf8]{inputenc}

\usepackage{array}
\usepackage{bm}
\usepackage{booktabs}
\usepackage{caption}
\usepackage{multirow}
\usepackage{subcaption}
\usepackage{todonotes}
\usepackage{tikz}
\usepackage{afterpage}

\usepackage{hyperref}
\usepackage[nameinlink,capitalize]{cleveref}

\makeatletter{}
\g@addto@macro\bfseries{\boldmath}
\renewcommand{\@email}[1]{\texttt{#1}}
\def\NAT@sort{\z@}
\makeatother

\let\originalleft\left
\let\originalright\right
\renewcommand*{\left}{\mathopen{}\mathclose\bgroup\originalleft}
\renewcommand*{\right}{\aftergroup\egroup\originalright}

\newcommand*{\seccoord}[1]{\ensuremath\hat{#1}} % Notation for a coordinate of the section
\newcommand*{\secx}{\ensuremath\seccoord{x}} % x component
\newcommand*{\secy}{\ensuremath\seccoord{y}} % y component
\newcommand*{\secz}{\ensuremath\seccoord{z}} % z component
\newcommand*{\secw}{\ensuremath\seccoord{w}} % 3x^2 + f z^4 component
\newcommand*{\ratsec}[1]{\ensuremath\hat{#1}}
\newcommand*{\zerosec}{\ensuremath\ratsec{o}}
\newcommand*{\sechomol}[1]{\ensuremath\mathcal{#1}}
\newcommand*{\zerohomol}{\ensuremath\sechomol{Z}}

\newcommand*{\chargeval}[1]{q = \pm #1}
\newcommand*{\height}[0]{\ensuremath\tilde{b}}
\newcommand*{\shioda}[0]{\ensuremath\sigma}

\DeclareMathOperator{\SO}{SO}
\DeclareMathOperator{\Sp}{Sp}
\DeclareMathOperator{\Spin}{Spin}
\DeclareMathOperator{\SU}{SU}
\DeclareMathOperator{\U}{U}
\DeclareMathOperator{\gE}{E}
\DeclareMathOperator{\gF}{F}
\DeclareMathOperator{\gG}{G}

\DeclareMathOperator{\tr}{tr}
\newcommand*{\asu}{\ensuremath\mathfrak{su}}
\newcommand*{\au}{\ensuremath\mathfrak{u}}
\newcommand*{\aso}{\ensuremath\mathfrak{so}}
\newcommand*{\asp}{\ensuremath\mathfrak{sp}}
\newcommand*{\aE}{\ensuremath\mathfrak{e}}
\newcommand*{\aF}{\ensuremath\mathfrak{f}}
\newcommand*{\aG}{\ensuremath\mathfrak{g}}
\newcommand*{\aH}{\ensuremath\mathfrak{h}}

\newcommand*{\cN}{\mathcal{N}}
\newcommand*{\cO}{\mathcal{O}}
\newcommand*{\cR}{\mathcal{R}}

\newcommand*{\C}{\mathbb{C}}

\newcommand*{\bP}{\mathbb{P}}
\newcommand*{\Q}{\mathbb{Q}}
\newcommand*{\R}{\mathbb{R}}
\newcommand*{\Z}{\mathbb{Z}}

\newcommand*{\univcover}[0]{\ensuremath G}

\newcommand*{\canonclass}[0]{\ensuremath K_B}
\newcommand*{\vertdiv}[0]{B}
\newcommand*{\vertdivindex}[0]{\alpha}

\newcommand*{\locus}[1]{\ensuremath\{#1\}}

\newcommand*{\divclass}[1]{\ensuremath [#1]}

\newcommand*{\representation}[0]{\bm{R}}
\newcommand*{\conjrep}[0]{\overline{\bm{R}}}
\newcommand*{\fullrepresentation}[0]{\bm{\mathrm{R}}}
\newcommand*{\fullconjrep}[0]{\overline{\bm{\mathrm{R}}}}

\DeclareMathOperator*{\ord}{ord}
\DeclareMathOperator{\ordalt}{ord}
\newcommand*{\ordvanish}[2]{\ensuremath \ord_{#1}(#2)}
\newcommand*{\ordvanishone}[1]{\ensuremath\ordalt_{1}(#1)}
\newcommand*{\ordvanishtwo}[1]{\ensuremath\ordalt_{2}(#1)}
\newcommand*{\valuation}[2]{\ensuremath v_{#1}(#2)}
\newcommand*{\residue}[2]{\ensuremath\overline{#1}_{#2}}
\newcommand*{\elltrouble}[4][]{\ensuremath R^{#1}_{#2}(#3,#4)}
\DeclarePairedDelimiter\ceil{\lceil}{\rceil}
\DeclarePairedDelimiter\floor{\lfloor}{\rfloor}
\DeclarePairedDelimiter\abs{\lvert}{\rvert}
\newcommand*{\pabs}[2]{\ensuremath\abs{#2}_{#1}}
\newcommand*{\moddist}[2]{u_{#1}(#2)}
\newcommand*{\pscharge}[0]{pseudo-charge}

\newcommand*{\eds}[0]{\ensuremath W}
\newcommand*{\ecurvex}[0]{\ensuremath \phi}
\newcommand*{\ecurvey}[0]{\ensuremath \omega}

\newcommand*{\divpoly}[0]{\ensuremath \Psi}

\newcommand*{\quotparam}[0]{\ensuremath\nu}
\newcommand*{\quotparamtwo}[0]{\ensuremath\mu}

\newcommand*{\scorr}[0]{\ensuremath S}

\newcommand*{\quotfunction}[2]{T_{#1}(#2)}
\newcommand*{\quottriplet}[2]{\vec{\tau}_{#1}(#2)}

\newcommand*{\matterfibcurve}[0]{\ensuremath c}
\newcommand*{\invcartan}[3]{\left(\mathcal{C}^{-1}_{#1}\right)_{#2#3}}
\newcommand*{\invcartanmat}[1]{\mathcal{C}^{-1}_{#1}}
\newcommand*{\cartanmat}[1]{\mathcal{C}_{#1}}

\newcommand*{\numabelian}[0]{\ensuremath L} % Number of u(1) algebras

\newcommand*{\ellipticPlus}[0]{\ensuremath [+]}

\newcommand*{\singtype}[1]{\ensuremath \text{#1}}

\newcolumntype{C}[1]{>{\centering\let\newline\\\arraybackslash\hspace{0pt}}m{#1}}
\newcolumntype{M}{>{$}c<{$}}

\title{Orders of Vanishing and U(1) Charges in F-theory}

\author[1]{Nikhil Raghuram}
\author[2]{and Andrew P. Turner}

\affiliation[1]{Department of Physics \\ Robeson Hall, 0435 \\ Virginia Tech \\ 850 West Campus Drive \\ Blacksburg, VA 24061, USA}
\affiliation[2]{Department of Physics and Astronomy, \\ University of Pennsylvania, \\ Philadelphia,  PA 19104, USA}
\emailAdd{raghuram.nikhil at gmail.com}
\emailAdd{turnerap at sas.upenn.edu}

\abstract{Many interesting questions about F-theory models, including several concerning the F-theory swampland, involve massless matter charged under U(1) gauge symmetries. It is therefore important to better understand the geometric properties of F-theory models realizing various U(1) charges. We propose that, for F-theory models described by elliptic fibrations in Weierstrass form, the U(1) charge of light matter is encoded in the orders of vanishing of the section components corresponding to the U(1) gauge symmetry. We give specific equations relating the U(1) charges to the orders of vanishing that seem to hold for both U(1)-charged singlets and for matter additionally charged under a simply-laced nonabelian gauge algebra. Our formulas correctly describe properties of F-theory models in the prior literature, and we give an argument that they should describe the orders of vanishing for arbitrarily high U(1) charges.
They also resemble formulas for the $p$-adic valuations of elliptic divisibility sequences developed by Stange~\cite{StangeEllTrouble}. These proposals could serve as a U(1) analogue of the Katz--Vafa method, allowing one to determine U(1) charges without resolution. Additionally, they predict geometric information about F-theory models with general U(1) charges, which may be useful for exploring the F-theory landscape and swampland.}

\begin{document}
\maketitle
\flushbottom

\section{Introduction}
\label{sec:intro}

A major goal of the string theory program is understanding how to construct a compactification of string theory realizing a desired massless spectrum. This problem is important for both characterizing the landscape of string compactifications and finding stringy realizations of our observed universe. F-theory~\cite{VafaF-theory, MorrisonVafaI, MorrisonVafaII} has emerged as a powerful tool for constructing string compactifications in large part because it geometrizes several physical features. Much of the information about an F-theory model's massless spectrum, particularly regarding its unbroken gauge symmetry and its light charged matter, is encoded in the mathematical properties of its corresponding elliptic fibration. To be more specific, an F-theory compactification down to $12-2d$ real dimensions is described by a complex $d$-dimensional elliptically fibered Calabi--Yau manifold. Massless nonabelian gauge bosons are supported along divisors in the base with singular elliptic fibers. After resolution, the components in the fibers at codimension one intersect in the pattern of an affine Dynkin diagram, in line with the Kodaira classification of singularity types~\cite{KodairaII}. When supplemented with information about monodromy~\cite{TateMath}, this affine Dynkin diagram tells us which nonabelian gauge algebra is supported along this locus~\cite{MorrisonVafaI,MorrisonVafaII,BershadskyEtAlSingularities}.
Light charged matter, meanwhile, is typically localized at codimension-two loci in the base where the singularity type enhances. The representation of this charged matter can be determined by resolving the singularities and determining how the introduced fiber components at codimension two intersect the resolution divisors for the codimension-one singularities.

While properly determining the nonabelian gauge algebra and matter content\footnote{Since this paper focuses on massless matter, we typically omit the word ``massless'' and use the term ``matter'' to refer to massless matter.} with this approach requires resolving singularities,\footnote{The string junction program offers a valid method for determining nonabelian gauge groups and matter representations that does not require resolutions. (See, for instance, \cite{GrassiHalversonShanesonPhysics,GrassiHalversonShanesonMath}.) We do not significantly discuss this approach here, although it would be interesting to investigate how the results obtained here tie into these methods.} one can often read off singularity types and their associated physical data through easier methods. If the elliptic fibration is in the Weierstrass form
\begin{equation}
y^2 = x^3 + f x z^4 + g z^6\,,
\end{equation}
the Kodaira table (given in \cref{tab:kodaira}) relates the singularity types and their corresponding Lie algebras to the orders of vanishing of $f$, $g$, and $\Delta=4f^3+27g^2$ at a particular divisor in the base. Furthermore, even though the Kodaira classification strictly holds only at codimension one, we can often heuristically use the Kodaira table to find singularity types at codimension two in the base as well. One can then determine matter representations with the Katz--Vafa method~\cite{KatzVafa}, in which one breaks the adjoint of the enhanced singularity type's corresponding Lie algebra to representations of the nonabelian gauge algebra. These techniques allow one to calculate the gauge group and charged matter content of a model simply by considering orders of vanishing, making the process of constructing and analyzing nonabelian F-theory models significantly easier. In addition to letting us quickly determine the massless spectrum of a given model, they can guide the process of constructing an F-theory model with a desired nonabelian gauge algebra and charged matter spectrum by describing expected features of the elliptic fibration.

However, many  interesting questions in the F-theory program involve abelian gauge algebras and light matter charged under them. Clearly, $\au(1)$ algebras are important for phenomenological model building in F-theory: the Standard Model gauge algebra has a $\au(1)$ factor, and many proposals for extending the Standard Model include extra $\au(1)$ algebras~\cite{RizzoZprime,LangackerZprime,GrimmWeigand,HalversonTASI}. They also offer an interesting arena for questions about the landscape and swampland~\cite{VafaSwamp} of F-theory models. As an example, consider 6D F-theory models with a $\U(1)$ gauge group. It has not yet been definitively determined which $\U(1)$ charges of massless matter can and cannot occur in such models, even though there has been significant progress on this problem and more general questions regarding constraints on $\U(1)$ theories~\cite{ParkTaylor,ParkIntersection,MorrisonParkU1,KleversEtAlToric,MonnierMoorePark,RaghuramTaylorLargeCharge,CianciMayorgaPenaValandroHighU1,LeeWeigandAbelianSwamp,CollinucciEtAlHighCharge,AndersonGrayOehlmannQuotients}. However, there are infinite families of charged matter spectra with unbounded $\U(1)$ charges that are consistent with the known 6D low-energy constraints~\cite{ParkTaylor,TaylorTurnerSwamp}. Only a finite number of these matter spectra can be realized in F-theory models~\cite{KumarMorrisonTaylorGlobalAspects}, presenting the possibility of an infinite swampland of such models. (See~\cite{RaghuramTaylorTurnerEnhancement} for recent developments regarding some of these infinite families.)

For these reasons, much work has focused on $\au(1)$ charged matter in F-theory constructions, both for matter charged only under $\au(1)$ algebras and matter additionally charged under nonabelian algebras~\cite{GrimmWeigand,DolanMarsanoSaulinaSchaferNameki,MarsanoSaulinaSchaferNamekiU1,MorrisonParkU1,CveticGrimmKlevers,MayrhoferPaltiWeigand,BorchmannSU5TopSummary,CveticKleversPiraguaMultU1,GrimmKapferKeitelRational,BraunGrimmKeitel,CveticGrassiKleversPiragua,BorchmannSU5Top,CveticKleversPiraguaSong,KrippendorfEtAlGUT,BraunCollinucciValandro,AntoniadisLeontaris,KuntzlerTateTrees,KleversEtAlToric,EsoleKangYau,LawrieSacco,LawrieEtAlRational,CveticKleversPiraguaTaylor,GrimmKapferKleversArithmetic,CveticEtAlHetU1,MorrisonParkTall,MorrisonParkTaylor,WangU1s,CveticLinU1,MayorgaPenaValandro,BuchmullerEtAlSO10,BaumeEtAlTorsion,Raghuram34,KimuraK3U1a,LeeRegaladoWeigand,CianciMayorgaPenaValandroHighU1,TaylorTurnerGeneric,KimuraK3U1b,LeeWeigandAbelianSwamp,CollinucciEtAlHighCharge,KimuraZ4,KimuraHalf3Fold,KimuraHalfFourfold,OehlmannSchimannek,RaghuramTaylorTurnerSM,KnappScheideggerSchimannek}. However, there are still open questions regarding $\au(1)$ algebras in F-theory, in part because, unlike nonabelian algebras, they are not associated with
codimension-one singularities of the fibration.
Instead, they are associated with extra rational sections of the elliptic fibration~\cite{MorrisonVafaII}. In more concrete terms, a rational section of a fibration in Weierstrass form is described by a solution
\begin{equation}
[x:y:z] = [\secx:\secy:\secz]
\end{equation}
of the Weierstrass equations, where $\secx$, $\secy$, $\secz$ are sections of appropriate line bundles on the base. An elliptic fibration may admit several or even an infinite number of such sections. They form a finitely generated group under an operation known as elliptic curve addition. There is a $\au(1)$ algebra corresponding to each generating section of this group with infinite order. Matter charged under a $\au(1)$ algebra still occurs at codimension-two loci in the base with singular fibers, but the $\au(1)$ charge is determined by how the section intersects the resolved singular fiber at the matter locus. Since $\au(1)$ algebras involve sections of the elliptic fibration, much of the technology for nonabelian gauge algebras does not carry over. The Katz--Vafa method, for instance, cannot be used to read off $\au(1)$ charges, as matter with different $\au(1)$ charges can occur at codimension-two loci with the same singularity type. This can be seen, for instance, in the Morrison--Park model~\cite{MorrisonParkU1}, which has a single $\au(1)$ gauge factor: in this model, the $\bm{1}_1$ loci and $\bm{1}_2$ loci are all of type $\singtype{I}_2$, where $(f,g,\Delta)$ vanish to orders $(0,0,2)$.

To the authors' knowledge, the prior F-theory literature does not present any systematic procedure analogous to the Katz--Vafa method for reading off $\au(1)$ charges from simple features such as orders of vanishing. However, there were tentative indications in~\cite{Raghuram34} that, at least for singlets, the $\au(1)$ charge is encoded in orders of vanishing of the $\secx$, $\secy$, and $\secz$ section components at a matter locus. It was noted there that in Weierstrass models with a $\U(1)$ gauge symmetry admitting $\bm{1}_{3}$ and $\bm{1}_{4}$ matter, the section components, particularly $\secz$, vanish to orders larger than 1 at the corresponding matter loci. The work also argued that the complicated non-UFD structure~\cite{KleversEtAlExotic} of these models directly reflects these higher orders of vanishing. These observations naturally suggest some correlation between the singlet charge and the orders of vanishing of the section components; in fact, \cite{Raghuram34} hypothesized a specific relation between these quantities.

Our goal here is to develop these observations into a systematic framework that could serve as a Katz--Vafa analogue for $\au(1)$ charges. Specifically, we aim to find rules relating $\au(1)$ charges to the orders of vanishing of the $\secx$, $\secy$, $\secz$, and $\secw = 3 \secx^2+f \secz^4$ components\footnote{When an elliptic fibration is in Weierstrass form, singularities occur at points on the fiber where $y=3x^2+fz^4=0$. Thus, $\secw$ provides valuable information about where and how a section hits a singular point on a fiber.} of the generating section.  In addition to singlet matter, we consider $\au(1)$-charged matter that is also charged under a semi-simple nonabelian gauge algebra. To make the scope of the analysis more manageable, we assume that the nonabelian part of the gauge algebra is simply-laced.\footnote{Some speculative thoughts on non-simply-laced situations are presented in \cref{sec:conclusions}.} Additionally, we only consider matter in generic~\cite{TaylorTurnerGeneric} representations of the nonabelian gauge factors, and we focus on matter loci where the elliptic fiber singularity type undergoes a rank-one enhancement.\footnote{There are notable examples even for $\au(1)$-charged singlet matter where the singularity type undergoes higher-rank enhancement, such as~\cite{Grassi:2021wii}, where the $\au(1)$-charged singlets are all supported at codimension-two $\singtype{II} \to \singtype{IV}$ loci.} Even with these simplifying assumptions, we consider several matter representations commonly found in F-theory models:\footnote{We do not significantly analyze matter supported along type $\singtype{III}$ or $\singtype{IV}$ loci. It would be interesting to investigate these situations in future work.}
\begin{itemize}
\item the singlet representation occurring at codimension-two $\singtype{I}_{1}\to \singtype{I}_{2}$ loci;
\item the fundamental and antisymmetric representations of $\asu(n)$ occurring, respectively, at codimension-two $\singtype{I}_{n}^{s}\to \singtype{I}_{n+1}$ and $\singtype{I}_{n}^{s}\to \singtype{I}^{*}_{n-4}$ loci;
\item the vector representations of $\aso(2n)$ occurring at codimension-two $\singtype{I}_{n-4}^{*s}\to \singtype{I}_{n-3}^{*}$ loci;
\item the spinor representations of $\aso(8)$, $\aso(10)$, $\aso(12)$, and $\aso(14)$ occurring, respectively, at codimension-two $\singtype{I}_{0}^{*s}\to \singtype{I}_{1}^{*}$, $\singtype{I}_{1}^{*s}\to \singtype{IV}^{*}$, $\singtype{I}_{2}^{*s}\to \singtype{III}^{*}$, and  $\singtype{I}_{3}^{*s}\to \singtype{II}^{*}$ loci;
\item the $\bm{27}$ representation of $\aE_6$ occurring at codimension-two $\singtype{IV}^{*s}\to \singtype{III}^*$ loci;
\item and the $\bm{56}$ representation of $\aE_7$ occurring at codimension-two $\singtype{III}^*\to \singtype{II}^*$ loci.
\end{itemize}

We propose a set of formulas, described in \cref{sec:summary}, that seem to correctly relate $\au(1)$ charges and orders of vanishing for all of these cases. The most important of these formulas describes the order of vanishing of $\secz$ at a codimension-two matter locus. To illustrate the basic idea, consider matter that occurs at the intersection of a gauge divisor supporting a simple nonabelian gauge algebra $\aG$ with the residual $\singtype{I}_1$ discriminant locus. Suppose that, at this matter locus, the singularity type enhances from one associated with $\aG$ to one associated with an enhanced Lie algebra $\aH$. Then, if $G$ and $H$ are the universal covering groups of $\aG$ and $\aH$, the order of vanishing for $\secz$ at the matter locus, $\ordvanishtwo{\secz}$, is schematically given by
\begin{equation}
\ordvanishtwo{\secz} = \frac{1}{2}\left(\frac{d_{G}}{d_{H}} q^2 + \invcartan{G}{\mathcal{I}}{\mathcal{I}} - \invcartan{H}{\mathcal{J}}{\mathcal{J}}\right)\,,
\end{equation}
where $q$ is the $\au(1)$ charge, $\invcartan{G}{\mathcal{I}}{\mathcal{I}}$ and $\invcartan{H}{\mathcal{J}}{\mathcal{J}}$ are particular diagonal elements of the inverse Cartan matrices for $G$ and $H$, and $d_G$ and $d_H$ are the orders (number of elements) of the centers of $G$ and $H$.

We present three lines of evidence in support of these proposals. First, they are satisfied by independently derived F-theory models from the prior literature with $\au(1)$ gauge algebras, as discussed in \cref{sec:explicitconstructions}.  Yet these previous F-theory models support only relatively small $\au(1)$ charges, and we would like to test the formulas over a large range of charges. Fortunately, a previously used strategy~\cite{MorrisonParkU1, Raghuram34}, which is reviewed in \cref{sec:strategy},  allows us to probe the behavior of models supporting large charges without explicitly constructing them. Roughly, the $\au(1)$ charge of matter is determined by how a generating section $\ratsec{s}$ of an elliptic fibration behaves at a matter locus in the base. One can also consider sections $m\ratsec{s}$ that are multiples of the generating section under elliptic curve addition. If a model admits matter with $\au(1)$ charge $q$, then at the matter locus, the non-generating section $m\ratsec{s}$ behaves as though it were a generating section supporting charge $mq$. Therefore, if we wish to obtain the orders of vanishing for generating section components in models supporting large charges, we can start with a model admitting smaller charges and examine multiples of the generating section. When we perform this analysis for the various types of charged matter mentioned above, the resulting orders of vanishing agree exactly with the proposed formulas. Finally, similar formulas describe the $p$-adic valuations of elliptic divisibility sequences (EDSs)  associated with singular elliptic curves~\cite{StangeEllTrouble}, although these formulas are written in a somewhat different format. Elliptic divisibility sequences are closely related to the procedure above involving multiples of generating sections, suggesting that the formulas from~\cite{StangeEllTrouble} should resemble our proposals, as observed.

These three points provide strong evidence in favor of the proposals, but they do not constitute a formal proof. We will not attempt to give such a proof here. One might in fact expect that the proposals hold only heuristically, given that the analogous Katz--Vafa method (at least as described above) is itself somewhat heuristic~\cite{KatzVafa,GrassiMorrisonGroupReps,MorrisonTaylorMaS,GrassiMorrison, EsoleKangSpell,Grassi:2018rva}. In particular, both methods involve applying the Kodaira classification at codimension two in the base; even though this often gives correct results, the Kodaira classification strictly holds only at codimension one. It would be important in future work to more properly establish our proposals and determine their exact range of validity.

However, even if they only hold heuristically, the proposals offer many potential benefits. Just as the Katz--Vafa method allows one to quickly determine nonabelian representations without resolution, these proposals would allow one to easily determine $\au(1)$ charges. Moreover, if one wishes to find an F-theory model realizing particular $\au(1)$ charges, these formulas predict properties of the model's generating section. Said another way, the formulas provide properties to aim for when constructing an F-theory model with desired $\au(1)$ charges. Since the proposals give the orders of vanishing for arbitrary charges, they could be an invaluable tool for exploring the F-theory landscape and swampland. Of course, they are intrinsically interesting for the mathematics of elliptic fibrations, particularly given their connection to $p$-adic valuations of elliptic divisibility sequences.

The formulas also exemplify the general theme that simple representations of light matter are easier to realize in string models than more complicated ones~\cite{IbanezUrangaSimpleReps, TaylorTurnerGeneric}. This contrasts with the situation in quantum field theory: while one must still satisfy conditions such as anomaly cancellation, the process of writing down a quantum field theory with complicated matter representations is not significantly more difficult than that for simpler representations. One can observe these ideas when working with nonabelian gauge symmetries, for which the more complicated representations tend to have larger dimensions. F-theory models with $\asu(2)$ algebras, for instance, almost automatically support light matter in the fundamental ($\bm{2}$) and adjoint ($\bm{3}$) representations. By contrast, the $\bm{4}$ representation only occurs when the $\asu(2)$ algebra is tuned in an intricate way~\cite{KleversTaylor,KleversEtAlExotic}. Even though matter representations with larger $\au(1)$ charges do not necessarily have larger dimensions, one would still expect that it is easier to obtain small $\au(1)$ matter charges in F-theory models than large ones. Indeed, our proposals suggest that the orders of vanishing for the section components increase with $\au(1)$ charge. According to the observations in~\cite{Raghuram34}, these larger orders of vanishing are associated with complicated structures in the Weierstrass models. The proposals therefore give a concrete explanation as to how large $\au(1)$ charges are more difficult to realize than small $\au(1)$ charges.

The rest of this paper is organized as follows. \Cref{sec:summary} summarizes our notations and results and shows how to apply the proposals in a particular example. Given the length of this paper, we have attempted to make this section as self-contained as possible. As such, a reader with sufficient background knowledge of F-theory should be able to understand and apply our formulas after reading only \cref{sec:summary}, at least at a mechanical level. In \cref{sec:review}, we review those aspects of $\au(1)$ gauge algebras in F-theory that are used in this paper. Because the centers of compact Lie groups are important for our results, \cref{sec:centers} discusses these centers and their connection to $\au(1)$ charges in more detail. This section reviews ideas in~\cite{CveticLinU1}, but it also explains in more detail our notations for the centers first mentioned in \cref{sec:summary}. Additionally, we list the allowed $\au(1)$ charges for the various matter representations considered here. \Cref{sec:strategy} describes the general strategy for our analysis, particularly the tactic of using multiples of generating sections to derive information about models supporting large $\au(1)$ charges. \Cref{sec:signs} discusses how the signs of $\au(1)$ charges fit into our strategies and our proposed formulas.

\Cref{sec:singlets,sec:su,sec:so,sec:e6,sec:e7,sec:multgaugealgebras}, which are the bulk of this paper, contain the detailed investigations of the specific types of charged matter we consider here. \Cref{sec:singlets} focuses on singlet matter with $\au(1)$ charges, while \cref{sec:su,sec:so,sec:e6,sec:e7} focus on $\au(1)$-charged matter that is additionally charged in representations of $\asu(n)$, $\aso(2n)$, $\aE_6$, and $\aE_7$, respectively. For each type of matter, we list explicit order-of-vanishing data for various $\au(1)$ charges and show that they satisfy the proposed formulas. We also relate the expressions for each case to similar formulas for EDS valuations in~\cite{StangeEllTrouble}.  In \cref{sec:multgaugealgebras}, we consider matter charged under multiple $\au(1)$ and simple Lie algebras, although we do not perform an exhaustive analysis of these situations.  While these sections may not be necessary for readers simply interested in the final results, they provide important evidence in favor of the formulas in \cref{sec:summary}. In \cref{sec:explicitconstructions}, we discuss how previous models in the F-theory literature follow our proposed formulas. \Cref{sec:conclusions} provides some concluding thoughts and future directions.

\section{Summary of results}
\label{sec:summary}
This section summarizes our results and describes our conventions. We also provide an example of how to use these results to determine the $\au(1)$ charges in a particular model. We have attempted to make this section as self-contained as possible, such that a reader solely interested in the final results can consult this section without referring to the rest of this paper. Nevertheless, such readers may benefit from \cref{sec:review,sec:centers}, which review the background material underlying these results. \Cref{sec:signs}, which discusses signs of $\au(1)$ charges, may also be helpful. Finally, \cref{app:specificreps} summarizes how the results given in this section are specialized for particular gauge algebras and representations. While \cref{app:specificreps} does not contain any results that cannot be obtained using the material in this section, some readers may find the specialized expressions given there to be useful.

\subsection{Notations and conventions}
\label{sec:summnotations}
\paragraph{Weierstrass form} We typically work with elliptic fibrations in the global Weierstrass form
\begin{equation}
\label{eq:globalWeierstrass}
y^2 = x^3 + f x z^4 + g z^6\,.
\end{equation}
Here, $[x:y:z]$ are the homogeneous coordinates of a $\mathbb{P}^{2,3,1}$ space, and $f$, $g$ are holomorphic sections of line bundles on the base. The discriminant of this elliptic fibration is
\begin{equation}
\label{eq:discriminant}
\Delta \equiv 4f^3 + 27 g^2\,.
\end{equation}
Since we focus on elliptically fibered Calabi--Yau manifolds, $f$ and $g$ are respectively sections of $\mathcal{O}(-4\canonclass)$ and $\mathcal{O}(-6\canonclass)$, where $\canonclass$ is the canonical class of the base $B$ of the elliptic fibration. We may also write elliptic fibrations in the local Weierstrass form
\begin{equation}
\label{eq:affineWeierstrass}
y^2 = x^3 + f x + g
\end{equation}
more commonly found in the F-theory literature. The global Weierstrass form and the local form are related by going to a chart where $z=1$. However, the zero section, which occurs at $[x:y:z]=[1:1:0]$, is more clearly visible in the global Weierstrass form.

\paragraph{Section components} Rational sections of the elliptic fibration are described as $[\secx:\secy:\secz]$, where the section components  $\secx$, $\secy$, and $\secz$ solve the Weierstrass form above. In line with the embedding of the elliptic fiber in $\mathbb{P}^{2,3,1}$, we are free to rescale the section components as
\begin{equation}
[\secx:\secy:\secz] \cong [\lambda^2\secx:\lambda^3\secy:\lambda\secz]\,.
\end{equation}
Thus, even though the section components can in principle be rational, we can clear denominators through a rescaling. Throughout this work, we assume that the section components have been rescaled to clear denominators and to remove any common factors that can be scaled away.
Therefore, we take the section components $\secx$, $\secy$, and $\secz$ to be holomorphic sections of line bundles on the base that solve the global Weierstrass form above. We also define the section component
\begin{equation}
\secw = 3\secx^2 +f \secz^4\,.
\end{equation}
Under the rescaling above, $\secw$ becomes $\lambda^4\secw$.

\paragraph{Singularity types} As described more fully in \cref{sec:review}, each singularity type in the Kodaira classification is associated with an ADE Lie algebra,\footnote{As we are interested in simply-laced gauge algebras in this paper, we focus on the split versions of the singularity types.} and in turn, a universal covering group for the ADE algebra. Therefore, in an abuse of language, we often refer to the singularity types by their associated universal covering groups. For instance, we may refer to a split $\singtype{I}_n$ singularity as an $\SU(n)$ singularity. While this can be done unambiguously for most singularity types, there are a few cases that require clarifications. The $\singtype{I}_2$ and $\singtype{III}$ singularity types are both associated with $\SU(2)$, while the $\singtype{I}_3$ and $\singtype{IV}$ singularity types are both associated with $\SU(3)$. In this paper, the terms ``$\SU(2)$ singularity'' and ``$\SU(3)$ singularity'' will correspond to $\singtype{I}_2$ and $\singtype{I}_3$ singularities, respectively; we always refer to the $\singtype{III}$ and $\singtype{IV}$ singularity types using the Kodaira notation. Meanwhile, we take the group for the $\singtype{I}_1$ and $\singtype{II}$ singularity types to be $\SU(1)$.

\paragraph{Shioda map} The Shioda map is defined to be~\cite{ParkIntersection,MorrisonParkU1,GrimmKapferKeitelRational}
\begin{equation}
\label{eq:shiodamap}
\shioda(\ratsec{s}) = \sechomol{S} - \zerohomol - \pi^*\left(D_B\right) + \sum_{\kappa,I,J}\left(\sechomol{S}\cdot \alpha_{\kappa,I}\right)\invcartan{\kappa}{I}{J}\mathcal{T}_{\kappa,J}\,.
\end{equation}
Here, $\sechomol{S}$ is the divisor corresponding to the section $\ratsec{s}$, $\zerohomol$ is the divisor corresponding to the zero section, and $\pi^*(D_B)$ is the pullback of a divisor $D_B$ in the base $B$. For a 6D F-theory model, described by an elliptically fibered threefold, the $\pi^*(D_B)$ term can be written as
\begin{equation}
\sum_{\vertdivindex}\left((\sechomol{S}-\zerohomol)\cdot\zerohomol\cdot\vertdiv^{\vertdivindex}\right)\vertdiv_\vertdivindex\,,
\end{equation}
where the $\vertdiv_\vertdivindex$ are pullbacks of the basis divisors of $H_2(B)$ and the $\vertdivindex$ indices are lowered and raised using the symmetric bilinear form $\Omega_{\alpha\beta}$ for $H_2(B)$. The index $\kappa$ labels the simple nonabelian gauge factors making up the gauge algebra, and $I,J$ run from $1$ to the rank of the $\kappa$th gauge factor. Finally, $\invcartan{\kappa}{}{}$ is the inverse Cartan matrix for the $\kappa$th gauge factor. Each nonabelian gauge factor is associated with a codimension-one locus in the base where the fiber (after resolution) consists of irreducible curves forming an affine Dynkin diagram. The $\alpha_{\kappa,I}$ are the irreducible curves corresponding to the simple roots of the $\kappa$th Lie algebra, and the  $\mathcal{T}_{\kappa,I}$ are the fibral divisors formed by fibering the $\alpha_{\kappa,I}$ over the codimension-one locus. Note that we will always assume the zero section is holomorphic; our primary focus is on elliptic fibrations in Weierstrass form, for which this assumption is valid.

\paragraph{$\au(1)$ charge units} We choose units for charges consistent with the definition of the Shioda map in \cref{eq:shiodamap}, as proposed in  \cite{CveticLinU1}. In these units, singlet $\au(1)$ charges are integers, and the lattice of singlet charges has unit spacing. The $\au(1)$ charge of matter that is also charged under a nonabelian gauge algebra may be fractional, owing to the term in Shioda map involving the inverse Cartan matrix. However, our formulas can easily be adapted for alternative charge normalizations: if the lattice of singlet charges has a spacing of $n$, one simply replaces $q$ in \cref{eq:ordvanishtwozgen,eq:ordvanishtwozgensingle} with $q/n$.

\paragraph{Matter representations} When we refer to matter in some representation $\fullrepresentation$ of a gauge algebra, we often implicitly mean the matter supported at a particular codimension-two locus in the F-theory base, which may involve matter fields in the representations $\fullrepresentation$ and its conjugate $\fullconjrep$. In 6D F-theory models, which have $\cN = (1, 0)$ supersymmetry, codimension-two loci typically support full hypermultiplets\footnote{If $\fullrepresentation$ is a pseudoreal representation, matter can occur in half-hypermultiplets. However, we are most interested in representations that have nonzero $\au(1)$ charges, so cases where the entire representation (including the $\au(1)$ charge) is pseudoreal are not too important here.} of $\fullrepresentation$ matter, which contain fields in both $\fullrepresentation$ and $\fullconjrep$. In 4D F-theory models, a chiral multiplet in a representation $\fullrepresentation$ contains fields only in the $\fullrepresentation$ representation, but there must be an accompanying CPT-conjugate antichiral multiplet in the $\fullconjrep$ representation. Both of these multiplets are supported at the same codimension-two locus. Thus, while we often describe matter as being in a representation $\fullrepresentation$ as a shorthand, one should remember that, due to supersymmetry considerations, there actually may be fields in both $\fullrepresentation$ and $\fullconjrep$.

\paragraph{Residues} We denote the least non-negative residue of $a$ modulo $b$ as $\residue{a}{b}$:
\begin{equation}
\residue{a}{b} = b\left(\frac{a}{b} - \floor*{\frac{a}{b}}\right)\,.
\end{equation}
For instance,
\begin{equation}
\residue{11}{6} = 5\,, \quad \residue{12}{6} = 0\,, \quad \residue{13}{6} = 1\,.
\end{equation}
We also define $\moddist{b}{a}$ as
\begin{equation}
\moddist{b}{a} = \min(\residue{a}{b}, b-\residue{a}{b})\,. \label{eq:moddistdef}
\end{equation}
Roughly, $\moddist{b}{a}$ gives the distance from an integer $a$ to the nearest multiple of $b$. For instance,
\begin{equation}
\moddist{6}{11} = 1\,, \quad \moddist{6}{12}= 0\,, \quad \moddist{6}{13}= 1\,.
\end{equation}

\paragraph{Orders of vanishing} We denote the order of vanishing of an expression $y$ at a locus $\locus{x=0}$ as
\begin{equation}
    \ordvanish{x=0}{y}\,.
\end{equation}
We also denote the order of vanishing of $y$ at a codimension-two locus $\locus{x_1=x_2=0}$ as
\begin{equation}
    \ordvanish{x_1,x_2}{y}\,.
\end{equation}
Since we are giving general proposals about the behavior at loci supporting $\au(1)$ charges, we often want to describe the orders of vanishing at a codimension-one or codimension-two locus supporting a gauge group or charged matter without specifying a particular locus. Therefore, we denote the order of vanishing of $y$ at an unspecified codimension-one locus as $\ordvanishone{y}$ and the order of vanishing at an unspecified codimension-two locus as $\ordvanishtwo{y}$.

\paragraph{Gauge group centers} At various points, we discuss elements of the center $Z(G)$ of a simple Lie group $G$. We refer to the order of $Z(G)$ (the number of elements in the center of $G$) as $d_{G}$. This number also equals the determinant of the Cartan matrix of $G$. We label elements of $Z(G)$ with the integer $\quotparam$; we also use $\quotparam$ to denote the discrete quotient involving the $\quotparam$ element of the center. For most of the gauge groups discussed here, the center is $\Z_m$ for some $m$, and we let $\quotparam$ run from $0$ to $m-1$. For $\Spin(4k)$, however, the center is $\Z_2\times\Z_2$. Since there are four elements in this center, we let $\quotparam$ run from 0 to 3, with $\quotparam=0$ referring to the identity element of the center. At codimension-two loci in the base supporting charged matter, the singularity type enhances, and one can associate a Lie group with the enhanced singularity type. It is useful to consider the center of this enhanced group, even though the enhanced group does not represent a physical gauge group. We label elements of the enhanced group's center with the integer $\quotparamtwo$.

It is also useful to define the function $\quotfunction{G}{\quotparam}$ and the triplet of functions $\quottriplet{G}{\quotparam}$, where $G$ is a simple Lie group. The expressions for $\quotfunction{G}{\quotparam}$ and $\quottriplet{G}{\quotparam}$, which are given in \cref{tab:quotfunctions}, depend on the group $G$ in question.
As discussed in \cref{sec:centers}, the $\quotfunction{G}{\quotparam}$ are essentially diagonal elements of the inverse Cartan matrix of $G$. The $\quottriplet{G}{\quotparam}$ provide information about the orders of vanishing of the $\secx$, $\secy$, and $\secw$ section components.

\begin{table}
    \centering

    \begin{tabular}{*{5}{c}} \toprule
        Singularity Type&  $G$ & $\quotfunction{G}{\quotparam}$  &  $\quottriplet{G}{\quotparam}$ & Valid Values of $\quotparam$ \\ \midrule
        $\singtype{I}_{1}$ & --- & $0$ & $(0,0,0)$ & 0\\[1em]
        $\singtype{II}$ & --- & $0$ & $\left(0,0,0\right)$ & 0 \\[1em]
        $\singtype{I}_{n}$ & $\SU(n)$ & $\frac{\quotparam(n-\quotparam)}{n}$ & $\left(0,1,1\right)\moddist{n}{\quotparam}$ & $0,1,\ldots,n-1$ \\[1em]
        $\singtype{III}$ & $\SU(2)$ & $\frac{\quotparam(2-\quotparam)}{2}$ &  $\left(1,1,1\right)\moddist{2}{\quotparam}$ & $0,1$ \\[1em]
        $\singtype{IV}$ & $\SU(3)$ & $\frac{\quotparam(3-\quotparam)}{3}$ &  $\left(1,1,2\right)\moddist{3}{\quotparam}$ & $0,1,2$ \\[1em]
        $\singtype{I}_{n-4}^*$ & $\Spin(2n)$ & $\begin{cases} 0 & \quotparam = 0 \\ 1 & \quotparam=2 \\ \frac{n}{4} & \quotparam=1,3 \end{cases}$ & $\begin{cases}\left(0,0,0\right) &  \quotparam = 0 \\ \left(1,2,2\right) & \quotparam = 2 \\\left(1,\floor*{\frac{n}{2}},\ceil*{\frac{n}{2}}\right) & \quotparam=1,3\end{cases}$ & $0,1,2,3$ \\[3em]
        $\singtype{IV}^{*}$ & $\gE_6$ & $\frac{2\quotparam(3-\quotparam)}{3}$ & $\left(2,2,3\right)\moddist{3}{\quotparam}$ & $0,1,2$ \\[1em]
        $\singtype{III}^*$ & $\gE_7$ & $\frac{3}{2}\quotparam$ & $\left(2,3,3\right)\quotparam$ & $0,1$ \\[1em]
        $\singtype{II}^*$ & $\gE_8$ & $0$ & $\left(0,0,0\right)$ & $0$ \\ \bottomrule
    \end{tabular}

    \caption{Expressions for $\quotfunction{G}{\quotparam}$ and $\quottriplet{G}{\quotparam}$. While the singularity types $\singtype{I}_1$ and $\singtype{II}$ do not have listed gauge group, $G$ can be thought of as $\SU(1)$ in these situations. In this paper, we focus on the split versions of these singularity types.}
    \label{tab:quotfunctions}
\end{table}

\subsection{Proposal}
\label{sec:proposals}
Suppose a codimension-one locus $\locus{\sigma=0}$ in the base supports a simply-laced gauge algebra $\aG$ whose universal covering group is $G$. The elliptic fibers along this locus are singular, and the singularity type is that corresponding to $G$ in the Kodaira classification. If there is also a $\au(1)$ gauge algebra, the section components of its associated generating section should vanish to orders
\begin{equation}
\ordvanishone{\secz} = 0\,, \quad \left(\ordvanishone{\secx},\ordvanishone{\secy}, \ordvanishone{\secw}\right) = \quottriplet{G}{\quotparam} \label{eq:ordvanishonegen}
\end{equation}
for some value of $\quotparam$ denoting an element of the center of $G$. As described in more detail in \cref{sec:centers}, $\quotparam$ encodes information about which component of the resolved singular fibers along  $\locus{\sigma=0}$ is hit by the generating section. This in turn (at least partially) describes the global structure of the gauge group~\cite{CveticLinU1}.

Now suppose there is some matter with $\au(1)$ charge $q$ occurring at a codimension-two locus in the base with an enhanced singularity type. At least locally, this codimension-two locus can be thought of as the intersection of the codimension-one loci $\locus{\sigma_{i}=0}$,\footnote{Some situations involve matter localized at nodal singularities of irreducible codimension-one loci in the base. The most notable examples are $\au(1)$ charged singlets, which occur at double points of the discriminant locus $\locus{\Delta=0}$. Locally, such situations still look like the intersection of multiple codimension-one loci, even though these loci may be identified globally. For the case of singlets, the singularity type can be thought of as enhancing from $\singtype{I}_1\times \singtype{I}_1$ to $\singtype{I}_2$, or from $\SU(1)\times\SU(1)$ to $\SU(2)$.} with $i$ running from $1$  to $N$.\footnote{In almost all cases of interest, $N$ will be 2. However, we leave open the possibility that $N$ can be greater than 2 to account for more exotic matter types, such as trifundamental representations.}  Each of these loci has singular fibers of type $G_i$, where we are referring to singularity types by their corresponding ADE groups in the Kodaira classification.\footnote{For the singularity types $\singtype{I}_1$ and $\singtype{II}$, whose corresponding group can roughly be thought of as $\SU(1)$, the only allowed value of $\quotparam$ is 0. As indicated in \cref{tab:quotfunctions}, $\quotfunction{}{\quotparam}$ and $\quottriplet{}{\quotparam}$ are 0 for these situations.} Each $\locus{\sigma_{i}=0}$ locus also has a corresponding $\quotparam_{i}$ consistent with the codimension-one orders of vanishings described above. At the codimension-two locus, the singularity type enhances from $G_{1}\times\ldots\times G_{N}$ to some singularity type $H$, where $H$ is again an ADE group. The generating section components for the $\au(1)$ should then vanish to orders
\begin{equation}
\ordvanishtwo{\secz} =\frac{1}{2}\left(\frac{\prod_{i}d_{G_{i}}}{d_H}q^2 + \sum_{i}\quotfunction{G_i}{\quotparam_{i}} - \quotfunction{H}{\quotparamtwo} \right)\label{eq:ordvanishtwozgen}
\end{equation}
and
\begin{equation}
\left(\ordvanishtwo{\secx},\ordvanishtwo{\secy},\ordvanishtwo{\secw}\right) = (2,3,4)\times \ordvanishtwo{\secz} + \quottriplet{H}{\quotparamtwo}\,. \label{eq:ordvanishtwoothergen}
\end{equation}
at the codimension-two locus. Here $d_{G}$ represents the order (the number of elements) of the center of $G$, while $\quotparamtwo$ is an integer representing a particular element of the center of $H$.

An important special case is matter charged under the gauge algebra $\aG\oplus\au(1)$, where $\aG$ is a simple Lie algebra. We again assume that $\aG$ is simply-laced. There is a codimension-one locus $\locus{\sigma=0}$ along which the elliptic fibers are singular with singularity type $G$. The orders of vanishing at $\locus{\sigma=0}$ are still given by \cref{eq:ordvanishonegen} for some $\quotparam$ in the center of $G$. Now suppose that matter charged under $\aG\oplus\au(1)$ occurs at the intersection of $\locus{\sigma=0}$ with the residual discriminant locus, which has singularity type $\singtype{I}_1$.  For the residual discriminant locus, whose ADE group can be thought of as $\SU(1)$, the associated $\quotparam$, $\quotfunction{}{\quotparam}$, and $\quottriplet{}{\quotparam}$ are 0. If the enhanced singularity type at the codimension-two locus is $H$, the orders of vanishing at the codimension-two locus are
\begin{equation}
\ordvanishtwo{\secz} =\frac{1}{2}\left(\frac{d_{G}}{d_H}q^2 + \quotfunction{G}{\quotparam} - \quotfunction{H}{\quotparamtwo} \right)\label{eq:ordvanishtwozgensingle}
\end{equation}
and
\begin{equation}
\left(\ordvanishtwo{\secx},\ordvanishtwo{\secy},\ordvanishtwo{\secw}\right) = (2,3,4)\times \ordvanishtwo{\secz} + \quottriplet{H}{\quotparamtwo}\,. \label{eq:ordvanishtwoothergensingle}
\end{equation}

These expressions provide a procedure for determining the $\au(1)$ charge of some matter given the orders of vanishing of the section components, at least up to sign. From $f$, $g$, and the discriminant $\Delta$, one can determine the $G_i$ and $H$, which in turn determine the $d_{G_i}$ and $d_H$. One can also use Equations \labelcref{eq:ordvanishonegen} and \labelcref{eq:ordvanishtwoothergen} to read off $\quotparam$ and $\quotparamtwo$ for the matter in question.\footnote{There may be two values of $\quotparam$ (or $\quotparamtwo$) corresponding to a particular value of $\quottriplet{G}{\quotparam}$ (or $\quottriplet{H}{\quotparamtwo}$). However, $\quotfunction{G}{\quotparam}$ will be the same for these two values of $\quotparam$, so this ambiguity does not cause any problems when using \cref{eq:ordvanishtwozgen}.} This information can then be plugged into \cref{eq:ordvanishtwozgen}, and one can solve for $q$ up to a sign.

To use these relations to predict the orders of vanishing for matter with a particular $\au(1)$ charge, one must determine the value of $\quotparamtwo$ corresponding to $\quotparam$ and $q$. The relations between these parameters, which depend on the specific groups and representations in question, are described in \cref{tab:quotparamtworelations}. Alternatively, one can try all values of $\quotparamtwo$ allowed for the codimension-two singularity type $H$ and determine which values of $\quotparamtwo$ give sensible orders of vanishing for a desired $\quotparam$ and $q$. However, note that even if the formulas predict reasonable orders of vanishing for a choice of $\quotparam$ and $q$, it does not imply that one can actually find an F-theory model that realizes the appropriate matter with this $\au(1)$ charge. The proposals should be thought of as giving the orders of vanishing assuming that a model supporting the appropriate matter exists. In fact, one likely could use these formulas to argue that certain $\au(1)$ charges cannot be realized in F-theory. One would first calculate the orders of vanishing predicted by the formulas for the $\au(1)$ charge in question. If attempts to construct F-theory model with a rational section admitting these orders of vanishing always lead to obstructions, it would hint that the corresponding $\au(1)$ charge cannot be realized.

For all the models discussed here, including several examples previously seen in the literature, these formulas are satisfied exactly. However, these formulas may more generally give the lowest orders of vanishing of the section components. As a helpful analogy, consider the Kodaira table in \cref{tab:kodaira}, which lists the orders of vanishing of $f$, $g$, and $\Delta$ for various singularity types. In some elliptic fibrations, the orders of vanishing at a locus supporting a particular singularity type may exceed those listed in the table, so long as these larger orders of vanishing do not correspond to an enhanced singularity type. For instance, while  $f$, $g$, and $\Delta$ typically vanish to orders $(3,4,8)$ at an $\singtype{IV}^*$ locus, they could alternatively vanish to orders $(4,4,8)$. However, if they vanish to orders $(3,5,9)$, the singularity type would be $\singtype{III}^*$ rather than $\singtype{IV}^*$. One might expect that the orders of vanishing of the section components behave in a similar way. As an example, our formulas predict that, for an $\asu(5)\oplus\au(1)$ model with $\quotparam=1$, $(\secx,\secy,\secz,\secw)$ should vanish to orders $(0,1,0,1)$ at the codimension-one $\asu(5)$ locus. An $\asu(5)\oplus\au(1)$ model with $\quotparam=1$ and codimension-one orders of vanishing such as $(0,1,0,2)$ may still be consistent with our proposals. However, orders of vanishing such as $(0,2,0,2)$ would correspond to different value of $\quotparam$, so the proposals would suggest $\asu(5)\oplus\au(1)$ models with such codimension-one orders of vanishing would have a different global gauge group structure. Since we do not analyze F-theory models exhibiting such behaviors here, these thoughts on models with higher orders of vanishing are somewhat speculative. It would therefore be important to more properly establish whether such situations can occur and how to handle them in future work.

\begin{table}
    \centering

    \begin{tabular}{*{5}{c}} \toprule
        Gauge Factor & Enhancement & Representation & Allowed $q$ & $\quotparamtwo$ \\ \midrule
        --- & $\singtype{I}_1 \to \singtype{I}_2$ &$\bm{1}_q$& $\Z$ & $\residue{q}{2}$ \\
        $\SU(n)$ & $\singtype{I}_n^{(s)}\to \singtype{I}_{n+1}$ & $\bm{n}_q$& $\frac{\quotparam}{n}+\Z$ & $\residue{nq}{n+1}$ \\
        $\SU(n)$, odd $n$ & $\singtype{I}_n^{(s)}\to \singtype{I}^*_{n-4}$  & $\bm{\frac{n}{2}(n-1)}_q$& $\frac{2\quotparam}{n}+\Z$ & $\residue{nq}{4}$\\
        $\SU(n)$, even $n$ & $\singtype{I}_n^{(s)}\to \singtype{I}^*_{n-4}$  & $\bm{\frac{n}{2}(n-1)}_q$& $\frac{2\quotparam}{n}+\Z$ & $\residue{(q-\frac{2}{n}\quotparam)}{2}+\residue{2\quotparam}{4}$ \\
        $\Spin(2n)$ & ${\singtype{I}_{n-4}^{*(s)}}\to \singtype{I}^{*}_{n-3}$ & $\bm{2n}_q$& $\frac{\quotparam}{2}+\Z$ & $\residue{(2q-\quotparam)}{4} + \residue{2q}{2}$\\
        $\Spin(8)$ & ${\singtype{I}_0^*}^{(s)}\to \singtype{I}^{*}_{1}$ & ${\bm{8_\text{s}}}_{,q}$& $\begin{cases}\frac{1}{2}+\Z & \quotparam=1,2\\\Z & \quotparam=0,3 \end{cases}$& $\residue{(2q-2\quotparam)}{4}$\\
        $\Spin(8)$ & ${\singtype{I}_0^*}^{(s)}\to \singtype{I}^{*}_{1}$ & ${\bm{8_\text{c}}}_{,q}$& $\begin{cases}\frac{1}{2}+\Z & \quotparam=2,3\\\Z & \quotparam=0,1 \end{cases}$&$\residue{(2q-2\quotparam)}{4}$ \\
        $\Spin(10)$ & ${\singtype{I}_1^*}^{(s)}\to {\singtype{IV}^*}$ & $\bm{16}_q$& $\frac{4-\quotparam}{4}+\Z$& $\residue{4q}{3}$\\
        $\Spin(12)$ & ${\singtype{I}_2^*}^{(s)}\to {\singtype{III}}^*$ & $\bm{32}_q\oplus\bm{1}_{2q}$&$\begin{cases}\frac{1}{2}+\Z & \quotparam=2,3\\\Z & \quotparam=0,1 \end{cases}$ & $\residue{(2q+\quotparam)}{2}$\\
        $\Spin(12)$ & ${\singtype{I}_2^*}^{(s)}\to {\singtype{III}}^*$ & $\bm{32^\prime}_q\oplus\bm{1}_{2q}$& $\begin{cases}\frac{1}{2}+\Z & \quotparam=1,2\\\Z & \quotparam=0,3 \end{cases}$ & $\residue{(2q+\quotparam)}{2}$\\
        $\Spin(14)$ & ${\singtype{I}_3^*}^{(s)}\to {\singtype{II}}^*$ & $\bm{64}_q\oplus\bm{14}_{2q}$& $\frac{\quotparam}{4}+\Z$ & 0\\
        $\gE_6$ & ${\singtype{IV}^*}^{(s)}\to {\singtype{III}}^*$ & $\bm{27}_q$ & $\frac{\quotparam}{3}+\Z$ & $\residue{3q}{2}$ \\
        $\gE_7$ & ${\singtype{III}^*}\to {\singtype{II}}^*$ & $\bm{56}_{q}\oplus\bm{1}_{2q}$ & $\frac{\quotparam}{2}+\Z$ & 0 \\ \bottomrule
    \end{tabular}

    \caption{Relations between $\quotparam$, $\quotparamtwo$, and the $\au(1)$ charge $q$ for various representations. Note that, in our conventions, the $\bm{32}$ representation of $\Spin(12)$ has highest weight $[0,0,0,0,0,1]$, while the $\bm{32^\prime}$ representation has highest weight $[0,0,0,0,1,0]$. This convention differs from~\cite{Slansky} but agrees with~\cite{YamatsuGroupTheory}.}
    \label{tab:quotparamtworelations}
\end{table}

\subsection{Example}
As an example of how to read off $\au(1)$ charges, consider the F-theory model described by a Weierstrass equation of the form
\begin{equation}
\begin{aligned}
    y^2 &= x^3 + \left(c_1 c_3 - b^2 c_0 - \frac{1}{3}c_2^2\right) x z^4 \\
    &\qquad\quad + \left(c_0 c_3^2 -\frac{1}{3} c_1 c_2 c_3 +\frac{2}{27}c_2^3 -\frac{2}{3}b^2 c_0 c_2+\frac{1}{4}b^2 c_1^2\right) z^6
\end{aligned}
\end{equation}
with
\begin{equation}
\begin{aligned}
    c_0 &= \sigma\Big[8 c_{1,1} c_{3,0}^3+\sigma  \left(2 c_{3,0}^2 \left(2 b^2 c_{1,1}^2+c_{1,2}\right)-8 b c_{1,1} c_{3,0} c_{2,1}+c_{2,1}^2\right) \\
   &\qquad\quad +\sigma ^2 \left(c_{2,1} \left(4 b^3
   c_{1,1}^2-b c_{1,2}\right)+2 c_{3,0} \left(-4 b^4 c_{1,1}^3+b^2 c_{1,1}
   c_{1,2}+c_{1,3}\right)\right)-\sigma ^3 c_{0,4}\Big]\,,\\
c_1 &= \sigma\Big[2 c_{3,0} \left(2 b c_{1,1} c_{3,0}+c_{2,1}\right)+b \sigma  \left(c_{1,2} c_{3,0}-2 b c_{1,1} c_{2,1}\right)+b \sigma ^2 c_{1,3}+\sigma ^3 c_{1,4}\Big]\,, \\
c_2 &= c_{3,0}^2+b \sigma  c_{2,1}+\sigma ^4 c_{2,4}\,,\\
c_3 &= b c_{3,0}+\sigma ^4 c_{3,4}\,.
\end{aligned}
\end{equation}
We assume that the base of the elliptic fibration is complex two-dimensional, giving us a 6D F-theory model. The parameters $b$ and $c_{i,j}$ are holomorphic sections of line bundles over the base, with the line bundles chosen such that $f$ and $g$ are holomorphic sections of $\mathcal{O}(-4K_B)$ and $\mathcal{O}(-6K_B)$. This model is a tuned version of the Morrison--Park form~\cite{MorrisonParkU1}, and it admits a generating section with components
\begin{equation}
\begin{aligned}
    \secx &= \frac{1}{3} b^2 c_{3,0}^2-\frac{2}{3} \sigma  \left(b^3 c_{2,1}\right)+\sigma ^4 \left(2
    b c_{3,0} c_{3,4}-\frac{2}{3} b^2 c_{2,4}\right)+\sigma ^8 c_{3,4}^2\,, \\
    \secy &= -2 \sigma  \left(b^5 c_{1,1} c_{3,0}^2\right)+\sigma ^2 \left(b^6 c_{1,1}
    c_{2,1}-\frac{1}{2} b^5 c_{1,2} c_{3,0}\right)-\frac{1}{2} \sigma ^3 \left(b^5
    c_{1,3}\right)\\
    &\qquad -\frac{1}{2} \sigma ^4 \left(b^2 \left(b^2 c_{1,4}-2 b c_{2,4} c_{3,0}+4
    c_{3,0}^2 c_{3,4}\right)\right)+b^3 \sigma ^5 c_{2,1} c_{3,4}\\
    &\qquad +b \sigma ^8 c_{3,4} \left(b c_{2,4}-3 c_{3,0} c_{3,4}\right)-\sigma^{12}c_{3,4}^3\,, \\
    \secz &= b\,, \\
    \secw &= -4 \sigma  \left(b^6 c_{1,1} c_{3,0}^3\right)+b^6 \sigma ^2 c_{3,0} \left(6 b c_{1,1}
    c_{2,1}-c_{3,0} \left(4 b^2 c_{1,1}^2+c_{1,2}\right)\right)\\
    &\qquad +b^6 \sigma ^3 \left(b
    c_{2,1} \left(c_{1,2}-4 b^2 c_{1,1}^2\right)+c_{3,0} \left(8 b^4 c_{1,1}^3-2 b^2
    c_{1,1} c_{1,2}-c_{1,3}\right)\right)\\
    &\qquad +\sigma ^4 \left(b^6 c_{0,4}+b^3 c_{3,0}
    \left(b^2 c_{1,4}-2 b c_{2,4} c_{3,0}+4 c_{3,0}^2 c_{3,4}\right)\right)\\
    &\qquad +2 b^4 \sigma
    ^5 \left(b c_{2,1} c_{2,4}+c_{3,0} c_{3,4} \left(2 b c_{1,1} c_{3,0}-3
    c_{2,1}\right)\right)\\
    &\qquad +b^5 \sigma ^6 c_{3,4} \left(c_{1,2} c_{3,0}-2 b c_{1,1}
    c_{2,1}\right) +b^5 \sigma ^7 c_{1,3} c_{3,4}\\
    &\qquad +b^2 \sigma ^8 \left(b^2 c_{2,4}^2+b
    c_{3,4} \left(b c_{1,4}-8 c_{2,4} c_{3,0}\right)+14 c_{3,0}^2 c_{3,4}^2\right) -4 \sigma ^9 b^3 c_{2,1} c_{3,4}^2 \\
    &\qquad - \sigma^{12}c_{3,4}^2\left(4 b^2 c_{2,4} - 12 b c_{3,0} c_{3,4} - 3 c_{3,4}^2 \sigma^4\right)\,.
\end{aligned}
\label{eq:examplesec}
\end{equation}
This indicates that the model has a $\au(1)$ gauge algebra. Additionally, the discriminant is proportional to $\sigma^5$:
\begin{equation}
\Delta = 2 b c_{1,1} c_{3,0}^4 \Delta^{\prime} \sigma^5 + \mathcal{O}(\sigma^6)\,,
\end{equation}
where
\begin{equation}
    \begin{aligned}
        \Delta^\prime &= -112 b^8 c_{1,1}^3 c_{2,1} c_{3,0}^2-4 b^4 c_{1,3} c_{2,1} c_{3,0}^2+4 b^3 c_{0,4} c_{3,0}^3+128 b^9 c_{1,1}^4 c_{3,0}^3+8 b^2 c_{1,4} c_{3,0}^4 \\
        &\qquad -16 b c_{2,4} c_{3,0}^5-4 b^6 c_{1,1} c_{2,1} \left(c_{2,1}^2-3 c_{1,2} c_{3,0}^2\right)+12 b^7 c_{1,1}^2 c_{3,0} \left(3 c_{2,1}^2-2 c_{1,2} c_{3,0}^2\right)\\
        &\qquad +b^5 \left(-2 c_{1,2} c_{2,1}^2 c_{3,0}+\left(c_{1,2}^2+8 c_{1,1} c_{1,3}\right) c_{3,0}^3\right)+32 c_{3,0}^6 c_{3,4}.
    \end{aligned}
\end{equation}
One can verify that the split condition~\cite{GrassiMorrison}
\begin{align}
\frac{9g}{2f}\Bigg|_{\sigma=0} = -\psi^2
\end{align}
is satisfied, indicating that the singularity type along $\locus{\sigma=0}$ is $\singtype{I}_5^s$ and that the supported gauge algebra is $\asu(5)$. There are no other codimension-one loci supporting nonabelian gauge algebras in the model, and the total gauge algebra is $\asu(5)\oplus\au(1)$.

There are two sources of charged matter in this model. If $\locus{\sigma=0}$ is a curve of genus $g$, there are $g$ hypermultiplets of adjoint ($\bm{24}$) matter not localized at a codimension-two locus. The $\au(1)$ charge of this matter is 0, and this matter is therefore not of significant interest to us; however, we must account for it to properly satisfy the 6D anomaly cancellation conditions. The remaining matter is localized at codimension-two loci. Along $\locus{\sigma=0}$, the singularity type enhances to type $\singtype{I}_{1}^{*}$ (or $\Spin(10)$) at $\locus{\sigma=c_{3,0}=0}$ and to type $\singtype{I}_{6}$ (or $\SU(6)$) at $\locus{\sigma=b=0}$, $\locus{\sigma=c_{1,1}=0}$, and $\locus{\sigma=\Delta^\prime=0}$. By the Katz--Vafa method, the $\singtype{I}_1^*$ locus supports hypermultiplets of two-index antisymmetric ($\bm{10}$) matter, while the $\singtype{I}_6$ loci support hypermultiplets of fundamental ($\bm{5}$) matter. There are also codimension-two loci not along $\locus{\sigma=0}$ where the singularity type enhances to $\singtype{I}_2$: $\locus{b=c_{3,4}=0}$ and
\begin{equation}
    \begin{aligned}
        V_{q=1} &= \locus{\secy / \sigma = \secw / \sigma = 0} \setminus \\
        &\qquad \Big(\locus{\sigma=c_{3,0}=0}\cup\locus{\sigma=b=0}\cup\locus{\sigma=c_{1,1}=0}\cup\locus{b=c_{3,4}=0}\Big)\,.
    \end{aligned}
\end{equation}
These loci support singlets uncharged under the $\asu(5)$ algebra.

To determine the $\au(1)$ charges, we should look at the orders of vanishing of the section components. First, along $\locus{\sigma=0}$, the section components, listed in \cref{eq:examplesec}, vanish to orders
\begin{equation}
\ordvanish{\sigma=0}{\secz} = 0\,, \quad \left(\ordvanish{\sigma=0}{\secx},\ordvanish{\sigma=0}{\secy},\ordvanish{\sigma=0}{\secw}\right) = \left(0,1,1\right)\,.
\end{equation}
Comparing to \cref{eq:ordvanishonegen} and noting that
\begin{equation}
\quottriplet{\SU(5)}{\quotparam} = (0,1,1)\times \moddist{5}{\quotparam}\,,
\end{equation}
we see that $\quotparam$ is either 1 or 4. Either can be chosen without affecting the end results of our calculation, so we take $\quotparam$ to be 1. We additionally know that the center of $\SU(5)$ is $\Z_5$, so $d_{SU(5)} = 5$.

The codimension-two orders of vanishing at the matter loci are listed in \cref{tab:exampleenhancements}. This information allows us to determine the $\au(1)$ charges up to sign, which we demonstrate for three of the matter loci.
\begin{itemize}
\item First, consider the locus $\locus{\sigma=c_{3,0}=0}$ locus, which supports $\bm{10}$ matter. From \cref{tab:exampleenhancements}, we see that
\begin{equation}
\left(\ordvanishtwo{\secx},\ordvanishtwo{\secy},\ordvanishtwo{\secw}\right) - (2,3,4)\times\ordvanishtwo{\secz} = \quottriplet{\Spin(10)}{\quotparamtwo}=(1,2,3)\,.
\end{equation}
According to \cref{tab:quotfunctions}, $\quotparamtwo$ is therefore either $1$ or $3$ for this matter locus. Additionally, the center of $\Spin(10)$ is $\Z_4$, and $d_{\Spin(10)}$ is 4. Plugging this information into either \cref{eq:ordvanishtwozgen} or \cref{eq:ordvanishtwozgensingle} leads to
\begin{equation}
q^2 = \frac{d_{\Spin(10)}}{d_{\SU(5)}}\Big(2\ordvanishtwo{\secz} + \quotfunction{\Spin(10)}{\quotparamtwo} - \quotfunction{\SU(5)}{\quotparam}\Big) = \frac{4}{5}\Big(2\times0 + \frac{5}{4} - \frac{4}{5} \Big) = \frac{9}{25}\,.
\end{equation}
Therefore, $\abs{q}$ is $\frac{3}{5}$ for the matter supported here.
\item Second, consider the locus $\locus{\sigma=b=0}$, which supports $\bm{5}$ matter. From \cref{tab:exampleenhancements},
\begin{equation}
\left(\ordvanishtwo{\secx},\ordvanishtwo{\secy},\ordvanishtwo{\secw}\right) - (2,3,4)\times\ordvanishtwo{\secz} = \quottriplet{\SU(6)}{\quotparamtwo}=(0,3,3)\,.
\end{equation}
According to \cref{tab:quotfunctions}, $\quotparamtwo$ is $3$ for this matter locus, and since the center of $\SU(6)$ is $\Z_6$, $d_{\SU(6)}$ is 6.
Plugging this information into either \cref{eq:ordvanishtwozgen} or \cref{eq:ordvanishtwozgensingle} leads to
\begin{equation}
q^2 = \frac{d_{\SU(6)}}{d_{\SU(5)}}\Big(2\ordvanishtwo{\secz} + \quotfunction{\SU(6)}{\quotparamtwo} - \quotfunction{\SU(5)}{\quotparam}\Big) = \frac{6}{5}\Big(2\times1 + \frac{9}{6} - \frac{4}{5} \Big) = \frac{81}{25}\,.
\end{equation}
Therefore, $\abs{q}$ is $\frac{9}{5}$ for the matter supported here.
\item Third, consider the locus $\locus{b=c_{3,4}=0}$, which supports singlet matter, uncharged under $\SU(5)$, with a potentially nonzero $\au(1)$ charge. This locus consists of double points of the discriminant curve. The singularity type enhances from $\singtype{I}_1\times \singtype{I}_1$ to $\singtype{I}_2$, or, in terms of ADE groups, from $\SU(1)\times\SU(1)$ to $\SU(2)$. For the $\SU(1)$ loci, $\quotparam$ is 0, and there are no nontrivial contributions from either $\quottriplet{\SU(1)}{\quotparam}$ or $\quotfunction{\SU(1)}{\quotparam}$. We still need to determine $\quotparamtwo$, however. According to the codimension-two orders of vanishing,
\begin{equation}
\left(\ordvanishtwo{\secx},\ordvanishtwo{\secy},\ordvanishtwo{\secw}\right) - (2,3,4)\times\ordvanishtwo{\secz} = \quottriplet{\SU(2)}{\quotparamtwo} = (0,0,0)\,,
\end{equation}
implying that $\quotparamtwo$ is 0. Additionally, $d_{SU(2)}$ is 2. Plugging this information into either \cref{eq:ordvanishtwozgen} or \cref{eq:ordvanishtwozgensingle} leads to
\begin{equation}
q^2 = d_{SU(2)}\Big(2\ordvanishtwo{\secz} + \quotfunction{\SU(2)}{\quotparamtwo}\Big) =2 \Big(2\times1 + 0 \Big) = 4\,.
\end{equation}
Therefore, $\abs{q}$ is $2$ for the matter supported here.
\end{itemize}
The values of $\abs{q}$ for the other loci can be found by similar procedures, and the results are summarized in the penultimate column of \cref{tab:exampleenhancements}.

\begin{table}
    \centering

    \begin{tabular}{*{5}{c}} \toprule
        Locus & Enhancement  & $\ordvanishtwo{\secx,\secy,\secz,\secw}$ & $\abs{q}$ &  $\asu(5)\oplus\au(1)$ Rep.\\ \midrule
        $\locus{\sigma=c_{3,0}=0}$ & $\singtype{I}_5\times \singtype{I}_1 \rightarrow \singtype{I}_1^*$ & (1,2,0,3) & $\frac{3}{5}$& $\bm{10}_{-\frac{3}{5}}$ \\
        $\locus{\sigma=b=0}$ & $\singtype{I}_5\times \singtype{I}_1 \rightarrow \singtype{I}_6$ & (2,6,1,7) & $\frac{9}{5}$ & $\bm{5}_{-\frac{9}{5}}$  \\
        $\locus{\sigma=c_{1,1}=0}$ & $\singtype{I}_5\times \singtype{I}_1 \rightarrow \singtype{I}_6$& (0,2,0,2)& $\frac{4}{5}$ & $\bm{5}_{-\frac{4}{5}}$\\
        $\locus{\sigma=\Delta^{\prime}=0}$ & $\singtype{I}_5\times \singtype{I}_1 \rightarrow \singtype{I}_6$  & (0,1,0,1) & $\frac{1}{5}$ & $\bm{5}_{\frac{1}{5}}$ \\
        $\locus{b=c_{3,4}=0}$ & $\singtype{I}_1\times \singtype{I}_1 \rightarrow \singtype{I}_2$ & (2,3,1,4) & $2$ & $\bm{1}_2$ \\
        $V_{q=1}$ & $\singtype{I}_1\times \singtype{I}_1 \rightarrow \singtype{I}_2$ & (0,1,0,1) & $1$ & $\bm{1}_1$ \\ \bottomrule
    \end{tabular}

    \caption{Matter loci for $\asu(5)\oplus\au(1)$ example along with the codimension-two orders of vanishing of the section components. For typographical reasons, we describe the singularity types using the Kodaira notation rather than the associated ADE gauge groups.}
    \label{tab:exampleenhancements}
\end{table}

We have not yet found the signs of the charges, which cannot be directly determined from our relations. For singlets, the sign of the charge is irrelevant, as the hypermultiplets contain fields in both the $\bm{1}_{q}$ and $\bm{1}_{-q}$ representations. For the other representations, the relative sign of the $\au(1)$ charges must be consistent with the global structure of the gauge group, as described in more detail in~\cite{CveticLinU1} and in \cref{sec:centers}. In particular, the $\au(1)$ charges for matter in a specific representation of $\asu(5)$ must be separated by integers, even if the charges themselves are fractional. Therefore, if we take the $\au(1)$ charge at $\locus{\sigma=\Delta^{\prime}=0}$ to be $+\frac{1}{5}$, the $\au(1)$ charges at $\locus{\sigma=b=0}$ and $\locus{\sigma=c_{1,1}=0}$ should be $-\frac{9}{5}$ and $-\frac{4}{5}$, respectively. By similar arguments, the charge at $\locus{\sigma=c_{3,0}=0}$ should be $-\frac{3}{5}$.\footnote{A quick way of seeing this is to consider the decomposition $\bm{5}\otimes\bm{5} = \bm{15}\oplus\bm{10}$. If $\bm{5}_{\frac{1}{5}}$ and $\bm{5}_{-\frac{4}{5}}$ are valid representations, then $\bm{5}_{\frac{1}{5}}\otimes\bm{5}_{-\frac{4}{5}}$ would have $\au(1)$ charge $-\frac{3}{5}$.}

This analysis leads to the representations listed in the last column of \cref{tab:exampleenhancements}. If one calculates the multiplicities of the matter loci and includes the $\bm{24}_0$ adjoint matte, one can verify that this matter spectrum satisfies the 6D gauge and gauge--gravitational anomaly cancellation conditions~\cite{ErlerAnomaly,ParkTaylor,ParkIntersection} for the generating section height
\begin{equation}
\height = -2\canonclass + 2\divclass{b} - \frac{4}{5} \divclass{\sigma}\,.
\end{equation}
Note that we have determined the spectrum without performing any resolutions of the singular Weierstrass model, illustrating the power of this approach.

\section{Review of F-theory}
\label{sec:review}

In this section, we will review some basic aspects of F-theory that will be important for the following discussion. Further information on F-theory can be found in~\cite{DenefLesHouches,TaylorTASI,WeigandTASI,ParkIntersection,CveticLinReview}.

F-theory can be thought of most straightforwardly as a non-perturbative version of Type IIB string theory that allows for consistent compactifications in the presence of 7-branes. These 7-branes cause the axiodilaton field to vary over the compactification space, leading the theory to be strongly-coupled in some regions. We track the variation of the axiodilaton by encoding it as the complex structure of an elliptic curve; this leads to the structure of an elliptic fibration $X$ over the compactification space $B$. We assume here that the base $B$ is smooth.

The elliptic fibration associated with an F-theory model can be described mathematically by a Weierstrass model \labelcref{eq:globalWeierstrass},
    \begin{equation}
        \label{eq:reviewGlobalWeierstrass}
        y^2 = x^3 + f x z^4 + g z^6\,,
    \end{equation}
 where $[x : y : z]$ are the homogeneous coordinates of $\bP^{2, 3, 1}$ and $f, g$ are holomorphic sections of line bundles on the base $B$. We will always consider elliptically fibered Calabi--Yau manifolds, in which case $f$ and $g$ are respectively sections of the line bundles $\cO(-4 \canonclass)$ and $\cO(-6 K_B)$, with $\canonclass$ the canonical class of the base.

 The elliptic fiber of this fibration may become singular over certain loci in the base. Fiber singularities at codimension one in the base signal that nonabelian gauge groups are supported at these loci (in the context of Type IIB, this indicates the presence of 7-branes wrapping these loci), while fiber singularities at codimension two in the base herald the presence of localized charged matter. The possible codimension-one fiber singularities were classified by Kodaira and N\'{e}ron~\cite{KodairaI,KodairaII,NeronClassification}, and are listed in \cref{tab:kodaira}. Using the Kodaira classification, one can read off the singularity type at a codimension-one locus by examining the orders of vanishing of $f$, $g$, and the discriminant $\Delta = 4 f^3 + 27 g^2$.

 \begin{table}[!ht]
    \centering

    \begin{tabular}{C{5.7em}*{3}{M}C{4.4em}C{8.3em}} \toprule
        Singularity \newline Type & \ord(f) & \ord(g) & \ord(\Delta) & Dynkin \newline Diagram & Universal Covering \newline Group \\ \midrule
        $\singtype{I}_0$                     & \ge 0   & \ge 0   & 0            & ---                     & ---                               \\
        $\singtype{I}_1$                     & 0       & 0       & 1            & ---                     & ---                               \\
        $\singtype{II}$                      & \ge 1   & 1       & 2            & ---                     & ---                               \\
        $\singtype{III}$                     & 1       & \ge 2   & 3            & $A_1$                   & $\SU(2)$                          \\
        $\singtype{IV}$                      & \ge 2   & 2       & 4            & $A_2$                   & $\SU(3)$                          \\
        $\singtype{I}_n$                     & 0       & 0       & N            & $A_{n - 1}$             & $\SU(n)$                          \\
        $\singtype{I}^*_0$                   & \ge 2   & \ge 3   & 6            & $D_4$                   & $\Spin(8)$                        \\
        $\singtype{I}^*_n$                   & 2       & 3       & N + 6        & $D_{n + 4}$             & $\Spin(2 n + 8)$                  \\
        $\singtype{IV}^*$                    & \ge 3   & 4       & 8            & $\gE_6$                 & $\gE_6$                           \\
        $\singtype{III}^*$                   & 3       & \ge 5   & 9            & $\gE_7$                 & $\gE_7$                           \\
        $\singtype{II}^*$                    & \ge 4   & 5       & 10           & $\gE_8$                 & $\gE_8$                           \\
        Non-minimal               & \ge 4   & \ge 6   & \ge 12       & \multicolumn{2}{c}{Not valid in F-theory at codim.-one}     \\ \bottomrule
    \end{tabular}

    \caption{The Kodaira classification of codimension-one singular fibers. The orders of vanishing of $f, g, \Delta$ at the singular locus in the base are given. The Dynkin diagram entry gives the intersection pattern of the exceptional $\bP^1$s introduced by the resolution procedure; these are potentially subject to Tate monodromy, though we will only consider ADE cases here. The gauge algebra entries list the universal cover of the gauge algebras associated with the given singularity type; which gauge algebra is actually realized by a given singularity is determined via the Tate split conditions mentioned in the text. Note that the final row lists non-minimal singularities, which cannot be resolved crepantly such that the fibration remains flat.}
    \label{tab:kodaira}
\end{table}

If one resolves the codimension-one fiber singularities via a sequence of blowups, new $\bP^1$ curves are introduced in the fiber over the codimension-one locus. The $\bP^1$ curves (including the $\bP^1$ hit by the zero section that exists prior to resolution) intersect in the pattern of an affine ADE Dynkin diagram, with the $\bP^1$ hit by the zero section serving as the affine node. Thus, each singularity type can be associated with an ADE Lie algebra. In compactifications to six or fewer dimensions, the $\bP^1$ curves can undergo monodromy when traversing a closed path along the discriminant locus, identifying them as being associated with the same exceptional divisor. This corresponds to a folding of the Dynkin diagram and thus a reduction of the associated Lie algebra to one of non-ADE type. Whether this occurs can be determined by considering certain algebraic conditions on $f, g, \Delta$ known as the Tate split conditions, which we will not elaborate on further here. In this paper, we will always consider split fibers, in which case the associated Lie algebra is always the ADE algebra corresponding to the given Kodaira singularity. This Lie algebra is precisely the nonabelian gauge algebra supported at the given codimension-one locus in the F-theory model.\footnote{It is worth noting that, in some situations, the true gauge algebra of the theory can differ from that indicated by the geometry due to T-branes~\cite{CecottiCordovaHeckmanVafaTBranes,AndersonHeckmanVafaTBranes}. In this paper, we assume that T-brane effects can be ignored and that the geometry accurately reflects the gauge algebra and light charged matter of the model.} In the discussion that follows, it is important to consider the exceptional divisors found by fibering the new $\bP^1$ curves over the codimension-one locus, which we denote by the symbol $\mathcal{T}$. The number of such exceptional divisors associated with a codimension-one locus equals the rank of the supported gauge algebra.

As mentioned above, matter is supported at codimension-two loci where the singularity type enhances. In the Type IIB picture, this occurs at the intersection of the 7-brane stacks associated with the codimension-one singular loci. While the Kodaira classification provides a complete classification of all possible codimension-one singularities, there is as yet no complete classification of the singularity types at codimension two; nonetheless, we often refer to codimension-two singularities using the Kodaira singularity type associated to the orders of vanishing of $(f, g, \Delta)$ at the given codimension-two locus.

In most situations, the matter representations supported at a given codimension-two locus can be read off without performing a full resolution to codimension two using what is known as the Katz--Vafa method~\cite{KatzVafa}. Consider a given codimension-one singular locus $\locus{\sigma = 0}$ supporting a gauge algebra $\aG$. At a codimension-two locus where the singularity enhances further, we can naively associate a larger gauge algebra $\aG' \supset \aG$ with the enhanced singularity type using the Kodaira classification (note, however, that there is not actually an enhanced gauge algebra supported at the codimension-two locus). We can then consider the branching rule for the adjoint representation of $\aG'$ to representations of $\aG$. This branching rule will always include the adjoint representation of $\aG$ as well as some number of singlets; the remaining representations, which always appear in conjugate pairs, are the Katz--Vafa prediction for the charged light matter supported at the codimension-two locus.

The Katz--Vafa method, while convenient, does not always produce the correct result; in order to properly determine the matter representations supported at a codimension-two locus, one must carry out a resolution of the singular Calabi--Yau manifold to codimension-two. The charged matter representations can then be read off from the dual M-theory picture. In this context, nonabelian charged matter arises from M2 branes wrapping particular combinations of the $\bP^1$ curves making up the resolved fiber over the codimension-two locus; the group theoretic weight associated with a particular combination of curves is determined by computing its intersection with the various exceptional divisors. More concretely, suppose we have a model with a semisimple gauge algebra $\aG$.
We can expand $\aG$ as a direct sum of $K$ simple Lie algebras labeled by the index $\kappa$:
\begin{equation}
\aG = \bigoplus_{\kappa = 1}^K \aG_\kappa\,.
\end{equation}
If a codimension-two locus supports matter in a representation
$\representation$ of $\aG$, then the resolved fiber at that locus consists of a set of irreducible curves.
A (possibly reducible) curve $\matterfibcurve$ that supports the weight $\vec{w}$ for the representation $\representation$ satisfies
\begin{equation}
\mathcal{T}_{\kappa, I}\cdot\matterfibcurve = w_{\kappa, I}\,.
\end{equation}
Here, $\mathcal{T}_{\kappa,I}$ is the exceptional divisor corresponding to the $I$th simple root of $\aG_{\kappa}$, while $w_{\kappa,I}$ is the  element of $\vec{w}$ corresponding to the $I$th Cartan generator of $\aG_{\kappa}$ in the Dynkin basis (the basis of fundamental weights).

\subsection{Abelian gauge algebras in F-theory and the Mordell--Weil group}

Abelian $\au(1)$ gauge factors arise in F-theory in a different fashion, and are not associated with codimension-one fiber singularities. Rather, abelian gauge factors are associated with additional rational sections of the elliptic fibration~\cite{MorrisonVafaII}.

F-theory constructions described by a Weierstrass model \labelcref{eq:reviewGlobalWeierstrass} will always have at least one rational section, the zero section $\zerosec$, given by $[\secx : \secy : \secz] = [1 : 1 : 0]$. However, an elliptic fibration may have additional rational sections. These rational sections form a group, known as the Mordell--Weil group. We discuss the addition operation of this group in the following section. According to the Mordell--Weil theorem, this group is finitely generated, taking the form~\cite{LangNeron}
    \begin{equation}
        \Z^r \oplus \Gamma\,.
    \end{equation}
The group $\Gamma$ is the torsion subgroup, which is associated in the F-theory context with the global structure of the gauge group; we will not discuss the torsional part of the Mordell--Weil group further here. The integer $r$ is known as the Mordell--Weil rank, and the $r$ independent sections other than the zero section are known as generating sections.

An F-theory model associated with an elliptic fibration that has Mordell--Weil rank $r$ has an abelian sector including a $\au(1)^r$ gauge algebra. The motivation for this most naturally comes from the dual M-theory picture. To each generating section $\ratsec{s}$ there is associated a divisor class $\sigma(\ratsec{s})$ whose Poincar\'{e} dual acts as an additional zero mode for the M-theory 3-form $C_3$, indicating the presence of an additional $\U(1)$ gauge boson associated with $\ratsec{s}$. The divisor $\sigma(\ratsec{s})$ is given by the Shioda map, which takes the form \labelcref{eq:shiodamap}:
    \begin{equation}
        \shioda(\ratsec{s}) = \sechomol{S} - \zerohomol - \sum_{\vertdivindex}\left((\sechomol{S}-\zerohomol)\cdot\zerohomol\cdot\vertdiv^{\vertdivindex}\right)\vertdiv_\vertdivindex + \sum_{\kappa,I,J}\left(\sechomol{S}\cdot \alpha_{\kappa,I}\right)\invcartan{\kappa}{I}{J}\mathcal{T}_{\kappa,J}\,.
    \end{equation}
This is a homomorphism from the Mordell--Weil group to the N\'{e}ron--Severi group with rational coefficients. As discussed above, weights of nonabelian matter representations are associated with particular combinations of fibral curves in the resolved threefold; the $\U(1)$ charges of the matter associated with a fibral curve $\matterfibcurve$ are given by
    \begin{equation}
        q = \sigma(\ratsec{s}_i) \cdot \matterfibcurve\,,
    \end{equation}
with $\ratsec{s}_i$ being the generating sections of the Mordell--Weil group.

One final important ingredient is the N\'{e}ron--Tate height pairing between rational sections $\ratsec{s}_i, \ratsec{s}_j$ (with $i, j$ possibly equal), given by
    \begin{equation}
        \begin{aligned}
            \height_{i j} &= -\pi_*(\sigma(\ratsec{s}_i) \cdot \sigma(\ratsec{s}_i)) \\
            &= -\pi_*(\sechomol{S}_i \cdot \sechomol{S}_j) - \canonclass + n_i + n_j - \sum_\kappa \left(\sechomol{S}_i \cdot \alpha_{\kappa, I}\right) \left(\cR_\kappa^{-1}\right)_{I J} \left(\sechomol{S}_j \cdot \alpha_{\kappa, J}\right) b_\kappa\,,
        \end{aligned}
    \end{equation}
with $\pi_*$ the pushforward to the homology lattice of the base, $n_i = \pi_*(\sechomol{S}_i \cdot \zerohomol)$ the locus along which the section $\ratsec{s}_i$ intersects the zero section, and $\cR_\kappa$ the normalized root matrix for the $\kappa$th nonabelian gauge factor. The height matrix $\height_{i j}$ plays the role of the 6D anomaly coefficients for the abelian gauge factors.

\subsection{The elliptic curve group law}

As we have seen, the rational sections of an elliptic fibration form a group, called the Mordell--Weil group. The addition operation of rational sections is a fiberwise extension of the group law for elliptic curves, which we will summarize here. For more information, see~\cite{Silverman:1338326}.

The addition operation of the Mordell--Weil group of rational points of an elliptic curve is induced by the standard addition in $\C$ via the definition of the elliptic curve as a complex torus $\C / \Lambda$. Under the map from this description of the elliptic curve to that given by the Weierstrass equation, the induced operation $\ellipticPlus$ is a well-defined addition on rational points of the elliptic curve with identity element $Z$, typically chosen to be the point $[1 : 1 : 0]$ when the elliptic curve is written in global Weierstrass form. This addition operation is defined by the property that if $P, Q, R$ are the three points of intersection of the elliptic curve with a line (counted with multiplicity), then $P \ellipticPlus Q \ellipticPlus R = Z$. Algorithmically, given two points $P$ and $Q$, we can find their sum $P \ellipticPlus Q$ by forming the line that passes through both $P$ and $Q$ (if $P = Q$, then we instead use the tangent line to the elliptic curve at $P$). This line intersects the elliptic curve in a third point $R$; the line passing through $R$ and $Z$ again intersects the elliptic curve in a third point, which is the result $P \ellipticPlus Q$.

As we will make frequent use of elliptic curve addition in this paper, we include here explicit expressions for the addition law when the elliptic curve is written in global Weierstrass form, with identity $Z = [1 : 1 : 0]$. Given two distinct points $P_1 = [x_1 : y_1 : z_1]$ and $P_2 = [x_2 : y_2 : z_2]$, $P_3 = P_1 \ellipticPlus P_2$ has coordinates
    \begin{equation}
        \begin{aligned}
            x_3 &= x_1 z_1^2 \left(x_2^2 + f z_2^4\right) + x_2 z_2^2 \left(x_1^2 + f z_1^4\right) - 2 z_1 z_2 (y_1 y_2 - g z_1^3 z_2^3)\,, \\
            y_3 &= -y_1^2 y_2 z_2^3 - 3 x_2 x_1^2 y_2 z_2 z_1^2 + 3 x_1 x_2^2 y_1 z_1 z_2^2 + y_2^2 y_1 z_1^3 - 3 g z_1^3 z_2^3 \left(y_2 z_1^3 - y_1 z_2^3\right) \\
            &\qquad - f z_1 z_2 \left(x_2 y_2 z_1^5 + 2 x_1 y_2 z_1^3 z_2^2 - 2 x_2 y_1 z_2^3 z_1^2 - x_1 y_1 z_2^5\right)\,, \\
            z_3 &= x_2 z_1^2 - x_1 z_2^2\,.
        \end{aligned}
    \end{equation}
Given a point $P = [x : y : z]$, the point $2 P = P \ellipticPlus P$ has coordinates
    \begin{equation}
        \begin{aligned}
            x_{2 P} &= \left(3 x^2 + f z^4\right)^2 - 8 x y^2\,, \\
            y_{2 P} &= -\left(3 x^2 + f z^4\right)^3 + 12 x y^2 \left(3 x^2 + f z^4\right) - 8 y^4\,, \\
            z_{2 P} &= 2 y z\,.
        \end{aligned}
    \end{equation}
The inverse of a point $P = [x : y : z]$ is simply $-P = [x : -y : z]$.

\section{Gauge group centers and allowed charges}
\label{sec:centers}
As a first step towards relating $\au(1)$ charges to the orders of vanishing of section components, we should examine the types of $\au(1)$ charges that can occur in F-theory models. The allowed $\au(1)$ charges of singlet matter are quantized. In line with the definition of the Shioda map in \cref{eq:shiodamap}, it is natural to normalize the charges such that the singlet charges are integral and that the lattice of singlet charges has a spacing of 1~\cite{CveticLinU1}. However, matter that is also charged under a nonabelian gauge factor may have fractional charges in this normalization. The allowed fractional charges in an F-theory model depend on the global structure of the gauge group, the description of which involves the center of the gauge group's universal cover.

Since it plays a key role in this paper, let us review the relation between fractional charges and elements of the center first described in~\cite{CveticLinU1}. Suppose we have a model with a gauge algebra $\aG \oplus \au(1)$, where $\aG$ is a semisimple Lie algebra given by
\begin{equation}
\aG = \bigoplus_{\kappa=1}^{K}\aG_{\kappa}\,.
\end{equation}
Recall from  \cref{sec:review}  that matter in a representation $\representation_q$ of $\aG\oplus\au(1)$ comes from M2 branes wrapping curves in the resolved fiber of a codimension-two locus in the base. Specifically, a weight $\vec{w}$ of the representation $\representation_{q}$ comes from wrapping an M2 brane on a curve $\matterfibcurve$ satisfying,
\begin{equation}
\label{eq:intweightequality}
\mathcal{T}_{\kappa, I}\cdot\matterfibcurve = w_{\kappa, I}\,,
\end{equation}
where $\mathcal{T}_{\kappa,I}$ is the fibral divisor corresponding to the $I$th simple root of $\aG_{\kappa}$ and $w_{\kappa,I}$ is the element of $\vec{w}$ corresponding to the $I$th Cartan generator of $\aG_{\kappa}$
The charge of this matter under the $\au(1)$ algebra is
\begin{equation}
q = \shioda(\ratsec{s})\cdot\matterfibcurve\,.
\end{equation}
\Cref{eq:shiodamap} states that  the Shioda map is given by
\begin{equation}
\shioda(\ratsec{s}) =\sechomol{S} - \zerohomol - \pi^*\left(D_B\right) + \sum_{\kappa,I} l_{\kappa,I} \mathcal{T}_{\kappa,I}\,,
\end{equation}
where
\begin{equation}
l_{\kappa,I}  = \sum_{J}(\sechomol{S}\cdot \alpha_{\kappa,J})\invcartan{\kappa}{I}{J}\,.
\end{equation}
In rough terms, the numbers $l_{\kappa,I}$ describe which of the components in the ADE fibers along codimension-one loci are hit by the section. They therefore depend on the behavior of the section at codimension one. However, these numbers are allowed to be fractional, as the inverse Cartan matrix $\invcartanmat{\kappa}$ can have fractional entries. In fact, the only fractional contributions to the charge $q$ come from the $l_{\kappa,I}$ term in the Shioda map, implying that
\begin{equation}
\label{eq:intlatticecondition}
q - \sum_{\kappa,I} l_{\kappa,I} \mathcal{T}_{\kappa,I}\cdot \matterfibcurve  \in \Z\,.
\end{equation}
This statement is true for any weight of any $\representation_q$ representation realized in the F-theory model. Note that $\sum_{\kappa,I} l_{\kappa,I} \mathcal{T}_{\kappa,I}\cdot \matterfibcurve$ may differ by integers for different weights in an irreducible representation $\representation_q$.

If $\mathcal{Q}$ is the generator of the $\au(1)$ algebra and $E_{\kappa,I}$ are the elements of the Cartan sub-algebra of $\aG_{\kappa}$, then $\Upsilon = \mathcal{Q} - \sum_{\kappa,I} l_{\kappa,I} E_{\kappa,I}$ is an element of the Cartan subalgebra of $\aG\oplus\au(1)$. Exponentiating this gives us an element $C=\exp[2\pi i \Upsilon]$ of $\univcover\times\U(1)$, the universal covering group for the algebra $\aG\oplus\au(1)$. $\univcover$ is the simply connected Lie group associated with $\aG$. For any representation $\representation_q$ of $\aG\oplus\au(1)$, $C$ acts on a weight $\vec{w}$ of the representation as
\begin{equation}
C \vec{w} = \exp\left[2\pi i \left(q - \sum_{\kappa,I}l_{\kappa,I} w_{\kappa,I}\right)\right]\vec{w}\,.
\end{equation}
The group element $C$ essentially multiplies each weight by a complex number of magnitude 1. This complex number is the same for all weights in an irreducible representation $\representation_{q}$, as $\sum_{\kappa,I}l_{\kappa,I}w_{\kappa,I}$ differs only by integers for the different weights of this irreducible representation.\footnote{One way of seeing this is to note that two different weights for the same irreducible representation differ by a root. The term $\sum_{\kappa,I}l_{\kappa,I}w_{\kappa,I}$ essentially takes the form $\sum_{\kappa,I,J}n_{J}\invcartan{\kappa}{I}{J} w_{\kappa, I}$, where the $n_{J}$ are integers. The inverse Cartan matrices convert weight and root vectors from the Dynkin basis to the basis of simple roots, and root vectors have integral elements in the latter basis. Therefore, the difference in $\sum_{\kappa,I,J}n_{J}\invcartan{\kappa}{I}{J} w_{\kappa, I}$ for two different weights, which differ by a root, should be integral.} Thus, $C$ is proportional to the identity element and commutes with every element of $\univcover\times\U(1)$. In other words, $C$ is in the center of $\univcover\times\U(1)$. Moreover, \cref{eq:intweightequality,eq:intlatticecondition} imply that $C$ acts trivially on any representation $\representation_q$ realized in the F-theory model. The gauge group of the F-theory model is therefore $(\univcover\times\U(1))/\Xi$, where $\Xi$ is a subgroup of the center of $\univcover\times\U(1)$ that includes $C$.

In summary, the $l_i$ coefficients in $\shioda(\ratsec{s})$, which control the allowed fractional charges in the F-theory model and affect the global structure of the gauge group, can be associated with an element of the gauge group's center. It is therefore worth describing the elements of the center in more detail. If the gauge algebra is $\aG\oplus\au(1)$ as before, we are interested in the center $Z(\univcover\times\U(1))$, where $\univcover$ is the simply connected Lie group associated with $\aG$. Since $\aG$ is a sum of simple Lie algebras $\aG_1\oplus\aG_2\oplus\ldots\oplus\aG_{K}$, the group $\univcover$ is a product $\univcover_1\times\univcover_2\times\ldots\univcover_{K}$ of the universal covers of the $\aG_{\kappa}$. The center of $\univcover\times\U(1)$ is then
\begin{equation}
Z\left(\univcover\times\U(1)\right) = Z(\univcover)\times Z\left(\U(1)\right) = \left(\prod_{\kappa}Z(\univcover_{\kappa})\right)\times\U(1)\,.
\end{equation}
An element of $Z\left(\univcover\times\U(1)\right)$ can be written as $(\quotparam_1,\quotparam_2,\ldots,\quotparam_{K}, \gamma)$, where $\quotparam_{\kappa}$ is an element $Z(\univcover_{\kappa})$ and $\gamma$ is an element of $\U(1)$. We want elements of $Z\left(\univcover\times\U(1)\right)$ analogous to $C$ above, so we let $\gamma$ be $\exp[2\pi i \mathcal{Q}]$. The $l_{\kappa,I}$ are then encoded in a choice of the $\quotparam_{\kappa}$. The centers of some of the compact simple Lie groups are given in \cref{tab:centers}. Many of these centers are $\Z_n$ groups, so it is convenient to let the $\quotparam_\kappa$ be integers corresponding to elements of the $\Z_n$. Even for the $\Z_2\times\Z_2$ center of $\Spin(4k)$, we label the elements of the center with integers ranging from 0 to 3. The notations for each group are discussed in more detail in the following subsections.

\begin{table}
    \centering

    \begin{tabular}{*{3}{c}}\toprule
        Algebra      & Group         & Center \\ \midrule
        $\asu(n)$    & $\SU(n)$      & $\Z_n$ \\
        $\aso(2n+1)$ & $\Spin(2n+1)$ & $\Z_2$ \\
        $\asp(n)$    & $\Sp(n)$      & $\Z_2$ \\
        $\aso(4k)$   & $\Spin(4k)$   & $\Z_2\times\Z_2$ \\
        $\aso(4k+2)$ & $\Spin(4k+2)$ & $\Z_4$ \\
        $\aE_6$      & $\gE_6$       & $\Z_3$ \\
        $\aE_7$      & $\gE_7$       & $\Z_2$ \\
        $\aE_8$      & $\gE_8$       & --- \\
        $\aF_4$      & $\gF_4$       & --- \\
        $\aG_2$      & $\gG_2$       & --- \\ \bottomrule
    \end{tabular}

    \caption{Centers of various groups. If a center is not specified, the center of the group is trivial. $n$ and $k$ are integers.}
    \label{tab:centers}
\end{table}

\begin{figure}
\centering
\includegraphics[scale=0.6]{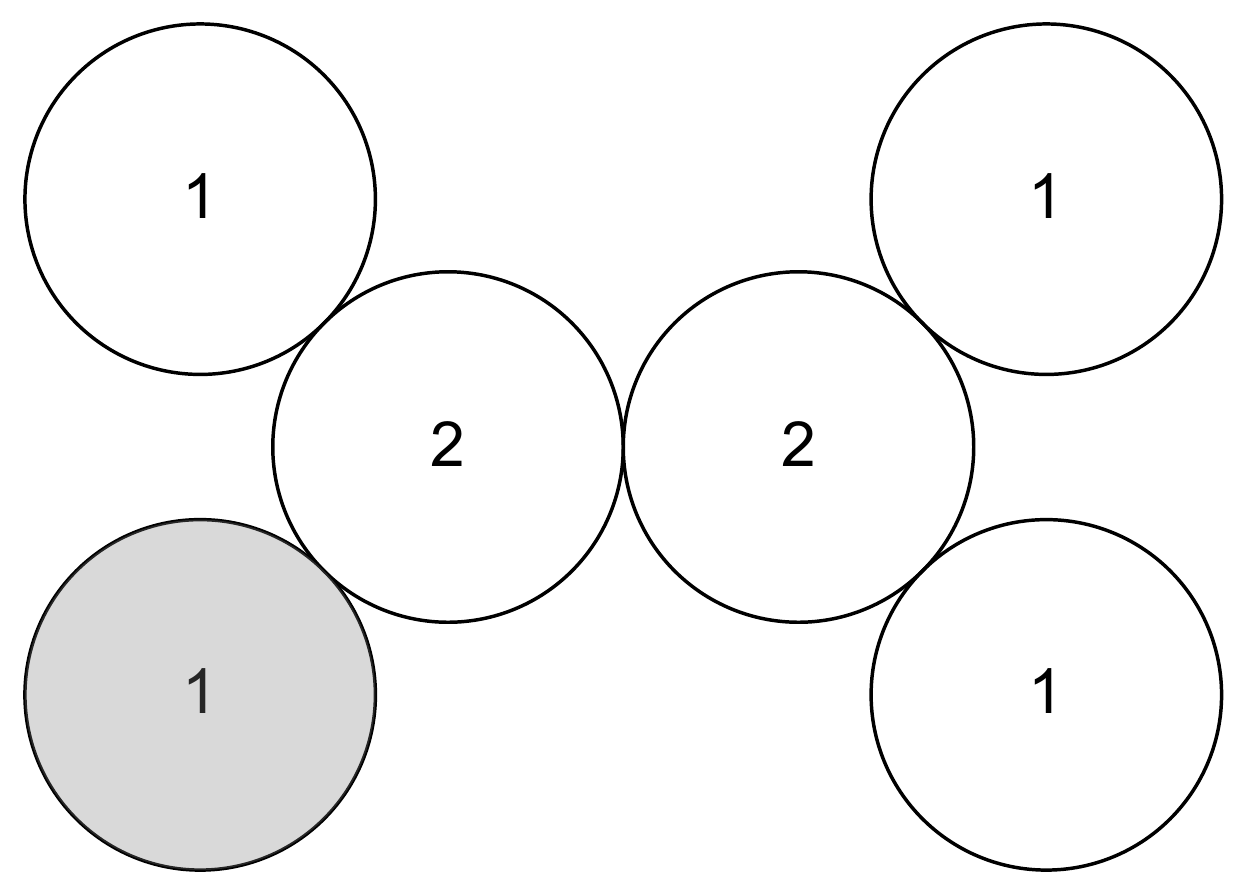}
\caption{Fiber at generic points along a codimension-one locus supporting an $\aso(10)$ algebra. The shaded curve corresponds to the affine node of the $\hat{D}_5$ Dynkin diagram. The numbers within each circle denote the multiplicity of the curve.}
\label{fig:so10fiber}
\end{figure}

We can describe the relation between the $\quotparam_{\kappa}$ and the $l_{\kappa,I}$ more explicitly: because
\begin{equation*}
l_{\kappa,I} = \sum_{J}\left(\sechomol{S}\cdot\alpha_{J}\right)\invcartan{\kappa}{I}{J}\,,
\end{equation*}
the $l_{\kappa,I}$ are determined by the behavior of the section along the codimension-one loci supporting the $\aG_\kappa$ gauge algebras. Let us first focus on the situation where $K = 1$, such that the gauge algebra is $\aG_1\oplus\au(1)$. We additionally assume that $\aG_1$ is simply-laced, as we primarily consider only simply-laced algebras in this paper. At the codimension-one locus supporting $\aG_1$,  the fiber consists of irreducible curves forming an affine ADE diagram corresponding to $\aG_1$; again, the curve representing the affine node is the one hit by the zero section. \cref{fig:so10fiber} illustrates this fiber for an example where $\aG_{1}=\aso(10)$. The generating section for the $\au(1)$ should intersect one of these curves transversally at generic points along the codimension-one locus.\footnote{We should not expect the section to exhibit more extreme behaviors, such as wrapping a curve, at generic points along the codimension-one locus. Such behavior would require all the section components to vanish simultaneously, at least prior to resolution. The section components would then be proportional to appropriate powers of the codimension-one locus, and these factors can be scaled away to remove the wrapping behavior.} The numbers $l_{\kappa,I}$ essentially indicate which of these curves is intersected by the section. Each element of the center should in turn correspond to a curve in the codimension-one fiber that may be hit by the section, and vice versa.\footnote{Along these lines, note that the different values of $\quotparam$ distinguish between, for instance, the $\singtype{I}_n^{(01)}$, $\singtype{I}_n^{(0|1)}$, $\singtype{I}_n^{(0||1)}$, \textellipsis{} models in the language of~\cite{KuntzlerTateTrees,LawrieEtAlRational}.} This statement naively seems to be problematic in the $\aso(10)$ example: while the codimension-one fiber has six irreducible curves, there are only four elements of $Z(\Spin(10)) = \Z_4$. However, the irreducible curves in the fiber may have multiplicities greater than one; these multiplicities are given by the dual Kac labels (or comarks) of the highest root. Because a section should intersect the fiber only once, it can only intersect one of the irreducible curves with multiplicity $1$~\cite{KuntzlerTateTrees, LawrieEtAlRational}. For the $\aso(10)\oplus\au(1)$ example, only the four curves at the ends of the fiber have multiplicity $1$. These four curves exactly match the four elements of $Z(\Spin(10)) = \Z_4$, in line with the expectation that the curve hit by the section at codimension one should specify an element of the center. If the section hits the affine component, we say that $\quotparam =0$, the identity element of the center. If the section hits the upper-left component in \cref{fig:so10fiber}, then we say that $\quotparam = 2$, and if the section hits the upper-right or lower-right curves, then we say that $\quotparam$ is $1$ or $3$, respectively.

The $l_{\kappa,I}$ numbers also provide a way to determine the allowed fractional charges for an F-theory model. Again, suppose that the gauge algebra is $\aG_1\oplus\au(1)$. If the section hits the affine node of the fiber at codimension one, the $l_{1,I}$ are $0$ and the $\au(1)$ charges for matter in any representation of $\aG_1$ must be integral. If the section hits a curve corresponding to the simple root $J$ of $\aG_1$, the $l_{1,I}$ are given by $\invcartan{\kappa}{I}{J}$. To find the allowed $\au(1)$ charges for matter in an irreducible representation $\representation_{q}$ of $\aG_1\oplus\au(1)$, we consider any weight $\vec{w}$ of $\representation_{q}$ in the Dynkin basis. The allowed $\au(1)$ charges are then given by
\begin{equation}
\left(\sum_{I}l_{1,I}w_{I}\right) + j\,, \quad j \in \Z\,.\label{eq:allowedcharges}
\end{equation}
Of course, we can relate the curve hit by the section to an element of $Z(\univcover_1)$, providing us with a way to determine the $\au(1)$ charges consistent with the global structure of the gauge group. \cref{eq:allowedcharges} shows that the allowed $\au(1)$ charges for $\conjrep$ are the negative of those for $\representation$, since the weights of  $\conjrep$ are the negative of those of $\representation$. This ensures that we can form $\representation_q$ hypermultiplets by combining fields in the $\representation_q$ and $\conjrep_{-q}$ representations.

To illustrate these ideas, consider the $\aso(10)\oplus\au(1)$ example above and assume that, at codimension-one, the section hits the upper-right curve in \cref{fig:so10fiber} corresponding to $\quotparam=1$. The Cartan and inverse Cartan matrices for $\aso(10)$ are
\begin{equation}
\cartanmat{\aso(10)} = \begin{pmatrix}2 & -1 & 0 & 0 & 0 \\ -1 & 2 & -1 & 0 & 0\\ 0 & -1 & 2 & -1 & -1\\ 0 & 0 & -1 & 2 & 0\\ 0 & 0 & -1 & 0 & 2\end{pmatrix}\,, \qquad \invcartanmat{\aso(10)} = \begin{pmatrix}1 & 1 & 1 & \frac{1}{2} & \frac{1}{2} \\1 & 2 & 2 & 1 & 1 \\1 & 2 & 3 & \frac{3}{2} & \frac{3}{2} \\\frac{1}{2} & 1 & \frac{3}{2} & \frac{5}{4} & \frac{3}{4} \\ \frac{1}{2} & 1 & \frac{3}{2} & \frac{3}{4} & \frac{5}{4}\end{pmatrix}\,,
\end{equation}
and the upper-right curve corresponds to $J=4$. Therefore,
\begin{equation}
l_{1,I} = \invcartan{\aso(10)}{I}{4}\,.
\end{equation}
For the $\bm{10}$ representation of $\aso(10)$, the highest weight is $[1,0,0,0,0]$, and
\begin{equation}
\sum_{I=1}^{5}\invcartan{\aso(10)}{I}{4}w_{I} = \frac{1}{2}\,.
\end{equation}
The allowed $\bm{10}$ charges are therefore $j+\frac{1}{2}$ for integer $j$. The $\bm{16}$ representation, meanwhile, has a highest weight of $[0,0,0,0,1]$, and
\begin{equation}
\sum_{I=1}^{5}\invcartan{\aso(10)}{I}{4}w_{I} = \frac{3}{4}\,.
\end{equation}
The allowed $\bm{16}$ charges are therefore $j-\frac{1}{4}$ for integer $j$.

It is useful to define a few functions of $\quotparam$. While the specific form of these functions depends on the simple gauge algebra in question, as described in detail below, we can state the general idea behind these functions without reference to a particular gauge algebra. As mentioned above, each possible value of $\quotparam$ is identified with a curve hit by the section at generic points along the codimension-one locus supporting a simple gauge algebra $\aG$. We define
\begin{equation}
\quotfunction{\univcover}{\quotparam} = \invcartan{\aG}{I}{I}\,,
\end{equation}
where $I$ denotes the curve identified with $\quotparam$. If $\quotparam$ is 0, we take $\quotfunction{\univcover}{\quotparam}$ to be 0 as well. In the $\aso(10)$ example, $\quotparam=1$ corresponds to the upper-right curve in \cref{fig:so10fiber}, which is specified by $I=4$. Since
\begin{equation}
\invcartan{\aso(10)}{4}{4}= \frac{5}{4}\,,
\end{equation}
we have
\begin{equation}
\quotfunction{\Spin(10)}{1} = \invcartan{\aso(10)}{4}{4} = \frac{5}{4}\,.
\end{equation}
Along similar lines,
\begin{equation}
    \begin{aligned}
        \quotfunction{\Spin(10)}{0} &= 0\,, \\
        \quotfunction{\Spin(10)}{2} &= \invcartan{\aso(10)}{1}{1} = \frac{1}{2}\,, \\
        \quotfunction{\Spin(10)}{3} &= \invcartan{\aso(10)}{5}{5}  = \frac{5}{4}\,.
    \end{aligned}
\end{equation}
We also define a function $\quottriplet{\univcover}{\quotparam}$ whose output is a triplet of integers. Roughly, it gives the values of
\begin{equation*}
\left(\ordvanishone{\secx},\ordvanishone{\secy},\ordvanishone{\secw}\right)
\end{equation*}
such that the section hits the fibral curve associated with $\quotparam$, at least at generic points along the codimension-one locus supporting $\aG$.\footnote{Intriguingly, the values of $\quottriplet{G}{\quotparam}$ for various $G$, $\quotparam$ also appear in the Tate's algorithm table~\cite{BershadskyEtAlSingularities} as the orders of vanishing of $a_2+\frac{1}{4}a_1^2$, $a_3$, and $a_4$ for various singularity types. This observation may be related to the shifts in $x$ and $y$ performed when converting the elliptic fibration from Tate form to Weierstrass form.} We will describe $\quottriplet{\univcover}{\quotparam}$ in more detail when we individually discuss the different gauge algebras. Even though we have developed these ideas based on the codimension-one behavior of the sections, it will be useful to use these same functions at codimension two as well. When these expressions are used at codimension two, $\quotparam$ is replaced with $\quotparamtwo$.

These ideas generalize naturally for models where $\aG$ consists of more than one simple algebra. In particular, let $\aG$ be the direct sum of simple algebras $\aG_{\kappa}$. Each $\aG_{\kappa}$ simple algebra is supported on a codimension-one locus in the base. The fiber at this locus consists of a set of curves forming the affine Dynkin diagram of $\aG_{\kappa}$. We have a set of numbers $l_{\kappa,I}$ for each $\aG_{\kappa}$ that describe which of the fibral curves is hit by the section, at least at generic points along the codimension-one locus for $\aG_{\kappa}$. If the section hits the affine node, then $l_{\kappa,I}=0$. If the section hits one of the other curves, then
\begin{equation}
l_{\kappa,I} = \sum_{J}\left(\mathcal{S}\cdot\alpha_{\kappa,J}\right)\invcartan{\kappa}{I}{J}\,,
\end{equation}
where $\invcartanmat{\kappa}$ is the inverse Cartan matrix for $\aG_{\kappa}$. Of course, the section can only hit those curves with multiplicity 1 in the fiber. As before, the $l_{\kappa,I}$ can alternatively by described using a specific element of $Z(\univcover\times\U(1))$. For each $\kappa$, we can associate $l_{\kappa,I}$ with an element $\quotparam_{\kappa}$ of the center of $\univcover_{\kappa}$, the universal covering group for $\aG_{\kappa}$. Then, the gauge group has a quotient by a group containing the element $(\quotparam_1,\ldots,\quotparam_{K},\gamma)$ of $Z(\univcover\times\U(1))$, where $K$ is the total number of simple, nonabelian gauge algebras. We can then find $\quotfunction{\univcover_{\kappa}}{\quotparam_\kappa}$ and $\quottriplet{\univcover_{\kappa}}{\quotparam_\kappa}$ for each $\aG_{\kappa}$.

We now describe how the ideas above are applied for the types of gauge algebras discussed here. We focus only on the simply-laced algebras. For each gauge algebra, we describe how the possible values of $\quotparam$ match with curves in the resolved fibers and give explicit expressions for $\quotfunction{\univcover}{\quotparam}$ and $\quottriplet{\univcover}{\quotparam}$. We also discuss the allowed $\au(1)$ charges for matter in representations of these gauge algebras.

\subsection{$\asu(n)$}
\label{sec:suncenters}
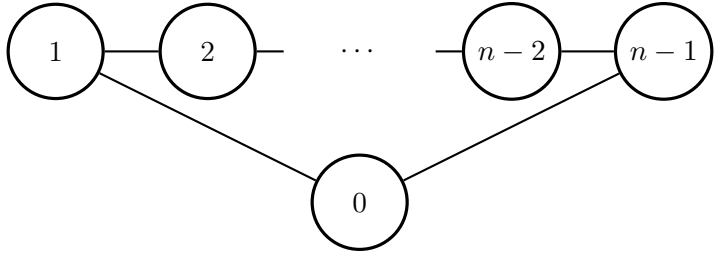
\begin{figure}
\centering
\begin{tikzpicture}[
multnode/.style={circle, draw=black, fill=gray!30, very thick, minimum size=1.25cm},
singnode/.style={circle, draw=black, fill=white, very thick, minimum size=1.25cm},
]

\node[singnode] (reg1) at (-4,0) {1};
\node[singnode] (reg2) at (-2,0) {2};
\node[singnode] (reg3) at (2,0) {$n-2$};
\node[singnode] (reg4) at (4,0) {$n-1$};
\node[singnode] (aff1) at (0,-2) {0};
\node (cont) at (0,0) {$\cdots$};

\draw[thick] (reg1)--(aff1);
\draw[thick] (reg1)--(reg2);
\draw[thick] (reg3)--(reg4);
\draw[thick] (aff1)--(reg4);
\draw[thick] (reg2)--(-1,0);
\draw[thick] (reg3)--(1,0);
\end{tikzpicture}
\caption{Dynkin diagram for $\asu(n)$. The numbers give the value of $I$ corresponding to each node; the affine node has $I=0$.}
\label{fig:sundynkindiagram}
\end{figure}

The affine Dynkin diagram for $\asu(n)$, shown in \cref{fig:sundynkindiagram}, has $n$ nodes. The curves are labeled by $I$, which runs from 0 to $n-1$ with the $I=0$ curve serving as the affine node of the diagram. All of the curves in the resolved $\asu(n)$ fiber have multiplicity $1$, so any of them can be hit by a section. This agrees with the fact that the center of $\SU(n)$ is $\Z_n$, and each curve should correspond to an element of the center. We take $\quotparam$ to be $n-I$, where $I$ refers to the curve hit by the section, or 0 if the section hits the $I=0$ affine node curve. In the language of~\cite{KuntzlerTateTrees, LawrieEtAlRational}, $\quotparam=0$ corresponds to the $\singtype{I}_{n}^{(01)}$ fiber, $\quotparam=1$ corresponds to the $\singtype{I}_{n}^{(0|1)}$ fiber, and in general $\quotparam$ corresponds to an $\singtype{I}_{n}^{(0|\ldots|1)}$ fiber with $\moddist{\quotparam}{n}$ bars between the 0 and the 1.

The inverse Cartan matrix for $\SU(n)$ takes the form~\cite{YamatsuGroupTheory}
\begin{equation}
\invcartan{\asu(n)}{I}{J} = \frac{1}{n}\min(I,J)\left(n-\max(I,J)\right)\,.
\end{equation}
Therefore, we define
\begin{equation}
\quotfunction{\SU(n)}{\quotparam} =\invcartan{\asu(n)}{n-\quotparam,}{n-\quotparam} = \frac{\quotparam(n-\quotparam)}{n}\,. \label{eq:quotfunctionsun}
\end{equation}
Since the highest weight of the fundamental representation is $[1,0,\ldots,0]$, \cref{eq:allowedcharges} implies that the allowed $\au(1)$ charges for the fundamental representation are
\begin{equation}
\frac{\quotparam}{n}+j\text{ for }j\in\Z\,.
\end{equation}
The two-index antisymmetric representation has a highest weight of $[0,1,0,\ldots,0]$, and its allowed $\au(1)$ charges are
\begin{equation}
\frac{2\quotparam}{n} + j\text{ for }j\in\Z\,.
\end{equation}
Note that the allowed charges for $\quotparam^\prime = n-\quotparam$ are the negative of those for $\quotparam$. This fact agrees with the group structure of the center, as the inverse of $\quotparam$ in $\Z_n$ is $n-\quotparam$.

For the antisymmetric representation of $\asu(n)$, the same charge can occur for different values of $\quotparam$ when $n$ is even. When the gauge algebra is $\asu(6)\oplus\au(1)$, for instance, $\bm{15}_1$ matter can occur when $\quotparam=0$ or when $\quotparam=3$. Because $\quotparam$ describes the global structure of the gauge group, a situation involving $\bm{15}_1$ with $\quotparam=0$ is distinct from a situation involving $\bm{15}_1$ with $\quotparam=3$. We should therefore expect that the orders of vanishing for $\bm{15}_1$ should be different for $\quotparam=0$ and $\quotparam=3$. This does not pose any issue if one is reading off the charges from a model, as one can determine $\quotparam$ from the codimenson-one orders of vanishing. However, when trying to predict the orders of vanishing for a certain charge in a spectrum, one may also need to choose a value of $\quotparam$. The need for this extra information is not a shortcoming of our approach; rather, it reflects the unavoidable fact that models with different global gauge groups are distinct.

We also define
\begin{equation}
\quottriplet{\SU(n)}{\quotparam} = \left(0,1,1\right)\times\moddist{n}{\quotparam}\label{eq:quottriplesun}
\end{equation}
for  $n\ge 2$, where $\moddist{n}{\quotparam}$ is defined in \cref{sec:summnotations}. These numbers describe the orders of vanishing for $\secx$, $\secy$, and $\secw$ such that the section hits the appropriate curve in the $\singtype{I}_n$ fiber. However, $\SU(2)$ and $\SU(3)$ can also be realized by type $\singtype{III}$ and type $\singtype{IV}$ fibers. To account for these special cases, we also define
\begin{equation}
\quottriplet{\singtype{III}}{\quotparam} = \left(1,1,1\right)\times\moddist{2}{\quotparam}\,, \quad \quottriplet{\singtype{IV}}{\quotparam} = \left(1,1,2\right)\times\moddist{3}{\quotparam}\,.
\end{equation}

\subsection{$\aso(2n)$}
\label{sec:soncenters}
\begin{figure}
\centering
\begin{tikzpicture}[
twonode/.style={circle, double, draw=black, fill=white, very thick, minimum size=1.25cm},
singnode/.style={circle, draw=black, fill=white, very thick, minimum size=1.25cm},
]

\node[singnode] (reg1) at (-4,2) {1};
\node[twonode] (reg2) at (-2,0) {2};
\node[twonode] (reg3) at (2,0) {$n-2$};
\node[singnode] (reg4) at (4,2) {$n-1$};
\node[singnode] (reg5) at (4,-2) {$n$};
\node[singnode] (aff1) at (-4,-2) {0};
\node (cont) at (0,0) {$\cdots$};

\draw[thick] (aff1)--(reg2);
\draw[thick] (reg1)--(reg2);
\draw[thick] (reg3)--(reg4);
\draw[thick] (reg3)--(reg5);
\draw[thick] (reg2)--(-1,0);
\draw[thick] (reg3)--(1,0);
\end{tikzpicture}
\caption{Dynkin diagram for $\aso(2n)$. The numbers give the value of $I$ corresponding to each node, with the affine node having $I=0$. Nodes marked with two concentric circles occur with multiplicity 2 in the fiber.}
\label{fig:so2ndynkindiagram}
\end{figure}
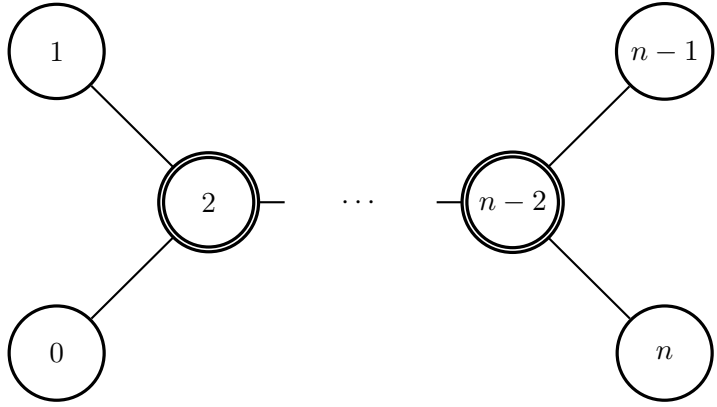

While the affine Dynkin diagram for $\aso(2n)$, shown in \cref{fig:so2ndynkindiagram}, has $n$ nodes, only four of them occur with multiplicity 1 in the resolved fiber. These are the nodes at the ends of the diagram, namely the affine node and the three nodes labeled by $I=1$, $n-1$, and $n$. The center $Z(\Spin(2n))$ is $\Z_4$ for odd $n$ and $\Z_2\times\Z_2$ for even $n$. In either case,  $Z(\Spin(2n))$ has four elements, in agreement with the number of multiplicity-one curves in the resolved fiber. For odd $n$, we label the elements of $\Z_4$ by $\quotparam=0,1,2,3$. We then identify $\quotparam=0$ with the affine node, $\quotparam=1$ with the $I=n-1$ node, $\quotparam=2$ with the $I=1$ node, and $\quotparam=3$ with the $I=n$ node. For even $n$, we still label the elements of $\Z_2\times\Z_2$ by $\quotparam=0,1,2,3$. We also identify $\quotparam=0$ with the affine node, $\quotparam=1$ with the $I=n-1$ node, $\quotparam=2$ with the $I=1$ node, and $\quotparam=3$ with the $I=n$ node. In the language of~\cite{KuntzlerTateTrees, LawrieEtAlRational}, $\quotparam=0$ corresponds to the ${\singtype{I}^*_{n-4}}^{(01)}$ fiber, $\quotparam=2$ corresponds to the ${\singtype{I}^*_{n-4}}^{(0|1)}$ fiber, and $\quotparam=1,3$ correspond to the ${\singtype{I}^*_{n-4}}^{(0||1)}$ fiber.

The inverse Cartan matrix for $\aso(2n)$ takes the form~\cite{YamatsuGroupTheory}
\begin{equation}
\invcartanmat{\aso(2n)}= \frac{1}{2}\begin{pmatrix}2 & 2 & 2 & \cdots & 2 & 1 & 1\\ 2 & 4 & 4 & \cdots & 4 & 2 & 2\\ 2& 4 & 6 & \cdots & 6 & 3 & 3\\\vdots & \vdots & \vdots & \ddots & \vdots & \vdots & \vdots \\2 & 4 & 6 & \cdots & 2(n-2) & (n-2) & (n-2) \\1 & 2 & 3 & \cdots & (n-2) & n/2 & (n-2)/2 \\ 1 & 2 & 3 & \cdots & (n-2) & (n-2)/2 & n/2  \end{pmatrix}\,.
\end{equation}
Therefore,
\begin{equation}
\quotfunction{\Spin(2n)}{\quotparam} = \begin{cases} 0 & \quotparam = 0 \\ 1& \quotparam = 2 \\ \frac{n}{4} & \quotparam = 1,3\end{cases}\,.
\end{equation}
The allowed charges for various representations depend on the value of $n$ modulo 4. The vector representation of $\aso(2n)$ has a highest weight of $[1,0,\ldots,0]$. There are also two spinor representations for all $n$, which have highest weights $[0,0,\ldots,0,1]$ and $[0,0,\ldots,1,0]$. For odd $n$, these representations are conjugates of each other, and spinor hypermultiplets contain fields in both of these representations. However, for even $n$, these two spinors are self-conjugate and can therefore be supported at distinct loci (
and can belong to distinct hypermultiplets in 6D contexts). The allowed $\au(1)$ charges for these representations are summarized in \cref{tab:so2ncharges}. As for the antisymmetric representation of $\asu(n)$, matter in particular representations can occur with the same charge for different values of $\quotparam$. Note that, for odd $n$, the allowed charges for $\quotparam=1$ and $\quotparam=3$ are negatives of each other, in line with the fact that $\quotparam=1$ and $\quotparam=3$ are each other's inverses in $\Z_4$.  Finally, we define
\begin{equation}
\quottriplet{\Spin(2n)}{\quotparam} = \begin{cases} (0,0,0) & \quotparam = 0\\ (1,2,2) & \quotparam = 2\\ (1,\floor*{\frac{n}{2}},\ceil*{\frac{n}{2}}) & \quotparam=1,3\end{cases}
\end{equation}

\begin{table}
    \centering

    \begin{tabular}{*{6}{c}}\toprule
        \multirow{2}{*}{Group}           & \multirow{2}{*}{Irrep} & \multicolumn{4}{c}{$\quotparam$} \\ \cline{3-6}
                                         &                        & 0 & 1             & 2             & 3 \\ \midrule
        \multirow{3}{*}{$\aso(8\ell)$}   & \textbf{V}             & 0 & $\frac{1}{2}$ & 0             & $\frac{1}{2}$ \\
                                         & \textbf{S}             & 0 & $\frac{1}{2}$ & $\frac{1}{2}$ & 0 \\
                                         & \textbf{C}             & 0 & 0             & $\frac{1}{2}$ & $\frac{1}{2}$ \\ \midrule
        \multirow{3}{*}{$\aso(8\ell+2)$} & \textbf{V}             & 0 & $\frac{1}{2}$ & 0             & $\frac{1}{2}$ \\
                                         & \textbf{S}             & 0 & $\frac{3}{4}$ & $\frac{1}{2}$ & $\frac{1}{4}$ \\
                                         & \textbf{C}             & 0 & $\frac{1}{4}$ & $\frac{1}{2}$ & $\frac{3}{4}$ \\ \midrule
        \multirow{3}{*}{$\aso(8\ell+4)$} & \textbf{V}             & 0 & $\frac{1}{2}$ & 0             & $\frac{1}{2}$ \\
                                         & \textbf{S}             & 0 & 0             & $\frac{1}{2}$ & $\frac{1}{2}$ \\
                                         & \textbf{C}             & 0 & $\frac{1}{2}$ & $\frac{1}{2}$ & 0 \\ \midrule
        \multirow{3}{*}{$\aso(8\ell+6)$} & \textbf{V}             & 0 & $\frac{1}{2}$ & 0             & $\frac{1}{2}$ \\
                                         & \textbf{S}             & 0 & $\frac{1}{4}$ & $\frac{1}{2}$ & $\frac{3}{4}$ \\
                                         & \textbf{C}             & 0 & $\frac{3}{4}$ & $\frac{1}{2}$ & $\frac{1}{4}$ \\ \bottomrule
    \end{tabular}

    \caption{Fractional parts of the allowed $\au(1)$ charges for various representations of $\aso(2n)$. \textbf{V} refers to the vector representation with highest weight $[1,0,\ldots,0]$. \textbf{S} refers to the spinor representation with highest weight $[0,\ldots,0,1]$. \textbf{C} refers to the spinor representation with highest weight $[0,\ldots,1,0]$.}
    \label{tab:so2ncharges}
\end{table}

\subsection{$\aE_6$, $\aE_7$, and $\aE_8$}
\label{sec:excenters}

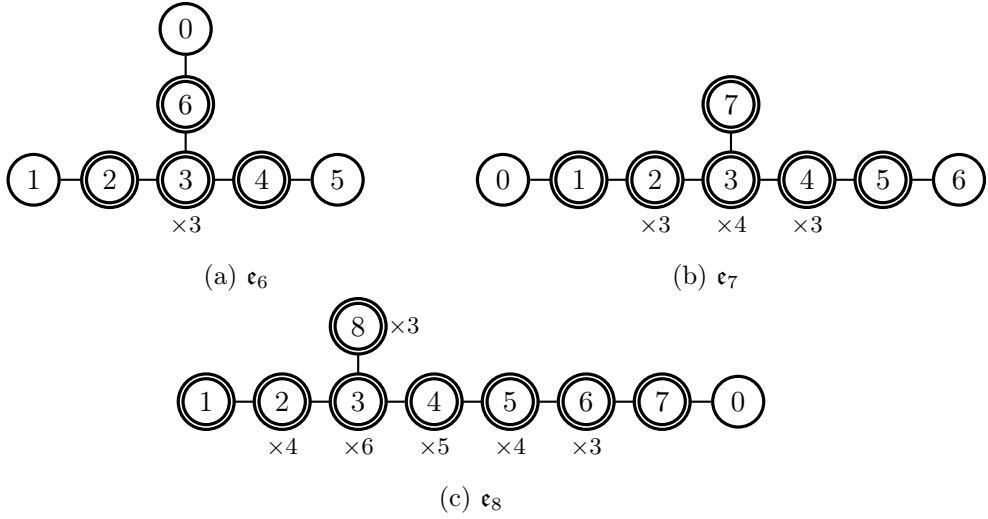
\begin{figure}
\centering
\begin{subfigure}[b]{0.4\textwidth}
\begin{tikzpicture}[
twonode/.style={circle, double, draw=black, fill=white, very thick, minimum size=.6cm},
singnode/.style={circle, draw=black, fill=white, very thick, minimum size=.6cm},
]

\node[singnode] (reg1) at (-2,0) {1};
\node[twonode] (reg2) at (-1,0) {2};
\node[twonode] (reg3) at (0,0) {3};
\node (label3) at (0,-0.6) {\footnotesize$\times 3$};
\node[twonode] (reg4) at (1,0) {$4$};
\node[singnode] (reg5) at (2,0) {$5$};
\node[twonode] (reg6) at (0,1) {$6$};
\node[singnode] (aff1) at (0,2) {0};

\draw[thick] (aff1)--(reg6)--(reg3);
\draw[thick] (reg1)--(reg2)-- (reg3)--(reg4)--(reg5);
\end{tikzpicture}
\caption{$\aE_6$}
\label{fig:e6dynkindiagram}
\end{subfigure}
\begin{subfigure}[b]{0.4\textwidth}
\begin{tikzpicture}[
twonode/.style={circle, double, draw=black, fill=white, very thick, minimum size=.6cm},
singnode/.style={circle, draw=black, fill=white, very thick, minimum size=.6cm},
]

\node[singnode] (aff1) at (-3,0) {0};
\node[twonode] (reg1) at (-2,0) {1};
\node[twonode] (reg2) at (-1,0) {2};
\node (label2) at (-1,-0.6) {\footnotesize$\times 3$};
\node[twonode] (reg3) at (0,0) {3};
\node (label3) at (0,-0.6) {\footnotesize$\times 4$};
\node[twonode] (reg4) at (1,0) {$4$};
\node (label4) at (1,-0.6) {\footnotesize$\times 3$};
\node[twonode] (reg5) at (2,0) {$5$};
\node[singnode] (reg6) at (3,0) {$6$};
\node[twonode] (reg7) at (0,1) {$7$};

\draw[thick] (reg7)--(reg3);
\draw[thick] (aff1)--(reg1)--(reg2)-- (reg3)--(reg4)--(reg5)--(reg6);
\end{tikzpicture}
\caption{$\aE_7$}
\label{fig:e7dynkindiagram}
\end{subfigure}

\begin{subfigure}[b]{0.6\textwidth}
\centering
\begin{tikzpicture}[
twonode/.style={circle, double, draw=black, fill=white, very thick, minimum size=.6cm},
singnode/.style={circle, draw=black, fill=white, very thick, minimum size=.6cm},
]

\node[twonode] (reg1) at (-2,0) {1};
\node[twonode] (reg2) at (-1,0) {2};
\node (label2) at (-1,-0.6) {\footnotesize$\times 4$};
\node[twonode] (reg3) at (0,0) {3};
\node (label3) at (0,-0.6) {\footnotesize$\times 6$};
\node[twonode] (reg4) at (1,0) {$4$};
\node (label4) at (1,-0.6) {\footnotesize$\times 5$};
\node[twonode] (reg5) at (2,0) {$5$};
\node (label5) at (2,-0.6) {\footnotesize$\times 4$};
\node[twonode] (reg6) at (3,0) {$6$};
\node (label6) at (3,-0.6) {\footnotesize$\times 3$};
\node[twonode] (reg7) at (4,0) {$7$};
\node[singnode] (aff1) at (5,0) {0};
\node[twonode] (reg8) at (0,1) {$8$};
\node (label6) at (0.6,1) {\footnotesize$\times 3$};

\draw[thick] (reg8)--(reg3);
\draw[thick] (reg1)--(reg2)--(reg3)--(reg4)--(reg5)--(reg6)--(reg7)--(aff1);
\end{tikzpicture}
\caption{$\aE_8$}
\label{fig:e8dynkindiagram}
\end{subfigure}

\caption{Dynkin diagrams for $\aE_6$,$\aE_7$ and $\aE_8$. The numbers within each circle give the value of $I$ corresponding to each node, with the affine node having $I=0$. Nodes marked with two concentric circles occur with multiplicity greater than 1 in the fiber. If the multiplicity is greater than 2, the multiplicity is indicated just outside the node. Nodes with concentric circles without a given multiplicity have multiplicity 2.}
\label{fig:se6dynkindiagram}
\end{figure}

The affine Dynkin diagram for $\aE_6$, shown in \cref{fig:e6dynkindiagram}, has seven nodes. Three of these nodes---the $I=0$, $1$, and $5$ nodes---occur with multiplicity 1 in the resolved fiber, in agreement with the three elements of $Z(\gE_{6}) = \Z_3$. Labeling the elements of $\Z_3$ with $\quotparam=0,1,2$, we associate $\quotparam=0$ with the $I=0$ affine node, $\quotparam=1$ with the $I=1$ curve, and $\quotparam=2$ with the $I=5$ curve. In the language of~\cite{KuntzlerTateTrees, LawrieEtAlRational}, $\quotparam=0$ corresponds to the ${\singtype{IV}^*}^{(01)}$ fiber, and $\quotparam=1,2$ correspond to the ${\singtype{IV}^*}^{(0|1)}$ fiber. The inverse Cartan matrix of $\aE_6$ is
\begin{equation}
\invcartanmat{\aE_6} = \frac{1}{3}\begin{pmatrix} 4 & 5 & 6 & 4 &  2 & 3\\ 5 & 10 & 12 & 8 & 4 & 6\\6 & 12 & 18 & 12 & 6 & 9\\ 4 & 8 & 12 & 10 & 5 & 6\\2 & 4 & 6 & 5 & 4 & 3\\3 & 6 & 9 & 6 & 3 & 6\end{pmatrix}\,.
\end{equation}
Based on its diagonal entries for $I=1$ and $I=5$, we define
\begin{equation}
\quotfunction{\gE_6}{\quotparam} = \frac{2}{3}\quotparam(3-\quotparam)\,.
\end{equation}
We also define
\begin{equation}
\quottriplet{\gE_6}{\quotparam} = \left(2,2,3\right)\times\moddist{3}{\quotparam}
\end{equation}
In this paper, we are primarily interested in matter in the $\bm{27}$ representation, with highest weight $[1,0,0,0,0,0]$, and its conjugate. From the form of the inverse Cartan matrix, the allowed $\au(1)$ charges for the $\bm{27}$ representation are
\begin{equation}
\frac{\quotparam}{3} + j \text{ for }j\in\Z\,.
\end{equation}
The allowed charges for the $\overline{\bm{27}}$ representation are the negatives of those for the $\bm{27}$ representation.

The affine Dynkin diagram for $\aE_7$, shown in \cref{fig:e7dynkindiagram}, has two nodes that occur with multiplicity $1$: the $I=0$ affine node and the $I=6$ node. The center of $\gE_7$ is $\Z_2$, which has two elements labeled by $\quotparam=0$ and $\quotparam=1$. We associate $\quotparam=0$ with the affine $I=0$ node and $\quotparam=1$ with the $I=6$ node. In the language of~\cite{KuntzlerTateTrees, LawrieEtAlRational}, $\quotparam=0$ corresponds to the ${\singtype{III}^*}^{(01)}$ fiber, and $\quotparam=1,2$ correspond to the ${\singtype{III}^*}^{(0|1)}$ fiber. Since the inverse Cartan matrix of $\aE_7$ has
\begin{equation}
\invcartan{\aE_7}{I}{6} = \frac{1}{2}\left(2,4,6,5,4,3,3\right)\,,
\end{equation}
we define
\begin{equation}
\quotfunction{\gE_7}{\quotparam} = \frac{3}{2}\quotparam\,.
\end{equation}
We also define
\begin{equation}
\quottriplet{\gE_7}{\quotparam} = \left(2,3,3\right)\times\moddist{2}{\quotparam} =  \left(2,3,3\right)\times\quotparam\,.
\end{equation}
We are primarily interested in matter in the pseudoreal $\bm{56}$ representation of $\gE_7$, which has highest weight $[0,0,0,0,0,1,0]$. The allowed $\au(1)$ charges for $\bm{56}$ matter are therefore
\begin{equation}
\frac{\nu}{2} + j\text{ for }j\in\Z\,.
\end{equation}

Finally, the affine Dynkin diagram for $\aE_8$, shown in \cref{fig:e8dynkindiagram} has only one node that occurs with multiplicity 1: the affine node. The center of $\gE_8$ is trivial, so we identify the affine node with $\quotparam=0$. We will not consider matter charged under an $\aE_8$ gauge algebra, as it is difficult if not impossible to find localized $\aE_8$ matter without $f$ and $g$ vanishing to orders $(4,6)$ at codimension two.\footnote{Such loci do not admit an easy interpretation in terms of charged matter, and they are in fact associated with tensionless non-critical strings and superconformal field theories. See the review article~\cite{HeckmanRudeliusSCFT} and references contained therein for more information.} However, we will encounter situations where the singularity type enhances to $\gE_8$ at codimension two. As evidenced by \cref{eq:ordvanishtwozgen} and \labelcref{eq:ordvanishtwoothergen}, the formulas we develop here use $\quotfunction{}{\quotparam}$ and $\quottriplet{}{\quotparam}$ for the group corresponding to the codimension-two singularity type. Therefore, we define
\begin{equation}
\quotfunction{\gE_8}{\quotparam} = 0\,, \quad \quottriplet{\gE_8}{\quotparam} = \left(0,0,0\right)\,.
\end{equation}

\section{General strategy}
\label{sec:strategy}
The ultimate goal of this work is to argue that information about the $\au(1)$ charge spectrum of an F-theory model is encoded in the orders of vanishing of the section components. In particular, we would like to establish explicit relations between the orders of vanishing at a codimension-two locus and the charge supported at that locus. Ideally, one would probe this problem by examining the generating sections in F-theory models that realize various types of charges. However, the currently known F-theory constructions realize only a small subset of the possible charges. As an example, consider F-theory models with just a $\U(1)$ gauge group. Explicit constructions of this type have realized charges up to $\chargeval{6}$~\cite{MorrisonParkU1,KleversEtAlToric,Raghuram34,CollinucciEtAlHighCharge,OehlmannSchimannek,KnappScheideggerSchimannek}, even though indirect arguments suggest F-theory $\U(1)$ models should be able to admit charges at least as large as~$\chargeval{21}$ \cite{RaghuramTaylorLargeCharge}. Since only a limited number of charges have actually been observed in F-theory models, one might imagine it would be difficult to make statements about arbitrary charges.

Fortunately, as first pointed out in~\cite{MorrisonParkU1}, there is a way to at least conjecture about charges not seen in the currently known F-theory constructions. To simplify the discussion, we assume that we are working with a 6D F-theory model, although the arguments should carry over to lower-dimensional models as well. Suppose our F-theory model has a $\au(1)$ gauge algebra and that there is a codimension-two locus in the base supporting matter with $\au(1)$ charge $q$. At the geometric level, the singularity type of the elliptic fiber enhances at this locus. After resolution, there is at least one exceptional curve at this locus supporting charged matter, which we refer to as $\matterfibcurve$. Since the locus supports matter with charge $q$, the generating section  $\ratsec{s}$ behaves at this locus in a way such that
\begin{equation}
    \sigma(\ratsec{s})\cdot \matterfibcurve = q\,.
\end{equation}

The elliptic fibration also admits sections that are multiples of the generating section. Since the Shioda map is a homomorphism, a section $m\ratsec{s}$ satisfies
\begin{equation}
    \sigma(m\ratsec{s})\cdot \matterfibcurve = m\sigma(\ratsec{s})\cdot \matterfibcurve = m q
\end{equation}
at this same codimension-two locus in the base. It naively appears as if the matter has ``charge'' $mq$ according to the section $m\ratsec{s}$. In the rest of this paper, we will refer to the quantity $\sigma(\ratsec{s}^\prime)\cdot \matterfibcurve$ as the \emph{\pscharge{}} when $\ratsec{s}^\prime$ is not a generating section. More specifically, we will say that a section $\ratsec{s}^\prime$, which may not necessarily be a generating section, ``realizes a \pscharge{} $q^\prime$'' at a particular locus when $\sigma(\ratsec{s}^\prime)\cdot \matterfibcurve=q^\prime$. Of course, since $m\ratsec{s}$ is not a generating section, it does not truly correspond to a $\au(1)$ gauge symmetry, and the model does not genuinely have charge $mq$ matter supported at this locus. Nevertheless, we expect that the section $m\ratsec{s}$ behaves as a generating section would in a model that genuinely supported charge $mq$ matter, at least near the codimension-two locus.

Thus, our strategy is to use non-generating sections realizing \pscharge{} $q$ to glean information about the generating sections in models genuinely admitting charge $q$ matter. Specifically, we start with a model admitting some set of relatively small charges, which we refer to as the ``seed'' model. Then, we consider multiples of the generating section, allowing us to find the orders of vanishing for the section components for non-generating sections admitting higher \pscharge{}s. These data can then be used to establish and test the expressions in \cref{sec:proposals} relating $\au(1)$ charges to the orders of vanishing. While we described this strategy in the context of a model with just a $\au(1)$ gauge algebra, it is equally applicable whenever the gauge algebra includes a $\au(1)$ factor, even if there are other nonabelian or abelian gauge factors. In addition to admitting larger \pscharge{}s than the generating section, a non-generating section can be associated with a different element $\quotparam$ of the center. If a generating section $\ratsec{s}$ is associated with a center element $\quotparam$, then  $\ratsec{s}^\prime = m\ratsec{s}$ is associated with the center element $\quotparam^{\prime}=m\quotparam$; in other words, the center element for $m\ratsec{s}$ is found by adding together $m$ copies of $\quotparam$ according to the group law of the center. This ensures that the \pscharge{}s realized by $m\ratsec{s}$ are allowed for $\quotparam^\prime=m\quotparam$ according to the rules in \cref{sec:centers}.

Key to this strategy is the assertion that a non-generating section realizing \pscharge{} $q$ and a generating section admitting charge $q$ behave the same way at the loci in question. This is a reasonable claim, since the behavior near a codimension-two locus is a local property of the section. The question of whether a section is a generating section, on the other hand, involves global properties of the model; the local behavior of the section near a particular locus would presumably not depend on such global properties. We will also see more concrete, albeit circumstantial, evidence in favor of this assertion. First, we can verify this claim in cases where we have an explicit F-theory model genuinely admitting charge $q$. Additionally, there are often multiple ways of obtaining a \pscharge{} $q$. For instance, a \pscharge{} $q=4$ can be obtained from the section $4\ratsec{s}$ in a model admitting $q=1$ matter or from the section $2\ratsec{s}$ in a model admitting $q=2$ matter. If our assertion is true, then we should see the same local behavior for a particular \pscharge{} regardless of how we obtain a particular \pscharge{}. In fact, our analysis gives the same orders of vanishing for each given \pscharge{} even when that \pscharge{} is produced in independent ways.\footnote{More precisely, we obtain the same orders of vanishing whenever we produce a \pscharge{} with the same value of $\quotparam$. See \cref{sec:centers}, particularly \cref{sec:suncenters}, for more details.} This suggests that our expressions relating the charge to the orders of vanishing hold more broadly.

\subsection{Signs}
\label{sec:signs}

At this point, we should discuss how the sign of the charge fits into our proposal. Using only the orders of vanishing of the $\secx$, $\secy$, $\secz$, and $\secw$ section components, one should not be able to determine the sign of the charge. When an elliptic fibration has a generating section $\ratsec{s}$, we are equally free to choose the inverse section $-\ratsec{s}$ as the generating section. If we take the generating section to be $-\ratsec{s}$, the $\au(1)$ charges should be the negative of those when $\ratsec{s}$ is the generating section. However, for an elliptic fibration in Weierstrass form, one can find the inverse of a section by flipping the sign of the $\secy$ component while leaving the other section components unchanged. As a result, the orders of vanishing of the section components at a particular locus should be the same for $\ratsec{s}$ and $-\ratsec{s}$. A formula relating the orders of vanishing to the charge should therefore be insensitive to the sign of the charge. Additionally, the question of whether to use $-\ratsec{s}$ instead of $\ratsec{s}$ as the generating section is equivalent to question of whether to use $\quotparam$ or its inverse in the center. Therefore, formulas for the orders of vanishing should not distinguish between $\quotparam$ (or $\quotparamtwo$) and its inverse. These ideas are demonstrated by the proposed formulas in \cref{sec:proposals}: they only depend on the square of the charge $q$, and the $\quotfunction{G}{\quotparam}$ and $\quottriplet{G}{\quotparam}$ are the same for $\quotparam$ and its inverse element.

This ambiguity between $\ratsec{s}$ and $-\ratsec{s}$ reflects the idea that the overall sign of the charges is not important physically. We should be able to flip the sign of all the $\au(1)$ charges without changing the physics. The more relevant property is the sign of a charge relative to other charges in the theory. There are a few different types of relative signs we should consider. First, if the gauge algebra has multiple $\au(1)$ factors, matter in a representation of the gauge algebra has a (possibly zero) charge for each $\au(1)$ factor. The relative sign of the charges in this representation can be important. In a 6D model with a $\au(1)^2$ gauge algebra, for example, one might wish to determine if the representation of some hypermultiplet is $\bm{1}_{1,1}$  or $\bm{1}_{1,-1}$. These representations are not conjugates of each other, and  $\bm{1}_{1,1}$ hypermultiplets should be distinguished from $\bm{1}_{1,-1}$ hypermultiplets. One can determine these sorts relative signs by applying the proposed rules for the orders of vanishing to linear combinations of the generating sections. For the $\au(1)^2$ example, suppose the generating sections are $\ratsec{s}_1$ and $\ratsec{s}_2$. Then, $\bm{1}_{1,-1}$ hypermultiplets would appear to be uncharged under the section $\ratsec{s}_1+\ratsec{s}_2$, whereas $\bm{1}_{1,1}$ hypermultiplets would have \pscharge{} $2$ according to $\ratsec{s}_1+\ratsec{s}_2$. One could therefore distinguish between $\bm{1}_{1,1}$ and $\bm{1}_{1,-1}$ hypermultiplets by applying the formulas in \cref{sec:proposals} to the section components of $\ratsec{s}_1+\ratsec{s}_2$.

Second, one may want to determine the sign of one matter multiplet's charge relative to that of another matter multiplet. For example, let us consider a model with a $\aG\oplus\au(1)$ algebra, where $\aG$ is a nonabelian Lie algebra. Suppose that the model has two matter loci supporting matter in the representations $\representation_{q_1}$ and  $\representation_{q_2}$ for some irreducible representation $\representation$ of $\aG$. The overall sign of $q_1$ and $q_2$ is unimportant: we are free to flip the overall sign of the $\au(1)$ charges, after which the matter representations would be $\representation_{-q_1}$ and $\representation_{-q_2}$. However, the relative sign between $q_1$ and $q_2$ is unchanged by this flip and may therefore be a meaningful property of the model. The equations in \cref{sec:proposals}, particularly \cref{eq:ordvanishtwozgen}, are insensitive to this relative sign.

While the inability to distinguish this relative sign is admittedly a shortcoming of the proposals, this information can, in many cases, be obtained relatively easily by alternative means. It is first important to recall that matter fields in a representation $\representation_q$ are typically accompanied by matter fields in the representation $\conjrep_{-q}$.  In 6D, for example, full hypermultiplets in a representation $\representation_q$ have fields transforming in both representations $\representation_q$ and $\conjrep_{-q}$. If $\representation$ is either a real or pseudoreal representation of $\aG$, the $\representation$ representation is isomorphic to the $\conjrep$ representation, and a hypermultiplet of $\representation_q$ matter can equivalently be viewed as a hypermultiplet of $\representation_{-q}$ matter. Even the relative signs of the charges are therefore unimportant when $\representation$ is either real or pseudoreal. An important example is singlet matter in the $\bm{1}_q$ representation, which is uncharged under the nonabelian gauge algebra. The singlet representation is real, and thus even the relative sign of the $\au(1)$ charge is unimportant.

This leaves us with the situations where $\representation$ is not self-conjugate. For the charge normalization conventions used here, the allowed $\au(1)$ charges for matter in the $\representation_q$ representation, which may be fractional, are separated by integers. Therefore, if one knows $\abs{q_1}$ and $\abs{q_2}$, one can often use this condition to find the relative sign difference between $q_1$ and $q_2$. As an example, suppose that $\aG$ is $\asu(5)$ and $\representation$ is the fundamental ($\bm{5}$) representation. If $\abs{q_1}$ is $\frac{1}{5}$ and $\abs{q_2}$ is $\frac{4}{5}$, one can automatically deduce that $q_1$ and $q_2$ must have opposite signs.

This trick fails, however, when the allowed charges are either integral or half-integral.  Depending on the context, there are other strategies that may be used to determine the relative signs. In 6D models with $\aG\oplus\au(1)$ gauge algebras, the massless spectrum must satisfy the anomaly condition~\cite{ErlerAnomaly,ParkTaylor,ParkIntersection}
\begin{equation}
    \sum_{i} x_{\representation, q_i} E_{\representation} q_{i} = 0\,,
\end{equation}
where $x_{\representation, q_i}$ is the number of hypermultiplets in the representation $\representation_{q_i}$. The anomaly coefficient $E_{\representation}$ is defined by
\begin{equation}
    \tr_{\representation}F^3 = E_{\representation} \tr F^3\,,
\end{equation}
where $F$ is the field strength for $\aG$, $\tr_{\representation}$ is the trace in the $\representation$ representation, and $\tr$ is the trace in the fundamental representation.\footnote{For $\asu(n)$, $E_{\representation}$ is $1$ for the fundamental representation and $n-4$ for the two-index antisymmetric representation. For all the other representations considered in this paper, $E_{\representation}$ is 0.}  Since this formula is clearly sensitive to the sign of $q_i$, one can sometimes glean information about the relative signs of the charges from this condition. In 4D models, Yukawa couplings correspond to codimension-three loci in the base where the singularity type enhances, which can often be viewed as the intersection of codimension-two loci supporting the matter participating in the Yukawa interaction. The Yukawa term in the Lagrangian must be invariant under gauge transformations, so the $\au(1)$ charges of matter fields participating in the interaction must sum to 0.  Thus, one can determine some information about the relative signs by knowing that the matter at certain loci admit a Yukawa interaction. In fact, one can often construct 6D and 4D models with the same Weierstrass tuning, potentially allowing both of these strategies to be used.

To summarize, the proposed formulas in \cref{sec:proposals} are not sensitive to the sign of the charge, but the physically relevant signs can often be determined without too much difficulty. There are some cases where certain signs are both meaningful and difficult to determine, however. While one may need to perform resolutions to fully determine the charges in such cases, the proposed formulas still provide an easy method of determining the absolute value of the charge, making them an invaluable tool. Of course, when attempting to find the orders of vanishing corresponding to a given charge, one can use the formulas in \cref{sec:proposals} for either choice of sign and still obtain the same result.

\section{Charged singlets}
\label{sec:singlets}
The primary focus for the rest of this paper is applying the strategy outlined in \cref{sec:strategy} to models admitting various types of matter. For each example, we generate order-of-vanishing data and show that they satisfy the proposals. Let us first consider loci supporting charged singlets. We can narrow our focus to F-theory models where the total gauge algebra is simply $\au(1)$, although our results appear to be valid for singlets in more general contexts. The matter in such models is supported at codimension-two loci in the base where the elliptic curve singularity type enhances to $\singtype{I}_2$; in 6D models, each such irreducible locus supports a single hypermultiplet of matter. The corresponding $\au(1)$ charges for this matter are integral, as implied by the form of the Shioda map in \cref{eq:shiodamap}.
Typically, one would determine the charge of the matter supported at a particular locus by resolving the $\singtype{I}_2$ singularities and examining the behavior of the section in the smooth model. In line with the proposals of \cref{sec:proposals}, however, we now ask whether one can determine these $\au(1)$ charges from orders of vanishing of the section components without performing a resolution.

This question was already discussed in~\cite{Raghuram34}, and the analysis there suggested a particular relation between the charge and the orders of vanishing in Weierstrass form. The work employed the strategy in \cref{sec:strategy} for an F-theory construction described by the Weierstrass model~\cite{MorrisonParkU1}
\begin{equation}
    \label{eq:singweier}
    y^2 = x^3 + \left(\hat{f}_{12}-3f_6^2\right)xz^4 + \left(f_9^2+2f_6^3-\hat{f}_{12}f_6\right)z^6\,.
\end{equation}
This elliptic fibration has Mordell--Weil rank $1$, and the generating section $\ratsec{s}$ can be written as
\begin{equation}
    \label{eq:singgensec}
    \ratsec{s}\colon [\secx:\secy:\secz] = \left[f_6:f_9:1\right]\,.
\end{equation}
There are no additional nonabelian gauge symmetries in this model, and the gauge algebra for this model is therefore $\au(1)$. The only matter in this model, supported at $\locus{f_9=\hat{f}_{12}=0}$, has charge $q=1$, as confirmed by resolving the $\singtype{I}_2$ singularities. Prior to resolution, the generating section hits the singular point in the fiber at $\locus{f_9=\hat{f}_{12}=0}$. The resolution introduces a new exceptional curve\footnote{To ease the discussion, we are using language that assumes that the elliptic fibration is a threefold, as is the case for 6D F-theory models.} $\matterfibcurve$, and the generating section hits $\matterfibcurve$ at $\locus{f_9=\hat{f}_{12}=0}$.
More specifically, we have that
\begin{equation}
    \sigma(\ratsec{s})\cdot \matterfibcurve = 1
\end{equation}
at $\locus{f_9=\hat{f}_{12}=0}$.

One can then examine the non-generating sections $q\ratsec{s}$, which realize \pscharge{} $q$ at  $\locus{f_9=\hat{f}_{12}=0}$. We are particularly interested in the orders of vanishing of the $\secx$, $\secy$, $\secz$, and $\secw=3\secx^2+f\secz^4$ section components at $\locus{f_9=\hat{f}_{12}=0}$; since the singlet loci occur along a codimension-one $\singtype{I}_1$ locus, the orders of vanishing at codimension one are unimportant. Some of these orders of vanishing are listed in \cref{tab:singorders}. As noticed in~\cite{Raghuram34}, these numbers follow a curious pattern. For a section $q\ratsec{s}$, the orders of vanishing for $(\secx,\secy,\secz,\secw)$ seem to be described by the expressions
\begin{alignat}{2}
(2,3,1,4) &\times \frac{q^2}{4}\,, && \qquad \text{for even }q\,, \label{eq:singletorderseven}\\
(2,3,1,4) &\times \frac{q^2-1}{4} +(0,1,0,1)\,, && \qquad \text{for odd }q\,.\label{eq:singletordersodd}
\end{alignat}
According to the arguments in \cref{sec:signs}, these patterns should hold regardless of the sign of $q$. These patterns have been verified up to $q=28$, as seen in \cref{tab:singorders}, and as discussed in \cref{sec:explicitconstructions}, these same orders of vanishing occur at singlet loci in models from the prior literature, suggesting that the orders of vanishing described by \cref{eq:singletorderseven,eq:singletordersodd} are universal features of the Weierstrass models at loci admitting charged singlets.

\begin{table}
    \centering

    \begin{tabular}{*{5}{>{$}c<{$}}}\toprule
        q  & \ordvanish{}{\secx} & \ordvanish{}{\secy} & \ordvanish{}{\secz} & \ordvanish{}{\secw} \\ \midrule
        1  & 0   & 1   & 0   & 1 \\
        2  & 2   & 3   & 1   & 4 \\
        3  & 4   & 7   & 2   & 9 \\
        4  & 8   & 12  & 4   & 16 \\
        5  & 12  & 19  & 6   & 25 \\
        6  & 18  & 27  & 9   & 36 \\
        7  & 24  & 37  & 12  & 49 \\
        8  & 32  & 48  & 16  & 64 \\
        9  & 40  & 61  & 20  & 81 \\
        10 & 50  & 75  & 25  & 100 \\
        11 & 60  & 91  & 30  & 121 \\
        12 & 72  & 108 & 36  & 144 \\
        13 & 84  & 127 & 42  & 169 \\
        14 & 98  & 147 & 49  & 196 \\
        15 & 112 & 169 & 56  & 225 \\
        16 & 128 & 192 & 64  & 256 \\
        17 & 144 & 217 & 72  & 289 \\
        18 & 162 & 243 & 81  & 324 \\
        19 & 180 & 271 & 90  & 361 \\
        20 & 200 & 300 & 100 & 400 \\
        21 & 220 & 331 & 110 & 441 \\
        22 & 242 & 363 & 121 & 484 \\
        23 & 264 & 397 & 132 & 529 \\
        24 & 288 & 432 & 144 & 576 \\
        25 & 312 & 469 & 156 & 625 \\
        26 & 338 & 507 & 169 & 676 \\
        27 & 364 & 547 & 182 & 729 \\
        28 & 392 & 588 & 196 & 784 \\ \bottomrule
    \end{tabular}

    \caption{Orders of vanishing for the $\secx$, $\secy$, $\secz$, and $\secw$ components of the $q\ratsec{s}$ sections at $\locus{f_9=\hat{f}_{12}=0}$. The generating section $\ratsec{s}$ is described by \cref{eq:singgensec}, and the Weierstrass model is given in \cref{sec:singlets}. This model supports singlets with charge $\chargeval{1}$ at $\locus{f_9=\hat{f}_{12}=0}$, and the section $q\ratsec{s}$ realizes \pscharge{} $q$ at this locus.}
    \label{tab:singorders}
\end{table}

The analysis in~\cite{Raghuram34} did not give an explanation for these expressions. One aspect of the
expressions, the $(0,1,0,1)$ term in \cref{eq:singletorderseven,eq:singletordersodd}, has a straightforward interpretation. When the unresolved elliptic fibration is written in Weierstrass form, the singular points occur at $y=w=0$ on the fiber. If the section hits the singular point on a fiber, the $\secy$ and $\secw$ components should go to 0. The  $(0,1,0,1)$ term therefore indicates that the section hits the singular point on the fibers at loci supporting odd charges. The $\frac{q^2}{4}$ and $\frac{q^2-1}{4}$ terms do not immediately present explanations, but two pieces of evidence indicate that \cref{eq:singletorderseven,eq:singletordersodd} are natural results. First, we can show that the formulas are specializations of those proposed in \cref{sec:proposals}. Second, these same expressions appear in previous work on elliptic divisibility sequences, suggesting that these formulas reflect deeper properties of elliptic fibrations.

\subsection{Connection to the proposed formulas}
To see the connection between the observed singlet patterns and the formulas in \cref{sec:proposals}, it is helpful to rewrite the formulas in \cref{eq:singletorderseven,eq:singletordersodd} as
\begin{equation}
\ordvanishtwo{\secz} = \frac{1}{2}\left(\frac{1}{2}q^2 - \frac{1}{2}\residue{q}{2}\right) = \floor*{\frac{q^2}{4}} \label{eq:singletorderrewritea}
\end{equation}
and
\begin{equation}
\left(\ordvanishtwo{\secx},\ordvanishtwo{\secy},\ordvanishtwo{\secw}\right) = \left(2,3,4\right)\times\ordvanishtwo{\secz} + \left(0,1,1\right)\residue{q}{2}\,.
\end{equation}
We have assumed that $q$ is an integer. Singlets occur at double points of the discriminant locus where the singularity type enhances from $\singtype{I}_1\times \singtype{I}_1$ at codimension one\footnote{Since the singlet locus is a double point of the discriminant locus, it resembles the intersection of two $\singtype{I}_1$ loci for a sufficiently small neighborhood around the singlet locus.} to $\singtype{I}_2$ at codimension two. If we refer to the singularity types by their corresponding ADE groups, the $\singtype{I}_1$ singularity type at codimension one can be associated with $\SU(1)$, while the $\singtype{I}_2$ singularity type at codimension two can be associated with $\SU(2)$. The center of $\SU(n)$ has $d_{\SU(n)} = n$ elements. For $\SU(1)$, there is only one element of the center denoted by $\quotparam=0$, and according to \cref{eq:quotfunctionsun}
\begin{equation}
\quotfunction{\SU(1)}{0} = 0\,, \quad \quottriplet{\SU(1)}{0} = (0,0,0)\,.
\end{equation}
As a result, all of the terms in the general proposals involving the codimension-one behavior of the section are trivial for singlets. For $\SU(2)$, there are two elements of the center, which are denoted by $\quotparamtwo=0$ and $\quotparamtwo=1$. According to \cref{eq:quotfunctionsun,eq:quottriplesun},
\begin{equation}
\quotfunction{\SU(2)}{\quotparamtwo} = \frac{1}{2}\quotparamtwo(2-\quotparamtwo) = \frac{1}{2}\quotparamtwo\,, \quad \quottriplet{\SU(2)}{\quotparamtwo} = (0,1,1)\times\moddist{2}{\quotparamtwo} = (0,1,1)\quotparamtwo\,.
\end{equation}
If we set $\quotparamtwo$ to $\residue{q}{2}$ and plug this information into the general formulas \labelcref{eq:ordvanishtwozgen} and \labelcref{eq:ordvanishtwoothergen} from \cref{sec:proposals}, we obtain
\begin{equation}
\ordvanishtwo{\secz} = \frac{1}{2}\left(\frac{1}{2}q^2 - \frac{1}{2}\residue{q}{2}\right)
\end{equation}
and
\begin{equation}
\left(\ordvanishtwo{\secx},\ordvanishtwo{\secy},\ordvanishtwo{\secw}\right) = \left(2,3,4\right)\times\ordvanishtwo{\secz} + \left(0,1,1\right)\moddist{2}{q}\,.
\end{equation}
This exactly matches the rewritten form of the singlet orders of vanishing above. \Cref{eq:singletorderseven,eq:singletordersodd} therefore agree with the general formulas proposed in \cref{sec:proposals}.

The fact that $\quotparamtwo = \residue{q}{2}$ demonstrates that elliptic curve addition respects the addition law for the centers of the groups in question, as discussed in \cref{sec:strategy}. In this case, we are interested in the $\Z_2$ center of $\SU(2)$, the group corresponding to the codimension-two singularity type. The generating section is associated with the $\quotparamtwo=1$ element of $\Z_2$. Based on the $\Z_2$ addition law, adding $\quotparamtwo=1$ to itself an even number of times should give the $\quotparamtwo=0$ identity element of $\Z_2$, while adding $\quotparamtwo=1$ to itself an odd number of times should give the $\quotparamtwo=1$ element of $\Z_2$. We in fact see that even multiples of the generating section are associated with $\quotparamtwo=0$ and odd multiples of the generating section are associated with $\quotparamtwo=1$. We observe similar behaviors in the examples considered later.

\subsection{The elliptic troublemaker sequence}
\label{sec:singleteds}
Remarkably, the sequence $\floor{\frac{q^2}{4}}$ seen in \cref{eq:singletorderrewritea} also appears in a different but somewhat related context: it is the elliptic troublemaker sequence $\elltrouble{q}{1}{2}$ defined in Stange's work~\cite{StangeEllTrouble} on the valuations of elliptic divisibility sequences. There in fact seem to be direct parallels between Stange's work and the problems explored in this paper, suggesting that Stange's results may guide us toward a firmer understanding of the observed orders of vanishing. Therefore, let us briefly review some of the relevant results from~\cite{StangeEllTrouble} along with some necessary background material.

Stange's work considers elliptic curves over finite extensions of the $p$-adic numbers $\Q_p$.  The $p$-adic numbers are a completion of the rational numbers $\Q$ just like the set of real numbers $\R$, but they have a different notion of the ``absolute value.'' Suppose we consider a nonzero rational number $x$, which we can uniquely write as $x = \frac{r}{s} p^n$ for a prime integer $p$ and coprime integers $r, s$ both coprime to $p$. We can then define the $p$-adic valuation of $x$ as
\begin{equation}
    \valuation{p}{x} = n\,.
\end{equation}
If $x$ is an integer, $\valuation{p}{x}$ gives the exponent of the largest power of $p$ that divides $x$. We define $v_p(0) = \infty$. This $p$-adic valuation satisfies the standard properties of a valuation:
\begin{itemize}
    \item $\valuation{p}{ab} = \valuation{p}{a}+\valuation{p}{b}$,
    \item $\valuation{p}{a+b} \ge \min(\valuation{p}{a},\valuation{p}{b})$, with exact equality if $\valuation{p}{a}\neq \valuation{p}{b}$,
    \item $\valuation{p}{a}=\infty$ if and only if $a=0$.
\end{itemize}
Note that the order of vanishing also satisfies these properties, a fact that will be exploited shortly. One can then define a $p$-adic absolute value $\pabs{p}{x}=p^{-\valuation{p}{x}}$.

There are two particular concepts regarding elliptic curves that are important in Stange's analysis. The first is the reduction of elliptic curves modulo a prime $p$. Consider an elliptic curve with integral coefficients over a field $K$ of characteristic 0 that is a finite extension of $\Q_p$.  To be concrete, one could consider this elliptic curve to be written in the Tate form\footnote{While this way of writing elliptic curves is known as the ``Tate form'' in the F-theory literature, it is also known as the ``Weierstrass form''  in other contexts. However, the F-theory literature typically reserves the term ``Weierstrass form'' for the form $y^2=x^3+fx+g$. We therefore refer to the form of the elliptic curve in \cref{eq:tateform} as the Tate form to be consistent with the F-theory conventions.}
\begin{equation}
    \label{eq:tateform}
    y^2 + a_1 x y + a_3 x = x^3 + a_2 x^2 + a_4 x + a_6\,,
\end{equation}
where the $a_i$ are integers. We can then form a reduced elliptic curve by reducing the $a_i$ coefficients modulo the prime number $p$. The new reduced elliptic curve is now over the residue field of $K$, where elements of $K$ that are equivalent modulo $p$ are identified. Even if the original curve is smooth, its reduction may be singular. If the elliptic curve remains smooth after reduction, one says that the original elliptic curve over $K$ has good reduction; if the elliptic curve has a nodal singularity after reduction, then the original elliptic curve is said to have multiplicative reduction; and if the reduced elliptic curve has a cusp singularity, the original elliptic curve is said to have additive reduction. Additionally, a rational point $P$ of the original elliptic curve maps onto a rational point of the reduced elliptic curve. We say that $P$ reduces to this new rational point.

The second important topic is that of elliptic divisibility sequences and division polynomials. Suppose we have an elliptic curve $E$ such as that in \cref{eq:tateform}. Then we can define a sequence of polynomials in $\Z[a_1,a_2,a_3,a_4,a_6,x,y]$ known as the division polynomials $\divpoly_m$ for $E$. This sequence can be constructed using the recursion relations
\begin{equation}
\begin{aligned}
    \divpoly_{2j+1} &= \divpoly_{j+2}\divpoly_j^3-\divpoly_{j-1}\divpoly_{j+1}^3\,, &  \text{for } j\ge2\,, \\
    \divpoly_{2j}\divpoly_{2} &= \divpoly_{j-1}^2\divpoly_{j}\divpoly_{j+2}-\divpoly_{j-2}\divpoly_{j}\divpoly_{j+1}^2\,, & \text{for } j\ge3\,,
\end{aligned}
\end{equation}
and the initial values
\begin{equation}
\begin{aligned}
    \divpoly_1 &= 1\,, \\
    \divpoly_2 &= 2y+a_1 x + a_3\,, \\
    \divpoly_3 &= 3 x^4 + b_2 x^3+ 3 b_4 x^2 + 3 b_6 x + b_8\,, \\
    \divpoly_4 &= \divpoly_2\left(2 x^6 + b_2 x^5 + 5 b_4 x^4 + 10 b_6 x^3 + 10 b_8 x^2 + (b_2 b_8-b_4 b_6) x + (b_4 b_8-b_6^2)\right)\,.
\end{aligned}
\end{equation}

The $b_i$ parameters in these initial values are the same as those familiar from, for instance, Tate's algorithm in F-theory~\cite{BershadskyEtAlSingularities}:
\begin{equation}
    \begin{aligned}
        b_2 &= a_1^2 + 4 a_2\,, \\
        b_4 &= a_1 a_3 + 2 a_4\,, \\
        b_6 &= a_3^2 + 4 a_6\,, \\
        b_8 &= b_2 a_6 - a_1 a_3 a_4 + a_2 a_3^2 - a_4^2\,.
    \end{aligned}
\end{equation}
The division polynomials can be used to describe the multiples of a point under elliptic curve addition. If we describe a generic point on the elliptic curve as $P=(x,y)$, then one can write the formula for a multiple of a point as
\begin{equation}
    \label{eq:pointdivpoly}
    mP=\left(\frac{\ecurvex_m}{\divpoly_m^2},\frac{\ecurvey_m}{\divpoly_m^3}\right)\,,
\end{equation}
where $\ecurvex_m$ and $\ecurvey_m$ are polynomials in $\Z[a_1,a_2,a_3,a_4,a_6,x,y]$. Now consider a rational point $P$. We can always scale the $a_i$ coefficients in \cref{eq:tateform} to convert $P$ into an integral point $P^\prime$. Evaluating the $\divpoly_m$ for this integral point $P^\prime$ and the scaled $a_i$ values gives a sequence of integers $W_n$ known as an elliptic divisibility sequence (EDS). This sequence, like all divisibility sequences, has the property that $W_m\mid W_n$ if $m\mid n$.

With these ingredients, we can now describe the theorem in Stange's work, Theorem~28, that is relevant for understanding the singlet charges. Consider an elliptic curve over an extension of $\Q_p$ with multiplicative reduction modulo $p$ and a rational
point $P$ on the elliptic curve that reduces to the singular point. Then, the valuations of the $W_m$ associated with $P$ are described by the formula
\begin{equation}
    \valuation{p}{\eds_m} = \elltrouble{m}{a_P}{l_P} + \begin{cases}\scorr_{m/n_P}(p,p,\valuation{}{p},0, s_{P}, w_{P}) & n_P \mid m\\
    0 & n_P \nmid m \end{cases}\,,
\end{equation}
where $\elltrouble{m}{a_P}{l_P}$ is the elliptic troublemaker sequence given by\footnote{In~\cite{StangeEllTrouble}, there are a variety of formulations of the elliptic troublemaker sequence that are equivalent for integer $m$.
}
\begin{equation}
    \elltrouble{m}{a_P}{l_P} = \frac{m^2-m}{2}a_P+\sum_{k=1}^{\floor{\frac{ma_P}{l_P}}}\left(k l_P - m a_P\right) - m^2 \sum_{k=1}^{\floor{\frac{a_P}{l_P}}}\left(k l_P -  a_P\right)\,.
\end{equation}
The integer $l_P$ is equivalent to $-\valuation{p}{j(E)}$, the negative of the valuation of the $j$-invariant of the elliptic curve. Meanwhile, $a_P$ is an integer determined by $P$, which we can take to satisfy $0 \le a_P < l_P$; it plays a role similar to our parameters $\quotparam$ and $\quotparamtwo$.\footnote{More specifically, $a_P$ indicates the component of the N\'{e}ron model special fiber containing $P$. It is zero if and only if $P$ has non-singular reduction, and since the theorem describes situations where $P$ has singular reduction, $a_P$ will often be nonzero when we use this theorem. See~\cite{StangeEllTrouble} for a more thorough explanation of $a_P$.} In fact, when $0\le a_P<l_P$ and $m$ is an integer, the elliptic troublemaker sequence can be written in the useful form
\begin{equation}
     \elltrouble{m}{a_P}{l_P} = \frac{a_P(l_P-a_P)}{2l_P}m^2 - \frac{\residue{(m a_P)}{l_P}(l_P-\residue{(m a_P)}{l_P})}{2l_P}\,.
\end{equation}
The integer $n_P$ is the smallest positive integer such that the reduction of $n_P P$ equals the reduction of the identity point of the elliptic curve. Finally $\scorr$ is a correction term that is described in more detail in \cref{app:scorr}.

Even though this theorem is formulated for elliptic curves over extensions of the $p$-adic numbers, it also seems give the correct orders of vanishing for the sections described above. This is likely explained by the underlying structure shared by both problems. The rational points of an elliptic curve are analogues of the rational sections of an elliptic fibration, and the $\eds_m$ elements of the elliptic divisibility sequence are roughly the analogues of the $\secz$ components of the sections considered above.\footnote{It is important to remember, however, that the $\eds_m$ are found from the division polynomials. For particular choices of $P$ and $m$, the $\divpoly_m$, $\ecurvex_m$, and $\ecurvey_m$ in \cref{eq:pointdivpoly} may share common factors that can be removed. In such cases, $W_m$ would be proportional to $\secz$ but may have additional factors. This possibility is more relevant for the cases with nonabelian gauge algebras described later.} Meanwhile, investigating the behavior of the sections at the locus $\locus{f_9=\hat{f}_{12}=0}$ requires analyzing expressions for $f$, $g$, and the section components after dropping terms proportional to $f_9$ and $\hat{f}_{12}$. This procedure parallels reducing an elliptic curve modulo $p$. Since the elliptic fibration has $\singtype{I}_2$ singularities at  $\locus{f_9=\hat{f}_{12}=0}$, the singular fibers there have nodal singularities, just as one would find after reducing an elliptic curve with multiplicative reduction. Additionally, the generating section hits the singular point at $\locus{f_9=\hat{f}_{12}=0}$, which corresponds to a point $P$ having singular reduction. Finally, the order of vanishing is itself a valuation and should play a role similar to the $p$-adic valuation.

These analogies suggest that the orders of vanishing for the $\secz$ components of the $q\ratsec{s}$ sections should be described by Stange's theorem. To make the connection more explicit, we can construct an elliptic curve analogue of \cref{eq:singweier} by setting $f_9$ and $\hat{f}_{12}$, the parameters corresponding to the matter locus, to an arbitrary prime integer $p$ (not equal to 3) while setting $f_6$ to a number such as 1. These substitutions lead to an elliptic curve
\begin{equation}
y^2 = x^3 + (p-3)x + (p^2-p+2)
\end{equation}
with a discriminant
\begin{equation}
\Delta = 4f^3+27g^2 = p^2 \left(27 p^2-50 p+99\right)
\end{equation}
and a rational point
\begin{equation}
(x,y) = (1,p)\,.
\end{equation}
This elliptic curve has multiplicative reduction modulo $p$. If we consider the EDS associated with $P$, $l_P$ (which is generally given by $-\valuation{p}{j(E)}$) equals 2 for this situation, as would be expected since the matter locus supports $\singtype{I}_2$ singularities. Since $P$ reduces to the singular point, we can take $a_P$ to be 1, in agreement with the value of $\quotparamtwo$ for the generating section. If one calculates several multiples of $P$, one finds that they never reduce to the identity point; evidently no multiples of $P$ reduce to the identity, and $n_P$ can be taken to be infinitely large. Stange's theorem would then suggest that
\begin{equation}
\valuation{p}{W_q} = \elltrouble{q}{1}{2} = \floor*{\frac{q^2}{4}}\,.
\end{equation}
In turn, one would expect the $\secz$ section components for $q\ratsec{s}$ to vanish to these same orders at the matter locus, as observed.

We therefore have a plausible explanation for the orders of vanishing observed in the non-generating sections for the Weierstrass model considered above. In turn, it is natural to expect that the orders of vanishing at loci genuinely supporting charged singlets should obey the same pattern. Of course, since Stange's theorem is formulated in a different context, one would ideally have a proof of a similar statement for elliptic fibrations. It would also be beneficial to have a rigorous proof that the behavior of the $q\ratsec{s}$ sections matches that of generating sections admitting charge-$q$ matter. We leave both of these questions to future work. Nevertheless, Stange's theorem seems to directly predict the observed orders of vanishing, and we will shortly see that similar formulas agree with the observed orders of vanishing when there are nonabelian gauge algebras as well.

\section{Representations of $\asu(n)\oplus\au(1)$}
\label{sec:su}
Now that we have seen the connection between the orders of vanishing and $\au(1)$ charges for singlets, we can start investigating what happens for matter simultaneously charged under a $\au(1)$ and a simple nonabelian gauge factor. We first focus on $\asu(n)\oplus\au(1)$ gauge algebras, restricting our attention to matter in the fundamental and antisymmetric representations of $\asu(n)$.
New complications arise when we include nonabelian gauge factors. The global structure of the total gauge group can have a nontrivial quotient involving both the $\au(1)$ factor and an element of the center of the nonabelian gauge factor. As a result, matter charged under the nonabelian gauge factor can have fractional $\au(1)$ charges in the units specified by the Shioda map in \cref{eq:shiodamap}. Additionally, there are now two types of orders of vanishing to consider: those at the codimension-two locus supporting the matter and those at the codimension-one locus supporting the nonabelian gauge factor.

Despite these complications, many of the ideas seen in the singlet analysis carry over to this new context. The proposals in \cref{sec:proposals} still seem to correctly describe the orders of vanishing of the section components supporting various $\au(1)$ charges. Furthermore, we will still find intriguing connections to the elliptic troublemaker sequences and the work in~\cite{StangeEllTrouble}.

\subsection{Fundamental representation of $\asu(n)$}
\label{sec:sunfundamentals}

As discussed in \cref{sec:suncenters}, an F-theory model with an $\asu(n)\oplus\au(1)$ gauge algebra and global structure described by the integer $\quotparam$ can support $\bm{n}_{q}$ matter with $\au(1)$ charge
\begin{equation}
    q = \frac{\quotparam}{n}+j \text{ for } j\in \Z\,.
\end{equation}
In other words, a model supporting $\bm{n}_{q}$ matter should have a global gauge group structure described by
\begin{equation}
    \quotparam = \residue{n q}{n}\,.
\end{equation}
These observations allow us to adapt the strategy of \cref{sec:strategy} to fundamental matter. We start with a seed model with an $\asu(n)\oplus\au(1)$ gauge algebra, a global structure described by $\quotparam=1$, and $\bm{n}_{1/n}$ matter. In F-theory, the $\asu(n)$ gauge algebra is supported along a codimension-one locus in the base with $\singtype{I}^{(s)}_n$ singularities, while the fundamental matter is supported at a codimension-two locus where the singularity type enhances to $\singtype{I}_{n+1}$. If this model has a generating section $\ratsec{s}$, the multiples $m\ratsec{s}$ would have $\quotparam=\residue{m}{n}$ and would realize \pscharge{} $\frac{m}{n}$ at the codimension-two locus. We can therefore obtain orders of vanishing corresponding to all allowed combinations of $\quotparam$ and the $\au(1)$ charge by examining multiples of this generating section.

To be more specific, we analyze three different seed models supporting $\asu(n)\oplus\au(1)$ algebras. For $n=2k+1$, $k\ge 1$, we consider the Weierstrass model
\begin{equation}
    \label{eq:suoddseedmodelfund}
    \begin{aligned}
        y^2 &= x^3 + \left[-\frac{1}{48} \left(b_{1,0}^2+
    4 c_{2,1}\sigma\right)^2+\frac{1}{2}
    b_{0,0} b_{1,0} c_{1,k} \sigma ^k+c_{3,1} c_{1,k}\sigma^{k+1}+b_{0,0}^2 \sigma ^{2k} c_{0,2 k}\right]x z^4 \\
    &\qquad\qquad\quad + \left[\frac{1}{864} \left(b_{1,0}^2+4c_{2,1} \sigma  \right)^3-\frac{1}{24}c_{1,k}
    \left(b_{1,0}^2+4 \sigma  c_{2,1}\right) \left(b_{0,0} b_{1,0}+2 \sigma
    c_{3,1}\right) \sigma ^k \right. \\
   &\qquad\qquad\qquad\qquad + \frac{1}{12} b_{0,0}^2  \left(3 c_{1,k}^2-b_{1,0}^2 c_{0,2
        k}\right)\sigma ^{2 k} \\
    &\qquad\qquad\qquad\qquad \left.-\frac{1}{3}c_{0,2 k} \left(-2 b_{0,0}^2 c_{2,1}+3 b_{1,0}
    b_{0,0} c_{3,1}+3 \sigma  c_{3,1}^2\right) \sigma ^{2 k+1} \right] z^6\,.
    \end{aligned}
\end{equation}
This Weierstrass model supports $\singtype{I}^{(s)}_{2k+1}$ singularities along $\locus{\sigma=0}$ and a generating section with components
\begin{equation}
    \label{eq:suoddseedfundseccomp}
    \begin{aligned}
    \secx &= \left(\frac{1}{2} b_{0,0} b_{1,0}+\sigma  c_{3,1}\right)^2-\frac{2}{3} b_{0,0}^2
    \left(\frac{b_{1,0}^2}{4}+\sigma  c_{2,1}\right)\,, \\
    \secy &= -\frac{\sigma}{2} \left(b_{0,0}^4 \sigma ^{k-1} c_{1,k}+\left(b_{0,0} b_{1,0}+2 \sigma
    c_{3,1}\right) \left(b_{1,0} b_{0,0} c_{3,1}-b_{0,0}^2 c_{2,1}+\sigma
    c_{3,1}^2\right)\right)\,, \\
    \secz &= b_{0,0}\,, \\
    \secw&= \frac{\sigma}{2} \Big[\left(b_{1,0} b_{0,0} c_{3,1}-b_{0,0}^2
    c_{2,1}+\sigma  c_{3,1}^2\right) \left(b_{0,0}^2
    \left(b_{1,0}^2-2 \sigma  c_{2,1}\right)+6 \sigma  b_{1,0} b_{0,0} c_{3,1}+6 \sigma ^2
    c_{3,1}^2\right) \\
    &\phantom{= \frac{\sigma}{2} \Big[}+2 b_{0,0}^6 \sigma^{2 k-1} c_{0,2 k}+b_{0,0}^4 \sigma^{k-1} c_{1,k}
    \left(b_{0,0} b_{1,0}+2 \sigma  c_{3,1}\right)\Big]\,,
    \end{aligned}
\end{equation}
implying that the gauge algebra is $\asu(2k+1)\oplus\au(1)$. The charged matter spectrum, summarized in \cref{tab:suoddseedfundmatter}, satisfies the gauge and gauge--gravitational anomaly conditions for $b=\divclass{\sigma}$ and $\height = -2\canonclass + 2\divclass{b_{0,0}} - \frac{2k}{2k+1}\divclass{\sigma}$. These models are equivalent to the $\mathcal{Q}(2k,k,1,1,0,0,k)$ models in~\cite{KuntzlerTateTrees}.

\begin{table}
    \centering

    \begin{tabular}{*{2}{>{$}c<{$}}}\toprule
        \text{Representation} & \text{Matter Locus} \\ \midrule
        \bm{\frac{n(n-1)}{2}}_{\mathrlap{2/n}} & V_a = \locus{\sigma = b_{1,0}=0} \\
        \bm{n}_{\mathrlap{(n+1)/n}} & V_{f,1} = \locus{\sigma = b_{0,0}=0} \\
        \bm{n}_{\mathrlap{(1-n)/n}} & V_{f,2} = \locus{\sigma = b_{0,0} c_{2,1}-b_{1,0} c_{3,1}=0} \\
        \bm{n}_{\mathrlap{1/n}} & V_{f,3} = \locus{\sigma = b_{1,0}^2 c_{0,2 k}+c_{1,k}^2=0} \\
        \bm{1}_{\mathrlap{2}} & V_{s,2} = \locus{b_{0,0} = c_{3,1}=0}\\
        \bm{1}_{\mathrlap{1}} & V_{s,1} = \locus{\frac{\secy}{\sigma}=\frac{\secw}{\sigma}=0}\setminus\left(V_{a}\cup V_{f,1}\cup V_{f,2} \cup V_{f,3} \cup V_{s,2}\right)\\\midrule
        \bm{(n^2-1)}_{\mathrlap{0}} & \locus{\sigma=0}\\\bottomrule
    \end{tabular}

    \caption{
    Matter spectrum for the $\asu(n)\oplus\au(1)$ fundamental seed models for $n = 2 k + 1$ and $k \ge 1$.
    Note that one can flip the sign of all $\au(1)$ charges and obtain an equally valid description of the matter spectrum (with a different value of $\quotparam$). For $n=3$ ($k=1$), the matter spectrum is similar but with two changes: the locus $V_{a}$ does not support any matter, while the $\bm{3}_{-2/3}$ locus is described by $\locus{\sigma=b_{0,0}^2 c_{1,k}-b_{1,0} b_{0,0} c_{2,1}+b_{1,0}^2 c_{3,1}=0}$.  }
    \label{tab:suoddseedfundmatter}
\end{table}

We can construct a seed model for $n=2k$, $k\ge2$ by taking \cref{eq:suoddseedmodelfund}, replacing $k$ with $k-1$, and setting
\begin{equation}
    \label{eq:sunfundevensubs}
    c_{1,k-1} \to c_{1,k}\sigma + \frac{1}{2} b_{2,k-1} b_{1,0}\,, \quad c_{0,2k-2} \to c_{0,2k-1}\sigma - \frac{1}{4}b_{2,k-1}^2\,.
\end{equation}
This Weierstrass model admits $\singtype{I}^{(s)}_{2k}$ singularities along $\locus{\sigma=0}$ and a generating section described by \cref{eq:suoddseedfundseccomp} with the above substitutions. As a result, the gauge algebra is $\asu(2k)\oplus\au(1)$. The matter spectrum, summarized in \cref{tab:suevenseedfundmatter}, satisfies the gauge and gauge--gravitational anomaly conditions for $b=\divclass{\sigma}$ and $\height = -2\canonclass + 2\divclass{b_{0,0}} - \frac{2k-1}{2k}\divclass{\sigma}$. These models are equivalent to the $\mathcal{Q}(2k-1,k,1,1,0,0,k-1)$ models in~\cite{KuntzlerTateTrees}.

\begin{table}
    \centering

    \begin{small}
    \begin{tabular}{*{2}{>{$}c<{$}}} \toprule
        \text{Representation} & \text{Matter Locus} \\\midrule
        \bm{\frac{n(n-1)}{2}}_{\mathrlap{2/n}} & V_a = \locus{\sigma = b_{1,0}=0} \\
        \bm{n}_{\mathrlap{(n+1)/n}} & V_{f,1} = \locus{\sigma = b_{0,0}=0} \\
        \bm{n}_{\mathrlap{(1-n)/n}} & V_{f,2} = \locus{\sigma = -2 b_{0,0} c_{2,1}+2 b_{1,0} c_{3,1}+b_{0,0}^2 \sigma^{k-2} b_{2,k-1}=0} \\
        \bm{n}_{\mathrlap{1/n}} & V_{f,3} = \locus{\sigma = 2 b_{1,0} \left(b_{1,0} c_{0,2 k-1}+b_{2,k-1} c_{1,k}\right)-2 c_{2,1}
            b_{2,k-1}^2+b_{0,0}  b_{2,k-1}^3\sigma^{k-2}=0} \\
        \bm{1}_{\mathrlap{2}} & V_{s,2} = \locus{b_{0,0} = c_{3,1}=0}\\
        \bm{1}_{\mathrlap{1}} & V_{s,1} = \locus{\frac{\secy}{\sigma}=\frac{\secw}{\sigma}=0}\setminus\left(V_{a}\cup V_{f,1}\cup V_{f,2} \cup V_{f,3} \cup V_{s,2}\right)\\\midrule
        \bm{(n^2-1)}_{\mathrlap{0}} & \locus{\sigma=0}\\\bottomrule
    \end{tabular}
    \end{small}

    \caption{Matter spectrum for the $\asu(n)\oplus\au(1)$ fundamental seed models for $n=2k$ and $k\ge2$. Note that one can flip the sign of all $\au(1)$ charges and obtain an equally valid description of the matter spectrum (with a different value of $\quotparam$).}
    \label{tab:suevenseedfundmatter}
\end{table}

The $n = 2$ case is somewhat unique, as one can obtain an $\asu(2)$ gauge factor (with $\singtype{I}_2$ fibers) without satisfying the split condition. We therefore consider a separate Weierstrass model just for the $n=2$ case:
\begin{equation}
    \label{eq:su2seedmodelfund}
    y^2 = x^3 + \left(-\frac{\phi ^2}{48}+f_1\sigma\right)x z^4 + \left(-\frac{\phi ^3}{864}+\frac{1}{12} f_1 \sigma  \phi+\gamma ^2 \sigma ^2 \right)z^6\,.
\end{equation}
This model admits $\singtype{I}_2$ singularities along $\locus{\sigma=0}$ and a generating section of the form
\begin{equation}
    \left[\secx:\secy:\secz\right] = \left[-\frac{\phi }{12}:\gamma  \sigma:1\right]\,, \quad \secw=f_1 \sigma\,,
\end{equation}
implying that the gauge algebra is $\asu(2)\oplus\au(1)$. Matter in the $\bm{2}_{1/2}$ representation is supported at $\locus{\sigma=f_1^2+\phi\gamma^2=0}$, and $\bm{1}_{1}$ matter occurs at $\locus{\gamma=f_1=0}$. Because both the $\bm{2}$ and $\bm{1}$ representations are self-conjugate, the sign of the $\au(1)$ charge is unimportant. (Additionally, there may be $\bm{3}_0$ matter that can propagate throughout the $\locus{\sigma=0}$ divisor.) For 6D models, the spectrum satisfies the anomaly cancellation conditions with $\height = -2\canonclass -\frac{1}{2}\divclass{\sigma}$.

As mentioned previously, we can calculate multiples of the models' generating sections and examine their behavior at the $\bm{n}_{1/n}$ loci, giving us orders of vanishing for various \pscharge{}s. In fact, our seed models also admit fundamental matter with $\au(1)$ charges larger than $1/n$, allowing us to strengthen the analysis. First, some \pscharge{}s can be obtained in multiple ways. In an $\asu(3)\oplus\au(1)$ model, for example, the orders of vanishing for \pscharge{} $4/3$ can be obtained from the behavior of $4\ratsec{s}$ at the $\bm{3}_{1/3}$ locus or from the behavior of $\ratsec{s}$ at the $\bm{3}_{4/3}$ locus. We always obtain the same orders of vanishing for a particular combination of $\quotparam$ and the \pscharge{} even if we obtain the \pscharge{} in different ways, suggesting that our strategy of looking at multiples of the generating section gives reasonable results. Second, these higher-charge fundamental loci allow us to probe some larger \pscharge{}s without calculating larger multiples of the generating section.

The orders of vanishing we obtain are summarized in \cref{tab:su2through5funddata,tab:su6through8funddata}. One can verify that the codimension-one orders of vanishing at the $\asu(n)$ locus are described by
\begin{equation}
    \left(\ordvanishone{\secx},\ordvanishone{\secy},\ordvanishone{\secw}\right) =  \quottriplet{\SU(n)}{\quotparam} = \left(0,\moddist{n}{\quotparam},\moddist{n}{\quotparam}\right)\,, \quad \ordvanishone{\secz} = 0
\end{equation}
at the codimension-one locus supporting the $\asu(n)$ gauge factor.  Additionally, the codimension-two orders of vanishing at the fundamental locus are given by
\begin{equation}
    \begin{aligned}
        \ordvanishtwo{\secz} &= \frac{1}{2}\left(\frac{n}{n+1}q^2 + \quotfunction{\SU(n)}{\quotparam} - \quotfunction{\SU(n+1)}{\quotparamtwo}\right) \\
        &= \frac{1}{2}\left(\frac{n}{n+1}q^2 + \frac{\quotparam(n-\quotparam)}{n} - \frac{\quotparamtwo(n+1-\quotparamtwo)}{n+1}\right)
    \end{aligned}
\end{equation}
and
\begin{equation}
    \begin{aligned}
        \left(\ordvanishtwo{\secx},\ordvanishtwo{\secy},\ordvanishtwo{\secw}\right) &= (2,3,4) \times\ordvanishtwo{\secz}+\quottriplet{\SU(n+1)}{\quotparamtwo} \\
         &= (2,3,4) \times\ordvanishtwo{\secz}+\left(0,\moddist{n+1}{\quotparamtwo},\moddist{n+1}{\quotparamtwo}\right)
    \end{aligned}
\end{equation}
for
\begin{equation}
    \quotparamtwo = \residue{nq}{n+1}\,.
\end{equation}
These formulas agree exactly with the proposals in \cref{sec:proposals}.

\begin{table}
    \begin{subtable}{0.5\linewidth}
        \centering

        \begin{tabular}{*{4}{>{$}c<{$}}}\toprule
            \abs{q}     & \quotparam & \ordalt_{(1)} & \ordalt_{(2)} \\ \midrule
            \frac{1}{2} & 1          & (0,1,0,1)     & (0,1,0,1) \\[0.2em]
            1           & 0          & (0,0,0,0)     & (0,1,0,1) \\[0.2em]
            \frac{3}{2} & 1          & (0,1,0,1)     & (2,3,1,4) \\[0.2em]
            2           & 0          & (0,0,0,0)     & (2,4,1,5) \\[0.2em]
            \frac{5}{2} & 1          & (0,1,0,1)     & (4,7,2,9) \\[0.2em]
            3           & 0          & (0,0,0,0)     & (6,9,3,12) \\[0.2em]
            \frac{7}{2} & 1          & (0,1,0,1)     & (8,13,4,17) \\[0.2em]
            4           & 0          & (0,0,0,0)     & (10,16,5,21) \\[0.2em]
            \frac{9}{2} & 1          & (0,1,0,1)     & (14,21,7,28) \\[0.2em]
            5           & 0          & (0,0,0,0)     & (16,25,8,33) \\ \bottomrule
        \end{tabular}

        \caption{$\bm{2}_{q}$ representation of $\asu(2)\oplus\au(1)$}
        \label{tab:su2funddata}
    \end{subtable}
    \begin{subtable}{0.5\linewidth}
        \centering

        \begin{tabular}{*{4}{>{$}c<{$}}}\toprule
            \abs{q}      & \quotparam      & \ordalt_{(1)} & \ordalt_{(2)} \\ \midrule
            \frac{1}{3}  & 1 \text{ or } 2 & (0,1,0,1)     & (0,1,0,1) \\[0.2em]
            \frac{2}{3}  & 1 \text{ or } 2 & (0,1,0,1)     & (0,2,0,2) \\[0.2em]
            1            & 0               & (0,0,0,0)     & (0,1,0,1) \\[0.2em]
            \frac{4}{3}  & 1 \text{ or } 2 & (0,1,0,1)     & (2,3,1,4) \\[0.2em]
            \frac{5}{3}  & 1 \text{ or } 2 & (0,1,0,1)     & (2,4,1,5) \\[0.2em]
            2            & 0               & (0,0,0,0)     & (2,5,1,6) \\[0.2em]
            \frac{8}{3}  & 1 \text{ or } 2 & (0,1,0,1)     & (6,9,3,12) \\[0.2em]
            \frac{10}{3} & 1 \text{ or } 2 & (0,1,0,1)     & (8,14,4,18) \\[0.2em]
            4            & 0               & (0,0,0,0)     & (12,18,6,24) \\[0.2em]
            \frac{16}{3} & 1 \text{ or } 2 & (0,1,0,1)     & (22,33,11,44) \\[0.2em]
            \frac{20}{3} & 1 \text{ or } 2 & (0,1,0,1)     & (34,51,17,68) \\[0.2em]
            8            & 0               & (0,0,0,0)     & (48,72,24,96) \\ \bottomrule
        \end{tabular}

        \caption{$\bm{3}_{q}$ representation of $\asu(3)\oplus\au(1)$}
        \label{tab:su3funddata}
    \end{subtable}

    \bigskip

    \begin{subtable}{0.5\linewidth}
        \centering

        \begin{tabular}{*{4}{>{$}c<{$}}}\toprule
            \abs{q}      & \quotparam      & \ordalt_{(1)} & \ordalt_{(2)} \\ \midrule
            \frac{1}{4}  & 1 \text{ or } 3 & (0,1,0,1)     & (0,1,0,1) \\[0.2em]
            \frac{1}{2}  & 2               & (0,2,0,2)     & (0,2,0,2) \\[0.2em]
            \frac{3}{4}  & 1 \text{ or } 3 & (0,1,0,1)     & (0,2,0,2) \\[0.2em]
            1            & 0               & (0,0,0,0)     & (0,1,0,1) \\[0.2em]
            \frac{5}{4}  & 1 \text{ or } 3 & (0,1,0,1)     & (2,3,1,4) \\[0.2em]
            \frac{3}{2}  & 2               & (0,2,0,2)     & (2,4,1,5) \\[0.2em]
            \frac{9}{4}  & 1 \text{ or } 3 & (0,1,0,1)     & (4,7,2,9) \\[0.2em]
            \frac{5}{2}  & 2               & (0,2,0,2)     & (6,9,3,12) \\[0.2em]
            3            & 0               & (0,0,0,0)     & (6,11,3,14) \\[0.2em]
            \frac{15}{4} & 1 \text{ or } 3 & (0,1,0,1)     & (12,18,6,24) \\[0.2em]
            \frac{9}{2}  & 2               & (0,2,0,2)     & (16,26,8,34) \\[0.2em]
            5            & 0               & (0,0,0,0)     & (20,30,10,40) \\[0.2em]
            \frac{25}{4} & 1 \text{ or } 3 & (0,1,0,1)     & (32,48,16,64) \\[0.2em]
            \frac{15}{2} & 2               & (0,2,0,2)     & (46,69,23,92) \\ \bottomrule
        \end{tabular}

        \caption{$\bm{4}_{q}$ representation of $\asu(4)\oplus\au(1)$}
        \label{tab:su4funddata}
    \end{subtable}
    \begin{subtable}{0.5\linewidth}
        \centering

        \begin{tabular}{*{4}{>{$}c<{$}}}\toprule
            \abs{q}      & \quotparam      & \ordalt_{(1)} & \ordalt_{(2)} \\ \midrule
            \frac{1}{5}  & 1 \text{ or } 4 & (0,1,0,1)     & (0,1,0,1) \\[0.2em]
            \frac{2}{5}  & 2 \text{ or } 3 & (0,2,0,2)     & (0,2,0,2) \\[0.2em]
            \frac{3}{5}  & 2 \text{ or } 3 & (0,2,0,2)     & (0,3,0,3) \\[0.2em]
            \frac{4}{5}  & 1 \text{ or } 4 & (0,1,0,1)     & (0,2,0,2) \\[0.2em]
            1            & 0               & (0,0,0,0)     & (0,1,0,1) \\[0.2em]
            \frac{6}{5}  & 1 \text{ or } 4 & (0,1,0,1)     & (2,3,1,4) \\[0.2em]
            \frac{8}{5}  & 2 \text{ or } 3 & (0,2,0,2)     & (2,5,1,6) \\[0.2em]
            \frac{12}{5} & 2 \text{ or } 3 & (0,2,0,2)     & (6,9,3,12) \\[0.2em]
            \frac{16}{5} & 1 \text{ or } 4 & (0,1,0,1)     & (8,14,4,18) \\[0.2em]
            \frac{18}{5} & 2 \text{ or } 3 & (0,2,0,2)     & (12,18,6,24) \\[0.2em]
            4            & 0               & (0,0,0,0)     & (12,20,6,26) \\[0.2em]
            \frac{24}{5} & 1 \text{ or } 4 & (0,1,0,1)     & (20,30,10,40) \\[0.2em]
            6            & 0               & (0,0,0,0)     & (30,45,15,60) \\[0.2em]
            \frac{36}{5} & 1 \text{ or } 4 & (0,1,0,1)     & (44,66,22,88) \\ \bottomrule
        \end{tabular}

        \caption{$\bm{5}_{q}$ representation of $\asu(5)\oplus\au(1)$}
        \label{tab:su5funddata}
    \end{subtable}

    \caption{Orders of vanishing of $(\secx,\secy,\secz,\secw)$ for various combinations of $\quotparam$ and \pscharge{} $q$ for fundamental matter of $\asu(n)\oplus\au(1)$ with $2 \le n \le 5$. The $\asu(2)\oplus\au(1)$ data are found using the model in \cref{eq:su2seedmodelfund}, the $\asu(3)\oplus\au(1)$ and $\asu(5)\oplus\au(1)$ data are found using the model in \cref{eq:suoddseedmodelfund}, and the $\asu(4)\oplus\au(1)$ data are found using the seed model for fundamentals with even $n$.}
    \label{tab:su2through5funddata}
\end{table}

\begin{table}
    \begin{subtable}{0.5\linewidth}
        \centering

        \begin{tabular}{*{4}{>{$}c<{$}}}\toprule
            \abs{q}      & \quotparam      & \ordalt_{(1)} & \ordalt_{(2)} \\ \midrule
            \frac{1}{6}  & 1 \text{ or } 5 & (0,1,0,1)     & (0,1,0,1) \\[0.2em]
            \frac{1}{3}  & 2 \text{ or } 4 & (0,2,0,2)     & (0,2,0,2) \\[0.2em]
            \frac{1}{2}  & 3               & (0,3,0,3)     & (0,3,0,3) \\[0.2em]
            \frac{2}{3}  & 2 \text{ or } 4 & (0,2,0,2)     & (0,3,0,3) \\[0.2em]
            \frac{5}{6}  & 1 \text{ or } 5 & (0,1,0,1)     & (0,2,0,2) \\[0.2em]
            \frac{7}{6}  & 1 \text{ or } 5 & (0,1,0,1)     & (2,3,1,4) \\[0.2em]
            \frac{5}{3}  & 2 \text{ or } 4 & (0,2,0,2)     & (2,6,1,7) \\[0.2em]
            \frac{7}{3}  & 2 \text{ or } 4 & (0,2,0,2)     & (6,9,3,12) \\[0.2em]
            \frac{5}{2}  & 3               & (0,3,0,3)     & (6,10,3,13) \\[0.2em]
            \frac{10}{3} & 2 \text{ or } 4 & (0,2,0,2)     & (10,16,5,21) \\[0.2em]
            \frac{7}{2}  & 3               & (0,3,0,3)     & (12,18,6,24) \\[0.2em]
            \frac{25}{6} & 1 \text{ or } 5 & (0,1,0,1)     & (14,24,7,31) \\[0.2em]
            \frac{14}{3} & 2 \text{ or } 4 & (0,2,0,2)     & (20,30,10,40) \\[0.2em]
            \frac{35}{6} & 1 \text{ or } 5 & (0,1,0,1)     & (30,45,15,60) \\ \bottomrule
        \end{tabular}

        \caption{$\bm{6}_{q}$ representation of $\asu(6)\oplus\au(1)$}
        \label{tab:su6funddata}
    \end{subtable}
    \begin{subtable}{0.5\linewidth}
        \centering

        \begin{tabular}{*{4}{>{$}c<{$}}}\toprule
            \abs{q}      & \quotparam      & \ordalt_{(1)} & \ordalt_{(2)} \\ \midrule
            \frac{1}{7}  & 1 \text{ or } 6 & (0,1,0,1)     & (0,1,0,1) \\[0.2em]
            \frac{2}{7}  & 2 \text{ or } 5 & (0,2,0,2)     & (0,2,0,2) \\[0.2em]
            \frac{3}{7}  & 3 \text{ or } 4 & (0,3,0,3)     & (0,3,0,3) \\[0.2em]
            \frac{4}{7}  & 3 \text{ or } 4 & (0,3,0,3)     & (0,4,0,4) \\[0.2em]
            \frac{5}{7}  & 2 \text{ or } 5 & (0,2,0,2)     & (0,3,0,3) \\[0.2em]
            \frac{6}{7}  & 1 \text{ or } 6 & (0,1,0,1)     & (0,2,0,2) \\[0.2em]
            \frac{8}{7}  & 1 \text{ or } 6 & (0,1,0,1)     & (2,3,1,4) \\[0.2em]
            \frac{12}{7} & 2 \text{ or } 5 & (0,2,0,2)     & (2,7,1,8) \\[0.2em]
            \frac{16}{7} & 2 \text{ or } 5 & (0,2,0,2)     & (6,9,3,12) \\[0.2em]
            \frac{18}{7} & 3 \text{ or } 4 & (0,3,0,3)     & (6,11,3,14) \\[0.2em]
            \frac{24}{7} & 3 \text{ or } 4 & (0,3,0,3)     & (12,18,6,24) \\[0.2em]
            \frac{30}{7} & 2 \text{ or } 5 & (0,2,0,2)     & (16,26,8,34) \\[0.2em]
            \frac{32}{7} & 3 \text{ or } 4 & (0,3,0,3)     & (20,30,10,40) \\[0.2em]
            \frac{36}{7} & 1 \text{ or } 6 & (0,1,0,1)     & (22,37,11,48) \\[0.2em]
            \frac{40}{7} & 2 \text{ or } 5 & (0,2,0,2)     & (30,45,15,60) \\[0.2em]
            \frac{48}{7} & 1 \text{ or } 6 & (0,1,0,1)     & (42,63,21,84) \\ \bottomrule
        \end{tabular}

        \caption{$\bm{7}_{q}$ representation of $\asu(7)\oplus\au(1)$}
        \label{tab:su7funddata}
    \end{subtable}

    \bigskip
    \centering

    \begin{subtable}{0.5\linewidth}
        \centering

        \begin{tabular}{*{4}{>{$}c<{$}}}\toprule
            \abs{q}      & \quotparam      & \ordalt_{(1)} & \ordalt_{(2)} \\ \midrule
            \frac{1}{8}  & 1 \text{ or } 7 & (0,1,0,1)     & (0,1,0,1) \\[0.2em]
            \frac{1}{4}  & 2 \text{ or } 6 & (0,2,0,2)     & (0,2,0,2) \\[0.2em]
            \frac{3}{8}  & 3 \text{ or } 5 & (0,3,0,3)     & (0,3,0,3) \\[0.2em]
            \frac{1}{2}  & 4               & (0,4,0,4)     & (0,4,0,4) \\[0.2em]
            \frac{7}{8}  & 1 \text{ or } 7 & (0,1,0,1)     & (0,2,0,2) \\[0.2em]
            \frac{9}{8}  & 1 \text{ or } 7 & (0,1,0,1)     & (2,3,1,4) \\[0.2em]
            \frac{7}{4}  & 2 \text{ or } 6 & (0,2,0,2)     & (2,7,1,8) \\[0.2em]
            \frac{9}{4}  & 2 \text{ or } 6 & (0,2,0,2)     & (6,9,3,12) \\[0.2em]
            \frac{21}{8} & 3 \text{ or } 5 & (0,3,0,3)     & (6,12,3,15) \\[0.2em]
            \frac{27}{8} & 3 \text{ or } 5 & (0,3,0,3)     & (12,18,6,24) \\[0.2em]
            \frac{7}{2}  & 4               & (0,4,0,4)     & (12,19,6,25) \\[0.2em]
            \frac{9}{2}  & 4               & (0,4,0,4)     & (20,30,10,40) \\ \bottomrule
        \end{tabular}

        \caption{$\bm{8}_{q}$ representation of $\asu(8)\oplus\au(1)$}
        \label{tab:su8funddata}
    \end{subtable}

    \caption{Orders of vanishing of $(\secx,\secy,\secz,\secw)$ for various combinations of $\quotparam$ and \pscharge{} $q$ for fundamental matter of $\asu(n)\oplus\au(1)$ with $6\le n \le 8$. The $\asu(7)\oplus\au(1)$  data are found using the model in \cref{eq:suoddseedmodelfund}, and the $\asu(6)\oplus\au(1)$ and $\asu(8)\oplus\au(1)$ data are found using the seed model for fundamentals with even $n$.}
    \label{tab:su6through8funddata}
\end{table}

\afterpage{\clearpage}

\subsection{Antisymmetric representation of $\asu(n)$ for odd $n$}
\label{sec:sunantisymmetricsodd}
In F-theory models, matter in the antisymmetric representation is allowed to have $\au(1)$-charges of the form
\begin{equation}
    \frac{2\quotparam}{n} + j \text{ for } j\in\Z\,.
\end{equation}
For odd $n$ (i.e., $n=2k+1$ with $k\ge2$), each allowed $\au(1)$ charge corresponds to a unique value of $\quotparam$:
\begin{equation}
    \quotparam = \residue{nq(k+1)}{n}\,.
\end{equation}
Therefore, suppose we have a seed model with an $\asu(2k+1)\oplus\au(1)$ algebra, a global structure described by $\quotparam=k+1$, and $\bm{\frac{n(n-1)}{2}}_{1/n}$ matter. The antisymmetric matter is supported at a codimension-two locus along the $\singtype{I}^{(s)}_{n}$ locus where the singularity type enhances to $\singtype{I}^{*}_{n-4}$. If the model has a generating section $\ratsec{s}$, the multiples $m\ratsec{s}$ would have $\quotparam=\residue{m(k+1)}{n}$ and would realize \pscharge{} $\frac{m}{n}$ at the codimension-two locus. Thus, multiples of the generating section would give us all allowed combinations of $\quotparam$ and the $\au(1)$ charge.

We can consider the Weierstrass model
\begin{equation}
    \label{eq:sunantiweierodd}
    \begin{aligned}
        y^2 &= x^3 + \left(-\frac{1}{48} \left(b_1^2-4 b_2 \sigma \right)^2+\frac{1}{2} \sigma ^k \left(b_1 c_1-2 c_0 \sigma \right)\right) x z^4 \\
        &\qquad + \frac{1}{864} \left(\left(b_1^2-4 b_2 \sigma \right)^3-36 \left(b_1^2-4 b_2 \sigma \right) \left(b_1 c_1-2 c_0 \sigma \right) \sigma ^k+216 c_1^2 \sigma ^{2 k}\right)z^6\,,
    \end{aligned}
\end{equation}
which admits $\singtype{I}^{(s)}_{2k+1}$ singularities at $\locus{\sigma=0}$ and a generating section $\ratsec{s}$ with components
\begin{equation}
    \label{eq:sunantisecodd}
    [\secx:\secy:\secz] =  \left[\frac{1}{12} \left(b_1^2-4 b_2 \sigma \right):\frac{1}{2} c_1 \sigma ^k:1\right]\,, \quad \secw= \frac{1}{2}\sigma^{k}\left(b_1 c_1-2c_0\sigma\right)\,.
\end{equation}
The gauge algebra for this Weierstrass model is therefore $\asu(2k+1)\oplus\au(1)$. The matter spectrum includes $\bm{\frac{n(n-1)}{2}}_{1/n}$ matter at $\locus{\sigma=b_1=0}$, $\bm{n}_{(k+1)/n}$ matter at $\locus{\sigma=c_1=0}$, $\bm{n}_{-k/n}$ matter at $\locus{\sigma=b_1 c_0-b_2 c_1=0}$, $\bm{1}_{1}$ matter at $\locus{c_0=c_1=0}$, and delocalized $\bm{(n^2-1)}_{0}$ matter allowed to propagate along $\locus{\sigma=0}$. (We can freely flip the overall sign of the $\au(1)$ charges while replacing $\quotparam$ with $n-\quotparam$.) For 6D models, this spectrum satisfies the gauge and gauge--gravitational anomaly conditions for $\height = -2\canonclass - \frac{k(k+1)}{n}\divclass{\sigma}$.

The multiples $m\ratsec{s}$ of the generating section support \pscharge{} $\frac{m}{n}$ at the antisymmetric locus with $\quotparam=\residue{m(k+1)}{n}$, so we can use these multiples to find the orders of vanishing corresponding to $\bm{\frac{n(n-1)}{2}}_{m/n}$ matter. The order of vanishing data are summarized in \cref{tab:sunantiodddata}. According to the formulas in \cref{sec:proposals}, we expect that the codimension-one orders of vanishing should be described by
\begin{equation}
    \left(\ordvanishone{\secx},\ordvanishone{\secy},\ordvanishone{\secw}\right) = \quottriplet{\SU(n)}{\quotparam} = \left(0,\moddist{n}{\quotparam},\moddist{n}{\quotparam}\right)\,, \quad \ordvanishone{\secz}=0\,.
\end{equation}
The codimension-two orders of vanishing, meanwhile, should be given by
\begin{equation}
    \ordvanishtwo{\secz} = \frac{1}{2}\left(\frac{n}{4}q^2 +\frac{\quotparam(n-\quotparam)}{n} - \quotfunction{\SO(2n)}{\quotparamtwo}\right)
\end{equation}
and
\begin{equation}
    \left(\ordvanishone{\secx},\ordvanishone{\secy},\ordvanishone{\secw}\right) = (2,3,4)\times\ordvanishtwo{\secz} + \quottriplet{\SO(2n)}{\quotparamtwo}
\end{equation}
for some value of $\quotparamtwo$. Indeed, the data satisfy these relations with
\begin{equation}
    \quotparamtwo = \residue{n q}{4}\,.
\end{equation}

\begin{table}
    \begin{subtable}{0.5\textwidth}
        \centering

        \begin{tabular}{*{4}{>{$}c<{$}}}\toprule
            \abs{q}      & \quotparam      & \ordalt_{(1)} & \ordalt_{(2)} \\ \midrule
            \frac{1}{5}  & 2 \text{ or } 3 & (0, 2, 0, 2)  & (1, 2, 0, 3) \\[0.2em]
            \frac{2}{5}  & 1 \text{ or } 4 & (0, 1, 0, 1)  & (1, 2, 0, 2) \\[0.2em]
            \frac{3}{5}  & 1 \text{ or } 4 & (0, 1, 0, 1)  & (1, 2, 0, 3) \\[0.2em]
            \frac{4}{5}  & 2 \text{ or } 3 & (0, 2, 0, 2)  & (2, 3, 1, 4) \\[0.2em]
            1            & 0               & (0, 0, 0, 0)  & (1, 2, 0, 3) \\[0.2em]
            \frac{6}{5}  & 2 \text{ or } 3 & (0, 2, 0, 2)  & (3, 5, 1, 6) \\[0.2em]
            \frac{7}{5}  & 1 \text{ or } 4 & (0, 1, 0, 1)  & (3, 5, 1, 7) \\[0.2em]
            \frac{8}{5}  & 1 \text{ or } 4 & (0, 1, 0, 1)  & (4, 6, 2, 8) \\[0.2em]
            \frac{9}{5}  & 2 \text{ or } 3 & (0, 2, 0, 2)  & (5, 8, 2, 11) \\[0.2em]
            2            & 0               & (0, 0, 0, 0)  & (5, 8, 2, 10) \\[0.2em]
            \frac{11}{5} & 2 \text{ or } 3 & (0, 2, 0, 2)  & (7, 11, 3, 15) \\[0.2em]
            \frac{12}{5} & 1 \text{ or } 4 & (0, 1, 0, 1)  & (8, 12, 4, 16) \\ \bottomrule
        \end{tabular}

        \caption{$\bm{10}_{q}$ representation of $\asu(5)\oplus\au(1)$}
        \label{tab:su5antidata}
    \end{subtable}
    \begin{subtable}{0.5\textwidth}
        \centering

        \begin{tabular}{*{4}{>{$}c<{$}}}\toprule
            \abs{q}      & \quotparam      & \ordalt_{(1)} & \ordalt_{(2)} \\ \midrule
            \frac{1}{7}  & 3 \text{ or } 4 & (0, 3, 0, 3)  & (1, 3, 0, 4) \\[0.2em]
            \frac{2}{7}  & 1 \text{ or } 6 & (0, 1, 0, 1)  & (1, 2, 0, 2) \\[0.2em]
            \frac{3}{7}  & 2 \text{ or } 5 & (0, 2, 0, 2)  & (1, 3, 0, 4) \\[0.2em]
            \frac{4}{7}  & 2 \text{ or } 5 & (0, 2, 0, 2)  & (2, 3, 1, 4) \\[0.2em]
            \frac{5}{7}  & 1 \text{ or } 6 & (0, 1, 0, 1)  & (1, 3, 0, 4) \\[0.2em]
            \frac{6}{7}  & 3 \text{ or } 4 & (0, 3, 0, 3)  & (3, 5, 1, 6) \\[0.2em]
            1            & 0               & (0, 0, 0, 0)  & (1, 3, 0, 4) \\[0.2em]
            \frac{8}{7}  & 3 \text{ or } 4 & (0, 3, 0, 3)  & (4, 6, 2, 8) \\[0.2em]
            \frac{9}{7}  & 1 \text{ or } 6 & (0, 1, 0, 1)  & (3, 6, 1, 8) \\[0.2em]
            \frac{10}{7} & 2 \text{ or } 5 & (0, 2, 0, 2)  & (5, 8, 2, 10) \\[0.2em]
            \frac{11}{7} & 2 \text{ or } 5 & (0, 2, 0, 2)  & (5, 9, 2, 12) \\[0.2em]
            \frac{12}{7} & 1 \text{ or } 6 & (0, 1, 0, 1)  & (6, 9, 3, 12) \\[0.2em]
            \frac{13}{7} & 3 \text{ or } 4 & (0, 3, 0, 3)  & (7, 12, 3, 16) \\[0.2em]
            2            & 0               & (0, 0, 0, 0)  & (7, 11, 3, 14) \\ \bottomrule
        \end{tabular}

        \caption{$\bm{21}_{q}$ representation of $\asu(7)\oplus\au(1)$}
        \label{tab:su7antidata}
    \end{subtable}

   \bigskip
   \centering

   \begin{subtable}{0.5\textwidth}
        \centering
        \begin{tabular}{*{4}{>{$}c<{$}}}\toprule
            \abs{q}      & \quotparam      & \ordalt_{(1)} & \ordalt_{(2)} \\ \midrule
            \frac{1}{9}  & 4 \text{ or } 5 & (0, 4, 0, 4)  & (1, 4, 0, 5) \\[0.2em]
            \frac{2}{9}  & 1 \text{ or } 8 & (0, 1, 0, 1)  & (1, 2, 0, 2) \\[0.2em]
            \frac{1}{3}  & 3 \text{ or } 6 & (0, 3, 0, 3)  & (1, 4, 0, 5) \\[0.2em]
            \frac{4}{9}  & 2 \text{ or } 7 & (0, 2, 0, 2)  & (2, 3, 1, 4) \\[0.2em]
            \frac{5}{9}  & 2 \text{ or } 7 & (0, 2, 0, 2)  & (1, 4, 0, 5) \\[0.2em]
            \frac{2}{3}  & 3 \text{ or } 6 & (0, 3, 0, 3)  & (3, 5, 1, 6) \\[0.2em]
            \frac{7}{9}  & 1 \text{ or } 8 & (0, 1, 0, 1)  & (1, 4, 0, 5) \\[0.2em]
            \frac{8}{9}  & 4 \text{ or } 5 & (0, 4, 0, 4)  & (4, 6, 2, 8) \\[0.2em]
            1            & 0               & (0, 0, 0, 0)  & (1, 4, 0, 5) \\[0.2em]
            \frac{10}{9} & 4 \text{ or } 5 & (0, 4, 0, 4)  & (5, 8, 2, 10) \\[0.2em]
            \frac{11}{9} & 1 \text{ or } 8 & (0, 1, 0, 1)  & (3, 7, 1, 9) \\[0.2em]
            \frac{4}{3}  & 3 \text{ or } 6 & (0, 3, 0, 3)  & (6, 9, 3, 12) \\ \bottomrule
        \end{tabular}

        \caption{$\bm{36}_{q}$ representation of $\asu(9)\oplus\au(1)$}
        \label{tab:su9antidata}
   \end{subtable}

   \caption{Orders of vanishing of $(\secx,\secy,\secz,\secw)$ for various combinations of $\quotparam$ and \pscharge{} $q$ for antisymmetric matter of $\asu(2k+1)\oplus\au(1)$. The data are found using the model in \cref{eq:sunantiweierodd}.}
   \label{tab:sunantiodddata}
\end{table}

\afterpage{\clearpage}

\subsection{Antisymmetric representation of $\asu(n)$ for even $n$}
\label{sec:sunantisymmetricseven}

Recall that matter in the antisymmetric representation of $\asu(n)$ is allowed to have $\au(1)$ charges of the form $\frac{2\quotparam}{n} + j$ for integers $j$. When $n$ is even (i.e., when $n=2k$), the allowed $nq$ are always even. More importantly, the same value of $q$ can occur for two different values of $\quotparam$: $\quotparam = \residue{kq}{n}$ or $\quotparam = \residue{kq+k}{n}$. Because of this fact, our strategy of using multiples of the generating section becomes more difficult. Suppose we start with a  $\asu(n)\oplus\au(1)$ seed model with $\quotparam=1$ and $\bm{\frac{n(n-1)}{2}}_{2/n}$ matter. By considering multiples $m\ratsec{s}$ of the generating section $\ratsec{s}$, we can obtain the \pscharge{}s $q=\frac{2m}{n}$ with
\begin{equation}
    \quotparam = \residue{m}{n} = \residue{kq}{n}\,.
\end{equation}
To obtain data for situations where $\quotparam = \residue{kq+k}{n}$, we need a different seed model. If we start with one admitting $\bm{\frac{n(n-1)}{2}}_{2/n}$ matter with $\quotparam=k+1$, multiples $m\ratsec{s}$ of the generating section $\ratsec{s}$ would give us \pscharge{}s $\frac{2m}{n}$ with
\begin{equation}
    \quotparam = \residue{m k + m}{n}\,.
\end{equation}
We can therefore use this seed model to obtain information about the $\quotparam = \residue{kq+k}{n}$ situations when $kq$ is odd. However, we would not learn anything about the $\quotparam = \residue{kq+k}{n}$ situations when $kq$ is even. If we were to start with a seed model admitting $\bm{\frac{n(n-1)}{2}}_{4/n}$ matter with $\quotparam=k+2$, we could obtain information about the $\quotparam = \residue{kq+k}{n}$ situations for some even values of $kq$, but we would miss those values of $kq$ that are multiples of $4$.

In order to get information about all of the $\quotparam = \residue{kq+k}{n}$ situations, we would need an infinite number of seed models. This would clearly be impractical, so we will consider only a limited number of $(\quotparam,q)$ combinations. Specifically, we will focus on the situations where $\quotparam = \residue{kq}{n}$ and the situations where $\quotparam = \residue{kq+k}{n}$ with odd $kq$. Thus, we need two seed models. For the $\quotparam = \residue{kq}{n}$ situations, we can consider the same Weierstrass model used for the fundamentals of $\asu(2k)\oplus\au(1)$, which is described in \cref{sec:sunfundamentals}. This model has $\quotparam=1$ and admits the locus $\locus{\sigma=b_{1,0}=0}$ supporting $\bm{\frac{n(n-1)}{2}}_{2/n}$ matter.

For the $\quotparam = \residue{kq+k}{n}$ cases with odd $kq$, we need a seed model with $\quotparam=k+1$ and $\bm{\frac{n(n-1)}{2}}_{2/n}$ matter, at least for $k\ge3$.\footnote{The $\asu(4)\oplus\au(1)$ with $k=2$ is somewhat special, as $\quotparam=3$ is the inverse of $\quotparam=1$ in $\Z_4$. Thus, the $\quotparam=3$ seed model would be equivalent to the $\quotparam=1$ seed model discussed directly above, implying that we do not need a second seed model to obtain the $\quotparam = \residue{kq+k}{n}$ \pscharge{}s with odd $kq$.}  We can use the Weierstrass construction
\begin{equation}
    \label{eq:sunevenantiseedmodelb}
    \begin{aligned}
        y^2 &= x^3 + \left[-\frac{1}{48} \left(b_1^2-4b_2 \sigma-4c_{1,0} \sigma ^{k-1}
        \right)^2+\sigma ^k \left(\frac{1}{3} b_2 c_{1,0}-\frac{1}{2} b_1 c_{1,1}+\sigma
        c_{0,1}\right)\right]x z^4 \\
        &\phantom{{}= x^3} + \Bigg[\frac{1}{864} \left(b_1^2-4 \sigma  b_2-4 \sigma ^{k-1} c_{1,0}\right)^3 \\
        &\qquad\qquad\quad-\frac{1}{72} \sigma^k \left(b_1^2-4 \sigma  b_2-4 \sigma ^{k-1} c_{1,0}\right) \left(6 \sigma
        c_{0,1}+2 b_2 c_{1,0}-3 b_1 c_{1,1}\right)\\
        &\qquad\qquad\quad+ \sigma^{2 k} \left(\frac{c_{1,1}^2}{4}-\frac{1}{3} c_{0,1} c_{1,0}\right)+\frac{2}{27} c_{1,0}^3
        \sigma^{3 k-3}\Bigg]z^6\,.
    \end{aligned}
\end{equation}
It admits $\singtype{I}^{(s)}_{2k}$ singularities along $\locus{\sigma=0}$ and a generating section $\ratsec{s}$ with components
\begin{equation}
\begin{aligned}
    [\secx:\secy:\secz] &= \left[\frac{1}{12} \left(b_1^2-4 b_2 \sigma \right):\frac{1}{6} \sigma^{k-1} \left(3 \sigma  c_{1,1}-b_1 c_{1,0}\right):1\right]\,, \\
    \secw &= -\frac{1}{6} \sigma^{k-1} \left(-b_1^2 c_{1,0}+\sigma  \left(2 b_2 c_{1,0}+3 b_1 c_{1,1}\right)-6 \sigma ^2 c_{0,1}+2 \sigma ^{k-1} c_{1,0}^2\right)\,.
    \label{eq:sunevenantiseedmodelbgensec}
\end{aligned}
\end{equation}
In addition to possible delocalized adjoint matter along $\locus{\sigma=0}$, the model supports $\bm{\frac{n(n-1)}{2}}_{2/n}$ matter at $\locus{\sigma=b_1=0}$, $\bm{n}_{(k+1)/n}$ matter at $\locus{\sigma=c_{1,0}=0}$, $\bm{n}_{-(k-1)/n}$ matter at
\begin{equation}
    \locus{\sigma=-\frac{1}{2} b_1^2 c_{0,1}-\frac{1}{6} b_2^2 c_{1,0}+\frac{1}{9} \sigma ^{k-3} b_1^2
        c_{1,0}^2+\frac{1}{2} b_1 b_2 c_{1,1}=0}\,,
\end{equation}
and $\bm{1}_{1}$ matter at
\begin{equation}
\left\{\frac{\secy}{\sigma^{k-1}} = \frac{\secw}{\sigma^{k-1}}=0\right\}\setminus\left(\locus{\sigma=b_{1}=0}\cup\locus{\sigma=c_{1,0}=0}\right)\,.
\end{equation}
(Of course, one can flip the signs of the $\au(1)$ charges to obtain an equally valid description of the spectrum with $\quotparam=k-1$.) One can verify that, for 6D models, this spectrum satisfies the gauge and gauge--gravitational anomalies with $\height = -2\canonclass -\frac{(k-1)(k+1)}{n}\divclass{\sigma}$.

\Cref{tab:sunantievendata} summarizes the orders of vanishing obtained from these two seed constructions. The codimension-one orders of vanishing satisfy the formulas
\begin{equation}
    \left(\ordvanishone{\secx},\ordvanishone{\secy},\ordvanishone{\secw}\right) = \quottriplet{\SU(n)}{\quotparam}\,, \quad \ordvanishone{\secz} = 0\,.
\end{equation}
The codimension-two orders of vanishing are given by
\begin{equation}
    \ordvanishtwo{\secz} = \frac{1}{2}\left(\frac{n}{4}q^2 + \quotfunction{\SU(n)}{\quotparam} - \quotfunction{\SO(2n)}{\quotparamtwo}\right)
\end{equation}
and
\begin{equation}
        \left(\ordvanishtwo{\secx},\ordvanishtwo{\secy},\ordvanishtwo{\secw}\right) = \left(2,3,4\right)\times\ordvanishtwo{\secz} + \quottriplet{\SO(2n)}{\quotparamtwo}
\end{equation}
for
\begin{equation}
    \quotparamtwo = \residue{\left(q-\frac{2}{n}\quotparam\right)}{2} +\residue{2\quotparam}{4}\,.
\end{equation}
These formulas exactly match those expected from the proposals in \cref{sec:proposals}.

\begin{table}
    \begin{small}
        \begin{subtable}{0.5\textwidth}
            \centering

            \begin{tabular}{*{4}{>{$}c<{$}}}\toprule
                \abs{q}      & \quotparam      & \ordalt_{(1)} & \ordalt_{(2)} \\ \midrule
                \frac{1}{2}  & 1 \text{ or } 3 & (0, 1, 0, 1)  & (1, 2, 0, 2) \\[0.2em]
                1            & 2               & (0, 2, 0, 2)  & (2, 3, 1, 4) \\[0.2em]
                \frac{3}{2}  & 1 \text{ or } 3 & (0, 1, 0, 1)  & (3, 5, 1, 6) \\[0.2em]
                2            & 0               & (0, 0, 0, 0)  & (4, 6, 2, 8) \\[0.2em]
                \frac{5}{2}  & 1 \text{ or } 3 & (0, 1, 0, 1)  & (7, 11, 3, 14) \\[0.2em]
                3            & 2               & (0, 2, 0, 2)  & (10, 15, 5, 20) \\[0.2em]
                \frac{7}{2}  & 1 \text{ or } 3 & (0, 1, 0, 1)  & (13, 20, 6, 26) \\[0.2em]
                4            & 0               & (0, 0, 0, 0)  & (16, 24, 8, 32) \\[0.2em]
                \frac{9}{2}  & 1 \text{ or } 3 & (0, 1, 0, 1)  & (21, 32, 10, 42) \\[0.2em]
                5            & 2               & (0, 2, 0, 2)  & (26, 39, 13, 52) \\[0.2em]
                \frac{11}{2} & 1 \text{ or } 3 & (0, 1, 0, 1)  & (31, 47, 15, 62) \\[0.2em]
                6            & 0               & (0, 0, 0, 0)  & (36, 54, 18, 72) \\[0.2em]
                \frac{13}{2} & 1 \text{ or } 3 & (0, 1, 0, 1)  & (43, 65, 21, 86) \\[0.2em]
                7            & 2               & (0, 2, 0, 2)  & (50, 75, 25, 100) \\ \bottomrule
            \end{tabular}

            \caption{$\bm{6}_{q}$ representation of $\asu(4)\oplus\au(1)$}
            \label{tab:su4antisymmetrics}
        \end{subtable}
        \begin{subtable}{0.5\textwidth}
            \centering

            \begin{tabular}{*{4}{>{$}c<{$}}}\toprule
                \abs{q}      & \quotparam      & \ordalt_{(1)} & \ordalt_{(2)} \\ \midrule
                \frac{1}{3}  & 1 \text{ or } 5 & (0, 1, 0, 1)  & (1, 2, 0, 2) \\[0.2em]
                \frac{1}{3}  & 2 \text{ or } 4 & (0, 2, 0, 2)  & (1, 3, 0, 3) \\[0.2em]
                \frac{2}{3}  & 2 \text{ or } 4 & (0, 2, 0, 2)  & (2, 3, 1, 4) \\[0.2em]
                1            & 0               & (0, 0, 0, 0)  & (1, 3, 0, 3) \\[0.2em]
                1            & 3               & (0, 3, 0, 3)  & (3, 5, 1, 6) \\[0.2em]
                \frac{4}{3}  & 2 \text{ or } 4 & (0, 2, 0, 2)  & (4, 6, 2, 8) \\[0.2em]
                \frac{5}{3}  & 2 \text{ or } 4 & (0, 2, 0, 2)  & (5, 9, 2, 11) \\[0.2em]
                \frac{5}{3}  & 1 \text{ or } 5 & (0, 1, 0, 1)  & (5, 8, 2, 10) \\[0.2em]
                2            & 0               & (0, 0, 0, 0)  & (6, 9, 3, 12) \\[0.2em]
                \frac{7}{3}  & 1 \text{ or } 5 & (0, 1, 0, 1)  & (9, 14, 4, 18) \\[0.2em]
                \frac{7}{3}  & 2 \text{ or } 4 & (0, 2, 0, 2)  & (9, 15, 4, 19) \\[0.2em]
                \frac{8}{3}  & 2 \text{ or } 4 & (0, 2, 0, 2)  & (12, 18, 6, 24) \\[0.2em]
                3            & 0               & (0, 0, 0, 0)  & (13, 21, 6, 27) \\[0.2em]
                3            & 3               & (0, 3, 0, 3)  & (15, 23, 7, 30) \\[0.2em]
                \frac{10}{3} & 2 \text{ or } 4 & (0, 2, 0, 2)  & (18, 27, 9, 36) \\[0.2em]
                \frac{11}{3} & 2 \text{ or } 4 & (0, 2, 0, 2)  & (21, 33, 10, 43) \\[0.2em]
                \frac{11}{3} & 1 \text{ or } 5 & (0, 1, 0, 1)  & (21, 32, 10, 42) \\[0.2em]
                4            & 0               & (0, 0, 0, 0)  & (24, 36, 12, 48) \\[0.2em]
                \frac{13}{3} & 2 \text{ or } 4 & (0, 2, 0, 2)  & (29, 45, 14, 59) \\[0.2em]
                \frac{14}{3} & 2 \text{ or } 4 & (0, 2, 0, 2)  & (34, 51, 17, 68) \\ \bottomrule
            \end{tabular}

            \caption{$\bm{15}_{q}$ representation of $\asu(6)\oplus\au(1)$}
            \label{tab:su6antisymmetrics}
        \end{subtable}

        \bigskip
        \centering

        \begin{subtable}{0.5\textwidth}
            \centering

            \begin{tabular}{*{4}{>{$}c<{$}}}\toprule
                \abs{q}     & \quotparam      & \ordalt_{(1)} & \ordalt_{(2)} \\ \midrule
                \frac{1}{4} & 1 \text{ or } 7 & (0, 1, 0, 1)  & (1, 2, 0, 2) \\[0.2em]
                \frac{1}{4} & 3 \text{ or } 5 & (0, 3, 0, 3)  & (1, 4, 0, 4) \\[0.2em]
                \frac{1}{2} & 2 \text{ or } 6 & (0, 2, 0, 2)  & (2, 3, 1, 4) \\[0.2em]
                \frac{3}{4} & 3 \text{ or } 5 & (0, 3, 0, 3)  & (3, 5, 1, 6) \\[0.2em]
                \frac{3}{4} & 1 \text{ or } 7 & (0, 1, 0, 1)  & (1, 4, 0, 4) \\[0.2em]
                1           & 4               & (0, 4, 0, 4)  & (4, 6, 2, 8) \\[0.2em]
                \frac{5}{4} & 1 \text{ or } 7 & (0, 1, 0, 1)  & (3, 7, 1, 8) \\[0.2em]
                \frac{5}{4} & 3 \text{ or } 5 & (0, 3, 0, 3)  & (5, 8, 2, 10) \\[0.2em]
                \frac{3}{2} & 2 \text{ or } 6 & (0, 2, 0, 2)  & (6, 9, 3, 12) \\[0.2em]
                \frac{7}{4} & 3 \text{ or } 5 & (0, 3, 0, 3)  & (7, 13, 3, 16) \\[0.2em]
                \frac{7}{4} & 1 \text{ or } 7 & (0, 1, 0, 1)  & (7, 11, 3, 14) \\[0.2em]
                2           & 0               & (0, 0, 0, 0)  & (8, 12, 4, 16) \\[0.2em]
                \frac{9}{4} & 1 \text{ or } 7 & (0, 1, 0, 1)  & (11, 17, 5, 22) \\[0.2em]
                \frac{9}{4} & 3 \text{ or } 5 & (0, 3, 0, 3)  & (11, 19, 5, 24) \\[0.2em]
                \frac{5}{2} & 2 \text{ or } 6 & (0, 2, 0, 2)  & (14, 21, 7, 28) \\ \bottomrule
            \end{tabular}

            \caption{$\bm{28}_{q}$ representation of $\asu(8)\oplus\au(1)$}
            \label{tab:su8antisymmetrics}
        \end{subtable}

    \end{small}

    \caption{Orders of vanishing of $(\secx,\secy,\secz,\secw)$ or various combinations of $\quotparam$ and \pscharge{} $q$ for antisymmetric matter of $\asu(2k)\oplus\au(1)$. }
    \label{tab:sunantievendata}
\end{table}

\afterpage{\clearpage}

\subsection{Connection to EDS valuations}
\label{sec:suneds}

So far, we have seen that the orders of vanishing obtained using the strategy of \cref{sec:strategy} agree with the proposed formulas in \cref{sec:proposals}. Just as was the case with singlets, the appearance of these orders of vanishing can at least be partially explained by the results in~\cite{StangeEllTrouble} regarding the $p$-adic valuations of elliptic divisibility sequences. However, there is one new complication for the non-singlet cases. The formula for $\ordvanishtwo{\secz}$ appears to be related to the difference in the $p$-adic valuations of two different elliptic divisibility sequences: one corresponding to the codimension-two behavior and the other corresponding to the codimension-one behavior. Despite this additional difficulty, we can see a clear connection between the proposals and the results in  \cite{StangeEllTrouble}.

\paragraph{Fundamental representation of $\asu(n)$}
As described in \cref{sec:sunfundamentals}, we could obtain all the allowed \pscharge{}s by starting with a seed model admitting $\bm{n}_{1/n}$ matter. In the F-theory constructions we consider, this matter occurs at codimension-two loci where the singularity type enhances from $\singtype{I}^{(s)}_{n}$ to $\singtype{I}_{n+1}$. Moreover, the generating section components $(\secx,\secy,\secz,\secw)$ vanish to orders $(0,1,0,1)$ at this locus, indicating that $\quotparamtwo=1$. Multiples $m\ratsec{s}$ of the generating section support \pscharge{} $\frac{m}{n}$.

If we write this codimension-two locus as $\locus{\sigma=\Delta_{1/n}=0}$ where $\locus{\sigma=0}$ is the $\asu(n)$ locus, we could imagine an elliptic curve found by setting $\sigma$ and $\Delta_{1/n}$ to some prime $p$ and setting other parameters to some arbitrary integer values. Because the elliptic fibration supports singularity type $\singtype{I}_{n+1}$ at the codimension-two locus, the elliptic curve analogue would have multiplicative reduction modulo $p$. Meanwhile, consider the rational point $P$ on the elliptic curve analogue found by making the appropriate substitutions into the generating section expressions. As discussed in \cref{sec:singleteds}, the result in~\cite{StangeEllTrouble} states that the elliptic divisibility sequence $\eds^{(2)}_{m}$ associated with $P$ should have $p$-adic valuations given by an elliptic troublemaker sequence:\footnote{We are assuming that no multiples of $P$ reduce to the identity point, but this seems to agree with the observed behaviors.}
\begin{equation}
    \valuation{p}{\eds^{(2)}_{m}} = \elltrouble{m}{a}{l}\,.
\end{equation}
The integer $l$ should equal $\valuation{p}{\Delta}$, which in this case is $n+1$. Meanwhile, $a$, which describes the component of the Neron fiber hit by $P$, corresponds to $\quotparamtwo$, which in this case is $1$. We would therefore expect that
\begin{equation}
    \valuation{p}{\eds^{(2)}_{m}} = \elltrouble{m}{1}{n+1} = \frac{1}{2}\left(\frac{n}{n+1}m^2 - \frac{\residue{m}{n+1}\left(n+1-\residue{m}{n+1}\right)}{n+1}\right)\,.
\end{equation}

However, $\valuation{p}{\eds^{(2)}_{m}}$ overcounts $\ordvanishtwo{\secz}$. Remember that if $\secx$, $\secy$, and $\secz$ are proportional to $A^2$, $A^3$ and $A$ for some factor $A$, one can scale the section components and remove this factor of $A$. We always get rid of such factors when calculating the $\ordvanishtwo{\secz}$ values. However, the recursion formulas for the elliptic divisibility sequences, which are analogous to formulas for the $\secz$ components of multiples of the generating section, do not remove such factors. In turn, the formulas from~\cite{StangeEllTrouble} count these extra factors as well. This did not cause any issues for the case of singlets, in part because there was no special codimension-one locus in the model. For the $\asu(n)\oplus\au(1)$ models, the $\secy$ and $\secw$ components of the generating section are proportional to $\sigma$. As a result, the formulas for the elliptic divisibility sequences would lead to extra factors of $\sigma$ that can be scaled away. To recover $\ordvanishtwo{\secz}$, we need to count these extra factors and subtract the result from $\valuation{p}{W^{(2)}_{m}}$.

We can in fact count these extra factors by considering a second EDS, $W^{(1)}_{m}$, just for the factors of $\sigma$. Imagine we found an elliptic curve by taking our seed model with $\bm{n}_{1/n}$ matter and setting $\sigma$ (but not $\Delta_{1/n}$) to a prime $p$. The resulting elliptic curve would have multiplicative reduction modulo $p$. If we consider a point found by making the appropriate substitutions into the generating section components, the corresponding EDS valuations should be given by
\begin{equation}
    \valuation{p}{\eds^{(1)}_{m}} = \elltrouble{m}{a}{l}\,.
\end{equation}
The integer $l$ should equal $\valuation{p}{\Delta} = n$, while $a$ should be given by the value of $\quotparam$ for the generating section, which is $1$. In the end, we have
\begin{equation}
        \valuation{p}{\eds^{(1)}_{m}} = \elltrouble{m}{1}{n} = \frac{1}{2}\left(\frac{n-1}{n}m^2 - \frac{\residue{m}{n}\left(n-\residue{m}{n}\right)}{n}\right)\,.
\end{equation}

The difference of these two sequences of EDS valuations is given by
\begin{equation}
    \valuation{p}{W^{(2)}_{m}} - \valuation{p}{W^{(1)}_{m}} = \frac{1}{2}\left(\frac{1}{n(n+1)}m^2 + \frac{\residue{m}{n}\left(n-\residue{m}{n}\right)}{n} - \frac{\residue{m}{n+1}\left(n+1-\residue{m}{n+1}\right)}{n+1}\right)
\end{equation}
Substituting $m=n q$, $\quotparam=\residue{m}{n}$, and $\quotparamtwo=\residue{m}{n+1}$, we can rewrite this as
\begin{equation}
    \begin{aligned}
    \valuation{p}{W^{(2)}_{m}} - \valuation{p}{W^{(1)}_{m}} &= \frac{1}{2}\left(\frac{n^2}{n(n+1)}q^2 + \frac{\quotparam\left(n-\quotparam\right)}{n} - \frac{\quotparamtwo\left(n+1-\quotparamtwo\right)}{n+1}\right)\\
    &= \frac{1}{2}\left(\frac{d_{\SU(n)}}{d_{\SU(n+1)}} q^2 + \quotfunction{\SU(n)}{\quotparam} - \quotfunction{\SU(n+1)}{\quotparamtwo} \right)\,.
    \end{aligned}
\end{equation}
This result agrees exactly with observed patterns in $\ordvanishtwo{\secz}$ and with the proposals in \cref{sec:proposals}.

\paragraph{Antisymmetric representation of $\asu(n)$ for odd $n$} We saw that all the allowed antisymmetric \pscharge{}s for odd $n>4$ can be obtained from multiples of the generating section in a model supporting antisymmetric matter with charge $\frac{1}{n}$. Such matter occurs in models with $\quotparam = \ceil*{\frac{n}{2}}=k+1$. We should therefore be able to obtain the formula for $\ordvanishtwo{\secz}$ from those in  \cite{StangeEllTrouble} describing the EDS valuations. As with fundamental matter, we expect $\ordvanishtwo{\secz}$ to be given by the difference of EDS valuations corresponding to the codimension-two and codimension-one behavior. For the codimension-one $\singtype{I}_n^{(s)}$ singularity type, consider the Weierstrass model in \cref{eq:sunantiweierodd}, which admits an $\asu(n)\oplus\au(1)$ model with $n=2k+1$. The locus $\locus{\sigma=0}$ supports the $\asu(n)$ factor, and the locus $\locus{\sigma=b_1=0}$ supports antisymmetric matter with $\au(1)$ charge $\frac{1}{n}$ (up to sign). To construct the EDS associated with the codimension-one behavior, we construct an analogous elliptic curve by setting $\sigma$ to a prime integer $p$ and the remaining parameters to numbers such that $f$ and $g$ of the resulting elliptic curve are integers. For instance, we can consider the elliptic curve given by the following Weierstrass equation (with $z$ set to 1):
\begin{equation}
y^2 = x^3 -(p-3) \left(2 p^k+3 p-9\right)x -\left(2(p-3)^3-p^{2 k}+2 (p-3)^2 p^k\right)\,.
\end{equation}
This elliptic curve has multiplicative reduction modulo $p$ with $\valuation{}{\Delta} = 2k+1$, and it admits the rational point
\begin{equation}
(x,y) = (3-p,p^k)\,.
\end{equation}
The rational point reduces to the singular point modulo $p$, and no multiples of this point reduce to the identity point. The EDS valuations associated with this elliptic curve are given by
\begin{equation}
\valuation{}{W_m^{(1)}}=\elltrouble{m}{k}{2k+1} = \frac{1}{2}\left(\frac{k(k+1)}{n} m^2 - \frac{\residue{mk}{n}(n-\residue{m k}{n})}{n}\right)\,.
\end{equation}

For the codimension-two behavior, we need to find the EDS valuations $\valuation{}{\eds_{m}^{(2)}}$ corresponding to $\singtype{I}^*_{n-4}$ loci with at $\locus{\sigma=b_1=0}$ with $\quotparamtwo = 1$ or $3$. The fibers at $\singtype{I}^*$ loci have cusp singularities, so the analogous elliptic curves should have additive reduction, unlike those with multiplicative reduction we encountered previously.
Fortunately, \cite{StangeEllTrouble} gives formulas for the valuations of EDSs for elliptic curves of additive reduction. The key insight is that, after a field extension by an element $p^\prime$, elliptic curves with additive reduction can be converted to a more minimal form with either good or multiplicative reduction. As an example, consider an elliptic curve $E$ over $\Q_p$ of the form
\begin{equation}
y^2 = x^3 + p^2 f^\prime x + p^3 g^\prime\,, \label{eq:arithredex}
\end{equation}
where $p$ is a prime integer and $f^\prime$, $g^\prime$ are integers not proportional to $p$ that satisfy
\begin{equation}
\Delta = 4\left(p^2 f^\prime\right)^3 + 27\left(p^3 g^\prime\right)^2 \propto  p^{6+(n-4)}
\end{equation}
for an integer $n\ge 4$. This elliptic curve, the analogue of $\singtype{I}^*_{n-4}$ singular fibers, has additive reduction modulo $p$. Suppose we extend $\Q_p$ by $p^\prime = \sqrt{p}$, such that the elliptic curve can be written as
\begin{equation}
y^2 = x^3 + (p^\prime)^4 f^\prime x + (p^\prime)^6 g^\prime\,.
\end{equation}
The elliptic curve is now in a non-minimal form, and we can remove the factors of ${p^\prime}$ by defining
\begin{equation}
x = (p^\prime)^2 x^{\prime}\,, \quad y = (p^\prime)^3 y^{\prime}\,.
\end{equation}
After plugging in these redefinitions and removing the common factor of  $(p^\prime)^6$, we obtain a new elliptic curve $E^\prime$ of the form
\begin{equation}
{y^\prime}^2 = {x^\prime}^3 +  f^\prime x^\prime + g^\prime\,.
\end{equation}
with a discriminant
\begin{equation}
\Delta_{E^\prime} = \frac{\Delta_{E}}{(p^\prime)^{12}}\,.
\end{equation}
$E^\prime$ no longer has additive reduction modulo $p^\prime$. If $E^\prime$ has good reduction, we say that the original curve $E$ has potential good reduction. If $E^\prime$ has multiplicative reduction, the original curve $E$ has potential multiplicative reduction.

Now consider the elliptic divisibility sequence $W_m$ associated with a point $P$ on $E$. The point $P$ maps onto a point $P^\prime$ on $E^\prime$, and we can form an EDS $W_m^\prime$ associated with $P^\prime$ and $E^\prime$.  We can also have a valuation $v_1$ for the extended field that is essentially a valuation with respect to $p^\prime$: for an element $\zeta$ of the original field, $\valuation{1}{\zeta} = d\valuation{}{\zeta}$, where $d$ is the ramification degree of the extension. For the example in \cref{eq:arithredex}, the ramification degree is $2$ because we extend the field by $\sqrt{p}$.

The valuations $\valuation{}{W_m}$ can be related to the valuations $\valuation{1}{W_m^\prime}$. According to Theorem~19 of~\cite{StangeEllTrouble}, if $E$ has potential good reduction, the valuations of $W_m$ are given by the formula
\begin{equation}
d \valuation{}{W_m} = \frac{m^2-1}{12} d \valuation{}{\Delta_{E}} + \valuation{1}{W_m^\prime}\,. \label{eq:valWnpotgood}
\end{equation}
And according to Theorem~14 of~\cite{StangeEllTrouble}, the valuations $\valuation{1}{W_m^\prime}$ take the form
\begin{equation}
\label{eq:valWngood}
\valuation{1}{W_m^\prime} = \begin{cases}\frac{\valuation{1}{x^\prime_{P^\prime}}}{2}m^2 & n_{P^\prime} = 1 \\ 0 &  n_{P^\prime}\neq 1\end{cases} + \begin{cases}\scorr_{m/n_{{P^\prime}}}\left(p^\prime, b, \valuation{1}{p^\prime}, h, s, w\right) & m \mid n_{{P^\prime}} \\ 0 &  m \nmid n_{P^\prime}\end{cases}\,.
\end{equation}
Here, $x^\prime_{P^\prime}$ is the $x^\prime$ coordinate of $P^\prime$, and $n_{P^\prime}$ is the smallest integer such that $n_{P^\prime} P^\prime$ reduces to the identity of $E^\prime$ modulo $p^\prime$. The $\scorr$ term in described in more detail in \cref{app:scorr}, but as argued there it can essentially be replaced by the constant $s$ for the cases of interest in this paper.

If $E$ has potential multiplicative reduction, Theorem~29 of~\cite{StangeEllTrouble} states that the valuations of $W_m$ are given by
\begin{equation}
d \valuation{}{W_m} = \frac{m^2-1}{4} d \valuation{}{f} + \valuation{1}{W_m^\prime}\,, \label{eq:valWnpotmult}
\end{equation}
where $f$ can be read off from the equation for $E$.\footnote{If $E$ is written in Tate form, $f$ takes the standard form $-((a_1^2+4a_2)^2-24(a_1 a_3 + a_4))/48$. The quantity we refer to as $f$ would more commonly be referred to as $c_{4}(E)$ in the mathematics literature.} If $P^\prime$ reduces to a point other than the singular point, the $\valuation{1}{W_m^\prime}$ are still given by \cref{eq:valWngood}. However, if $P^\prime$ reduces to the singular point, the $ \valuation{1}{W_m^\prime}$ are given by the multiplicative reduction expressions, and the $\valuation{}{W_m}$ valuations take the form
\begin{equation}
\valuation{}{W_m} = \frac{m^2-1}{4} \valuation{}{f} + \frac{1}{d}\left(\elltrouble{m}{a}{l} +  \begin{cases}\scorr_{m/n_{P^\prime}}\left(p^\prime, p^\prime, \valuation{1}{p^\prime}, 0, s, w\right) &  n_{P^\prime} \mid m\\ 0 & n_{P^\prime} \nmid m \end{cases} \right) \label{eq:valWnpotmultsing}
\end{equation}
for integers $a$, $l$, and $n_{P^\prime}$.

To apply this to the codimension-two $\singtype{I}^*_{n-4}$ singularities, we consider an elliptic curve found by taking \cref{eq:sunantiweierodd}, setting $\sigma$ and $b_1$ to integers $p$ and $6p$ (where $p$ is a prime integer not equal to 2 or 3), and letting other parameters be appropriate numbers. For a particular choice of these numbers, we obtain the elliptic curve $E$
\begin{equation}
y^2 = x^3 + p^2 \left(-3 (1-3 p)^2+4 p^{k-1}\right)x +p^3 \left(2 (3 p-1)^3 -4 (3 p-1) p^{k-1} +p^{2 k-3}\right)\,,
\end{equation}
which admits a rational point
\begin{equation}
P\colon (x,y) = (p (3 p-1),p^k)\,.
\end{equation}
This elliptic curve has additive reduction with $\valuation{}{\Delta}=2k+3$. After following the procedure outlined above, the resulting elliptic curve $E^\prime$ has multiplicative reduction modulo $p^\prime$ with $l = \valuation{1}{\Delta^\prime}=4k-6$. Additionally, $P^\prime$ reduces to the singular point on $E^\prime$. However, $4P^\prime$ reduces to the identity point,  implying that $n_{P^\prime}=4$. The $y^\prime$ coordinate of $P^\prime$ is proportional to ${p^\prime}^{2k-3}$, and $a$ should be $2k-3$. Therefore, the EDS valuations for the codimension-two behavior are given by
\begin{equation}
    \label{eq:sooddvaluations}
    \begin{aligned}
        \valuation{}{\eds_{m}^{(2)}} &= \frac{m^2-1}{2} + \frac{1}{2}\left(\elltrouble{m}{2k-3}{4k-6}+\begin{cases}\scorr_{m/4}\left(p^\prime, p^\prime, \valuation{1}{p^\prime}, 0, s, w\right) & 4\mid m \\ 0 & 4\nmid m \end{cases}\right)\\
        &= \frac{n}{8}m^2 - \frac{1}{2}\begin{cases}1-s & 4\mid m\\ 1 & (2\mid m) \text{ and }(4\nmid m)\\\frac{n}{4} & (2\nmid m)\end{cases}\,.
    \end{aligned}
\end{equation}

Combining everything together, we expect that when the antisymmetric matter has charge or \pscharge{} $q=\pm\frac{m}{n}$, $\ordvanishtwo{\secz}$ should be given by
\begin{equation}
\valuation{}{\eds_{m}^{(2)}} - \valuation{}{\eds_{m}^{(1)}} = \frac{1}{8n}m^2 + \frac{\residue{mk}{n}(n-\residue{m k}{n})}{2 n} - \frac{1}{2}\begin{cases}1-s & 4\mid m\\ 1 & (2\mid m) \text{ and }(4\nmid m)\\\frac{n}{4} & (2\nmid m)\end{cases}\,.
\end{equation}
Demanding that this formula give the observed $\ordvanishtwo{\secz}$ of 1 for antisymmetric matter of $\asu(5)$ with $m=4$, we can set $s$ to 1, and if we plug in
\begin{equation}
    m = n q\,, \quad \quotparam = \residue{m (k + 1)}{n} = n - \residue{m k}{n}\,, \quad \quotparamtwo = \residue{m}{4} = \residue{n q}{4}\,,
\end{equation}
we find that
\begin{equation}
    \begin{aligned}
    \valuation{}{\eds_{m}^{(2)}} - \valuation{}{\eds_{m}^{(1)}} &= \frac{1}{2}\left(\frac{n}{4}q^2 + \frac{\quotparam(n-\quotparam)}{n} - \begin{cases}0 & \quotparamtwo=0\\ 1 & \quotparamtwo = 2\\\frac{n}{4} & \quotparamtwo = 1,3\end{cases}\right)\\
    &= \frac{1}{2}\left(\frac{d_{SU(n)}}{d_{SO(2n)}} q^2 + \quotfunction{\SU(n)}{\quotparam} - \quotfunction{\SO(2n)}{\quotparamtwo} \right)\,,
    \end{aligned}
\end{equation}
in exact agreement with the proposals in \cref{sec:proposals} and the observed order-of-vanishing data.

\paragraph{Antisymmetric representation of $\asu(n)$ for even $n$}

For the antisymmetric representation with even $n$ (i.e, $n=2k$), we obtained the order-of-vanishing data from two types of seed models admitting  $\bm{\frac{n(n-1)}{2}}_{2/n}$ matter: those with $\quotparam=1$ and those with $\quotparam=\frac{n}{2}+1$. In order to establish a connection to the EDS valuations, we need to analyze each seed model separately. Still, we can see that the formulas from~\cite{StangeEllTrouble} reproduce the observed order of vanishing data and the expressions expected from the proposals.

We begin with the  $\quotparam=1$ seed construction, which is described by the the Weierstrass model in \cref{eq:suoddseedmodelfund} with the substitutions in \cref{eq:sunfundevensubs}. The model's $\asu(2k)$ gauge factor is supported along the codimension-one locus $\locus{\sigma=0}$, which supports $\singtype{I}_{2k}^{(s)}$ singularities. The  $\bm{\frac{n(n-1)}{2}}_{2/n}$ matter is supported at the codimension-two locus $\locus{\sigma=b_{1,0}=0}$. As before, we need EDS valuations corresponding to both the codimension-one and the codimension-two behavior. The codimension-one EDS valuations, which were calculated above, are given by
\begin{equation}
\valuation{}{\eds^{(1)}_{m}} = \elltrouble{m}{1}{n} = \frac{1}{2}\left(\frac{n-1}{n}m^2 - \frac{\residue{m}{n}\left(n-\residue{m}{n}\right)}{n}\right)\,.
\end{equation}
We can calculate the codimension-two EDS valuations through a procedure similar to that used for antisymmetric matter with odd $n$. This analysis is performed in \cref{app:sonedsnu2}, so as not to clutter the discussion here. The end result is that
\begin{equation}
    \valuation{}{\eds^{(2)}_{m}} = \frac{1}{2}m^2 - \frac{1}{2}\begin{cases}0 & m\mid 2 \\ 1&  m\nmid 2\end{cases}\,.
\end{equation}
Combining these two expressions we find that
\begin{equation}
    \valuation{}{\eds_{m}^{(2)}} - \valuation{}{\eds_{m}^{(1)}} = \frac{1}{2}\left(\frac{1}{n}m^2 + \frac{\residue{m}{n}\left(n-\residue{m}{n}\right)}{n} - \begin{cases}0 & 2\mid m \\ 1&  2\nmid m\end{cases}\right)\,.
\end{equation}
If we use the relations
\begin{equation}
    q = \frac{2}{n}m\,, \quad \quotparam = \residue{m}{n}\,, \quad \quotparamtwo = \residue{\left(q-\frac{2\quotparam}{n}\right)}{2} + \residue{2\quotparam}{4} =2 \residue{m}{2}
\end{equation}
for this seed model, our result can be rewritten as
\begin{equation}
    \begin{aligned}
        \valuation{}{\eds_{m}^{(2)}} - \valuation{}{\eds_{m}^{(1)}} &= \frac{1}{2}\left(\frac{n}{4}q^2 + \frac{\quotparam\left(n-\quotparam\right)}{n} - \begin{cases}0 & \quotparamtwo= 0 \\ 1&  \quotparamtwo= 2\end{cases}\right)\\
        &= \frac{1}{2}\left(\frac{d_{\SU(n)}}{d_{\SO(2n)}}q^2 + \quotfunction{\SU(n)}{\quotparam} - \quotfunction{\SO(n)}{\quotparam}\right)\,.
    \end{aligned}
\end{equation}

The $\quotparam=\frac{n}{2}+1$ seed construction is described by the Weierstrass model in \cref{eq:sunevenantiseedmodelb}. The model supports $\singtype{I}^{(s)}_{2k}$ singularities along $\locus{\sigma=0}$ and a generating section with components given by \cref{eq:sunevenantiseedmodelbgensec}. The  $\bm{\frac{n(n-1)}{2}}_{2/n}$ matter is supported at the codimension-two locus $\locus{\sigma=b_{1}=0}$. To find the EDS valuations corresponding to the codimension-one behavior, we can construct an elliptic curve analogue by setting $\sigma$ to some prime $p$ and the other parameters to some arbitrary numbers. The resulting elliptic curve would have multiplicative reduction modulo $p$ with $\valuation{p}{\Delta} = 2k$, and the codimension-one EDS valuations should be given by
\begin{equation}
    \valuation{}{\eds^{(1)}_{m}} = \elltrouble{m}{k+1}{2k} = \frac{1}{2}\left(\frac{(k+1)(k-1)}{2k}m^2 - \frac{\residue{m(k+1)}{n}\left(n-\residue{m(k+1)}{n}\right)}{n}\right)\,.
\end{equation}
Calculating the codimension-two EDS valuations requires using the procedure for elliptic curves with additive reduction. The analysis is performed in \cref{app:sonedsnu1}, and the end result is
\begin{equation}
    \valuation{}{\eds^{(2)}_{m}} = \frac{1}{2}\left(\frac{n}{4}m^2 - \begin{cases}0 & 2\mid m \\ \frac{n}{4} &  2\nmid m\end{cases}\right)\,.
\end{equation}
Combining the results, we find that
\begin{equation}
    \valuation{}{\eds^{(2)}_{m}} - \valuation{}{\eds^{(1)}_{m}} =\frac{1}{2}\left( \frac{1}{n}m^2 +\frac{\residue{m(k+1)}{n}\left(n-\residue{m(k+1)}{n}\right)}{n} - \begin{cases}0 & 2\mid m \\ \frac{n}{4} &  2\nmid m\end{cases}\right)\,.
\end{equation}
If we use the relations
\begin{equation}
    q = \frac{2}{n}m\,, \quad \quotparam = \residue{(k+1)m}{n}\,, \quad \quotparamtwo = \residue{\left(q-\frac{2\quotparam}{n}\right)}{2} + \residue{2\quotparam}{4} =\left(1+2 \residue{(k+1)}{2}\right) \residue{m}{2}\,,
\end{equation}
for this seed model, our result can be rewritten as
\begin{equation}
    \begin{aligned}
    \valuation{}{\eds^{(2)}_{m}} - \valuation{}{\eds^{(1)}_{m}} &= \frac{n}{4}q^2 +\frac{\residue{\quotparam}{n}\left(n-\residue{\quotparam}{n}\right)}{n} - \begin{cases}0 & \quotparamtwo=0 \\ \frac{n}{4} &  \quotparamtwo = 1\text{ or }3\end{cases}\\
    &= \frac{1}{2}\left(\frac{d_{\SU(n)}}{d_{\SO(2n)}}q^2 + \quotfunction{\SU(n)}{\quotparam} - \quotfunction{\SO(n)}{\quotparam}\right).
    \end{aligned}
\end{equation}

To summarize, we see that, regardless of which seed model for $\asu(2k)\oplus\au(1)$ antisymmetric matter we use, the expressions for $\valuation{}{\eds^{(2)}_{m}} - \valuation{}{\eds^{(1)}_{m}}$ calculated using the formulas from~\cite{StangeEllTrouble} agree with the proposals in \cref{sec:proposals}. Moreover, these formulas reproduce the orders of vanishing observed by calculating multiples of the generating sections. This result gives us further confidence in the proposals.

\section{Representations of $\aso(2n)\oplus\au(1)$}
\label{sec:so}
We now turn to models with $\aso(2n)\oplus\au(1)$ gauge algebras, focusing on matter charged in the vector and spinor representations of $\aso(2n)$. Both these matter representations occur at codimension-two loci where a codimension-one locus with $\singtype{I}^{*s}_{n-4}$ singular fibers intersects the residual $\singtype{I}_1$ discriminant locus. The ideas and strategies used for the singlet and $\asu(n)\oplus\au(1)$ situations carry over here, but the analysis is more challenging due to the centers of the universal covering groups. $\Spin(4k)$, the universal covering group of $\aso(4k)$, has a $\Z_4$ center, but  $\Spin(4k+2)$, the universal covering group of $\aso(4k+2)$, has a $\Z_2\times\Z_2$ center. We therefore analyze the $\aso(4k)$ and $\aso(4k+2)$ representations separately. The $\Z_2\times\Z_2$ center poses the additional complication that it has more than one generating element, as seen above when investigating the antisymmetric representation of $\asu(n)$. We therefore need a larger number of seed models to generate a representative sample of \pscharge{}s.

\subsection{Spinor representations $\aso(4k+2)$}
We begin with matter charged in the spinor representations of $\aso(4k+2)$, the simplest of the $\aso(2n)\oplus\au(1)$ situations considered here. As discussed in \cref{sec:soncenters}, the two spinor representations of $\aso(4k+2)$ are conjugate to each other, so a codimension-two locus supporting $\aso(4k+2)$ spinor matter supports fields in both representations. Here, we restrict our attention to the spinor representations of $\aso(10)$ and $\aso(14)$, as attempts to realize spinor matter in F-theory for $\aso(2n)$ algebras beyond $\aso(14)$ produce constructions with codimension-two loci where $f$ and $g$ vanish to orders $(4,6)$. The spinors of $\aso(10)$ occur at codimension-two loci where the singularity type enhances from type $\singtype{I}^{*s}_{1}$ to type $\singtype{IV}^*$; in terms of the ADE groups, this is an enhancement from $\Spin(10)$ to $\gE_6$. The spinors of $\aso(14)$ meanwhile occur at codimension-two loci with an $\singtype{I}^{*s}_{3}\to \singtype{II}^*$ (or $\Spin(14)\rightarrow \gE_8$) enhancement.

Both $\Spin(10)$ and $\Spin(14)$ have $\Z_4$ centers, and $\quotparam$ can take a value from $0$ to $3$. For $\aso(10)$, the allowed $\au(1)$ charges for the spinor with highest weight $[0,0,0,0,1]$ are $\frac{4-\quotparam}{4}+\Z$. For $\aso(14)$, the allowed $\au(1)$ charges for the spinor with highest weight $[0,0,0,0,0,0,1]$ are $\frac{\quotparam}{4}+\Z$. Suppose we have a seed $\aso(10)\oplus\au(1)$ or $\aso(14)\oplus\au(1)$ model with a codimension-two locus supporting spinor matter with $\au(1)$ charge $\pm\frac{1}{4}$. This model would have either $\quotparam=1$ or $\quotparam=3$, and by the arguments in \cref{sec:signs}, we can freely choose our generating section such that we obtain either value of $\quotparam$. Then, we can find multiples $m\ratsec{s}$ of the generating section that would realize \pscharge{} $\pm \frac{m}{4}$ at the spinor locus with either $\quotparam = \residue{m}{4}$ (if the chosen generating section has $\quotparam=1$) or $\quotparam = \residue{3m}{4}$ (if the chosen generating section has $\quotparam=3$). We can therefore generate all the allowed combinations of \pscharge{} and $\quotparam$ from this seed model. Since the orders of vanishing of the section components are insensitive to the sign of the \pscharge{}, the orders of vanishing for \pscharge{} $-q$ are the same as those for \pscharge{} $+q$. Based on the discussion in \cref{sec:strategy}, we can use these data to establish formulas relating the orders of vanishing to the $\au(1)$ charge of the spinor matter.

\paragraph{$\aso(10)$} We consider the Weierstrass model given by
\begin{equation}
    \label{eq:so10spinorseed}
    \begin{aligned}
        y^2 &= x^3 -\frac{1}{3} \sigma ^2 \left(b_{0,0}^2 \left(b_{2,1}^2+3 \sigma  c_{0,3}\right)+\sigma b_{2,1} b_{0,0} c_{2,2}+\sigma  \left(\sigma  c_{2,2}^2-3 c_{1,2} c_{3,1}\right)\right)x z^4\\
        &\qquad + \frac{\sigma ^3}{108} \Big[\sigma  \left(27 c_{3,1}^2 \left(b_{2,1}^2+4 \sigma c_{0,3}\right)+8 \sigma ^2 c_{2,2}^3-36 \sigma  c_{1,2} c_{3,1} c_{2,2}\right) \\
        &\qquad\qquad\qquad - 4 b_{0,0}^3 \left(9 \sigma  b_{2,1} c_{0,3}+2 b_{2,1}^3\right) + 3 \sigma  b_{0,0}^2 \left(9 c_{1,2}^2-4 c_{2,2} \left(b_{2,1}^2+6 \sigma  c_{0,3}\right)\right) \\
        &\qquad\qquad\qquad + 6 \sigma b_{2,1} b_{0,0} \left(2 \sigma  c_{2,2}^2-3 c_{1,2} c_{3,1}\right)\Big]z^6\,.
    \end{aligned}
\end{equation}
This model is equivalent to the $\mathcal{Q}(3, 2, 2, 1, 0, 1, 1)$ model from~\cite{KuntzlerTateTrees}, which has an $\aso(10)\oplus\au(1)$ gauge algebra. The $\aso(10)$ algebra is tuned along $\locus{\sigma=0}$, while the Weierstrass model admits a generating section $\ratsec{s}$ with coordinates
\begin{equation}
\label{eq:so10spinorseedgensec}
\begin{aligned}
\secx &= \sigma ^2 c_{3,1}^2-\frac{1}{3} \sigma  b_{0,0}^2 \left(b_{0,0} b_{2,1}+2 \sigma c_{2,2}\right)\,, \\
\secy &= \frac{1}{2} \sigma ^2 \left(b_{0,0}^2 c_{3,1} \left(b_{0,0} b_{2,1}+2 \sigma c_{2,2}\right)-c_{1,2}b_{0,0}^4-2 \sigma  c_{3,1}^3\right)\,, \\
\secz &= b_{0,0}\,, \\
\secw &= \sigma ^3 \Big[b_{0,0}^4 \left(\sigma  c_{2,2}^2+c_{1,2} c_{3,1}\right)-4 \sigma b_{0,0}^2 c_{2,2} c_{3,1}^2-c_{0,3}b_{0,0}^6 \\
&\qquad\quad +b_{2,1} b_{0,0}^5 c_{2,2}-2 b_{2,1} b_{0,0}^3 c_{3,1}^2+3 \sigma  c_{3,1}^4\Big]\,.
\end{aligned}
\end{equation}
As can be verified by comparison with the listed matter spectrum in~\cite{KuntzlerTateTrees}, this model supports $\bm{16}_{- \frac{1}{4}}$ matter at $\locus{\sigma=b_{2,1}=0}$, $\bm{16}_{\frac{3}{4}}$ matter at $\locus{\sigma=b_{0,0}=0}$, and $\bm{10}_{\frac{1}{2}}$ matter at $\locus{\sigma=b_{0,0}c_{1,2}-b_{2,1}c_{3,1}=0}$. For this model, $\quotparam$ is $1$. (Alternatively, we can flip the signs of the $\bm{16}$ charges and take $\quotparam$ to be $3$; since the $\bm{10}$ representation is real, the sign of its charge is irrelevant as per the arguments in \cref{sec:signs}.)  After determining the spectrum of charged singlets and accounting for nonlocal adjoint matter uncharged under the $\au(1)$ algebra, one can verify that this matter spectrum satisfies the 6D anomaly cancellation conditions with $\height = -2\canonclass + 2\divclass{b_{0,0}} -\frac{5}{4}\divclass{\sigma}$.

We now calculate multiples of the generating section $\ratsec{s}$, examine their behavior at the $\bm{16}_{-\frac{1}{4}}$ locus $\locus{\sigma=b_{2,1}=0}$, and find the orders of vanishing corresponding to various \pscharge{}s. We can also find the orders of vanishing at the $\bm{16}_{\frac{3}{4}}$ locus, providing both an additional check on the data from the $\bm{16}_{- \frac{1}{4}}$ locus and the ability to probe larger \pscharge{}s without calculating larger multiples of $\ratsec{s}$. The resulting data are listed in \cref{tab:so10spinordata}. The observed codimension-one orders of vanishing satisfy the equations
\begin{equation}
\ordvanishone{\secz} = 0\,, \quad \left(\ordvanishone{\secx},\ordvanishone{\secy},\ordvanishone{\secw}\right) = \quottriplet{\Spin(10)}{\quotparam}\,,
\end{equation}
And if we let
\begin{equation}
\quotparamtwo = \residue{4q}{3}\,,
\end{equation}
where $q$ is the \pscharge{}, the codimension-two orders of vanishing satisfy the equations
\begin{equation}
\label{eq:ordtwozso10spinor}
\ordvanishtwo{\secz} = \frac{1}{2}\left(\frac{4}{3}q^2 + \quotfunction{\Spin(10)}{\quotparam} - \quotfunction{\gE_6}{\quotparamtwo}\right)\,.
\end{equation}
and
\begin{equation}
\left(\ordvanishtwo{\secx},\ordvanishtwo{\secy},\ordvanishtwo{\secw}\right) = (2,3,4)\times\ordvanishtwo{\secz} + \quottriplet{\gE_6}{\quotparamtwo}\,.
\end{equation}
Since  $\Spin(10)$ has a center with $d_{\Spin(10)} = 4$ elements and $\gE_6$ has a center with $d_{E_6}=3$ elements, the orders of vanishing satisfy the exact patterns proposed in \cref{sec:proposals} for a $\Spin(10)\to\gE_6$ enhancement locus.

\begin{table}
    \begin{subtable}{.5\linewidth}
        \centering

        \begin{tabular}{*{4}{>{$}c<{$}}}\toprule
            \abs{q}      & \quotparam      & \ordalt_{(1)} & \ordalt_{(2)} \\\midrule
            \frac{1}{4}  & 1 \text{ or } 3 & (1,2,0,3)     & (2,2,0,3) \\[0.2em]
            \frac{1}{2}  & 2               & (1,2,0,2)     & (2,2,0,3) \\[0.2em]
            \frac{3}{4}  & 1 \text{ or } 3 & (1,2,0,3)     & (2,3,1,4) \\[0.2em]
            1            & 0               & (0,0,0,0)     & (2,2,0,3) \\[0.2em]
            \frac{5}{4}  & 1 \text{ or } 3 & (1,2,0,3)     & (4,5,1,7) \\[0.2em]
            \frac{3}{2}  & 2               & (1,2,0,2)     & (4,6,2,8) \\[0.2em]
            \frac{7}{4}  & 1 \text{ or } 3 & (1,2,0,3)     & (6,8,2,11) \\[0.2em]
            2            & 0               & (0,0,0,0)     & (6,8,2,11) \\[0.2em]
            \frac{9}{4}  & 1 \text{ or } 3 & (1,2,0,3)     & (8,12,4,16) \\[0.2em]
            \frac{5}{2}  & 2               & (1,2,0,2)     & (10,14,4,19) \\[0.2em]
            \frac{11}{4} & 1 \text{ or } 3 & (1,2,0,3)     & (12,17,5,23) \\[0.2em]
            3            & 0               & (0,0,0,0)     & (12,18,6,24) \\[0.2em]
            \frac{13}{4} & 1 \text{ or } 3 & (1,2,0,3)     & (16,23,7,31) \\[0.2em]
            \frac{7}{2}  & 2               & (1,2,0,2)     & (18,26,8,35) \\[0.2em]
            \frac{15}{4} & 1 \text{ or } 3 & (1,2,0,3)     & (20,30,10,40) \\[0.2em]
            \frac{9}{2}  & 2               & (1,2,0,2)     & (28,42,14,56) \\[0.2em]
            \frac{21}{4} & 1 \text{ or } 3 & (1,2,0,3)     & (38,57,19,76) \\[0.2em]
            6            & 0               & (0,0,0,0)     & (48,72,24,96) \\ \bottomrule
        \end{tabular}

        \caption{$\bm{16}_{q}$ representation of $\aso(10)\oplus\au(1)$}
        \label{tab:so10spinordata}
    \end{subtable}
    \begin{subtable}{.5\linewidth}
        \centering

        \begin{tabular}{*{4}{>{$}c<{$}}}\toprule
            \abs{q}      & \quotparam      & \ordalt_{(1)} & \ordalt_{(2)} \\\midrule
            \frac{1}{4}  & 1 \text{ or } 3 & (1,3,0,4)     & (2,3,1,4) \\[0.2em]
            \frac{1}{2}  & 2               & (1,2,0,2)     & (2,3,1,4) \\[0.2em]
            \frac{3}{4}  & 1 \text{ or } 3 & (1,3,0,4)     & (4,6,2,8) \\[0.2em]
            1            & 0               & (0,0,0,0)     & (4,6,2,8) \\[0.2em]
            \frac{5}{4}  & 1 \text{ or } 3 & (1,3,0,4)     & (8,12,4,16) \\[0.2em]
            \frac{3}{2}  & 2               & (1,2,0,2)     & (10,15,5,20) \\[0.2em]
            \frac{7}{4}  & 1 \text{ or } 3 & (1,3,0,4)     & (14,21,7,28) \\[0.2em]
            2            & 0               & (0,0,0,0)     & (16,24,8,32) \\[0.2em]
            \frac{9}{4}  & 1 \text{ or } 3 & (1,3,0,4)     & (22,33,11,44) \\[0.2em]
            \frac{5}{2}  & 2               & (1,2,0,2)     & (26,39,13,52) \\[0.2em]
            \frac{11}{4} & 1 \text{ or } 3 & (1,3,0,4)     & (32,48,16,64) \\[0.2em]
            3            & 0               & (0,0,0,0)     & (36,54,18,72) \\[0.2em]
            \frac{13}{4} & 1 \text{ or } 3 & (1,3,0,4)     & (44,66,22,88) \\[0.2em]
            \frac{7}{2}  & 2               & (1,2,0,2)     & (50,75,25,100) \\[0.2em]
            \frac{15}{4} & 1 \text{ or } 3 & (1,3,0,4)     & (58,87,29,116) \\[0.2em]
            4            & 0               & (0,0,0,0)     & (64,96,32,128) \\[0.2em]
            \frac{17}{4} & 1 \text{ or } 3 & (1,3,0,4)     & (74,111,37,148) \\[0.2em]
            \frac{9}{2}  & 2               & (1,2,0,2)     & (82,123,41,164) \\[0.2em]
            \frac{19}{4} & 1 \text{ or } 3 & (1,3,0,4)     & (92,138,46,184) \\ \bottomrule
        \end{tabular}

        \caption{$\bm{64}_{q}\oplus\bm{14}_{2q}$ representation of $\aso(14)\oplus\au(1)$}
        \label{tab:so14spinordata}
    \end{subtable}

    \caption{Orders of vanishing for various combinations of $\quotparam$ and \pscharge{} $q$ and  for spinor matter in $\aso(4k+2)\oplus\au(1)$ models. The $\aso(10)\oplus\au(1)$ data are found using the model in \cref{eq:so10spinorseed}, while the $\aso(14)\oplus\au(1)$ data are found using the model in \cref{eq:so14spinorseed}.}
    \label{tab:oddsonspinordata}
\end{table}

\paragraph{$\aso(14)\oplus\au(1)$} Matter charged in the spinor ($\bm{64}$) representation of $\aso(14)$ occurs at loci where the singularity type enhances from $\singtype{I}_3^{*s}$ (or $\Spin(14)$) to $\singtype{II}^*$ (or $\gE_8$). These codimension-two loci also support matter in the vector $(\bm{14})$ representation. In a 6D model, for instance, an irreducible codimension-two $\singtype{I}_3^{*s}\to \singtype{II}^*$ locus supports a hypermultiplet of $\bm{64}$ matter and a hypermultiplet of $\bm{14}$ matter. We should therefore determine how the $\au(1)$ charge of the $\bm{64}$ matter  compares with that of the $\bm{14}$ matter at the same locus. The $\bm{14}$ representation has a weight $[0,0,0,0,0,-1,1]$, and the $\bm{64}$ representation has weights $[0,0,0,0,0,0,1]$ and $[0,0,0,0,0,-1,0]$.\footnote{We follow the notation of~\cite{Slansky} where the $\bm{64}$ representation has highest weight $[0,0,0,0,0,0,1]$. Other sources may use alternative conventions.} If the curves in the $\gE_8$ fiber supporting these two $\bm{64}$ weights are $\matterfibcurve_+$ and $\matterfibcurve_-$, then one can verify that $\matterfibcurve_+ + \matterfibcurve_-$ is one of the two curves in the fiber supporting the $\bm{14}$ weight.\footnote{Because the $\gE_8$ adjoint representation decomposes as $\bm{91}\oplus\bm{1}\oplus\bm{64}\oplus\overline{\bm{64}} \oplus\bm{14}\oplus\bm{14}$, there are two curves corresponding to the $[0,0,0,0,0,-1,1]$ weight of the $\bm{14}$ representation. We only need one of them for the purposes of this argument.} All the weights in the $\bm{64}$ representation should have the same $\au(1)$ charge $q$, implying that, for a section $\ratsec{s}$,
\begin{equation}
\shioda(\ratsec{s})\cdot \matterfibcurve_{+} = \shioda(\ratsec{s})\cdot\matterfibcurve_{-} = q\,.
\end{equation}
But the curve $\matterfibcurve_+ + \matterfibcurve_-$ should then satisfy
\begin{equation}
\shioda(\ratsec{s})\cdot \left(\matterfibcurve_{+}+\matterfibcurve_{-} \right)= 2q\,.
\end{equation}
The charge of the $[0,0,0,0,0,-1,1]$ weight should be the same as the charge of the other weights in the same $\bm{14}$ representation. Therefore, the charge of the $\bm{14}$ matter should be twice that of the $\bm{64}$ matter. In other words, the codimension-two locus should support matter fields in the $\bm{64}_{q}$ and $\bm{14}_{2q}$ representations, which are accompanied by fields in the $\overline{\bm{64}}_{-q}$ and $\bm{14}_{-2q}$ representations. In a 6D model, an irreducible $\singtype{I}_3^{*s}\to \singtype{II}^*$ locus would support a $\bm{64}_{q}$ hypermultiplet and and a $\bm{14}_{2q}$ hypermultiplet. When we say that an $\singtype{I}_3^{*s}\to \singtype{II}^*$ locus supports matter with $\au(1)$ charge $q$, we implicitly take $q$ to be the charge of the spinor matter; the charge of the vector matter can automatically be determined from this information.

We now consider the Weierstrass model
\begin{equation}
    \label{eq:so14spinorseed}
    \begin{aligned}
        y^2 &= x^3 -\frac{\sigma ^2 }{3} \Big[4 \beta _{2,1}^2 b_{0,0}^8+8 \sigma  \beta _{2,1} b_{0,0}^4 c_{2,2} \\
        &\qquad\qquad\qquad +3 \sigma b_{0,0}^2 \left(\sigma c_{0,4}-4 \beta _{2,1} c_{3,1}^2\right)+\sigma^2 \left(c_{2,2}^2-3 c_{1,3} c_{3,1}\right)\Big] x z^4 \\
        &\qquad + \frac{\sigma ^3}{108}  \Big[-64 \beta _{2,1}^3 b_{0,0}^{12}-24 \sigma ^2 \beta _{2,1} \Big(3 b_{0,0}^6 c_{0,4} \\
        &\qquad\qquad\qquad\qquad\qquad + b_{0,0}^4 \left(5 c_{2,2}^2-3 c_{1,3} c_{3,1}\right)-12 b_{0,0}^2 c_{2,2} c_{3,1}^2+9 c_{3,1}^4\Big) \\
        &\qquad\qquad\qquad +96 \sigma  \beta _{2,1}^2 b_{0,0}^6 \left(3 c_{3,1}^2-2 b_{0,0}^2 c_{2,2}\right)+\sigma ^3 \Big(9 b_{0,0}^2 \left(3 c_{1,3}^2-8 c_{0,4} c_{2,2}\right) \\
        &\qquad\qquad\qquad\qquad\qquad +4 \left(2 c_{2,2}^3-9 c_{1,3} c_{3,1} c_{2,2}+27 c_{0,4} c_{3,1}^2\right)\Big)\Big]z^6\,.
    \end{aligned}
\end{equation}
The $(f,g,\Delta)$ of this model vanish to orders $(2,3,9)$ at $\locus{\sigma=0}$, and the split condition for $\singtype{I}_{3}^*$ \cite{KatzEtAlTate,GrassiMorrison,MorrisonTaylorClusters} singularities is satisfied. The model therefore supports an $\aso(14)$ gauge algebra along $\locus{\sigma=0}$. It also admits a generating section $\ratsec{s}$ with coordinates
\begin{align}
\secx&=-\frac{1}{3} \sigma  \left(2 \beta _{2,1} b_{0,0}^6+2 \sigma  b_{0,0}^2 c_{2,2}-3 \sigma
   c_{3,1}^2\right)\,, \\
\secy &= -\frac{1}{2} \sigma ^3 \left(b_{0,0}^4 c_{1,3}-2 b_{0,0}^2 c_{2,2} c_{3,1}+2
   c_{3,1}^3\right)\,, \\
\secz &= b_{0,0}\,, \\
\secw &= \sigma ^4 \left(-b_{0,0}^6 c_{0,4}+b_{0,0}^4 \left(c_{2,2}^2+c_{1,3}
   c_{3,1}\right)-4 b_{0,0}^2 c_{2,2} c_{3,1}^2+3 c_{3,1}^4\right)\,,
\end{align}
indicating that the full gauge algebra is $\aso(14)\oplus\au(1)$. At $\locus{\sigma=b_{0,0}=0}$, the singularity type enhances from $\singtype{I}_3^{*s}$ to $\singtype{II}^*$ (or from $\Spin(14)$ to $\gE_8$). This codimension-two locus supports $\bm{64}_{\frac{1}{4}}\oplus\bm{14}_{\frac{1}{2}}$ matter, along with the appropriate conjugate matter.
At $\locus{\sigma=b_{0,0}^4 c_{1,3}-2 b_{0,0}^2 c_{2,2} c_{3,1}+2 c_{3,1}^3=0}$, the singularity type enhances to $\singtype{I}_{4}^*$ (or $\Spin(16)$), and the locus supports $\bm{14}_{\frac{1}{2}}$ matter. Based on these $\au(1)$ charges, $\quotparam$ is $1$. (Alternatively, we could say that the model has $\quotparam=3$ with $\bm{64}_{-\frac{1}{4}}\oplus\bm{14}_{\frac{1}{2}}$ matter at $\locus{\sigma=b_{0,0}=0}$. Because the $\bm{14}$ representation is real, the sign of its $\au(1)$ charge is not important as per the arguments in \cref{sec:signs}.) While $(f,g,\Delta)$ vanish to orders $(4,6,12)$ at $\locus{\sigma=\beta_{2,1}=0}$, the presence of such loci can be avoided by an appropriate choice of divisor classes. For instance, if we take $\divclass{\beta_{2,1}}$ to be trivial, there are no codimension-two $(4,6,12)$ loci. After calculating the spectrum of charged singlets and $\aso(14)$ adjoints uncharged under the $\au(1)$ algebra, one can show that this spectrum satisfies the 6D anomaly cancellation conditions, at least for situations without codimension-two $(4,6,12)$ loci.

As before, we can calculate multiples of $\ratsec{s}$ and find the orders of vanishing of the section components at $\locus{\sigma=b_{0,0}=0}$. These numbers are listed in \cref{tab:so14spinordata} along with the corresponding values of $\quotparam$ and of the spinor \pscharge{} $q$. The codimension-one orders of vanishing satisfy the equations
\begin{equation}
\ordvanishone{\secz} = 0\,, \quad \left(\ordvanishone{\secx},\ordvanishone{\secy},\ordvanishone{\secw}\right) = \quottriplet{\Spin(14)}{\quotparam}\,.
\end{equation}
The codimension-two orders of vanishing, meanwhile, satisfy the equations
\begin{gather}
\ordvanishtwo{\secz}= \frac{1}{2}\left(4q^2  + \quotfunction{\Spin(14)}{\quotparam} \right)\,, \label{eq:ordtwozspinorso14}\\
 \left(\ordvanishtwo{\secx},\ordvanishtwo{\secy},\ordvanishtwo{\secw}\right) = (2,3,4)\times\ordvanishtwo{\secz}\,.
\end{gather}
The center for $\gE_8$, the group associated with the codimension-two singularity type, is the trivial group. As such, the only allowed value of $\quotparamtwo$ is 0, and $\quotfunction{\gE_8}{\quotparamtwo}$ and $\quottriplet{\gE_8}{\quotparamtwo}$ are both trivial. Keeping this in mind, we see that the observed orders of vanishing for the $\aso(14)$ spinors agree with the proposals in \cref{sec:proposals}.

\subsection{Spinor representations of $\aso(4k)$}
We now turn to the spinor representations of $\aso(4k)$. We restrict our attention to the spinors of $\aso(8)$ and $\aso(12)$, as attempts to obtain spinors for larger values of $k$ produce models with codimension-two loci where $(f,g)$ vanish to orders $(4,6)$. Both of these algebras have two spinor representations. For $\aso(8)$, the two spinor representations are $\bm{8_\text{s}}$, with highest weight $[0,0,0,1]$, and $\bm{8}_c$, with highest weight $[0,0,1,0]$. Since these two representations are both real, they occur at different codimension-two loci in the F-theory model. For $\aso(12)$, the two spinor representations are $\bm{32}$, with highest weight $[0,0,0,0,0,1]$, and $\bm{32^\prime}$, with highest weight $[0,0,0,0,1,0]$.\footnote{While this convention agrees with~\cite{YamatsuGroupTheory}, the highest weights of the $\bm{32}$ and $\bm{32^\prime}$ representation are switched in the conventions of~\cite{Slansky}.} They are pseudoreal representations that occur at distinct codimension-two loci. While the $\bm{32}$ and $\bm{32^\prime}$ representations can in principle come  in half-hypermultiplets in 6D models, we are most interested in situations where matter in these representations also has a nonzero charge under an $\au(1)$ algebra. The total representation is no longer pseudoreal after including this charge, so $\aso(12)$ spinor matter with a nonzero $\au(1)$ charge must occur in full hypermultiplets. Because both the $\aso(8)$ and $\aso(12)$ spinor representations are self-conjugate, the sign of the $\au(1)$ charge is unimportant for these cases.

The universal covering groups $\Spin(8)$ and $\Spin(12)$ both have $\Z_2\times\Z_2$ centers. As discussed in \cref{sec:centers}, we label the four elements of $\Z_2\times\Z_2$ with an integer $\quotparam$ ranging from 0 to 3. The allowed $\au(1)$ charges are integral for all the spinor representations when $\quotparam=0$ and are half-integral for all the spinor representations when $\quotparam=2$. When $\quotparam=1$, the allowed charges for $\bm{8_\text{s}}$ and  $\bm{32^\prime}$ representations are half-integral, while those for the  $\bm{8_\text{c}}$ and  $\bm{32}$ representations are integral. When $\quotparam=3$, the roles are reversed: allowed charges for $\bm{8_\text{c}}$ and  $\bm{32}$ representations are half-integral, while those for the  $\bm{8_\text{s}}$ and  $\bm{32}^\prime$ representations are integral. Note that the $\aso(2n)$ fibers admit an automorphism that exchanges the two nodes at the end of the Dynkin diagram; these are the two nodes on the right hand side in \cref{fig:so2ndynkindiagram}. This automorphism exchanges the two spinor representations for a given algebra. In turn, the automorphism sends $\quotparam=1$ to $\quotparam=3$, and vice versa. The elements $\quotparam=2$ and $\quotparam=0$ are sent to themselves under the automorphism. In F-theory models, there is an ambiguity in the labeling of the curves exchanged under the automorphism, so we can perform the automorphism without changing the underlying geometry. As a result, we can often determine geometric information about one spinor representation from information about the other spinor representation. For instance, if we knew the orders of vanishing for $\bm{32}$ matter with some $\au(1)$ charge when $\quotparam=1$, we automatically know the orders of vanishing for $\bm{32}^\prime$ matter with the same charge when $\quotparam=3$.

With this in mind, we can now describe how to generate various combinations of the \pscharge{} and $\quotparam$. Starting with a $\quotparam=1$ model with ${\bm{8_\text{s}}}_{,\frac{1}{2}}$ or $\bm{32^\prime}_{\frac{1}{2}}$ matter, we can obtain all the half-integral $\quotparam=1$ \pscharge{}s and all the integral $\quotparam=0$ \pscharge{}s. If we start with a $\quotparam=2$ model with ${\bm{8_\text{s}}}_{,\frac{1}{2}}$ or $\bm{32^\prime}_{\frac{1}{2}}$ matter, we can obtain all the half-integral $\quotparam=2$ \pscharge{}s while reproducing all the integral $\quotparam=0$ \pscharge{}s. Obtaining the integral $\quotparam=3$ \pscharge{}s is more difficult. If we have a $\quotparam=3$ model with ${\bm{8_\text{s}}}_{,1}$ or $\bm{32^\prime}_{1}$ matter, we can generate the odd $\quotparam=3$ \pscharge{}s but not the even ones: the even multiples of the generating section, which would realize the even \pscharge{}s, would have $\quotparam=0$ according to the $\Z_2\times\Z_2$ group structure. We could generate some of the even $\quotparam=3$ \pscharge{}s from a model with ${\bm{8_\text{s}}}_{,2}$ or $\bm{32^\prime}_{2}$ matter, but we would miss those \pscharge{}s that are multiples of 4. Even if we attempt to fill in these gaps by starting with charges that are larger and larger powers of $2$, we always miss some of the even $\quotparam=3$ \pscharge{}s. We are therefore unable to practically obtain order-of-vanishing data for all of the even $\quotparam=3$ \pscharge{}s in a particular range. Nevertheless, we find that the order-of-vanishing data we are able to obtain follow the expected formulas in \cref{sec:proposals}, making it highly plausible that the missing \pscharge{}s follow these same patterns. Of course, once we have data for the $\bm{8_\text{s}}$ and $\bm{32^\prime}$ representations, we can use the automorphism to automatically obtain data for the $\bm{8_\text{c}}$ and $\bm{32}$ representations. This allows us to cut down on the number of seed models used. In fact, the triality of $\aso(8)$ models, discussed in more detail below, implies that we can read off orders of vanishing for the vector representation from the spinor data, and vice versa. As a result, we can obtain a large portion of the $\aso(8)$ data from just a single model, and the discussion of $\aso(8)$ spinors will also cover the $\bm{8_\text{v}}$ representation.

\paragraph{$\aso(8)\oplus\au(1)$} The two spinor representations of $\aso(8)$ and the vector representation $\bm{8_\text{v}}$ all occur at codimension-two loci where the singularity type enhances from $\singtype{I}_0^{*s}$ ($\Spin(8)$) to $\singtype{I}_1^*$ ($\Spin(10)$), a fact that reflects the triality of $\aso(8)$. As can be seen from the symmetries of its Dynkin diagram, the $\aso(8)$ algebra admits automorphisms that exchange the $\bm{8_\text{s}}$, $\bm{8_\text{c}}$, and the $\bm{8_\text{v}}$ representations. These automorphisms also exchange the $\quotparam=1$, $\quotparam=2$, and $\quotparam=3$ elements of the center in a way consistent with the allowed $\au(1)$ charges in \cref{tab:so2ncharges}. For example, ${\bm{8_\text{c}}}_{,1}$ matter (with $\au(1)$ charge $q=1$) in a $\quotparam=1$ model can alternatively be viewed as ${\bm{8_\text{v}}}_{,1}$ matter in a $\quotparam=2$ model or as ${\bm{8_\text{s}}}_{,1}$ matter in a $\quotparam=3$ model.

Let us consider the Weierstrass model
\begin{equation}
    \label{eq:so8model}
    \begin{aligned}
        y^2 &= x^3 + \frac{1}{3} \sigma ^2 \left(-3 b_{0,0}^2 b_{2,1}^2+3 \sigma  c_{1,2} c_{3,1}-c_{2,1}^2\right)x z^4\\
        &\qquad\quad +\frac{1}{108} \sigma ^3 \Big(27 \sigma  b_{0,0}^2 c_{1,2}^2+108 \sigma  b_{2,1}^2 c_{3,1}^2-72 b_{0,0}^2 b_{2,1}^2 c_{2,1}\\
        &\qquad\qquad\qquad\qquad\quad -36 \sigma  c_{1,2} c_{3,1} c_{2,1}+8 c_{2,1}^3\Big)z^6\,.
    \end{aligned}
\end{equation}
The elliptic fibration has $\singtype{I}_0^{*s}$ singularities along $\locus{\sigma=0}$, signaling the presence of an $\aso(8)$ algebra. It also admits a generating section $\ratsec{s}$ with coordinates
\begin{equation}
[\secx:\secy:\secz] = \left[\frac{1}{3} \sigma  \left(3 \sigma  c_{3,1}^2-2 b_{0,0}^2 c_{2,1}\right): -\frac{1}{2} \sigma ^2 \left(b_{0,0}^4 c_{1,2}-2 b_{0,0}^2 c_{2,1} c_{3,1}+2 \sigma c_{3,1}^3\right):b_{0,0}\right]
\end{equation}
and
\begin{equation}
\secw = \sigma ^2 \left(b_{0,0}^4 \left(\sigma  c_{1,2} c_{3,1}+c_{2,1}^2\right)-4 \sigma
   b_{0,0}^2 c_{2,1} c_{3,1}^2-b_{2,1}^2 b_{0,0}^6+3 \sigma ^2 c_{3,1}^4\right)\,.
\end{equation}
Thus the total gauge algebra is $\aso(8)\oplus\au(1)$. There are four codimension-two loci where the singularity type enhances to $\singtype{I}_1^*$: $\locus{\sigma=b_{0,0}=0}$, $\locus{\sigma=b_{2,1}=0}$, $\locus{\sigma=c_{2,1}-b_{0,0}b_{2,1}=0}$, and $\locus{\sigma=c_{2,1}+b_{0,0}b_{2,1}=0}$. The locus $\locus{\sigma=b_{0,0}=0}$ supports matter with $\au(1)$ charge $q=1$, the locus $\locus{\sigma=b_{2,1}=0}$ supports matter with $\au(1)$ charge $q=0$, and the remaining two loci support matter with $\au(1)$ charge $q=\frac{1}{2}$. Because of the triality of $\aso(8)$, there are multiple ways of describing the $\aso(8)$ representations supported at these loci, which are summarized in \cref{tab:repsforso8model}. Regardless of which description of the matter representations is used, the spectrum satisfies the 6D anomaly cancellation conditions with $\height = -2K_B + 2\divclass{b_{0,0}} - \divclass{\sigma}$.

\begin{table}
    \centering

    \begin{tabular}{*{7}{>{$}c<{$}}}\toprule
        \multirow{2}{*}{\text{Locus}}           & \multicolumn{6}{c}{$\quotparam$} \\ \cline{2-7}
                                                & 1                                & 1                                & 2                                & 2                                & 3                                & 3 \\ \midrule
        \locus{\sigma=b_{0,0}=0}                & {\bm{8_\text{c}}}_{,1}           & {\bm{8_\text{c}}}_{,1}           & {\bm{8_\text{v}}}_{,1}           & {\bm{8_\text{v}}}_{,1}           & {\bm{8_\text{s}}}_{,1}           & {\bm{8_\text{s}}}_{,1} \\
        \locus{\sigma=b_{2,1}=0}                & {\bm{8_\text{c}}}_{,0}           & {\bm{8_\text{c}}}_{,0}           & {\bm{8_\text{v}}}_{,0}           & {\bm{8_\text{v}}}_{,0}           & {\bm{8_\text{s}}}_{,0}           & {\bm{8_\text{s}}}_{,0} \\
        \locus{\sigma=c_{2,1}-b_{0,0}b_{2,1}=0} & {\bm{8_\text{v}}}_{,\frac{1}{2}} & {\bm{8_\text{s}}}_{,\frac{1}{2}} & {\bm{8_\text{s}}}_{,\frac{1}{2}} & {\bm{8_\text{c}}}_{,\frac{1}{2}} & {\bm{8_\text{c}}}_{,\frac{1}{2}} & {\bm{8_\text{v}}}_{,\frac{1}{2}} \\
        \locus{\sigma=c_{2,1}+b_{0,0}b_{2,1}=0} & {\bm{8_\text{s}}}_{,\frac{1}{2}} & {\bm{8_\text{v}}}_{,\frac{1}{2}} & {\bm{8_\text{c}}}_{,\frac{1}{2}} & {\bm{8_\text{s}}}_{,\frac{1}{2}} & {\bm{8_\text{v}}}_{,\frac{1}{2}} & {\bm{8_\text{c}}}_{,\frac{1}{2}} \\ \bottomrule
    \end{tabular}

    \caption{Matter representations supported at the four codimension-two $\singtype{I}_0^{*s}\rightarrow \singtype{I}_1^*$ loci in the elliptic fibration of \cref{eq:so8model}. Because of the triality of $\aso(8)$, there are multiple ways of describing the supported matter representations. Each column corresponds to one choice of matter representations and $\quotparam$.}
    \label{tab:repsforso8model}
\end{table}

To generate the order-of-vanishing data, we calculate multiples of $\ratsec{s}$ and find the orders of vanishing of the section components at the various loci. The results are listed in \cref{tab:so8data}. Note that these results do not include orders of vanishing for some combinations of $\quotparam$ and \pscharge{}. For instance, we have only given orders of vanishing for the odd integer \pscharge{}s when $\quotparam=1,2,3$ in light of the difficulties in generating the even integer \pscharge{}s described above. We have also included only the even integer \pscharge{}s for $\quotparam=0$ when $\abs{q}>8$; this simply reflects the data we chose to generate, and there is no fundamental difficulty in obtaining the odd \pscharge{}s for $\quotparam=0$. Despite these missing \pscharge{} values, one can see that the observed data follow the patterns proposed in \cref{sec:proposals}. The codimension-one orders of vanishing follow the pattern
\begin{equation}
\ordvanishone{\secz} = 0\,, \quad \left(\ordvanishone{\secx},\ordvanishone{\secy},\ordvanishone{\secw}\right) = \quottriplet{\Spin(8)}{\quotparam}\,,
\end{equation}
as expected, and if we let
\begin{equation}
\label{eq:so8mu}
\quotparamtwo = \begin{cases}\residue{(2q-2\quotparam)}{4} & \text{for }\bm{8_\text{s}}\text{ or }\bm{8_\text{c}}\\ \residue{(2q-\quotparam)}{4} +\residue{2q}{2}& \text{for }\bm{8_\text{v}}\end{cases}\,,
\end{equation}
the codimension-two orders of vanishing satisfy the expressions
\begin{equation}
\ordvanishtwo{\secz} = \frac{1}{2}\left(q^2 + \quotfunction{\Spin(8)}{\quotparam} - \quotfunction{\Spin(10)}{\quotparamtwo}\right) \label{eq:so8spinorord2z}
\end{equation}
and
\begin{equation}
\left(\ordvanishtwo{\secx},\ordvanishtwo{\secy},\ordvanishtwo{\secw}\right) = \left(2,3,4\right)\times\ordvanishtwo{\secz} + \quottriplet{\Spin(10)}{\quotparamtwo}\,.
\end{equation}
Since $d_{\Spin(8)}=d_{\Spin(10)}=4$, the codimension-two formulas agree with the expectations from \cref{sec:proposals}. Although we did not find orders of vanishing for the even \pscharge{}s when $\quotparam\neq 0$, \cref{eq:so8spinorord2z} is guaranteed to give us integer $\ordvanishtwo{\secz}$ for these cases. The expressions in \cref{eq:so8mu} give $\quotparamtwo=2$ whenever $q$ is even and $\quotparam$ takes the appropriate nonzero value, and plugging this information into  \cref{eq:so8spinorord2z} always results in an integer. These facts suggest that the expressions in \cref{sec:proposals} correctly relate the $\au(1)$ charge of $\aso(8)$ spinor or vector matter to the orders of vanishing of the section components.

\begin{table}
    \begin{subtable}{0.5\textwidth}
        \centering

        \begin{tabular}{*{3}{>{$}c<{$}}}\toprule
            \abs{q} & \ordalt_{(1)} & \ordalt_{(2)} \\ \midrule
            1       & (1,2,0,2)     & (2,3,1,4) \\[0.2em]
            3       & (1,2,0,2)     & (10,15,5,20) \\[0.2em]
            5       & (1,2,0,2)     & (26,39,13,52) \\[0.2em]
            7       & (1,2,0,2)     & (50,75,25,100) \\[0.2em]
            9       & (1,2,0,2)     & (82,123,41,164) \\[0.2em]
            11      & (1,2,0,2)     & (122,183,61,244) \\[0.2em]
            13      & (1,2,0,2)     & (170,255,85,340) \\[0.2em]
            15      & (1,2,0,2)     & (226,339,113,452) \\[0.2em]
            17      & (1,2,0,2)     & (290,435,145,580) \\ \bottomrule
        \end{tabular}

        \caption{$\quotparam=1$ for $\bm{8_\text{c}}$, $\quotparam=2$ for $\bm{8_\text{v}}$, $\quotparam=3$ for $\bm{8_\text{s}}$}
        \label{tab:so8datanu1int}
    \end{subtable}
    \begin{subtable}{0.5\textwidth}
        \centering

        \begin{tabular}{*{3}{>{$}c<{$}}}\toprule
            \abs{q}      & \ordalt_{(1)} & \ordalt_{(2)} \\ \midrule
            \frac{1}{2}  & (1,2,0,2)     & (1,2,0,3) \\[0.2em]
            \frac{3}{2}  & (1,2,0,2)     & (3,5,1,7) \\[0.2em]
            \frac{5}{2}  & (1,2,0,2)     & (7,11,3,15) \\[0.2em]
            \frac{7}{2}  & (1,2,0,2)     & (13,20,6,27) \\[0.2em]
            \frac{9}{2}  & (1,2,0,2)     & (21,32,10,43) \\[0.2em]
            \frac{11}{2} & (1,2,0,2)     & (31,47,15,63) \\[0.2em]
            \frac{13}{2} & (1,2,0,2)     & (43,65,21,87) \\[0.2em]
            \frac{15}{2} & (1,2,0,2)     & (57,86,28,115) \\[0.2em]
            \frac{17}{2} & (1,2,0,2)     & (73,110,36,147) \\ \bottomrule
        \end{tabular}

        \caption{$\quotparam=2,3$ for $\bm{8_\text{c}}$, $\quotparam=1,3$ for $\bm{8_\text{v}}$, $\quotparam=1,2$ for $\bm{8_\text{s}}$}
        \label{tab:so8datanu1frac}
    \end{subtable}

    \bigskip
    \centering

    \begin{subtable}{0.5\textwidth}
        \centering

        \begin{tabular}{*{3}{>{$}c<{$}}}\toprule
            \abs{q} & \ordalt_{(1)} & \ordalt_{(2)} \\ \midrule
            1       & (0,0,0,0)     & (1,2,0,2) \\[0.2em]
            2       & (0,0,0,0)     & (4,6,2,8) \\[0.2em]
            3       & (0,0,0,0)     & (9,14,4,18) \\[0.2em]
            4       & (0,0,0,0)     & (16,24,8,32) \\[0.2em]
            5       & (0,0,0,0)     & (25,38,12,50) \\[0.2em]
            6       & (0,0,0,0)     & (36,54,18,72) \\[0.2em]
            7       & (0,0,0,0)     & (49,74,24,98) \\[0.2em]
            8       & (0,0,0,0)     & (64,96,32,128) \\[0.2em]
            10      & (0,0,0,0)     & (100,150,50,200) \\[0.2em]
            12      & (0,0,0,0)     & (144,216,72,288) \\[0.2em]
            14      & (0,0,0,0)     & (196,294,98,392) \\[0.2em]
            16      & (0,0,0,0)     & (256,384,128,512) \\ \bottomrule
        \end{tabular}

        \caption{$\quotparam=0$ for $\bm{8_\text{s}}$, $\bm{8_\text{c}}$, and $\bm{8_\text{v}}$}
        \label{tab:so8datanu0}
    \end{subtable}

    \caption{Orders of vanishing of the section components for various combinations \pscharge{}s $q$ and $\quotparam$ for $\bm{8_\text{s}}$, $\bm{8_\text{c}}$, and $\bm{8_\text{v}}$ matter in $\aso(8)\oplus\au(1)$ models. The data are obtained using the $\aso(8)\oplus\au(1)$ model in \cref{eq:so8model}.}
    \label{tab:so8data}
\end{table}

\paragraph{$\aso(12)\oplus\au(1)$} The two spinor representations of $\aso(12)$ both occur at codimension-two loci where the singularity type enhances from $\singtype{I}_2^{*s}$ ($\Spin(12)$) to $\singtype{III}^*$ ($\gE_7$). After resolution, the codimension-two singularities can either be incompletely or completely resolved~\cite{MorrisonTaylorMaS, KanHalfHyper}. In 6D models, an incomplete resolution locus supports a half-hypermultiplet of spinor matter. A complete resolution locus supports a full hypermultiplet of spinor matter and an additional singlet, as can be seen from the decomposition of the $\gE_7$ adjoint representation:
\begin{equation}
\begin{aligned}
\gE_7 &\to \SO(12) \times \SU(2)\,, \\
\bm{133} &\to (\bm{66},\bm{1}) \oplus (\bm{32},\bm{2}) \oplus (\bm{1},\bm{3})\,.
\end{aligned}
\end{equation}
The $(\bm{32},\bm{2})$ degrees of freedom combine to form a $\bm{32}$ hypermultiplet, while the positive and negative roots of the $(\bm{1},\bm{3})$ degrees of freedom form the extra singlet hypermultiplet. By the automorphism of $\aso(12)$, there is a similar decomposition for the $\bm{32}^\prime$, and this spinor is also accompanied by a singlet at complete resolution loci.  Since spinor matter with $\au(1)$ charges must come in full hypermultiplets, we are primarily interested in the complete resolution loci. However, the spinor and singlet matter at these loci can both be charged under the $\au(1)$ algebra, raising the question of how the singlet charge compares to the spinor charge. The charge of the singlet matter is in fact twice that of the spinor matter.\footnote{Recall that, because the singlet representation is real, the sign of the singlet matter is unimportant.} Because the spinor matter occurs in full hypermultiplets, the resolved $\gE_7$ fiber contains two curves, $\gamma_+$ and $\gamma_-$, representing the two highest weights of the spinor representation in $(\bm{32},\bm{2})$ (or $(\bm{32^\prime},\bm{2})$). These two curves would belong to the two different half-hypermultiplets that combine to form the full hypermultiplet of spinor matter. The half-hypermultiplets should have opposite $\au(1)$ charges, implying that, for a generating section $\ratsec{s}$,
\begin{equation}
\sigma(\ratsec{s})\cdot\gamma_+ = -  \sigma(\ratsec{s})\cdot\gamma_{-} = q\,.
\end{equation}
But $\gamma_{+}$ and $\gamma_{-}$ also serve as the two weights of the fundamental of $\asu(2)$. The difference $\gamma_{+}-\gamma_{-}$ between these two curves is the root in the $\asu(2)$ adjoint, and it corresponds to the curve supporting the singlet. The charge of the singlet matter should be given by
\begin{equation}
\sigma(\ratsec{s})\cdot (\gamma_{+}-\gamma_{-}) = 2q\,.
\end{equation}
The $\au(1)$ charge of the singlet matter is therefore twice that of the spinor matter, as claimed above. While we gave this argument for the $\bm{32}$ spinor, the same argument carries through for the $\bm{32^\prime}$ spinor without significant modifications.

The $\aso(12)$ algebra does not admit the triality of the $\aso(8)$ algebra. We would therefore need multiple Weierstrass models to obtain a representative set of $q$ and $\quotparam$ combinations, but we have not been able to obtain models that support $\bm{32^\prime}_{1}$ matter with $\quotparam=3$ or $\bm{32^\prime}_{\frac{1}{2}}$ matter with $\quotparam=2$ (or the $\bm{32}$ matter related to this $\bm{32^\prime}$ matter by the automorphism). Whenever we attempt to construct such models, $(f, g, \Delta)$ end up vanishing to orders $(4,6,12)$ at the would-be spinor locus. However, there is the $\aso(12)\oplus\au(1)$ Weierstrass model
\begin{equation}
    \label{eq:so12spinormodel}
    \begin{aligned}
        y^2 &= x^3 -\frac{1}{3} \sigma ^2 \left[3 \sigma  b_{0,0}^2 c_{0,3}+\sigma  \left(\sigma c_{2,2}^2-3 c_{3,1} \left(b_{2,1} c_{3,1}+\sigma  c_{1,3}\right)\right)+b_{2,1}^2 b_{0,0}^4\right] x z^4 \\
        &\qquad\quad + \frac{\sigma ^3}{108}\Big[9 \sigma  b_{0,0}^2 \left(4 \sigma  b_{2,1}\left(c_{2,2}^2+c_{1,3} c_{3,1}\right)+4 b_{2,1}^2 c_{3,1}^2+3 \sigma ^2 c_{1,3}^2-8\sigma  c_{0,3} c_{2,2}\right)\\
        &\qquad\qquad\qquad\quad + 4 \sigma ^2 \left(-9 c_{3,1} c_{2,2} \left(2 b_{2,1}c_{3,1}+\sigma  c_{1,3}\right)+2 \sigma  c_{2,2}^3+27 c_{0,3} c_{3,1}^2\right)\\
        &\qquad\qquad\qquad\quad - 36\sigma  b_{2,1} b_{0,0}^4 c_{0,3}-8 b_{2,1}^3 b_{0,0}^6\Big]z^6\,,
    \end{aligned}
\end{equation}
which has $\singtype{I}_{2}^{*s}$ fibers along $\locus{\sigma=0}$ and a generating section $\ratsec{s}$ with components
\begin{equation}
\begin{aligned}
\secx &= -\frac{1}{3} \sigma  \left(2 \sigma  b_{0,0}^2 c_{2,2}+b_{2,1} b_{0,0}^4-3 \sigma c_{3,1}^2\right)\,, \\
\secy &= -\frac{1}{2} \sigma ^3 \left(b_{0,0}^4 c_{1,3}-2 b_{0,0}^2 c_{2,2} c_{3,1}+2c_{3,1}^3\right)\,, \\
\secz &=  b_{0,0}\,, \\
\secw &= \frac{\sigma ^3}{3} \Big[3 b_{0,0}^4 \left(c_{3,1} \left(\sigma  c_{1,3}-b_{2,1}c_{3,1}\right)+\sigma  c_{2,2}^2\right)\\
 &\qquad\qquad -12 \sigma  b_{0,0}^2 c_{2,2}c_{3,1}^2+b_{0,0}^6 \left(4 b_{2,1} c_{2,2}-3 c_{0,3}\right)+9 \sigma c_{3,1}^4\Big]\,.
\end{aligned}
\end{equation}
The model, which has $\quotparam=1$, supports $\bm{32^\prime}_{\frac{1}{2}} \oplus \bm{1}_{1}$ matter at $\locus{\sigma=b_{0,0}=0}$, $\bm{32}_{0}$ matter at $\locus{\sigma=b_{2,1}=0}$, and $\bm{12}_{\frac{1}{2}}$ matter at $\locus{\sigma=3 b_{0,0}^2 c_{0,3}-4 b_{2,1} b_{0,0}^2 c_{2,2}+3 b_{2,1} c_{3,1}^2=0}$. In 6D models, $\locus{\sigma=b_{0,0}=0}$ supports full hypermultiplets of $\bm{32^\prime}_{\frac{1}{2}}$ matter, while $\locus{\sigma=b_{2,1}=0}$ supports half-hypermultiplets of $\bm{32}_{0}$ matter. This same construction can alternatively be viewed as a $\quotparam=3$ model with $\bm{32}_{\frac{1}{2}} \oplus \bm{1}_{1}$ matter at $\locus{\sigma=b_{0,0}=0}$, $\bm{32^\prime}_{0}$ matter at $\locus{\sigma=b_{2,1}=0}$, and $\bm{12}_{\frac{1}{2}}$ matter at $\locus{\sigma=3 b_{0,0}^2 c_{0,3}-4 b_{2,1} b_{0,0}^2 c_{2,2}+3 b_{2,1} c_{3,1}^2=0}$. For either description, the massless charged spectrum, after accounting for additional singlets and $\aso(12)$ adjoints, satisfies the 6D anomaly cancellation conditions with $\tilde{b} = -2\canonclass + 2 \divclass{b_{0,0}} - \frac{3}{2}\divclass{\sigma}$.

We can therefore find the orders of vanishing for at least some of the spinor \pscharge{}s by calculating the components of multiples of $\ratsec{s}$ and finding their orders of vanishing at $\locus{\sigma=b_{0,0}=0}$. The resulting data are given in \cref{tab:so12spinordata}. The codimension-one orders of vanishing satisfy the expression
\begin{equation}
\ordvanishone{\secz} = 0\,, \quad \left(\ordvanishone{\secx},\ordvanishone{\secy},\ordvanishone{\secw}\right) = \quottriplet{\Spin(12)}{\quotparam}\,.
\end{equation}
And if we let\footnote{For all of the orders of vanishing in \cref{tab:so12spinordata}, $\quotparamtwo$ would be 0. While it might seem arbitrary to use this more complicated expression for $\quotparamtwo$, it can be motivated by the requirement that \cref{eq:so12spinorord2z} give integer $\ordvanishtwo{\secz}$ for the $(q,\quotparam)$ combinations that were not generated.}
\begin{equation}
\quotparamtwo = \residue{(2q+\quotparam)}{2}\,,
\end{equation}
the codimension-two orders of vanishing satisfy
\begin{equation}
\ordvanishtwo{\secz} = \frac{1}{2}\left(2q^2 + \quotfunction{\Spin(12)}{\quotparam} -  \quotfunction{\gE_7}{\quotparamtwo}\right) \label{eq:so12spinorord2z}
\end{equation}
and
\begin{equation}
 \left(\ordvanishtwo{\secx},\ordvanishtwo{\secy},\ordvanishtwo{\secw}\right) = \left(2,3,4\right)\times\ordvanishtwo{\secz} + \quottriplet{\gE_7}{\quotparamtwo}\,.
\end{equation}
Recalling that $d_{\Spin(12)}=4$ and $d_{\gE_7}=2$, we see that these expressions agree exactly with those expected from \cref{sec:proposals}. Note that we can rewrite the relationship between $q$, $\quotparam$, and $\quotparamtwo$ as
\begin{equation}
\quotparamtwo = \begin{cases}1 & \quotparam=2,3 \\ 0 & \quotparam=0,1\end{cases} \; \text{ for } \bm{32}_{q}\oplus\bm{1}_{2q}\,, \quad
\quotparamtwo = \begin{cases}1 & \quotparam=1,2 \\ 0 & \quotparam=0,3\end{cases} \; \text{ for } \bm{32^\prime}_{q}\oplus\bm{1}_{2q}\,.
\end{equation}

\begin{table}
    \centering

    \begin{subtable}{0.45\textwidth}
        \centering

        \begin{tabular}{*{3}{>{$}c<{$}}}\toprule
            \abs{q} & \ordalt_{(1)} & \ordalt_{(2)} \\ \midrule
            1       & (0,0,0,0)     & (2,3,1,4) \\
            2       & (0,0,0,0)     & (8,12,4,16) \\
            3       & (0,0,0,0)     & (18,27,9,36) \\
            4       & (0,0,0,0)     & (32,48,16,64) \\
            5       & (0,0,0,0)     & (50,75,25,100) \\
            6       & (0,0,0,0)     & (72,108,36,144) \\
            7       & (0,0,0,0)     & (98,147,49,196) \\
            8       & (0,0,0,0)     & (128,192,64,256) \\ \bottomrule
        \end{tabular}

        \caption{$\quotparam=0$ for both $\bm{32}_{q}\oplus\bm{1}_{2q}$ and  $\bm{32^\prime}_{q}\oplus\bm{1}_{2q}$ matter}
    \end{subtable}
    \begin{subtable}{0.45\textwidth}
        \centering

        \begin{tabular}{*{3}{>{$}c<{$}}}\toprule
            \abs{q}      & \ordalt_{(1)} & \ordalt_{(2)} \\ \midrule
            \frac{1}{2}  & (1,3,0,3)     & (2,3,1,4) \\[0.2em]
            \frac{3}{2}  & (1,3,0,3)     & (6,9,3,12) \\[0.2em]
            \frac{5}{2}  & (1,3,0,3)     & (14,21,7,28) \\[0.2em]
            \frac{7}{2}  & (1,3,0,3)     & (26,39,13,52) \\[0.2em]
            \frac{9}{2}  & (1,3,0,3)     & (42,63,21,84) \\[0.2em]
            \frac{11}{2} & (1,3,0,3)     & (62,93,31,124) \\[0.2em]
            \frac{13}{2} & (1,3,0,3)     & (86,129,43,172) \\[0.2em]
            \frac{15}{2} & (1,3,0,3)     & (114,171,57,228) \\[0.2em]
            \frac{17}{2} & (1,3,0,3)     & (146,219,73,292) \\ \bottomrule
        \end{tabular}

        \caption{$\quotparam=1$ for $\bm{32^\prime}_{q}\oplus\bm{1}_{2q}$ matter,  $\quotparam=3$ for $\bm{32}_{q}\oplus\bm{1}_{2q}$ matter}
    \end{subtable}

    \caption{Orders of vanishing of $(\secx, \secy, \secz, \secw)$ for various combinations of $\quotparam$ and \pscharge{}s $q$ for $\aso(12)$ spinor matter. This matter occurs at codimension-two loci with $\singtype{I}_{2}^{*s}\to \singtype{III}^*$ enhancements. The data are obtained using the $\aso(12)\oplus\au(1)$ model in \cref{eq:so12spinormodel}.}
    \label{tab:so12spinordata}
\end{table}

Of course, these conclusions for the $\aso(12)$ spinors were reached with only a subset of the allowed $(q,\quotparam)$ combinations. While we have not been able to find seed constructions for the missing combinations, it is unclear whether such models exist or whether some more fundamental constraint limits our ability to obtain these models. Perhaps some argument shows that, for instance, $\bm{32^\prime}_{1}$ matter cannot occur in a $\quotparam=3$ F-theory model without codimension-two (4,6,12) loci. It would be worthwhile to explore these questions in future work.

 \subsection{Vector representation of $\aso(4k+2)$}
We now consider the vector representation of $\aso(4k+2)$, which occurs at codimension-two loci where the singularity type enhances from $\singtype{I}^{*s}_{2k-3}$ (or $\Spin(4k+2)$) to $\singtype{I}^{*}_{2k-2}$ (or $\Spin(4k+4)$).  Specifically, an irreducible $\singtype{I}^{*s}_{2k-3}\rightarrow \singtype{I}^{*}_{2k-2}$ locus in a 6D model would support a hypermultiplet of vector ($\bm{4k+2}$) matter. As per the discussion in \cref{sec:soncenters}, the allowed $\au(1)$ charges are half-integral for $\quotparam=1$ and $3$ and integral for $\quotparam=0$ and $2$. Since the vector representation is real, the sign of the $\au(1)$ charge is unimportant. The automorphism previously mentioned that exchanges the two spinor representations is still applicable. It sends the vector representation to itself, but as before it sends $\quotparam$ to its inverse element in $\Z_4$. Thus, the automorphism essentially has no impact on matter in the vector representation, even though it may change the value of $\quotparam$ for the model.

We need multiple seed models to generate the \pscharge{}s corresponding to the allowed $\au(1)$ charges. If we start with a $\quotparam=1$ model with $\bm{(4k+2)}_{\frac{1}{2}}$ matter, we can generate the $\frac{1}{2}+2\Z$ \pscharge{}s for $\quotparam=1$, the $\frac{3}{2}+2\Z$ \pscharge{}s for $\quotparam=3$, the odd integral \pscharge{}s for $\quotparam=2$, and the even integral \pscharge{}s for $\quotparam=0$. By the automorphism, this same model also gives us the $\frac{3}{2}+2\Z$ \pscharge{}s for $\quotparam=1$ and the $\frac{1}{2}+2\Z$ \pscharge{}s for $\quotparam=3$. This leaves us with the odd $\quotparam=0$ and even $\quotparam=2$ \pscharge{}s. The odd $\quotparam=0$ \pscharge{}s can be obtained from a $\quotparam=0$ model admitting $\bm{(4k+2)}_{1}$ matter. In the process, we can also obtain the even $\quotparam=0$ \pscharge{}s, providing us an additional check on some of the data generated from the $\quotparam=1$ model.

Obtaining the even $\quotparam=2$ \pscharge{}s is more difficult. Starting with a $\quotparam=2$ model admitting $\bm{(4k+2)}_{2}$ matter, we could obtain the even $\quotparam=2$ \pscharge{}s that are not multiples of $4$. One could use a $\quotparam=2$ model with $\bm{(4k+2)}_{4}$ vector matter to obtain some \pscharge{}s that are multiples of 4, but we would still miss those \pscharge{}s that are multiples of 8. In order to obtain \pscharge{}s proportional to larger and larger powers of 2, we need seed models admitting vector matter with larger and larger $\au(1)$ charges. Even obtaining a model admitting $\bm{(4k+2)}_{2}$ matter is challenging, let alone models with higher charges.\footnote{If we take the formulas in \cref{sec:proposals} at face value, the $\secz$ component of the section should vanish to order 2 at the $\bm{(4k+2)}_{2}$ matter locus. The algebraic structure of such models is more complicated, since it makes use of non-UFD cancellations~\cite{Raghuram34}.} We therefore will not generate any of the even $\quotparam=2$ \pscharge{}s here.

For the $\quotparam=1$ seed models, we use Weierstrass constructions of the form
\begin{equation}
y^2 = x^3 - v^2\left(3\phi^2 + f_{k+1}v^{k-1}\right)x z^4 + v^3\left(2\phi^3+\phi f_{k+1} v^{k-1} + \gamma^2 v^{2k-3}\right)\label{eq:so4mp2vectormodela}
\end{equation}
and consider models where $k$ varies from 2 to 6. To avoid the additional complications of a spinor locus, we construct these seed models over an $\mathbb{F}_4$ base and let $\locus{v=0}$ be the curve of self-intersection $-4$. This model admits $\singtype{I}_{2k-3}^{*s}$ singularities along $\locus{v=0}$ and a generating section $\ratsec{s}$ with coordinates
\begin{equation}
[\secx:\secy:\secz] = [\phi v: \gamma v^{k}:1]\,, \quad \secw = -f_{k+1}v^{k+1}\,.
\end{equation}
The gauge algebra is therefore $\aso(4k+2)\oplus\au(1)$. The locus $\locus{v=\gamma=0}$ supports $\bm{(4k+2)}_{\frac{1}{2}}$ matter, while $\locus{f_{k+1}=\gamma=0}$ supports $\bm{1}_{1}$ singlet matter. Together, this matter spectrum satisfies the 6D anomaly cancellation conditions for $\height = -2\canonclass - \frac{2k+1}{4}\divclass{v}$. This model can be viewed as either a $\quotparam=1$ model or as a $\quotparam=3$ model; these two options are related by the automorphism discussed previously.

For the $\quotparam=0$ seed models, we use Weierstrass constructions of the form
\begin{equation}
    \label{eq:so4mp2vectormodelb}
    \begin{aligned}
        y^2 &= x^3 +\sigma^2\left(-\frac{1}{3} c_{2,1}^2+c_{3,0} \sigma^{k-1} c_{1,k+1}-\frac{1}{4} \sigma^{2 k-2} \left(b_{2,k}^2+4 c_{0,2k+1} \sigma\right)\right)x z^4 \\
        &\qquad\quad + \frac{\sigma^3}{108}\Big[8 c_{2,1}^3+9 \sigma^{2 k-3} \left(3\sigma^2 c_{1,k+1}^2-\left(2 \sigma c_{2,1}-3 c_{3,0}^2\right) \left(b_{2,k}^2+4 c_{0,2k+1}\sigma\right)\right)\\
        &\qquad\qquad\qquad\quad - 36 c_{2,1} c_{3,0} c_{1,k+1} \sigma^{k-1}\Big]z^6\,,
    \end{aligned}
\end{equation}
and we let $k$ vary from 2 to 6. We no longer assume that the base is $\mathbb{F}_4$.\footnote{In fact, if the base is $\mathbb{F}_4$ and $\locus{\sigma=0}$ has self-intersection -4, the parameter $c_{3,0}$ is forced to be proportional to $\sigma$, and the gauge algebra enhances.} Again, there are $\singtype{I}_{2k-3}^{*s}$ singularities along $\locus{\sigma=0}$, and the model admits a generating section $\ratsec{s}$ with coordinates
\begin{gather}
[\secx:\secy:\secz] = \left[c_{3,0}^2-\frac{2}{3} \sigma c_{2,1}:-c_{3,0}^3+\sigma c_{2,1} c_{3,0}-\frac{1}{2} \sigma^{k+1} c_{1,k+1}:1\right]\,, \\
\secw = 3 c_{3,0}^4-4 \sigma c_{2,1} c_{3,0}^2+\sigma^2 c_{2,1}^2+c_{3,0} \sigma^{k+1}c_{1,k+1}-\frac{1}{4} \sigma^{2 k} \left(b_{2,k}^2+4 c_{0,2k+1} \sigma\right)\,.
\end{gather}
The codimension-two locus $\locus{\sigma=c_{3,0}=0}$ supports $\bm{(4k+2)}_{1}$ matter, while the codimension-two locus $\locus{\sigma=b_{2,k}=0}$ supports $\bm{(4k+2)}_{0}$ matter. When $k=2$, the locus $\locus{\sigma=c_{2,1}=0}$ supports $\bm{16}_{0}$ matter, and after including the contributions from singlets and uncharged adjoints, the spectrum satisfies the 6D anomaly cancellation conditions with $\height=-2\canonclass$. When $m>2$, $(f, g,\Delta)$ vanish at least to orders $(4,6,12)$ at $\locus{\sigma=c_{2,1}=0}$, making it difficult to determine the matter associated with this locus and its contribution to the $\aso(4k+2)$ anomaly conditions. However, if we assume that any matter contributions at $\locus{\sigma=c_{2,1}=0}$ are uncharged under the $\au(1)$, just as was the case for $k=2$, the rest of the spectrum (including singlets) satisfies the $\au(1)$ anomaly cancellation conditions with $\height=-2\canonclass$. Since we are only interested in the $\locus{\sigma=c_{3,0}=0}$ vector locus, we proceed with this construction in spite of the codimension-two $(4,6,12)$ locus. As confirmation that this $(4,6,12)$ locus does not cause significant issues, we can verify that the results for the even \pscharge{}s found from the $\quotparam=0$ construction agree with those found from the $\quotparam=1$ construction.

We now use these constructions to generate $\aso(4k+2)\oplus\au(1)$ models for $k$ ranging from $2$ to $6$ and find the orders of vanishing associated with the various vector \pscharge{}s. We do not list the orders of vanishing at codimension one for the sake of brevity, but they are all given by
\begin{equation}
\ordvanishone{\secz} = 0\,, \quad \left(\ordvanishone{\secx},\ordvanishone{\secy},\ordvanishtwo{\secw}\right) = \quottriplet{\Spin(4k+2)}{\quotparam}\,.
\end{equation}
The orders of vanishing at codimension two are listed in \cref{tab:so4mp2vectordata}. If we let
\begin{equation}
\quotparamtwo = \residue{(2q-\quotparam)}{4} + \residue{2q}{2}\,,
\end{equation}
all of these numbers follow the formulas
\begin{equation}
\ordvanishtwo{\secz} = \frac{1}{2}\left(q^2 + \quotfunction{\Spin(4k+2)}{\quotparam} -  \quotfunction{\Spin(4k+4)}{\quotparamtwo}\right)
\end{equation}
and
\begin{equation}
\left(\ordvanishtwo{\secx},\ordvanishtwo{\secy},\ordvanishtwo{\secw}\right) = \left(2,3,4\right)\ordvanishtwo{\secz}+  \quottriplet{\Spin(4k+4)}{\quotparamtwo}\,.
\end{equation}
Since the center of $\Spin(2n)$ has $d_{\Spin(2n)}=4$ elements, these formulas exactly agree with the expectations from \cref{sec:proposals}.

\begin{table}
    \centering

    \begin{tiny}
    \begin{tabular}{*{7}{>{$}c<{$}}}\toprule
        \quotparam               & \abs{q}      & \aso(10)         & \aso(14)         & \aso(18)         & \aso(22)         & \aso(26) \\\midrule
        \multirow{12}{*}{0}      & 1            & (1,2,0,2)        & (1,2,0,2)        & (1,2,0,2)        & (1,2,0,2)        & (1,2,0,2) \\
                                 & 2            & (4,6,2,8)        & (4,6,2,8)        & (4,6,2,8)        & (4,6,2,8)        & (4,6,2,8) \\
                                 & 3            & (9,14,4,18)      & (9,14,4,18)      & (9,14,4,18)      & (9,14,4,18)      & (9,14,4,18) \\
                                 & 4            & (16,24,8,32)     & (16,24,8,32)     & (16,24,8,32)     & (16,24,8,32)     & (16,24,8,32) \\
                                 & 5            & (25,38,12,50)    & (25,38,12,50)    & (25,38,12,50)    & (25,38,12,50)    & (25,38,12,50) \\
                                 & 6            & (36,54,18,72)    & (36,54,18,72)    & (36,54,18,72)    & (36,54,18,72)    & (36,54,18,72) \\
                                 & 7            & (49,74,24,98)    & (49,74,24,98)    & (49,74,24,98)    & (49,74,24,98)    & (49,74,24,98) \\
                                 & 8            & (64,96,32,128)   & (64,96,32,128)   & (64,96,32,128)   & (64,96,32,128)   & (64,96,32,128) \\
                                 & 9            & (81,122,40,162)  & (81,122,40,162)  & (81,122,40,162)  & (81,122,40,162)  & (81,122,40,162) \\
                                 & 10           & (100,150,50,200) & (100,150,50,200) & (100,150,50,200) & (100,150,50,200) & (100,150,50,200) \\
                                 & 11           & (121,182,60,242) & (121,182,60,242) & (121,182,60,242) & (121,182,60,242) & (121,182,60,242) \\
                                 & 12           & (144,216,72,288) & (144,216,72,288) & (144,216,72,288) & (144,216,72,288) & (144,216,72,288) \\ \midrule
        \multirow{13}{*}{1 or 3} & \frac{1}{2}  & (1,3,0,3)        & (1,4,0,4)        & (1,5,0,5)        & (1,6,0,6)        & (1,7,0,7) \\[0.4em]
                                 & \frac{3}{2}  & (3,6,1,7)        & (3,7,1,8)        & (3,8,1,9)        & (3,9,1,10)       & (3,10,1,11) \\[0.4em]
                                 & \frac{5}{2}  & (7,12,3,15)      & (7,13,3,16)      & (7,14,3,17)      & (7,15,3,18)      & (7,16,3,19) \\[0.4em]
                                 & \frac{7}{2}  & (13,21,6,27)     & (13,22,6,28)     & (13,23,6,29)     & (13,24,6,30)     & (13,25,6,31) \\[0.4em]
                                 & \frac{9}{2}  & (21,33,10,43)    & (21,34,10,44)    & (21,35,10,45)    & (21,36,10,46)    & (21,37,10,47) \\[0.4em]
                                 & \frac{11}{2} & (31,48,15,63)    & (31,49,15,64)    & (31,50,15,65)    & (31,51,15,66)    & (31,52,15,67) \\[0.4em]
                                 & \frac{13}{2} & (43,66,21,87)    & (43,67,21,88)    & (43,68,21,89)    & (43,69,21,90)    & (43,70,21,91) \\[0.4em]
                                 & \frac{15}{2} & (57,87,28,115)   & (57,88,28,116)   & (57,89,28,117)   & (57,90,28,118)   & (57,91,28,119) \\[0.4em]
                                 & \frac{17}{2} & (73,111,36,147)  & (73,112,36,148)  & (73,113,36,149)  & (73,114,36,150)  & (73,115,36,151) \\[0.4em]
                                 & \frac{19}{2} & (91,138,45,183)  & (91,139,45,184)  & (91,140,45,185)  & (91,141,45,186)  & (91,142,45,187) \\[0.4em]
                                 & \frac{21}{2} & (111,168,55,223) & (111,169,55,224) & (111,170,55,225) & (111,171,55,226) & (111,172,55,227) \\[0.4em]
                                 & \frac{23}{2} & (133,201,66,267) & (133,202,66,268) & (133,203,66,269) & (133,204,66,270) & (133,205,66,271) \\[0.4em]
                                 & \frac{25}{2} & (157,237,78,315) & (157,238,78,316) & (157,239,78,317) & (157,240,78,318) & (157,241,78,319) \\ \midrule
        \multirow{6}{*}{2}       & 1            & (2,3,1,4)        & (2,3,1,4)        & (2,3,1,4)        & (2,3,1,4)        & (2,3,1,4) \\
                                 & 3            & (10,15,5,20)     & (10,15,5,20)     & (10,15,5,20)     & (10,15,5,20)     & (10,15,5,20) \\
                                 & 5            & (26,39,13,52)    & (26,39,13,52)    & (26,39,13,52)    & (26,39,13,52)    & (26,39,13,52) \\
                                 & 7            & (50,75,25,100)   & (50,75,25,100)   & (50,75,25,100)   & (50,75,25,100)   & (50,75,25,100) \\
                                 & 9            & (82,123,41,164)  & (82,123,41,164)  & (82,123,41,164)  & (82,123,41,164)  & (82,123,41,164) \\
                                 & 11           & (122,183,61,244) & (122,183,61,244) & (122,183,61,244) & (122,183,61,244) & (122,183,61,244) \\ \bottomrule
    \end{tabular}
    \end{tiny}

    \caption{Codimension-two orders of vanishing of $(\secx, \secy,\secz,\secw)$ for various \pscharge{}s $q$ of $\aso(4k+2)\oplus\au(1)$ vector matter. This matter is supported at $\singtype{I}_{2k-3}^{*s}\to \singtype{I}_{2k-2}^{*s}$ loci. The data are found using the constructions in \cref{eq:so4mp2vectormodela,eq:so4mp2vectormodelb}.}
    \label{tab:so4mp2vectordata}
\end{table}

\subsection{Vector representation of $\aso(4k)$}
We now turn to the vector representation of $\aso(4k)$. Previously, we used the triality of $\aso(8)$ to determine the orders of vanishing for $\bm{8_\text{v}}$ matter from those for $\bm{8_\text{s}}$ and $\bm{8_\text{c}}$ matter. We can no longer use this triality for $k>2$, so we must separately analyze vector matter for these cases. The vector representation of $\aso(4k)$ occurs at codimension-two loci where the singularity type enhances from $\singtype{I}^{*s}_{2k-4}$ (or $\Spin(4k)$) to $\singtype{I}^{*}_{2k-3}$ (or $\Spin(4k+2)$). In 6D models, an irreducible locus of this type supports a hypermultiplet of $\bm{4k}$ matter.  Just as for the $\aso(4k+2)$ vector representation, the allowed $\au(1)$ charges for the $\aso(4k)$ vector representation are half-integral for $\quotparam=1,3$ and integral for $\quotparam=0,2$.  And because the vector representation is real, the sign of the $\au(1)$ charge is unimportant for vector matter. The automorphism for $\aso(4k)$ does not affect the vector matter but exchanges $\quotparam=1$ and $\quotparam=3$.

We again need multiple models to generate a representative sample of \pscharge{}s. If we start with a $\quotparam=1$ model with $\bm{4k}_{\frac{1}{2}}$ matter, one can obtain all the half-integral $\quotparam=1$ \pscharge{}s and all the integral $\quotparam=0$ \pscharge{}s by examining multiples of the generating section. By the automorphism, this same model can be viewed as a $\quotparam=3$ model with $\bm{4k}_{\frac{1}{2}}$ matter, and the $\quotparam=1$ \pscharge{}s can equivalently be viewed as $\quotparam=3$ \pscharge{}s. Therefore, we can obtain all of the half-integral $\quotparam=3$ \pscharge{}s from this same construction. This leaves us with the $\quotparam=2$ \pscharge{}s. Starting with a $\quotparam=2$ seed model with $\bm{4k}_{1}$ matter, which can be constructed relatively easily, we can get all of the odd $\quotparam=2$ \pscharge{}s. This also gives us an alternative method of generating the even $\quotparam=0$ \pscharge{}s. Obtaining the even \pscharge{}s with $\quotparam=2$ is more difficult: in order to get a \pscharge{} proportional to $2^j$, we would need a seed model admitting $\bm{4k}_{2^j}$ matter. Therefore, we only generate the odd $\quotparam=2$ \pscharge{}s here.

For the $\quotparam=1$ seed models, we consider constructions over an $\mathbb{F}_4$ base given by the Weierstrass model
\begin{equation}
\label{eq:so4mvectormodela}
y^2 = x^3 - v^2\left(3 \phi ^2+f_k v^{k-2}\right)x z^4 + v^3\left(2 \phi ^3+\phi  f_k v^{k-2}+\gamma ^2 v^{2 k-3}\right)\,,
\end{equation}
where $\locus{v=0}$ is a curve in $\mathbb{F}_4$ of self-intersection $-4$. We let $k$ vary from 3 to 5.\footnote{Since we previously determined the orders of vanishing for $\aso(8)$ vector matter, we do not need to consider the $k=2$ version of this construction.} The model has $\singtype{I}^{*s}_{2k-4}$ singular fibers along $\locus{v=0}$ and admits a generating section of the form
\begin{equation}
[\secx:\secy:\secz] = \left[v\phi:v^{k}\gamma:1\right]\,, \quad \secw = -v^{k}f_{k}\,,
\end{equation}
implying that the gauge algebra is $\aso(4k)\oplus\au(1)$. The codimension-two locus $\locus{v=f_{k}=0}$ supports $\bm{4k}_{\frac{1}{2}}$ matter, while the codimension-two locus $\locus{f_{k}=\gamma=0}$ supports $\bm{1}_{1}$ matter. The resulting spectra satisfy the 6D anomaly cancellation conditions with $\height = -2\canonclass - \frac{k}{2}\divclass{v}$. By the automorphism, these constructions can alternatively be viewed as $\quotparam=3$ model with the same matter spectra.

For the $\quotparam=2$ seed models, we use constructions given by the Weierstrass model
\begin{equation}
    \label{eq:so4mvectormodelb}
    \begin{aligned}
        y^2 &= x^3 -\frac{v^2}{3}\left(c_{2,1}^2+3 b_{0,0}^2 c_{0,2 k-1} v^{2k-3} +3 b_{0,0}^2  b_{2,k-1}^2v^{2k-4}-3 c_{3,1}c_{1,k}v^{k-1} \right)x z^4\\
        &\qquad\quad -\frac{v^3}{108}\Big[-8 c_{2,1}^3-9 v^{2 k-4} \Big(b_{0,0}^2 \left(3 v c_{1,k}^2-8 c_{2,1} \left(b_{2,k-1}^2+v c_{0,2k-1}\right)\right) \\
        &\qquad\qquad\qquad\quad +12 v c_{3,1}^2 \left(b_{2,k-1}^2+v c_{0,2 k-1}\right)\Big)+36 c_{2,1} c_{3,1} v^{k-1} c_{1,k}\Big]z^6\,.
    \end{aligned}
\end{equation}
Again, we construct these elliptic fibrations over an $\mathbb{F}_4$ base and take $\locus{v=0}$ to be a curve of self-intersection $-4$. We also let $k$ run from $3$ to $5$. There are $\singtype{I}^{*s}_{2k-4}$ singularities along $\locus{v=0}$, and the Weierstrass model admits a generating section with components
\begin{equation}
\begin{aligned}
\secx &= \frac{ v }{3}\left(3 v c_{3,1}^2-2 b_{0,0}^2 c_{2,1}\right)\,, \\
\secy &= \frac{v^2}{2}\left(2 b_{0,0}^2 c_{2,1} c_{3,1}-b_{0,0}^4 c_{1,k} v^{k-2}-2 v c_{3,1}^3\right)\,, \\
\secz &= b_{0,0}\,, \\
\secw &=  -v^2 \Big(-b_{0,0}^4 c_{2,1}^2+4 v b_{0,0}^2 c_{2,1} c_{3,1}^2-3 v^2 c_{3,1}^4-b_{0,0}^4 c_{3,1} v^{k-1} c_{1,k} \\
&\qquad\quad\qquad +b_{0,0}^6 v^{2k-4} b_{2,k-1}^2+b_{0,0}^6 c_{0,2 k-1}v^{2k-3} \Big)\,.
\end{aligned}
\end{equation}
Thus, the gauge algebra is $\aso(4k)\oplus\au(1)$.\footnote{For $k=5$, $\divclass{c_{0,2m-1}}$ is ineffective. As a result, there is an extra generating section, and the gauge algebra is $\aso(20)\oplus\au(1)\oplus\au(1)$. However, the extra $\au(1)$ gauge factor does not affect the discussion and can be safely ignored.} The model supports $\bm{4k}_{1}$ matter at $\locus{v=b_{0,0}=0}$ and $\bm{4k}_{0}$ matter at $\locus{v= b_{2,k-1}=0}$. After accounting for charged singlets, this spectrum satisfies the anomaly cancellation conditions for $\height = -2\canonclass + 2\divclass{b_{0,0}} -\divclass{v}$.

We now construct multiples of the generating sections in these models and find the orders of vanishing for various \pscharge{}s $q$ of $\bm{4k}$ vector matter. We do not list the codimension-one orders of vanishing here for brevity, but the generated data for $k=2$ through $5$ satisfy the equations
\begin{equation}
\ordvanishone{\secz} = 0\,, \quad \left(\ordvanishone{\secx},\ordvanishone{\secy},\ordvanishone{\secw}\right) = \quottriplet{\Spin(4k)}{\quotparam}\,.
\end{equation}
The codimension-two orders of vanishing for $k=3$ through $5$ are listed in \cref{tab:so4mvectordata}; for $k=2$, we can use the previously determined $\aso(8)$ data in \cref{tab:so8data}. If we let
\begin{equation}
\quotparamtwo = \residue{(2q-\quotparam)}{4} + \residue{2q}{2}\,,
\end{equation}
all of the codimension-two orders of vanishing satisfy the equations
\begin{gather}
\ordvanishtwo{\secz} = \frac{1}{2}\left(q^2 + \quotfunction{\Spin(4k)}{\quotparam} - \quotfunction{\Spin(4k+2)}{\quotparamtwo} \right)\,, \\
\left(\ordvanishtwo{\secx},\ordvanishtwo{\secy},\ordvanishtwo{\secw}\right) = \left(2,3,4\right)\times \ordvanishtwo{\secz} + \quottriplet{\Spin(4k+2)}{\quotparamtwo}\,.
\end{gather}
Since $\Spin(2n)$ has a center with $d_{\Spin(2n)}=4$ elements, the observed orders of vanishing agree with the expectations from \cref{sec:proposals}.

\begin{table}
    \centering

    \begin{tabular}{*{5}{>{$}c<{$}}}\toprule
        \quotparam               & \abs{q}      & \aso(12)         & \aso(16)         & \aso(20) \\ \midrule
        \multirow{13}{*}{0}      & 1            & (1,2,0,2)        & (1,2,0,2)        & (1,2,0,2) \\
                                 & 2            & (4,6,2,8)        & (4,6,2,8)        & (4,6,2,8) \\
                                 & 3            & (9,14,4,18)      & (9,14,4,18)      & (9,14,4,18) \\
                                 & 4            & (16,24,8,32)     & (16,24,8,32)     & (16,24,8,32) \\
                                 & 5            & (25,38,12,50)    & (25,38,12,50)    & (25,38,12,50) \\
                                 & 6            & (36,54,18,72)    & (36,54,18,72)    & (36,54,18,72) \\
                                 & 7            & (49,74,24,98)    & (49,74,24,98)    & (49,74,24,98) \\
                                 & 8            & (64,96,32,128)   & (64,96,32,128)   & (64,96,32,128) \\
                                 & 9            & (81,122,40,162)  & (81,122,40,162)  & (81,122,40,162) \\
                                 & 10           & (100,150,50,200) & (100,150,50,200) & (100,150,50,200) \\
                                 & 11           & (121,182,60,242) & (121,182,60,242) & (121,182,60,242) \\
                                 & 12           & (144,216,72,288) & (144,216,72,288) & (144,216,72,288) \\
                                 & 13           & (169,254,84,338) & (169,254,84,338) & (169,254,84,338) \\ \midrule
        \multirow{13}{*}{1 or 3} & \frac{1}{2}  & (1,3,0,4)        & (1,4,0,5)        & (1,5,0,6) \\[0.2em]
                                 & \frac{3}{2}  & (3,6,1,8)        & (3,7,1,9)        & (3,8,1,10) \\[0.2em]
                                 & \frac{5}{2}  & (7,12,3,16)      & (7,13,3,17)      & (7,14,3,18) \\[0.2em]
                                 & \frac{7}{2}  & (13,21,6,28)     & (13,22,6,29)     & (13,23,6,30) \\[0.2em]
                                 & \frac{9}{2}  & (21,33,10,44)    & (21,34,10,45)    & (21,35,10,46) \\[0.2em]
                                 & \frac{11}{2} & (31,48,15,64)    & (31,49,15,65)    & (31,50,15,66) \\[0.2em]
                                 & \frac{13}{2} & (43,66,21,88)    & (43,67,21,89)    & (43,68,21,90) \\[0.2em]
                                 & \frac{15}{2} & (57,87,28,116)   & (57,88,28,117)   & (57,89,28,118) \\[0.2em]
                                 & \frac{17}{2} & (73,111,36,148)  & (73,112,36,149)  & (73,113,36,150) \\[0.2em]
                                 & \frac{19}{2} & (91,138,45,184)  & (91,139,45,185)  & (91,140,45,186) \\[0.2em]
                                 & \frac{21}{2} & (111,168,55,224) & (111,169,55,225) & (111,170,55,226) \\[0.2em]
                                 & \frac{23}{2} & (133,201,66,268) & (133,202,66,269) & (133,203,66,270) \\[0.2em]
                                 & \frac{25}{2} & (157,237,78,316) & (157,238,78,317) & (157,239,78,318) \\ \midrule
        \multirow{7}{*}{2}      & 1            & (2,3,1,4)        & (2,3,1,4)        & (2,3,1,4) \\
                                 & 3            & (10,15,5,20)     & (10,15,5,20)     & (10,15,5,20) \\
                                 & 5            & (26,39,13,52)    & (26,39,13,52)    & (26,39,13,52) \\
                                 & 7            & (50,75,25,100)   & (50,75,25,100)   & (50,75,25,100) \\
                                 & 9            & (82,123,41,164)  & (82,123,41,164)  & (82,123,41,164) \\
                                 & 11           & (122,183,61,244) & (122,183,61,244) & (122,183,61,244) \\
                                 & 13           & (170,255,85,340) & (170,255,85,340) & (170,255,85,340) \\ \bottomrule
    \end{tabular}

    \caption{Codimension-two orders of vanishing of $(\secx, \secy,\secz,\secw)$ for various \pscharge{}s $q$ of $\aso(4k)\oplus\au(1)$ vector matter. This matter is supported at $\singtype{I}_{2k-4}^{*s}\to \singtype{I}_{2k-3}^{*}$ loci. The data are found using the constructions in Equations \labelcref{eq:so4mvectormodela} and \labelcref{eq:so4mvectormodelb}.}
    \label{tab:so4mvectordata}
\end{table}

\subsection{Connection to EDS valuations}
\label{sec:soeds}

We can now show that the formulas for EDS valuations in~\cite{StangeEllTrouble} reproduce the patterns for $\aso(4k)$ and $\aso(4k+2)$ models proposed in \cref{sec:proposals} and observed in this section. As described in \cref{sec:suneds}, for each seed model, we should subtract the EDS valuations for the codimension-one behavior from those for the codimension-two behavior. The resulting numbers should agree with the observed $\ordvanishtwo{\secz}$ values for multiples of the seed model's generating section admitting the appropriate \pscharge{}s.

There are five singularity types that need to be considered for the $\aso$ representations discussed above. Some of the EDS valuations for $\singtype{I}^*_{2k-3}$ singularities were calculated in \cref{sec:suneds} when analyzing the antisymmetric representation of $\asu(2k+1)$; the remaining ones are calculated in \cref{app:sonoddedsnu0}. Those for the $\singtype{I}^*_{2k-4}$, $\singtype{IV}^*$, $\singtype{III}^*$, and $\singtype{II}^*$ singularity types are respectively calculated in \cref{app:sonedseven}, \cref{app:e6eds}, \cref{app:e7eds}, and \cref{app:e8eds}.
Having found the EDS valuations for these different singularity types, we can show how to reproduce the $\ordvanishtwo{\secz}$ formulas for the $\aso(2n)$ representations.

\paragraph{Vectors of $\aso(4k+2)$} We considered two different seed models for the $\bm{(4k+2)}_{q}$ representation of $\aso(4k+2)\oplus\au(1)$, so we must determine the EDS valuations for these two models separately. We start with the $\quotparam=1$ seed model. The generating section components $(\secx,\secy,\secz,\secw)$ vanish to orders $(1,k,0,k+1)$ along the codimension-one locus supporting the $\aso(4k+2)$. The EDS valuations corresponding to the codimension-one behavior are given by \cref{eq:sooddvaluations} with $n$ replaced with $2k+1$  and $s$ set to 1:
\begin{equation}
    \valuation{}{\eds_{m}^{(1)}} = \frac{2k+1}{8}m^2 - \frac{1}{2}\begin{cases}0 & 4\mid m\\ 1 & (2\mid m) \text{ and }(4\nmid m)\\\frac{2k+1}{4} & (2\nmid m)\end{cases}
\end{equation} At the codimension-two locus supporting the seed model's $\bm{(4k+2)}_{\frac{1}{2}}$ matter, the generating section components $(\secx,\secy,\secz,\secw)$ vanish to orders $(1,k+1,0,k+1)$. The EDS valuations corresponding to the codimension-two behavior are therefore given by \cref{eq:soevenvaluationsmu1} with $n$ replaced by $2k+2$:
\begin{equation}
    \valuation{}{\eds_{m}^{(2)}} = \frac{2k+2}{8}m^2 - \frac{1}{2}\begin{cases}0 &  2 \mid m\\ \frac{2k+2}{4} &  2 \nmid m \end{cases}\,.
\end{equation}
The difference between these two sequences of EDS valuations is
\begin{equation}
    \valuation{}{\eds_{m}^{(2)}}-\valuation{}{\eds_{m}^{(1)}} = \frac{1}{2}\left(\frac{1}{4}m^2+ \begin{cases}0 & 4\mid m\\ 1 & (2\mid m) \text{ and }(4\nmid m)\\\frac{2k+1}{4} & (2\nmid m)\end{cases} - \begin{cases}0 &  2 \mid m\\ \frac{2k+2}{4} &  2 \nmid m \end{cases}\right)
\end{equation}
Using the relations
\begin{equation}
    q = \frac{m}{2}\,, \quad \quotparam = \residue{m}{4}\,, \quad \quotparamtwo = \residue{(2q-\quotparam)}{4}+ \residue{2q}{2} = \residue{m}{2}\,,
\end{equation}
we can rewrite this as
\begin{equation}
    \valuation{}{\eds_{m}^{(2)}}-\valuation{}{\eds_{m}^{(1)}} = \frac{1}{2}\left(q^2 + \quotfunction{\SO(4k+2)}{\quotparam} - \quotfunction{\SO(4k+4)}{\quotparamtwo}\right)\,.
\end{equation}
The right-hand side of this expression exactly matches the expected formula for $\ordvanishtwo{\secz}$ from the proposals, and the formula agrees with the observed orders of vanishing.

For the $\quotparam=0$ seed model, the generating section components $(\secx, \secy, \secz, \secw)$ vanish to orders $(0,0,0,0)$ at the codimension-one locus supporting the $\aso(4k+2)$ factor. Because all the generating section components vanish to order 0, the EDS valuations $\valuation{}{\eds_{m}^{(1)}}$ are $0$. At the codimension-two locus supporting the $\bm{(4k+2)}_{1}$ matter, the generating section components $(\secx, \secy, \secz, \secw)$ vanish to orders $(1,2,0,2)$, and the corresponding EDS valuations $\valuation{}{\eds_{m}^{(2)}}$ are given by \cref{eq:soevenvaluationsmu2}. We therefore have
\begin{equation}
        \valuation{}{\eds_{m}^{(2)}}-\valuation{}{\eds_{m}^{(1)}} = \valuation{}{\eds_{m}^{(2)}} = \frac{m^2}{2} -\frac{1}{2}\begin{cases}0 & 2\mid m \\ 1 & 2\nmid m\end{cases}\,.
\end{equation}
Using the relations
\begin{equation}
    q = m\,, \quad \quotparam = 0\,, \quad  \quotparamtwo = \residue{(2q-\quotparam)}{4}+ \residue{2q}{2} = 2\residue{m}{2}\,,
\end{equation}
this can be rewritten as
\begin{equation}
    \valuation{}{\eds_{m}^{(2)}}-\valuation{}{\eds_{m}^{(1)}} = \frac{1}{2}\left(q^2 + \quotfunction{\SO(4k+2)}{\quotparam} - \quotfunction{\SO(4k+4)}{\quotparamtwo}\right)\,.
\end{equation}
Again, the right-hand side agrees with expression for $\ordvanishtwo{\secz}$ expected from the proposals and reproduces the observed orders of vanishing.

\paragraph{Vectors of $\aso(4k)$} We considered two different seed models for the $\bm{4k}_{q}$ representation of $\aso(4k)\oplus\au(1)$, so we need to perform separate EDS analyses for each of these models. We start with the $\quotparam=1$ seed model. Along the codimension-one locus supporting the $\aso(4k)$ gauge factor, the generating section components $(\secx,\secy,\secz,\secw)$ vanish to orders $(1,k,0,k)$. Therefore, the EDS valuations corresponding to the codimension-one behavior are given by \cref{eq:soevenvaluationsmu1} with $n$ set to $2k$:
\begin{equation}
\valuation{}{\eds^{(1)}_{m}} =  \frac{2k}{8}m^2 - \frac{1}{2}\begin{cases}0 &  2 \mid m\\ \frac{2k}{4} &  2 \nmid m \end{cases}\,.
\end{equation}
At the codimension-two locus supporting the $\bm{4k}_{1/2}$ matter, the $(\secx,\secy,\secz,\secw)$ vanish to orders $(1,k,0,k+1)$, and the EDS valuations corresponding to the codimension-two behavior are given by \cref{eq:sooddvaluations} with $n$ set to $2k+1$ and $s$ set to $1$:
\begin{equation}
    \valuation{}{\eds^{(2)}_{m}} = \frac{2k+1}{8}m^2 - \frac{1}{2}\begin{cases}0 & 4\mid m\\ 1 & (2\mid m) \text{ and }(4\nmid m)\\\frac{2k+1}{4} & (2\nmid m)\end{cases}\,.
\end{equation}
Using the relations\footnote{We are free to use either $\quotparam = \residue{m}{2}, \quotparamtwo=\residue{m}{4}$ or $\quotparam=3\residue{m}{2},\quotparamtwo=\residue{3m}{4}$, as the two options are related by the automorphism. One obtains similar results with either choice.}
\begin{equation}
    q = \frac{m}{2}\,, \quad \quotparam =  \residue{m}{2}\,, \quad \quotparamtwo = \residue{(2q-\quotparam)}{4}+ \residue{2q}{2} = \residue{m}{4}\,,
\end{equation}
one can write the difference of the two EDS valuations as
\begin{equation}
    \valuation{}{\eds^{(2)}_{m}} -  \valuation{}{\eds^{(1)}_{m}}  = \frac{1}{2}\left(q^2 + \quotfunction{\SO(4k)}{\quotparam} - \quotfunction{\SO(4k+2)}{\quotparamtwo}\right)\,.
\end{equation}
The right-hand side agrees with the expected formula for $\ordvanishtwo{\secz}$ from the proposals and reproduces the observed order of vanishing data.

For the $\quotparam=2$ seed model, the generating section components $(\secx,\secy,\secz,\secw)$ vanish to orders $(1,2,0,2)$ along the codimension-one locus supporting the $\aso(4k)$ gauge algebra. The EDS valuations for the codimension-one behavior are therefore given by \cref{eq:soevenvaluationsmu2}:
\begin{equation}
    \valuation{}{\eds_{m}^{(1)}} = \frac{m^2}{2} -\frac{1}{2}\begin{cases}0 & 2\mid m \\ 1 & 2\nmid m\end{cases}\,.
\end{equation}
At the codimension-two locus supporting the $\bm{4k}_{1}$ matter, $(\secx,\secy,\secz,\secw)$ vanish to orders $(2,3,1,4)$. The corresponding EDS valuations for this situations are given by \cref{eq:sooddedsnu0}:
\begin{equation}
    \valuation{}{\eds_{m}^{(2)}} = m^2\,.
\end{equation}
If we use the relations
\begin{equation}
    q = m\,, \quad \quotparam = 2\residue{m}{2}\,, \quad \quotparamtwo = \residue{(2q-\quotparam)}{4}+ \residue{2q}{2} = 0\,,
\end{equation}
the difference between these two EDS valuations can be written as
\begin{equation}
    \valuation{}{\eds_{m}^{(2)}}-\valuation{}{\eds_{m}^{(1)}} = \frac{1}{2}\left(q^2 + \quotfunction{\SO(4k)}{\quotparam} - \quotfunction{\SO(4k)}{\quotparamtwo}\right)\,,
\end{equation}
in agreement with the proposals.

\paragraph{Spinors of $\aso(8)$} While we considered only one seed model, given by \cref{eq:so8model}, for the spinors of $\aso(8)$, we generated data using two different matter loci with distinct charges. We need to perform the EDS analysis separately for each of these loci. The codimension-one EDS valuations are the same for both these loci and are given by \cref{eq:soevenvaluationsmu2}. The codimension-two EDS valuations, however, are different for the two cases.

At the loci $\locus{\sigma=c_{2,1}\pm b_{0,0} b_{2,1} = 0}$, which support matter with $\au(1)$ charge $\frac{1}{2}$, the generating section components $(\secx,\secy,\secz,\secw)$ vanish to orders $(1,2,0,3)$. Therefore, the codimension-two orders of vanishing are described by \cref{eq:sooddvaluations} with $n=5$ and $s=1$, and we have
\begin{equation}
    \valuation{}{\eds_{m}^{(2)}} - \valuation{}{\eds_{m}^{(1)}} = \frac{1}{2}\left(\frac{1}{4}m^2 - \begin{cases}0 & 4\mid m \\ 1 & (2\mid m)\text{ and }(4\nmid m)\\ \frac{5}{4} & m\nmid 2 \end{cases} + \begin{cases}0 & 2\mid m \\ 1 & (2\nmid m)\end{cases}\right)\,.
\end{equation}
If we note that $q=m/2$ and use the appropriate expressions for $\quotparam$ and $\quotparamtwo$, this formula reproduces the $\ordvanishtwo{\secz}$ given in \cref{tab:so8datanu1frac,tab:so8datanu0} and agrees with  the proposed formulas in \cref{sec:proposals}.

At the locus $\locus{\sigma=b_{0,0} = 0}$, which supports matter with $\au(1)$ charge $1$, the generating section components $(\secx,\secy,\secz,\secw)$ vanish to orders $(2,3,1,4)$. The EDS valuations for this situation, calculated in \cref{app:sonoddedsnu0}, are given by $\valuation{}{\eds_{m}^{(2)}} = m^2$. Therefore
\begin{equation}
    \valuation{}{\eds_{m}^{(2)}} - \valuation{}{\eds_{m}^{(1)}} = \frac{1}{2}\left(m^2 + \begin{cases}0 & 2\mid m \\ 1 & (2\nmid m)\end{cases}\right)\,.
\end{equation}
Noting that $q=m$ and that $\quotparamtwo=0$ for this seed model, this expression again agrees with the proposals and the observed $\ordvanishtwo{\secz}$ values.

\paragraph{Spinors of $\aso(10)$} For the spinors of $\aso(10)$, we considered a single seed model given by \cref{eq:so10spinorseedgensec}. The section components $(\secx,\secy,\secz,\secw)$ vanish to orders $(1,2,0,3)$ at the codimension-one locus supporting the $\singtype{I}_1^{*s}$ singularities and to orders $(2,2,0,3)$ at the codimension-two locus supporting the $\bm{16}_{-\frac{1}{4}}$ matter. To show the connection to EDS valuations, we should subtract the EDS valuations for $\singtype{I}^*_{1}$ singularities in \cref{eq:sooddvaluations} (with $n=5$, $s=1$) from those for $\singtype{IV}^*$ singularities in \cref{eq:edsvale6}:
\begin{equation}
    \begin{aligned}
    \valuation{}{W_{m}^{(2)}} - \valuation{}{W_{m}^{(1)}} &= \left(\frac{2}{3}m^2  - \begin{cases} 0 & 3\mid m \\  \frac{2}{3} & 3\nmid m\end{cases}\right) \\
    &\qquad\qquad - \left(\frac{5}{8}m^2 - \frac{1}{2}\begin{cases}0 & 4\mid m\\ 1 & (2\mid m) \text{ and }(4\nmid m)\\\frac{5}{4} & (2\nmid m)\end{cases}\right) \\
    &= \frac{1}{2}\left(\frac{1}{12}m^2 +\begin{cases}0& \residue{m}{4}=0\\ 1 & \residue{m}{4}=2\\\frac{5}{4} & \residue{m}{4}=1,3\end{cases} - \begin{cases} 0 & \residue{m}{3}=0\\  \frac{4}{3} & \residue{m}{3}=1,2\end{cases}\right)\,.
    \end{aligned}
\end{equation}
Then, if we let
\begin{equation}
m = -4q\,, \quad \quotparam = \residue{m}{4}\,, \quad \quotparamtwo = \residue{(-m)}{3} = \residue{4q}{3}\,,
\end{equation}
we find that
\begin{equation}
\valuation{}{W_{m}^{(2)}} - \valuation{}{W_{m}^{(1)}} = \frac{1}{2}\left(\frac{4}{3}q^2 + \begin{cases}0 &\quotparam=0\\ 1 & \quotparam=2\\\frac{5}{4} & \quotparam=1,3\end{cases} - \begin{cases} 0 & \quotparamtwo=0\\  \frac{4}{3} & \quotparamtwo=1,2\end{cases}\right)\,,
\end{equation}
exactly in agreement with the proposed formula for $\ordvanishtwo{\secz}$.

\paragraph{Spinors of $\aso(12)$} For the spinors of $\aso(12)$, we considered only one seed model given by \cref{eq:so12spinormodel}. At the codimension-one locus supporting the $\singtype{I}_{2}^{*}$ singularities, the section components $(\secx,\secy,\secz,\secw)$ vanish to orders $(1,3,0,3)$. Therefore, the EDS valuations corresponding to the codimension-one behavior are given by \cref{eq:soevenvaluationsmu1} with $n=6$:
\begin{equation}
    \valuation{}{\eds_{m}^{(1)}} = \frac{6}{8}m^2 - \frac{1}{2}\begin{cases}0 &  2 \mid m\\ \frac{6}{4} &  2 \nmid m \end{cases}\,.
\end{equation}
At the codimension-two $\singtype{III}^*$ locus supporting the $\bm{32^\prime}_{\frac{1}{2}} \oplus \bm{1}_{1}$ matter, the section components vanish to orders $(2,3,1,4)$. The EDS valuations corresponding to the codimension-two behavior, which are calculated in \cref{app:e7edsnu0}, are given by $\valuation{}{\eds_{m}^{(2)}} = m^2$. The difference of the EDS valuations is given by
\begin{equation}
    \valuation{}{\eds_{m}^{(2)}}-\valuation{}{\eds_{m}^{(1)}} = \frac{1}{2}\left(\frac{1}{2}m^2 +\begin{cases}0 &  2 \mid m\\ \frac{3}{2} &  2 \nmid m \end{cases}\right)\,.
\end{equation}
After noting that $q=\frac{m}{2}$, $\quotparam=\residue{m}{2}$ or $3\residue{m}{2}$, and $\quotparamtwo=0$, we can rewrite this expression as
\begin{equation}
    \valuation{}{\eds_{m}^{(2)}}-\valuation{}{\eds_{m}^{(1)}} = \frac{1}{2}\left(2q^2+\quotfunction{\SO(12)}{\quotparam} - \quotfunction{\gE_7}{\quotparamtwo}\right)\,.
\end{equation}
Thus, we have recovered the expected expression for $\ordvanishtwo{\secz}$ from \cref{sec:proposals}.

\paragraph{Spinors of $\aso(14)$} For the spinors of $\aso(14)$, we considered a single seed model described by \cref{eq:so14spinorseed}. The $(\secx,\secy,\secz,\secw)$ sections components vanish to orders $(1,3,0,4)$ at the codimension-one $\aso(14)$ locus and to orders $(2,3,1,4)$ at the codimension-two $\singtype{II}^*$ matter locus. We therefore subtract the EDS valuation for $\singtype{I}^*_{3}$ singularities given in \cref{eq:sooddvaluations} (with $n=7$ and $s=1$) from those for $\singtype{II}^*$ singularities in \cref{eq:edsvale8}:
\begin{equation}
    \begin{aligned}
        \valuation{}{W_{m}^{(2)}} - \valuation{}{W_{m}^{(1)}} &= m^2 - \left(\frac{7}{8}m^2 - \frac{1}{2}\begin{cases}0& 4\mid m\\ 1 & (2\mid m) \text{ and }(4\nmid m)\\\frac{7}{4} & (2\nmid m)\end{cases}\right) \\
        &= \frac{1}{8}m^2 + \frac{1}{2}\begin{cases}0 & \residue{m}{4}=0\\ 1 & \residue{m}{4}=2\\\frac{7}{4} & \residue{m}{4}=1,3\end{cases}\,.
    \end{aligned}
\end{equation}
Letting
\begin{equation}
m = 4q\,, \quad \quotparam = \residue{m}{4}\,, \quad \quotparamtwo = 0\,,
\end{equation}
we find that
\begin{equation}
\valuation{}{W_{m}^{(2)}} - \valuation{}{W_{m}^{(1)}} = \frac{1}{2}\left(4q^2 + \frac{1}{2}\begin{cases}0 & \quotparam=0\\ 1 &  \quotparam=2\\\frac{7}{4} &  \quotparam=1,3\end{cases}\right)\,,
\end{equation}
exactly in agreement with  the formula for $\ordvanishtwo{\secz}$ expected from the proposals.

\section{Representations of $\aE_6\oplus\au(1)$}
\label{sec:e6}
For models with $\aE_6\oplus\au(1)$ gauge algebras, we are primarily interested in matter in the $\bm{27}$ representation of $\aE_6$. Such matter occurs at the intersection of the codimension-one $\singtype{IV}^{*}$ locus supporting the $\aE_6$ gauge algebra with the residual $\singtype{I}_1$ discriminant locus. The singularity type enhances to $\singtype{III}^{*}$ at this codimension-two locus, or $\gE_7$ in the terms of ADE groups. The center of $\gE_6$ is $\Z_3$, whose elements we denote by $\quotparam=0,1,2$. For a gauge group with a quotient involving the $\quotparam$ element of the center, the allowed $\au(1)$ charges for  $\bm{27}$ matter are then
\begin{equation}
\frac{\quotparam}{3} + j \text{ for } j\in \Z\,.
\end{equation}
In other words, a model with an $\aE_6 \oplus \au(1)$ gauge algebra can support $\bm{27}_{q}$ matter when the gauge group quotient involves the element
\begin{equation}
\quotparam = \residue{3q}{3}\,.
\end{equation}
of the $\gE_6$ center. The allowed charges for $\quotparam=1$ and $\quotparam=2$ are equivalent up to a sign, as discussed in \cref{sec:excenters}.

To find the orders of vanishing for the $\bm{27}$ charges, we can start with the Weierstrass model
\begin{equation}
\label{eq:e6weiermodel}
y^2 = x^3 + \sigma ^3 f_ 3 x z^4 + \sigma ^4 \left(y_ 2^2-f_3 \sigma  x_2-x_2^3\sigma^2\right)z^6\,,
\end{equation}
which admits the section $\ratsec{s}$ with coordinates
\begin{equation}
\label{eq:e6gensec}
[\secx:\secy:\secz] =  \left[\sigma ^2 x_ 2:\sigma ^2 y_ 2:1\right]\,, \quad \secw = \sigma^3\left(f_3 + 3 x_2^2 \sigma\right)\,.
\end{equation}
The $f$ and $g$ of the Weierstrass model are respectively proportional to $\sigma^3$ and $\sigma^4$, and the order $\sigma^4$ term in $g$ is a perfect square. We therefore have an $\aE_6$ gauge factor tuned along $\locus{\sigma=0}$. Including the $\au(1)$ generated by the section above, the gauge algebra is $\aE_6 \oplus \au(1)$. This model is in fact equivalent to the $\mathcal{Q}(3,2,2,1,0,1,2)$ model of~\cite{KuntzlerTateTrees} with $b_{0,0}$ set to 1. By comparison to the spectrum given there, we can determine that the model supports
$\bm{27}_{1/3}$ matter at $\locus{\sigma=y_2=0}$ and that $\quotparam=1$. (Alternatively, we can take $\quotparam$ to be 2 and say that  $\locus{\sigma=y_2=0}$ supports $\bm{27}_{-1/3}$ matter.)
There is also $\bm{1}_{1}$ matter supported at $\locus{y_{2}=f_{3}+3 x_{2}^2\sigma=0}$. For a 6D F-theory model constructed from this elliptic fibration, this matter spectrum, together with the $1+\frac{1}{2}\divclass{\sigma}\cdot(\canonclass + \divclass{\sigma})$ hypermultiplets of $\bm{78}_0$ matter, satisfies the anomaly cancellation conditions with $\height = -2\canonclass -\frac{4}{3}\divclass{\sigma}$. The gauge group is therefore $(\gE_6\times\U(1))/\Z_3$.

The sections $m\ratsec{s}$ realize \pscharge{} $m/3$ at the $\bm{27}$ locus $\locus{\sigma=y_2=0}$ with $\quotparam=\residue{m}{3}$, giving us information about the orders of vanishing at $\bm{27}_q$ loci for all of the allowed charges.  \cref{tab:e6xu1orders} lists the orders of vanishing at $\locus{\sigma=0}$ at $\locus{\sigma=y_2=0}$ for some of these \pscharge{}s. The codimension-one orders of vanishing for a \pscharge{} $q$, which would correspond to a quotient involving the $\quotparam=\residue{3q}{3}$ element of the center, are described by the formula
\begin{equation}
\label{eq:e6codimone}
\ordvanishone{\secz} = 0\,, \quad \left(\ordvanishone{\secx}, \ordvanishone{\secy}, \ordvanishone{\secw}\right) = \quottriplet{\gE_6}{\quotparam}\,.
\end{equation}
And if we let
\begin{equation}
\quotparamtwo = \residue{3q}{2}\,,
\end{equation}
the codimension-two orders of vanishing follow the expressions
\begin{equation}
\ordvanishtwo{\secz} = \frac{1}{2}\left(\frac{3}{2}q^2 + \quotfunction{\gE_6}{\quotparam} - \quotfunction{\gE_7}{\quotparamtwo}\right)\label{eq:e6codimtwoz}
\end{equation}
and
\begin{equation}
\label{eq:e6codimtwoxy}
\left(\ordvanishtwo{\secx}, \ordvanishtwo{\secy}, \ordvanishtwo{\secw}\right) = (2,3,4)\times\ordvanishtwo{\secz} +\quottriplet{\gE_7}{\quotparamtwo}\,.
\end{equation}
Since $d_{\gE_6}=3$ and $d_{\gE_7}=2$, these formulas are in exact agreement with the proposals from \cref{sec:proposals}. We expect that the generating section components should vanish to these same orders at genuine $\bm{27}_q$ loci in $\aE_6\oplus\au(1)$ models.

\begin{table}
    \centering

    \begin{tabular}{*{4}{>{$}c<{$}}}\toprule
        \abs{q}      & \quotparam      & \text{Codimension One} & \text{Codimension Two} \\ \midrule
        \frac{1}{3}  & 1 \text{ or } 2 & (2,2,0,3)       & (2,3,0,3) \\[0.2em]
        \frac{2}{3}  & 1 \text{ or } 2 & (2,2,0,3)       & (2,3,1,4) \\[0.2em]
        1            & 0               & (0,0,0,0)       & (2,3,0,3) \\[0.2em]
        \frac{4}{3}  & 1 \text{ or } 2 & (2,2,0,3)       & (4,6,2,8) \\[0.2em]
        \frac{5}{3}  & 1 \text{ or } 2 & (2,2,0,3)       & (6,9,2,11) \\[0.2em]
        2            & 0               & (0,0,0,0)       & (6,9,3,12) \\[0.2em]
        \frac{7}{3}  & 1 \text{ or } 2 & (2,2,0,3)       & (10,15,4,19) \\[0.2em]
        \frac{8}{3}  & 1 \text{ or } 2 & (2,2,0,3)       & (12,18,6,24) \\[0.2em]
        3            & 0               & (0,0,0,0)       & (14,21,6,27) \\[0.2em]
        \frac{10}{3} & 1 \text{ or } 2 & (2,2,0,3)       & (18,27,9,36) \\[0.2em]
        \frac{11}{3} & 1 \text{ or } 2 & (2,2,0,3)       & (22,33,10,43) \\[0.2em]
        4            & 0               & (0,0,0,0)       & (24,36,12,48) \\[0.2em]
        \frac{13}{3} & 1 \text{ or } 2 & (2,2,0,3)       & (30,45,14,59) \\[0.2em]
        \frac{14}{3} & 1 \text{ or } 2 & (2,2,0,3)       & (34,51,17,68) \\[0.2em]
        5            & 0               & (0,0,0,0)       & (38,57,18,75) \\[0.2em]
        \frac{16}{3} & 1 \text{ or } 2 & (2,2,0,3)       & (44,66,22,88) \\[0.2em]
        \frac{17}{3} & 1 \text{ or } 2 & (2,2,0,3)       & (50,75,24,99) \\[0.2em]
        6            & 0               & (0,0,0,0)       & (54,81,27,108) \\[0.2em]
        \frac{19}{3} & 1 \text{ or } 2 & (2,2,0,3)       & (62,93,30,123) \\[0.2em]
        \frac{20}{3} & 1 \text{ or } 2 & (2,2,0,3)       & (68,102,34,136) \\[0.2em]
        7            & 0               & (0,0,0,0)       & (74,111,36,147) \\[0.2em]
        \frac{22}{3} & 1 \text{ or } 2 & (2,2,0,3)       & (82,123,41,164) \\[0.2em]
        \frac{23}{3} & 1 \text{ or } 2 & (2,2,0,3)       & (90,135,44,179) \\[0.2em]
        8            & 0               & (0,0,0,0)       & (96,144,48,192) \\ \bottomrule
    \end{tabular}

    \caption{Orders of vanishing of $(\secx, \secy, \secz,\secw)$ for combinations of $\quotparam$ and $\bm{27}$ \pscharge{}s $q$ at codimension one and codimension two. These data are found by calculating multiples of the generating section \labelcref{eq:e6gensec} for the $\aE_6\oplus\au(1)$ Weierstrass model in \cref{eq:e6weiermodel}.}
    \label{tab:e6xu1orders}
\end{table}

\subsection{Connection to EDS valuations}
Because all of the allowed combinations of $\quotparam$ and \pscharge{} can be obtained from multiples of the generating section $\ratsec{s}$, we expect that \cref{eq:e6codimtwoz} should agree with formulas in~\cite{StangeEllTrouble} describing the valuations of elliptic divisibility sequences. As with our previous examples, we must consider two sets of EDS valuations: $\valuation{}{\eds_{m}^{(1)}}$ for the codimension-one singularity type and $\valuation{}{\eds_{m}^{(2)}}$ for the codimension-two singularity type. The EDS valuations for the codimension-one $\singtype{IV}^*$ singularity type are calculated in \cref{app:e6eds} and are given by \cref{eq:edsvale6}:
\begin{equation}
    \valuation{}{\eds_{m}^{(1)}} = \frac{2}{3}m^2 - \begin{cases}0&3\mid m \\ \frac{2}{3} & 3\nmid m\end{cases}\,.
\end{equation}
The EDS valuations for the codimension-two $\singtype{III}^*$ singularities, meanwhile, are calculated in \cref{app:e7edsnu1}, and are given by \cref{eq:e7edsnu1final}:
\begin{equation}
    \valuation{}{\eds_{m}^{(2)}} = \frac{3}{4}m^2 - \begin{cases}0&2\mid m\\ \frac{3}{4} & 2\nmid m\end{cases}\,.
\end{equation}

The orders of vanishing of $\secz$ at codimension-two for multiples $m\ratsec{s}$ of the generating section $\ratsec{s}$ should then be given by $\valuation{}{\eds_m^{(2)}}- \valuation{}{\eds_m^{(1)}}$:
\begin{equation}
\ordvanishtwo{\secz} = \frac{3}{4}m^2 -\frac{2}{3}m^2 - \begin{cases} 0 & 2\mid m\\ \frac{3}{4} & 2\nmid m \end{cases} +  \begin{cases}0 & 3 \mid m \\ \frac{2}{3} & 3\nmid m\end{cases}\,.
\end{equation}
Since $m\ratsec{s}$ admits \pscharge{} $q=\frac{m}{3}$ for the $\bm{27}$ matter, this expression can be rewritten as
\begin{equation}
\ordvanishtwo{\secz} = \frac{3}{4}q^2 - \begin{cases} 0 & 2\mid (3q)\\ \frac{3}{4} & 2\nmid (3q) \end{cases} +  \begin{cases}0 & 3 \mid (3q) \\ \frac{2}{3} & 3\nmid (3q)\end{cases}\,.
\end{equation}
Recalling that $\quotparam=\residue{3q}{3}$ and that $\quotparamtwo=\residue{3q}{2}$, this formula exactly matches \cref{eq:e6codimtwoz}, as expected.

\section{Representations of $\aE_7\oplus\au(1)$}
\label{sec:e7}

For models with $\aE_7\oplus\au(1)$ gauge algebras, we are mostly interested in matter in the $\bm{56}$ representation. The center of $\gE_7$ is $\Z_2$, whose elements we denote as $\quotparam = 0,1$. As discussed in \cref{sec:excenters}, when the quotient involves the $\quotparam$ element of the center, the allowed $\au(1)$ charges for $\bm{56}$ matter are
\begin{equation}
\label{eq:e7possiblecharges}
\frac{\quotparam}{2} + j \text{ for }j\in\Z\,.
\end{equation}
In other words, a model with an $\aE_7\oplus\au(1)$ gauge algebra can support $\bm{56}_q$ matter when the quotient involves the element of the center corresponding to
\begin{equation}
\quotparam = \residue{2q}{2}\,.
\end{equation}

In F-theory models, $\bm{56}$ matter occurs at codimension-two loci where the singularity type enhances from from $\singtype{III}^*$ to $\singtype{II}^*$. Phrased in the ADE language, the singularity type enhances from $\gE_7$ to $\gE_8$. The $\bm{56}$ representation is pseudoreal, and if the supported $\bm{56}$ matter is uncharged under the $\au(1)$, it can be realized as half-hypermultiplets in 6D models. The fibers at such loci would take the shape of an incompletely resolved $\gE_8$ diagram~\cite{MorrisonTaylorMaS, KanHalfHyper}. However, if the $\bm{56}$ matter has a nonzero $\au(1)$ charge, the total representation, which includes the $\au(1)$ charge, is no longer pseudoreal. Therefore, $\bm{56}_q$ matter can only come in full hypermultiplets in 6D when $q$ is nonzero, and the fibers at the corresponding codimension-two loci form completely resolved $\gE_8$ diagrams. Since we are almost exclusively interested in $\au(1)$-charged $\bm{56}$ matter, we primarily deal with full hypermultiplets. The pseudoreal nature of the $\bm{56}$ representation also implies that, in line with the arguments in \cref{sec:signs}, the sign of the $\au(1)$ charge is unimportant for $\bm{56}$ matter.

The codimension-two $\singtype{II}^*$ loci admitting complete resolutions support both $\bm{56}$ and singlet matter. Specifically, in 6D models, an irreducible codimension-two locus with a completely resolvable $\singtype{II}^*$ singular fiber supports a full hypermultiplet of $\bm{56}$ matter and a full hypermultiplet of singlet matter. Just as was the case for the spinors of $\aso(12)$, the appearance of this extra singlet can be understood using the Katz--Vafa method and the following branching rule:
\begin{equation}
\begin{aligned}
\gE_8 &\to \gE_7 \times \SU(2)\,, \\
\bm{248} &\to (\bm{133},\bm{1}) \oplus (\bm{56},\bm{2}) \oplus (\bm{1},\bm{3})\,.
\end{aligned}
\end{equation}
The $(\bm{56},\bm{2})$ degrees of freedom combine to form the full hypermultiplet of $\bm{56}$ matter. The singlet hypermultiplet is formed by the degrees of freedom corresponding to the positive and negative roots of the $(\bm{1},\bm{3})$ in this decomposition.

This additional singlet can also have a $\au(1)$ charge, raising the question of how this singlet's charge compares to the charge of the $\bm{56}$ matter. In the resolved $\gE_8$ fiber at the codimension-two locus, there are two curves, $\gamma_+$ and $\gamma_-$, which represent the highest $\bm{56}$ weights in the $(\bm{56},\bm{2})$ of the decomposition. The $\au(1)$ charges for these $\bm{56}$ weights should be negatives of each other, implying that
\begin{equation}
 q_{\bm{56}} = \sigma(\ratsec{s})\cdot\gamma_+  = -\sigma(\ratsec{s})\cdot \gamma_-\,,
\end{equation}
where $\ratsec{s}$ is the generating section for the $\au(1)$. These two weights are also the highest and lowest weights of an $\asu(2)$ fundamental, so the curves are related by
\begin{equation}
\gamma_+ = \gamma_- + c\,.
\end{equation}
The curve $c$ represents a root of the $(\bm{1},\bm{3})$ part of the decomposition, and it corresponds to the singlet matter. The charge of this singlet is
\begin{equation}
q_{\bm{1}} =  \sigma(\ratsec{s})\cdot c  = \sigma(\ratsec{s})\cdot\gamma_{+} - \sigma(\ratsec{s})\cdot\gamma_{-} = 2  \sigma(\ratsec{s})\cdot\gamma_{+} = 2 q_{56}\,.
\end{equation}
In other words, the charge of the singlet supported at the $\singtype{II}^*$ locus should be twice that of $\bm{56}$ matter. Since the singlet representation is real, the arguments in \cref{sec:signs} suggest that the sign of the singlet charge does not matter.

We can now investigate how the orders of vanishing of the section components encode the charges of the $\bm{56}$ matter and the associated singlet. We start with the Weierstrass model
\begin{equation}
    \label{eq:e7weiermodel}
    \begin{aligned}
        y^2 &= x^3 -\frac{1}{3} \sigma^3 \left(3 b^2 c_{0,3}+\sigma  \left(c_{2,2}^2-3 c_{1,3} c_{3,1}\right)\right)x z^4 \\
        &\quad + \frac{1}{108} \sigma ^5 \left(9 b^2 \left(3 c_{1,3}^2 \sigma -8 c_{0,3} c_{2,2}\right)+4 \left(27 c_{0,3} c_{3,1}^2-9 c_{1,3} c_{2,2} c_{3,1} \sigma +2 c_{2,2}^3 \sigma \right)\right) z^6\,,
    \end{aligned}
\end{equation}
which admits a generating section $\ratsec{s}$ with components
\begin{equation}z
    \label{eq:e7gensec}
    \begin{gathered}
        [\secx:\secy:\secz] = \left[c_{3,1}^2 \sigma ^2-\frac{2}{3} b^2 c_{2,2} \sigma ^2:-\frac{1}{2} \sigma ^3 \left(b^4 c_{1,3}-2 b^2 c_{2,2} c_{3,1}+2 c_{3,1}^3\right):b\right]\,, \\
        \secw = -\sigma^3 \left(b^6 c_{0,3}-b^4 c_{2,2}^2 \sigma -b^4 c_{1,3} c_{3,1} \sigma +4 b^2 c_{2,2} c_{3,1}^2 \sigma -3 c_{3,1}^4 \sigma \right)\,.
    \end{gathered}
\end{equation}
This model is equivalent to the $\mathcal{Q}(3, 3, 2, 1, 0, 1, 2)$ model of~\cite{KuntzlerTateTrees} after trivial redefinitions of the parameters. It supports an $\aE_7$ gauge algebra along $\locus{\sigma=0}$. In 6D F-theory models constructed from this elliptic fibration, the codimension-two locus $\locus{\sigma=b=0}$ supports full hypermultiplets of $\bm{56}_{1/2} \oplus \bm{1}_1$ matter, in line with the observations in~\cite{KuntzlerTateTrees}.
 The model also supports $\bm{1}_2$ hypermultiplets at $\locus{b=c_{3,1}=0}$ and $\bm{1}_{1}$ hypermultiplets at
\begin{equation}
\left\locus{\frac{\secy}{\sigma^3} = \frac{\secw}{\sigma^3}=0\right}  \textbackslash \locus{b=c_{3,1}=0}\,.
\end{equation}
While $f$ and $g$ vanish to orders $4$ and $6$ at the codimension-two locus $\locus{\sigma=c_{0,3}=0}$, one can avoid the problems created by this locus by choosing the classes of the parameters such that $\divclass{\sigma}\cdot\divclass{c_{0,3}} = 0$. For instance, one can choose the divisor classes such that $\divclass{c_{0,3}}=0$, in which case $\divclass{b}=-2\canonclass{} -\frac{3}{2}\divclass{\sigma}$. In this situation, one can verify that the charged matter spectrum satisfies the gauge and mixed gauge--gravitational anomaly cancellation conditions with $\height = -2\canonclass{} + 2\divclass{b} - \frac{3}{2} \divclass{\sigma}$. From the $\au(1)$ charge of the $\bm{56}$ matter, one can see that the generating section, and therefore the model as a whole, has $\quotparam=1$.

We now consider multiples of the generating section $\ratsec{s}$. The $\bm{56}$ matter at $\locus{\sigma=b=0}$ would have \pscharge{} $q=\frac{1}{2}m$ for the section $m\ratsec{s}$ with $\quotparam=\residue{m}{2}$. The $m\ratsec{s}$ sections therefore give us analogues for all of the possible $\bm{56}$ charges described \cref{eq:e7possiblecharges}. \cref{tab:e7xu1orders} lists the orders of vanishing of the $m\ratsec{s}$ section components for various values of $m$, which follow a clear pattern. The codimension-one orders of vanishing are described by the equations
\begin{equation}
\ordvanishone{\secz} = 0\,, \quad \left(\ordvanishone{\secx},\ordvanishone{\secy},\ordvanishone{\secw}\right) = \quottriplet{\gE_7}{\quotparam}\,.
\end{equation}
The codimension-two orders of vanishing are given by
\begin{gather}
\ordvanishtwo{\secz} = \frac{1}{2}\left(2 q^2 + \quotfunction{\gE_7}{\quotparam} \right) \label{eq:e7ord2z56}\,, \\
  \left(\ordvanishtwo{\secx},\ordvanishtwo{\secy},\ordvanishtwo{\secw}\right) = \left(2,3,4\right)\times\ordvanishtwo{\secz}\,.
\end{gather}
The center of $\gE_{8}$ is trivial, and $\quotparamtwo$, $\quotfunction{\gE_{8}}{\quotparamtwo}$, and $\quottriplet{\gE_8}{\quotparamtwo}$ are all trivial as well. These formulas therefore agree with the expectations from \cref{sec:proposals}.

\begin{table}
    \centering

    \begin{tabular}{*{4}{>{$}c<{$}}}\toprule
        \abs{q}      & \quotparam & \text{Codimension One} & \text{Codimension Two} \\ \midrule
        \frac{1}{2}  & 1          & (2,3,0,3)              & (2,3,1,4) \\[0.2em]
        1            & 0          & (0,0,0,0)              & (2,3,1,4) \\[0.2em]
        \frac{3}{2}  & 1          & (2,3,0,3)              & (6,9,3,12) \\[0.2em]
        2            & 0          & (0,0,0,0)              & (8,12,4,16) \\[0.2em]
        \frac{5}{2}  & 1          & (2,3,0,3)              & (14,21,7,28) \\[0.2em]
        3            & 0          & (0,0,0,0)              & (18,27,9,36) \\[0.2em]
        \frac{7}{2}  & 1          & (2,3,0,3)              & (26,39,13,52) \\[0.2em]
        4            & 0          & (0,0,0,0)              & (32,48,16,64) \\[0.2em]
        \frac{9}{2}  & 1          & (2,3,0,3)              & (42,63,21,84) \\[0.2em]
        5            & 0          & (0,0,0,0)              & (50,75,25,100) \\[0.2em]
        \frac{11}{2} & 1          & (2,3,0,3)              & (62,93,31,124) \\[0.2em]
        6            & 0          & (0,0,0,0)              & (72,108,36,144) \\[0.2em]
        \frac{13}{2} & 1          & (2,3,0,3)              & (86,129,43,172) \\[0.2em]
        7            & 0          & (0,0,0,0)              & (98,147,49,196) \\[0.2em]
        \frac{15}{2} & 1          & (2,3,0,3)              & (114,171,57,228) \\[0.2em]
        8            & 0          & (0,0,0,0)              & (128,192,64,256) \\[0.2em]
        \frac{17}{2} & 1          & (2,3,0,3)              & (146,219,73,292) \\[0.2em]
        9            & 0          & (0,0,0,0)              & (162,243,81,324) \\ \bottomrule
    \end{tabular}

    \caption{Orders of vanishing of $(\secx, \secy, \secz,\secw)$ for combinations of $\quotparam$ and the $\bm{56}$ \pscharge{} $q$ at codimension one ($\locus{\sigma=0}$) and codimension two ($\locus{\sigma=b=0}$). These data are found by calculating multiples of the generating section \labelcref{eq:e7gensec} for the $\aE_7\oplus\au(1)$ Weierstrass model in \cref{eq:e7weiermodel}; \pscharge{} $q$ is realized by the section $2q\ratsec{s}$. One should observe the same orders of vanishing in Weierstrass models supporting $\bm{56}_q\oplus\bm{1}_{2q}$ matter.}
    \label{tab:e7xu1orders}
\end{table}

\subsection{Connection to EDS valuations}
Since all of the allowed combinations of $\quotparam$ and \pscharge{} $q$ can be obtained from multiples of the generating section $\ratsec{s}$, we expect that the formulas for the orders of vanishing above should agree with the formulas for EDS valuations in~\cite{StangeEllTrouble}. As before, we start by finding the EDS valuations corresponding to the codimension-one behavior and the codimension-two behavior of \cref{eq:e7weiermodel}. For the $\singtype{III}^*$ singularities at codimension-one, the generating section \labelcref{eq:e7gensec} vanishes to orders
$(2,2,0,3)$ at $\locus{\sigma=0}$. The corresponding EDS valuations are calculated by a procedure similar to that in \cref{app:e7edsnu1} and are given by \cref{eq:e7edsnu1final}:
\begin{equation}
    \valuation{}{\eds_{m}^{(1)}} = \frac{3}{4}m^2 - \begin{cases}0 & 2\mid m\\\frac{3}{4} & 2\nmid m\end{cases}\,.
\end{equation}
For the $\singtype{II}^*$ singularities at codimension-two, the generating section \labelcref{eq:e7gensec} vanishes to orders  $(2,3,1,4)$ at $\locus{\sigma=b=0}$. The codimension-two EDS valuations are calculated by a procedure similar to that in \cref{app:e8eds} and are given by \cref{eq:edsvale8}:
\begin{equation}
    \valuation{}{\eds_{m}^{(2)}} = m^2\,.
\end{equation}

The formula for $\ordvanishtwo{\secz}$ of the $m\ratsec{s}$ sections should be given by the difference
\begin{equation}
\valuation{}{\eds^{(2)}_{m}} - \valuation{}{\eds^{(1)}_{m}}  = \frac{1}{4}m^2 + \begin{cases} 0 & 2\mid m \\ \frac{3}{4} & 2\nmid m \end{cases}\,.
\end{equation}
Noting that the section $m\ratsec{s}$ realizes \pscharge{} $q=\frac{m}{2}$ for the $\bm{56}$ matter, we should expect that
\begin{equation}
\ordvanishtwo{\secz} = q^2 + \frac{3}{4} \residue{2q}{2} = \frac{1}{2} \left(2 q^2 + \frac{3}{2} \residue{2q}{2}\right)\,,
\end{equation}
exactly in agreement with the observed order of vanishing data, \cref{eq:e7ord2z56}, and the proposals in \cref{sec:proposals}.

\section{Matter charged under multiple gauge algebra factors}
\label{sec:multgaugealgebras}
Up to this point, we have only analyzed situations where the gauge algebra is either $\au(1)$ or $\aG\oplus\au(1)$, where $\aG$ is a simple Lie algebra. However, F-theory models can have more complicated gauge algebras consisting of multiple $\au(1)$ and nonabelian factors. Matter in such models can be charged under many of these factors. For instance, an F-theory model might have an $\asu(m)\oplus\asu(n)\oplus\au(1)\oplus\au(1)$ gauge algebra and matter in the $(\bm{m},\overline{\bm{n}})_{q_1,q_2}$ representation. This section focuses on these situations. We divide the analysis into two parts. We first discuss how to approach situations with matter charged under multiple $\au(1)$ factors. We then turn to matter charged under multiple simple nonabelian factors and (at least) one $\au(1)$. Despite the added complications in these examples, we will see that the connections between the $\au(1)$ charges and orders of vanishing proposed in \cref{sec:proposals} still hold.

\subsection{Multiple $\au(1)$ factors}
If there are $\numabelian$ $\au(1)$ factors, the Mordell--Weil rank of the corresponding elliptic fibration should be $\numabelian$, and each $\au(1)$ factor corresponds to a particular generating section. The formulas in \cref{sec:proposals} should hold without modification for each generating section. More specifically, if the nonabelian part of the gauge algebra is simple and nontrivial, each generating section has an associated integer $\quotparam$ that specifies which component of the resolved fiber at codimension-one is hit by the section.\footnote{If the nonabelian part of the gauge algebra is semisimple with $K$ simple nonabelian factors, there are a total of $K\times L$ integers playing the role of $\quotparam$, one for each combination of generating section and simple nonabelian factor.} Along the codimension-one locus supporting this nonabelian gauge algebra, the section components for a particular generating section should vanish to the orders dictated by its $\quotparam$ integer. If some matter has a certain charge under a particular $\au(1)$, the section components for the corresponding generating section should vanish to the orders proposed in \cref{sec:proposals} at the matter locus. To determine some matter field's charge under all of the $\numabelian$ $\au(1)$ factors, one would apply the procedures outlined in \cref{sec:proposals} to each generating section.

Note that there can be different choices for the basis of these $\au(1)$'s when there are multiple $\au(1)$ factors. The $\au(1)$ charges of matter can change if one uses a different $\au(1)$ basis. In the elliptic fibration, the basis of $\au(1)$ algebras corresponds to the basis of generating sections. If one changes the $\au(1)$ basis, the generating section basis should change as well, and the new $\au(1)$ charges should agree with the orders of vanishing for the new basis sections as per the rules in \cref{sec:proposals}. Thus, even though the charge spectrum changes with the $\au(1)$ basis, the corresponding change in the basis of sections should ensure that the formulas in \cref{sec:proposals} still hold. As long as one maintains a consistent relation between the $\au(1)$ basis and the basis of generating sections, changing the basis of $\au(1)$ algebras should not pose any problems.

Given that the formulas in \cref{sec:proposals} are essentially unchanged when there are multiple $\au(1)$ factors, we will not extensively discuss examples of this type here. However, \cref{sec:explicitconstructions} lists some models in the prior F-theory literature with multiple $\au(1)$ factors, all of which agree with the ideas discussed here and in \cref{sec:proposals}.

\subsection{Multiple nonabelian factors}
\label{sec:bifundamentals}
Now let us consider the case where the nonabelian part of the gauge algebra has multiple simple factors. Specifically, suppose the gauge algebra is
\begin{equation}
\bigoplus_{\kappa=1}^{K}\aG_{\kappa} \oplus \au(1)\,,
\end{equation}
where each $\aG_{\kappa}$ is a simple Lie algebra. As before, we assume that the $\aG_{\kappa}$ are all simply-laced. Because the argument above suggests that the presence of multiple $\au(1)$ factors does not significantly change the analysis, we assume for simplicity that there is only a single $\au(1)$ associated with a generating section $\ratsec{s}$. Each $\aG_{\kappa}$ is supported along a codimension-one locus $\locus{\sigma_{\kappa}=0}$ in the base with singular fibers of the appropriate type. We also have integers $\quotparam_{\kappa}$ describing which resolved fiber component along $\locus{\sigma_{\kappa}=0}$ is hit by $\ratsec{s}$. According to \cref{sec:proposals}, the section components for $\ratsec{s}$ should vanish to the orders described by \cref{eq:ordvanishonegen} at each of the $\aG_{\kappa}$ loci:
\begin{equation}
\ordvanish{\sigma_{k}=0}{\secz} = 0\,, \quad \left(\ordvanish{\sigma_{k}=0}{\secx},\ordvanish{\sigma_{k}=0}{\secy},\ordvanish{\sigma_{k}=0}{\secw}\right) = \quottriplet{G_{\kappa}}{\quotparam_{\kappa}}\,.
\end{equation}
Here, $G_{\kappa}$ is the universal covering group for $\aG_{\kappa}$. Now consider matter that is charged under multiple $\aG_{\kappa}$ and that is supported at the codimension-two intersection locus of the corresponding $\locus{\sigma_{\kappa}=0}$. If the singularity type enhances at this matter locus to one associated with the group $H$, the section components should vanish to the orders specified by Equations \labelcref{eq:ordvanishtwozgen} and \labelcref{eq:ordvanishtwoothergen} for some $\quotparamtwo$ in the center of $H$. The index $i$ in \cref{eq:ordvanishtwozgen} runs over those simple nonabelian algebras under which the matter is charged.

To illustrate these ideas, let us focus on models with $\asu(m)\oplus\asu(n)\oplus\au(1)$ gauge algebras and matter in the bifundamental $(\bm{m},\overline{\bm{n}})_{q}$ representation. At the codimension-two locus supporting the bifundamental matter, the singularity type enhances from $\singtype{I}_{m}^{s}\times \singtype{I}_{n}^{s}$ (or $\SU(m)\times \SU(n)$) to $\singtype{I}_{m+n}^{s}$ (or $\SU(m+n)$). If we have integers $(\quotparam_{m}, \quotparam_{n})$ describing the fiber components hit by $\ratsec{s}$, then the allowed $\au(1)$ charges for this bifundamental matter are given by
\begin{equation}
\frac{\quotparam_{m}}{m} - \frac{\quotparam_{n}}{n} + j \text{ for }j\in\Z\,.
\end{equation}
According to the formulas in \cref{sec:proposals}, the section components of $\ratsec{s}$ should vanish to the orders
\begin{equation}
\label{eq:bifundordersbegin}
\begin{gathered}
\ordvanish{\sigma_{m}=0}{\secz} =0\,, \\
\left(\ordvanish{\sigma_{m}=0}{\secx},\ordvanish{\sigma_{m}=0}{\secy},\ordvanish{\sigma_{m}=0}{\secw}\right) = \quottriplet{\SU(m)}{\quotparam_{m}} = (0,1,1)\times\moddist{m}{\quotparam_{m}}
\end{gathered}
\end{equation}
at the codimension-one locus $\locus{\sigma_{m}=0}$ supporting the $\asu(m)$ algebra and to the orders
\begin{equation}
\begin{gathered}
\ordvanish{\sigma_{n}=0}{\secz} =0\,, \\
\left(\ordvanish{\sigma_{n}=0}{\secx},\ordvanish{\sigma_{n}=0}{\secy},\ordvanish{\sigma_{n}=0}{\secw}\right) = \quottriplet{\SU(n)}{\quotparam_{n}} = (0,1,1)\times\moddist{n}{\quotparam_{n}}
\end{gathered}
\end{equation}
at the codimension-one locus $\locus{\sigma_{n}=0}$ supporting the $\asu(n)$ algebra. At the matter locus, the $\secz$ component of the generating section should vanish to order
\begin{equation}
    \begin{aligned}
        \ordvanishtwo{\secz} &= \frac{1}{2}\left(\frac{m n}{m+n}q^2 + \quotfunction{\SU(m)}{\quotparam_{m}} + \quotfunction{\SU(n)}{\quotparam_{n}} - \quotfunction{\SU(m+n)}{\quotparamtwo} \right) \\
        &= \frac{1}{2}\left(\frac{m n}{m+n}q^2 + \frac{\quotparam_m (m-\quotparam_{m})}{m} +  \frac{\quotparam_n (n-\quotparam_{n})}{n} -  \frac{\quotparamtwo (m+n-\quotparamtwo)}{m+n} \right)\,,
    \end{aligned}
\end{equation}
where $\quotparamtwo$ is an integer representing an element of the center of $\SU(m+n)$. Additionally, we would expect that
\begin{equation}
\left(\ordvanishtwo{\secx},\ordvanishtwo{\secy},\ordvanishtwo{\secw}\right) = (2,3,4)\times\ordvanishtwo{\secz} + \quottriplet{\SU(m+n)}{\quotparamtwo}
\end{equation}
One can verify that this formula gives integral $\ordvanishtwo{\secz}$ if
\begin{equation}
\label{eq:bifundordersend}
\quotparamtwo = \residue{\left(mq + \quotparam_{n} + \frac{m}{n}\quotparam_{n}\right)}{m+n}\,.
\end{equation}
This expression for $\quotparamtwo$ also reduces to the $\asu(m)$ fundamental $\quotparamtwo$ expression when $n=1$.

We can check these proposals for the bifundamentals by looking at a model with an $\asu(4)\oplus\asu(5)\oplus\au(1)$ gauge algebra. This example is given by the Weierstrass model
\begin{equation}
\label{eq:su4su5weier}
\begin{aligned}
    y^2 &= x^3 + \left(-\frac{c_2^2}{3}+c_1 c_3-c_0 b^2\right) x z^4 \\
    &\qquad\quad + \left(\frac{2 }{27}c_2^3-\frac{2}{3} b^2 c_0 c_2+\frac{1}{4} b^2 c_1^2-\frac{1}{3} c_1 c_3 c_2+c_0 c_3^2\right)z^6\,,
\end{aligned}
\end{equation}
where
\begin{equation}
\label{eq:su4su5tunings}
\begin{gathered}
c_0 = \frac{1}{4} \sigma_4^2 \sigma_5^4 \left({b_ 2^\prime}^2+4 {c_ 0^\prime} \sigma_4\right)\,, \\
c_1 = \frac{1}{2} \sigma_4 \sigma_5^2 ({b_ 1^\prime} {b_ 2^\prime}+2 {c_ 1^\prime} \sigma_4)\,, \\
c_2 = \frac{1}{4} \left(4 \sigma_4 \sigma_5 c_ 2^\prime+{b_ 1^\prime}^2\right)\,, \\
c_3 = \frac{b {b_1^\prime}}{2}+c_ 3^\prime \sigma_4 \sigma_5\,.
\end{gathered}
\end{equation}
Because the discriminant is proportional to $\sigma_4^4\sigma_5^5$ and the split conditions are satisfied, this model supports an $\asu(4)$ algebra on $\locus{\sigma_4 = 0}$ and an $\asu(5)$ algebra on $\locus{\sigma_5 = 0}$. Additionally, it admits a generating section $\ratsec{s}$ with section components
\begin{equation}
\begin{aligned}
\secx &= \left(\frac{b {b_1^\prime}}{2}+c_3^\prime \sigma_4 \sigma_5\right)^2-\frac{1}{6} b^2 \left({b_1^\prime}^2+4 c_2^\prime \sigma_4 \sigma_5\right)\,, \\
\secy &= -\frac{1}{4} \sigma_4 \sigma_5 \Big(b^4 {b_1^\prime} {b_2^\prime} \sigma_5+2 b^4 {c_1^\prime} \sigma_4 \sigma_5-2 b^3 {b_1^\prime} {c_2^\prime}+2 b^2 {b_1^\prime}^2 {c_3^\prime} \\
&\qquad\qquad\qquad\quad - 4 b^2 {c_2^\prime} {c_3^\prime} \sigma_4 \sigma_5+6 b {b_1^\prime} {c_3^\prime}^2 \sigma_4 \sigma_5+4 {c_3^\prime}^3 \sigma_4^2 \sigma_5^2\Big)\,, \\
\secz &= b\,, \\
\secw &= -\frac{1}{4} \sigma_4 \sigma_5 \Big(b^6 {b_2^\prime}^2 \sigma_4 \sigma_5^3+4 b^6 {c_0^\prime} \sigma_4^2 \sigma_5^3-b^5 {b_1^\prime}^2 {b_2^\prime} \sigma_5-2 b^5 {b_1^\prime} {c_1^\prime} \sigma_4 \sigma_5+2 b^4 {b_1^\prime}^2 {c_2^\prime} \\
&\qquad\qquad\qquad\quad - 2 b^4 {b_1^\prime} {b_2^\prime} {c_3^\prime} \sigma_4 \sigma_5^2-4 b^4 {c_1^\prime} {c_3^\prime} \sigma_4^2 \sigma_5^2-4 b^4 {c_2^\prime}^2 \sigma_4 \sigma_5 \\
&\qquad\qquad\qquad\quad - 2 b^3 {b_1^\prime}^3 {c_3^\prime}+16 b^3 {b_1^\prime} {c_2^\prime} {c_3^\prime} \sigma_4 \sigma_5-14 b^2 {b_1^\prime}^2 {c_3^\prime}^2 \sigma_4 \sigma_5 \\
&\qquad\qquad\qquad\quad + 16 b^2 {c_2^\prime} {c_3^\prime}^2 \sigma_4^2 \sigma_5^2-24 b {b_1^\prime} {c_3^\prime}^3 \sigma_4^2 \sigma_5^2-12 {c_3^\prime}^4 \sigma_4^3 \sigma_5^3\Big)\,.
\end{aligned}
\end{equation}
The complete gauge algebra is therefore $\asu(4)\oplus\asu(5)\oplus\au(1)$. The charged matter spectrum, which is summarized in \cref{tab:su5su4matterspectrum}, satisfies the anomaly cancellation conditions with $\height = -2\canonclass + 2\divclass{b} - \frac{3}{4}\divclass{\sigma_4} - \frac{4}{5}\divclass{\sigma_5}$. As can be seen from the $\au(1)$ charges, we can take $\quotparam_{4}$ and $\quotparam_{5}$ to both be 1;
this agrees with the orders of vanishing of $(\secx,\secy,\secz,\secw)$ at $\locus{\sigma_{4}=0}$ and $\locus{\sigma_{5}=0}$. The codimension-two orders of vanishing at the matter loci in \cref{tab:su5su4matterspectrum} also agree with the formulas discussed previously.

\begin{table}
    \centering

    \begin{tabular}{>{$}c<{$}@{\hskip 20pt}>{$}c<{$}>{$}c<{$}}\toprule
        \text{Matter} & \text{Locus} & \ordalt_{(2)} \\ \midrule
        (\bm{4},\overline{\bm{5}})_{\mathrlap{1/20}} & \locus{\sigma_{4}=\sigma_{5}=0} & (0,2,0,2) \\
        (\bm{4},\bm{1})_{\mathrlap{5/4}} & \locus{\sigma_{4}=b=0} & (2,3,1,4) \\
        (\bm{4},\bm{1})_{\mathrlap{-3/4}} &\locus{\sigma_{4}=b^2 {b_2^\prime} \sigma_5-2 b {c_2^\prime}+2 {b_1^\prime} {c_3^\prime}=0} & (0,2,0,2) \\
        (\bm{4},\bm{1})_{\mathrlap{1/4}} &  \locus{\sigma_{4}=b {b_2^\prime}^3 \sigma_5^2-2 {b_1^\prime}^2 {c_0^\prime}+2 {b_1^\prime} {b_2^\prime} {c_1^\prime}-2 {b_2^\prime}^2 {c_2^\prime} \sigma_5 = 0} & (0,1,0,1) \\
        (\bm{6},\bm{1})_{\mathrlap{1/2}} &  \locus{\sigma_{4}=b_1^\prime=0} & (1,2,0,2) \\
        (\bm{1},\bm{5})_{\mathrlap{6/5}} & \locus{\sigma_{5}=b=0} & (2,3,1,4) \\
        (\bm{1},\bm{5})_{\mathrlap{-4/5}} & \locus{\sigma_{5}=b {c_2^\prime}-{b_1^\prime} {c_3^\prime} = 0} & (0,2,0,2) \\
        (\bm{1},\bm{5})_{\mathrlap{1/5}} & \locus{\sigma_{5}={b_1^\prime}^2 {c_0^\prime}-{b_1^\prime} {b_2^\prime} {c_1^\prime}-{c_1^\prime}^2 \sigma_4=0} & (0,1,0,1) \\
        (\bm{1},\bm{10})_{\mathrlap{2/5}} & \locus{\sigma_{5}=b_1^\prime=0} & (1,2,0,2) \\
        (\bm{1},\bm{1})_{\mathrlap{2}} & \locus{b = c_3^\prime=0} & (1,2,3,4) \\ \bottomrule
    \end{tabular}

    \caption{Matter loci for the $\asu(4)\oplus\asu(5)\oplus\au(1)$ model, along with the codimension-two orders of vanishing of $(\secx,\secy,\secz,\secw)$. In addition to the matter loci listed here, there are $(\bm{1},\bm{1})_{1}$ singlets located at loci where $\secy/(\sigma_4\sigma_5)$ and $\secw/(\sigma_4\sigma_5)$ simultaneously vanish that are not part of the matter loci above. The model may also have delocalized $(\bm{6},\bm{1})_{0}$ and $(\bm{1},\bm{10})_{0}$ adjoint matter that can propagate along the $\locus{\sigma_{4}=0}$ and $\locus{\sigma_{5}=0}$ gauge divisors.}
    \label{tab:su5su4matterspectrum}
\end{table}

Importantly, the model admits $(\bm{4},\overline{\bm{5}})_{1/20}$ matter at the locus $\locus{\sigma_{4}=\sigma_{5}=0}$, where the section components vanish to orders $(0,2,0,2)$. As with the previous examples, we can use multiples of the generating section to predict the orders of vanishing for at least some combinations of $\quotparam_4$, $\quotparam_5$, and bifundamental $\au(1)$ charge $q$. Specifically, we can obtain the combinations of the form
\begin{equation}
\quotparam_{4} = \residue{r}{4}\,, \quad \quotparam_{5} = \residue{r}{5}\,, \quad q = \frac{r}{20}\,,
\end{equation}
where $r$ is an integer. The resulting orders of vanishing, which are listed in \cref{tab:su5su4data}, satisfy Equations \labelcref{eq:bifundordersbegin} through \labelcref{eq:bifundordersend}. Thus, the proposals in \cref{sec:proposals} hold for the $(\bm{4},\overline{\bm{5}})$ \pscharge{}s considered here, suggesting that they should hold generally for $(\bm{4},\overline{\bm{5}})_{q}$ matter with genuine $\au(1)$ charge $q$.

\begin{table}
    \centering

    \begin{tabular}{*{6}{>{$}c<{$}}}\toprule
        \abs{q}       & \quotparam_{4} & \quotparam_{5} & \ord_{\sigma_{4}=0} & \ord_{\sigma_{5}=0} & \ordalt_{(2)} \\ \midrule
        \frac{1}{20}  & 1              & 1              & (0,1,0,1)           & (0,1,0,1)           & (0,2,0,2) \\[0.2em]
        \frac{1}{10}  & 2              & 2              & (0,2,0,2)           & (0,2,0,2)           & (0,4,0,4) \\[0.2em]
        \frac{3}{20}  & 3              & 3              & (0,1,0,1)           & (0,2,0,2)           & (0,3,0,3) \\[0.2em]
        \frac{1}{5}   & 0              & 4              & (0,0,0,0)           & (0,1,0,1)           & (0,1,0,1) \\[0.2em]
        \frac{1}{4}   & 1              & 0              & (0,1,0,1)           & (0,0,0,0)           & (0,1,0,1) \\[0.2em]
        \frac{3}{10}  & 2              & 1              & (0,2,0,2)           & (0,1,0,1)           & (0,3,0,3) \\[0.2em]
        \frac{7}{20}  & 3              & 2              & (0,1,0,1)           & (0,2,0,2)           & (0,4,0,4) \\[0.2em]
        \frac{2}{5}   & 0              & 3              & (0,0,0,0)           & (0,2,0,2)           & (0,2,0,2) \\[0.2em]
        \frac{9}{20}  & 1              & 4              & (0,1,0,1)           & (0,1,0,1)           & (2,3,1,4) \\[0.2em]
        \frac{1}{2}   & 2              & 0              & (0,2,0,2)           & (0,0,0,0)           & (0,2,0,2) \\[0.2em]
        \frac{11}{20} & 3              & 1              & (0,1,0,1)           & (0,1,0,1)           & (0,4,0,4) \\[0.2em]
        \frac{3}{5}   & 0              & 2              & (0,0,0,0)           & (0,2,0,2)           & (0,3,0,3) \\[0.2em]
        \frac{13}{20} & 1              & 3              & (0,1,0,1)           & (0,2,0,2)           & (2,4,1,5) \\[0.2em]
        \frac{7}{10}  & 2              & 4              & (0,2,0,2)           & (0,1,0,1)           & (2,4,1,5) \\[0.2em]
        \frac{3}{4}   & 3              & 0              & (0,1,0,1)           & (0,0,0,0)           & (0,3,0,3) \\[0.2em]
        \frac{4}{5}   & 0              & 1              & (0,0,0,0)           & (0,1,0,1)           & (0,4,0,4) \\[0.2em]
        \frac{17}{20} & 1              & 2              & (0,1,0,1)           & (0,2,0,2)           & (2,5,1,6) \\ \bottomrule
    \end{tabular}

    \caption{Orders of vanishing for $(\secx,\secy,\secz,\secw)$ at codimension one and two for combinations of $\quotparam_{4}$, $\quotparam_{5}$, and \pscharge{} $q$ for $(\bm{4},\overline{\bm{5}})$ bifundamental matter. These data are found using the $\asu(4)\oplus\asu(5)\oplus\au(1)$ construction described by Equations \labelcref{eq:su4su5weier} and \labelcref{eq:su4su5tunings}.}
    \label{tab:su5su4data}
\end{table}

These same ideas should be applicable for representations beyond the $(\bm{m},\overline{\bm{n}})$ bifundamentals. One can also have $(\bm{m},\bm{n})_{q}$ bifundamentals of $\asu(m)\oplus\asu(n)\oplus\au(1)$, whose allowed $\au(1)$ charges are given by
\begin{equation}
\frac{\quotparam_m}{m} + \frac{\quotparam_{n}}{n} + j \text{ for }j\in\Z\,.
\end{equation}
These bifundamentals also occur at the intersection of codimension-one $\singtype{I}_m^{*s}$ and $\singtype{I}_n^{*s}$ loci where the singularity type enhances to $\singtype{I}_{m+n}^{*s}$. Therefore, we expect that, for instance,
\begin{equation}
\ordvanishtwo{\secz} = \frac{1}{2}\left(\frac{m n}{m+n}q^2 + \quotfunction{\SU(m)}{\quotparam_{m}} + \quotfunction{\SU(n)}{\quotparam_{n}} - \quotfunction{\SU(m+n)}{\quotparamtwo}\right)
\end{equation}
for some $\quotparamtwo$ representing an element of $\Z_{m+n}$, the center of $\SU(m+n)$. This formula gives integer $\ordvanishtwo{\secz}$ for
\begin{equation}
\quotparamtwo = \residue{\left(mq - \quotparam_{n} - \frac{m}{n}\quotparam_{n}\right)}{m+n}\,,
\end{equation}
and the expression expression for $\quotparamtwo$ also reduces to that for $\asu(m)$ fundamentals when $n=1$. The proposals in \cref{sec:proposals} should also cover other types of matter charged under multiple nonabelian gauge algebras, such as $(\bm{m},\bm{n})_{q}$ matter charged under $\aso(m)\oplus\asu(n)\oplus\au(1)$ algebras. However, there are also matter representations involving only a simple nonabelian algebra that occur at singularities of a gauge divisor. If the gauge algebra is $\asu(n)\oplus\au(1)$, for instance, one can have matter in the adjoint or symmetric representations of $\asu(n)$ localized at double point singularities of the gauge divisor~\cite{SadovGreenSchwarz,MorrisonTaylorMaS,CveticKleversPiraguaTaylor,AndersonGrayRaghuramTaylorMiT,KleversTaylor,KleversEtAlExotic}. If we focus on a sufficiently small neighborhood near the matter locus, these situations more closely resemble the bifundamental matter situations considered above: the matter locus locally appears to be the intersection of two codimension-one $\singtype{I}_n^{*s}$ loci, even though these two components are actually connected due to global properties of the gauge divisor. If such matter also has a $\au(1)$ charge, the orders of vanishing of the section components may be described by formulas akin to those for bifundamental matter. These more exotic matter representations lie outside of the scope of this paper, so we do not rigorously analyze them here. Nevertheless, they pose an interesting direction for future work.

\section{Examples in explicit constructions}
\label{sec:explicitconstructions}

We have so far performed somewhat indirect tests of the proposals in \cref{sec:proposals}. Specifically, we have used multiples of generating sections admitting relatively small $\au(1)$ charges as proxies for genuine generating sections admitting larger $\au(1)$ charges. Underlying this strategy is the assumption that both types of sections should behave similarly at the appropriate loci. Since the $\au(1)$ charge only depends on the local behavior of sections near a matter locus, this assumption is reasonable. This strategy of course lets us circumvent the difficulties in finding F-theory models admitting large charges.

However, we can also test the proposals by examining previous F-theory models with $\au(1)$ algebras. We have verified that, for the following models in the prior literature, the massless charged matter spectrum agrees with the formulas in \cref{sec:proposals}:
\begin{itemize}
\item The Morrison--Park construction in~\cite{MorrisonParkU1}, which has a $\U(1)$ gauge group and admits $\bm{1}_{1}$ and $\bm{1}_{2}$ matter.
\item The $\U(1)$ constructions in~\cite{KleversEtAlToric} and~\cite{Raghuram34} that admit $\bm{1}_{1}$, $\bm{1}_{2}$, and $\bm{1}_{3}$ matter.
\item The $\U(1)$ construction in~\cite{Raghuram34} that admits $\bm{1}_{1}$, $\bm{1}_{2}$, $\bm{1}_{3}$, and $\bm{1}_{4}$ matter.
\item The $\U(1)\times\U(1)$ models in~\cite{CveticKleversPiraguaMultU1} and~\cite{CveticKleversPiraguaTaylor}. By looking at linear combinations of the generating sections, we can also recover the relative signs of the charges under the two $\au(1)$ factors.
\item The $\U(1)\times\U(1)\times\U(1)$ model in~\cite{CveticKleversPiraguaSong}. Again, we can also recover the relative signs of charges under the three $\au(1)$ factors by considering linear combinations of the generating sections.
\item The models in~\cite{KleversEtAlExotic} with $\au(1)$ algebras, namely the $F_2$, $F_3$, $F_5$, $F_6$, $F_7$, $F_8$, $F_9$, $F_{12}$, $F_{14}$, and $F_{15}$ models. Note that the charge spectra are derived using non-Weierstrass descriptions of the elliptic fibrations, so when there are multiple $\au(1)$ gauge algebras, once must be careful in matching the basis of Weierstrass generating sections to the basis of $\au(1)$ charges.
\item The $\asu(5)\oplus\au(1)$, $\aso(10)\oplus\au(1)$, $\aE_6\oplus\au(1)$, and $\aE_7\oplus\au(1)$ models in~\cite{KuntzlerTateTrees}. While~\cite{KuntzlerTateTrees} includes models with other gauge algebras, matter spectra are not listed for these alternative models. Note that the $\au(1)$ charges are given in an alternative set of units where all charges, including those for matter also charged under a nonabelian algebra, are integral. The lattice spacing for singlet charges in this convention may be an integer larger than 1. While these models are tunings of elliptic fibrations with fibers embedded $\mathbb{P}^{1,1,2}$, we can form the equivalent Weierstrass models by performing the same tunings in the Morrison--Park Weierstrass model given in~\cite{MorrisonParkU1}.
\item The $\asu(5)\oplus\au(1)\oplus\au(1)$, $\asu(4)\oplus\au(1)\oplus\au(1)$, $\asu(5)\oplus\au(1)$, and $\asu(4)\oplus\au(1)$ models in~\cite{BorchmannSU5Top}. Again, the $\au(1)$ charges in these models are given in an alternative set of units where all charges are integral. The $\asu(5)\oplus\au(1)$ and $\asu(4)\oplus\au(1)$ models given there are either equivalent to or specializations of models in~\cite{KuntzlerTateTrees}.
\item The $\asu(3)\oplus\asu(2)\oplus\au(1)$ model in~\cite{RaghuramTaylorTurnerSM}, which is a generalization of the $\asu(3)\oplus\asu(2)\oplus\au(1)$ model in~\cite{KleversEtAlToric}. The $\asu(5)$ enhancement discussion in~\cite{RaghuramTaylorTurnerSM} also makes use of a ``would-be'' $(\bm{3},\bm{2})_{-5/6}$ matter locus. At this would-be locus, the section components vanish to the orders appropriate for $(\bm{3},\bm{2})_{-5/6}$ matter according to the equations in \cref{sec:proposals,sec:bifundamentals}.
\item The models involving $\au(1)$ factors in~\cite{RaghuramTaylorTurnerEnhancement}, including the $\asu(18)\oplus\au(1)$ and $\asu(19)\oplus\au(1)$ models. While these models were not analyzed using resolutions, the matter spectra given there, which can be recovered with the techniques presented here, agree with the anomaly cancellation conditions and the expectations from Higgsing related nonabelian models.
\end{itemize}
These models provide a broad set of gauge algebras, covering many of the representations discussed here. Many of them also have multiple $\au(1)$ factors and matter charged under multiple nonabelian factors, in further support of the ideas in \cref{sec:multgaugealgebras}. The fact that a variety of models in independent works follow the formulas in \cref{sec:proposals}, together with the evidence presented in previous sections, suggests that our proposals should hold more broadly.

\section{Conclusions}
\label{sec:conclusions}
To summarize, we proposed a series of formulas, outlined in \cref{sec:proposals}, that relate the $\au(1)$ charge of matter supported at a particular locus to the orders of vanishing of the generating section's components at that locus. As discussed in \cref{sec:explicitconstructions}, these formulas accurately reproduce the $\au(1)$ charge spectrum of various F-theory models from the prior literature. They also correctly describe the results of the indirect method of \cref{sec:strategy} for analyzing large $\au(1)$ charges, and they appear to be related to formulas from~\cite{StangeEllTrouble} for the $p$-adic valuations of elliptic divisibility sequences. In addition to offering a way of quickly determining the $\au(1)$ charges of light matter supported in an F-theory model, the proposals can aid explorations of the F-theory landscape by predicting the geometric properties of models admitting desired charge spectra.

There are several natural directions to pursue in future work. First, it is important to determine how these proposals would apply in models with gauge algebras and nonabelian matter representations beyond those covered here. For instance, our analysis focused on matter charged under abelian and simply-laced nonabelian gauge factors, but one might expect similar formulas to hold when non-simply-laced gauge factors are present as well. One possibility is that the proposals are essentially unchanged even when the gauge algebra is non-simply laced. Even if matter is charged under a $\au(1)$ factor and a non-simply-laced nonabelian factor, one would take the Lie group $G$ for the nonabelian factor to be the simply-laced Lie group associated with the codimension-one singularity type. The only practical difference would then be that monodromy effects would restrict the allowed values of $\quotparam$ and $\quotparamtwo$. This is just a tentative suggestion, however, and more work should be done to systematically investigate how the formulas may change for matter charged under non-simply-laced factors.

Furthermore, the proposals should be checked in models admitting $\au(1)$-charged matter additionally charged in exotic nonabelian representations, such as the adjoint,\footnote{While the adjoint representation by itself is not typically viewed as an exotic representations, it is difficult to find F-theory constructions with adjoint matter that also has a $\au(1)$ charge. Such situations would likely require adjoint matter localized at double point singularities~\cite{MorrisonTaylorMaS,KleversEtAlExotic}.} symmetric, and three-index antisymmetric representations of $\asu(n)$. Situations where the codimension-two singularity type enhances by a rank greater than one are also worthy of examination. An intriguing question is whether our proposals apply for $\au(1)$ models with superconformal matter loci, where $f$ and $g$ vanish to orders $(4,6)$ or greater at codimension two. If so, our formulas could provide interesting insights into $\U(1)$ flavor symmetries of 6D SCFTs~\cite{LeeRegaladoWeigand, ApruzzietalU1SCFT}.

A second important direction is clarifying the formulas' exact range of validity. At the same time, we should better understand why the proposals work from both mathematical and physical perspectives. The seemingly related formulas in~\cite{StangeEllTrouble} describing EDS valuations may provide some insights into geometric explanations for the proposals. In fact, the parallels between these two contexts suggest an intriguing connection between topics in physics and number theory.  This possibility makes understanding the physics behind the proposals all the more important. In particular, one might hope that a physical understanding of the proposals may translate to new perspectives on elliptic divisibility sequences and related mathematical topics. Alternatively, it may be possible to definitively establish the proposals by considering intersection numbers in the full elliptic fibration. The charge itself is given by $\shioda(\ratsec{s})\cdot c$, suggesting that the proposals could be understood in terms of similar intersection numbers. This approach would likely require a more concrete understanding of how the wrapping behavior of sections at codimension-two loci is encapsulated in the section components.

While the analysis here focused on models in Weierstrass form, there may be similar relations between $\au(1)$ charges and orders of vanishing for elliptic fibrations in alternative forms, such as those in Tate form~\cite{BershadskyEtAlSingularities, KatzEtAlTate} or those described as hypersurfaces in $\mathbb{P}^{1,1,2}$ \cite{MorrisonParkU1, KuntzlerTateTrees}. It is oftentimes more convenient to work with these alternative forms, and an easy method for reading off $\au(1)$ charges in such situations could simplify several analyses.

Finally, it would be interesting to use the proposals to characterize the landscape and swampland of F-theory $\au(1)$ models. A variety of important open problems involve understanding which models with ``large'' $\au(1)$ charges can occur in F-theory. Even for the simplest case of models with just a $\au(1)$ gauge algebra, determining a bound on the $\au(1)$ charges of singlet matter and constructing F-theory models with singlet charges larger than 6 (in appropriately scaled units) remain key goals in the F-theory program. Because they seem to work for arbitrarily large charges, the proposals would allow us to glean information about models with larger charges without directly constructing them. Thus, they may show the way towards constructions with new types of $\au(1)$ charges, such as singlet charges larger than $6$. Alternatively, the proposals may enable one to anticipate geometric obstructions to realizing particular $\au(1)$ charges. Such observations could, for instance, lead to new bounds on the $\au(1)$ charges and charge spectra that can occur in F-theory. A better physical understanding of the proposals may also reveal other constraints on $\au(1)$ algebras in F-theory. In general, a clearer knowledge of the geometric properties of $\au(1)$-charged matter can offer new insights into F-theory.

\acknowledgments We would like to thank Washington Taylor for helpful discussions, and Timo Weigand and Antonella Grassi for useful comments. The work of APT was supported by DOE grant DE-SC00012567 and DOE (HEP) Award DE-SC0013528. The work of NR was supported by NSF grant PHY-1720321.

\appendix

\section{Expressions for specific gauge algebras and representations}
\label{app:specificreps}
\subsection{Singlets}
Charged singlets occur at codimension-two loci in the base of the elliptic fibration with $\singtype{I}_2$ singular fibers. In units corresponding to the definition of Shioda map in \cref{eq:shiodamap}, charged singlets should have integer charges. At a codimension-two locus supporting singlets with charge $q$, the $\secz$ component vanishes to order
\begin{equation}
\ordvanishtwo{\secz} =\frac{q^2}{4}-\frac{\residue{q}{2}}{4} = \floor*{\frac{q^2}{4}}\,,
\end{equation}
while the $\secx$, $\secy$, and $\secw$ components vanish to orders
\begin{equation}
\left(\ordvanishtwo{\secx},\ordvanishtwo{\secy},\ordvanishtwo{\secw}\right) = \left(2,3,4\right)\times\ordvanishtwo{\secz} + \left(0,1,1\right)\residue{q}{2}
\end{equation}

Based on these equations, if one knows the orders of vanishing of the section components at a codimension-two singlet locus, one can determine the charge $q$ of the supported matter (up to sign) as
\begin{equation}
q^2 = \ordvanishtwo{\secy} + \ordvanishtwo{\secz} = \ordvanishtwo{\secw}\,.
\end{equation}

\subsection{$\asu(n)$}
For models with an $\asu(n)\oplus\au(1)$ gauge algebra, we label elements of the center of $\SU(n)$ with the parameter $\quotparam$, which can run from 0 to $n-1$. The section components in a model with a quotient corresponding to $\quotparam$ vanish to the following orders at the codimension-one locus supporting the $\asu(n)$ algebra:
\begin{equation}
\ordvanishone{\secz} = 0\,, \quad \left(\ordvanishone{\secx},\ordvanishone{\secy}, \ordvanishone{\secw}\right) = \left(0,1,1\right)\times\moddist{n}{\quotparam}\,.
\end{equation}

\paragraph{Fundamentals} An  $\asu(n)\oplus\au(1)$ model can have fundamental matter with charge $q=\frac{\quotparam}{n} + j$ for integers $j$. In other words, fundamental matter with charge $q$ can occur when $\quotparam$ is  $\residue{nq}{n}$. At the codimension-two locus supporting the $\bm{n}_{q}$ matter, the $\secz$ component should vanish to order
\begin{equation}
\ordvanishtwo{\secz} = \frac{1}{2}\left(\frac{n}{n+1}q^2 + \frac{\quotparam(n-\quotparam)}{n} - \frac{\quotparamtwo\left(n+1-\quotparamtwo\right)}{(n+1)}\right)\,,
\end{equation}
where
\begin{equation}
\quotparamtwo = \residue{nq}{n+1}\,.
\end{equation}
The other section components should vanish to orders
\begin{equation}
\left(\ordvanishtwo{\secx},\ordvanishtwo{\secy},\ordvanishtwo{\secw}\right) = \left(2,3,4\right)\times\ordvanishtwo{\secz} + \left(0,1,1\right)\times\moddist{n+1}{\quotparamtwo}\,.
\end{equation}

Based on these equations, if one knows the orders of vanishing of the section components at the codimension-one locus supporting the $\asu(n)$ factor and the codimension-two fundamental locus, one can determine the $\au(1)$ charge $q$ of the supported fundamental matter (up to sign) as
\begin{equation}
q^2 = {\frac{2(n+1)}{n}\left(\ordvanishtwo{\secz} + \frac{\tilde{\quotparamtwo}\left(n+1-\tilde{\quotparamtwo}\right)}{2(n+1)} - \frac{\tilde{\quotparam}\left(n-\tilde{\quotparam}\right)}{2n}\right)}\,,
\end{equation}
where
\begin{equation}
\begin{gathered}
\tilde{\quotparam} = \ordvanishone{\secy}= \ordvanishone{\secw}\,, \\
\tilde{\quotparamtwo} = \ordvanishtwo{\secy}  - 3\ordvanishtwo{\secz} =  \ordvanishtwo{\secw}  - 4\ordvanishtwo{\secz}\,.
\end{gathered}
\end{equation}

\paragraph{Antisymmetric matter for odd $n$}  An  $\asu(n)\oplus\au(1)$ model with discrete quotient $\quotparam$ can have antisymmetric matter with charge $q=\frac{2\quotparam}{n} +j$ for integers $j$. For odd $n$ (i.e., $n=2k+1$), this means that antisymmetric matter with charge $q$ can occur when
\begin{equation}
\quotparam = \residue{nq(k+1)}{n}\,.
\end{equation}
At the codimension-two locus supporting the antisymmetric matter, the $\secz$ component should vanish to order
\begin{equation}
\ordvanishtwo{\secz} = \frac{1}{2}\left(\frac{n}{4}q^2 +  \frac{\quotparam(n-\quotparam)}{n} -  \begin{cases}0 & \quotparamtwo=0 \\ 1 & \quotparamtwo=2 \\ \frac{n}{4} & \quotparamtwo=1,3\end{cases}\right)
\end{equation}
with
\begin{equation}
\quotparamtwo = \residue{nq}{4}\,.
\end{equation}
The other section components vanish to orders
\begin{equation}
\left(\ordvanishtwo{\secx},\ordvanishtwo{\secy},\ordvanishtwo{\secw}\right)\\ = \left(2,3,4\right)\times \ordvanishtwo{\secz} + \begin{cases}(0,0,0) & \quotparamtwo=0 \\ (1,2,2)& \quotparamtwo=2 \\ \left(1,\floor*{\frac{n}{2}}, \ceil*{\frac{n}{2}}\right)& \quotparamtwo=1,3\end{cases}
\end{equation}
If one knows the orders of vanishing at the codimension-one $\asu(n)$ locus and at the codimension-two antisymmetric locus, the $\au(1)$ charge of the supported matter can be determined (up to sign) as
\begin{equation}
q^2 = \frac{8}{n}\left(\ordvanishtwo{\secz} - \frac{\tilde{\quotparam}(n-\tilde{\quotparam})}{2n} + \frac{1}{2}\begin{cases}0 & \ordvanishtwo{\secw}-4\ordvanishtwo{\secz}=0 \\ 1 & \ordvanishtwo{\secw}-4\ordvanishtwo{\secz}=2 \\ \frac{n}{4} & \ordvanishtwo{\secw}-4\ordvanishtwo{\secz}=\ceil*{\frac{n}{2}}\end{cases}\right) \label{eq:summantiordtoq}
\end{equation}
where $\tilde{\quotparam} = \ordvanishone{\secy}$.\footnote{If we let $\tilde{\quotparamtwo} = \ordvanishtwo{\secy}  - 3\ordvanishtwo{\secz}$ and $ \tilde{\quotparamtwo}^\prime = \ordvanishtwo{\secw}  - 4\ordvanishtwo{\secz}$, the last term in parentheses in \cref{eq:summantiordtoq} can be written as $\frac{\tilde{\quotparamtwo}+\tilde{\quotparamtwo}^\prime}{8}$.}

\paragraph{Antisymmetric matter for even $n$} As before, an  $\asu(n)\oplus\au(1)$ model with discrete quotient $\quotparam$ can have antisymmetric matter with charge $q=\frac{2\quotparam}{n} + j$ for integers $j$. For even $n$ (i.e., $n=2k$), this means that antisymmetric matter with charge $q$ can occur when $\quotparam$ is either $\residue{(kq)}{n}$ or $\residue{(kq+k)}{n}$. At the codimension-two locus supporting the antisymmetric matter, the $\secz$ component vanishes to order
\begin{equation}
\ordvanishtwo{\secz} = \frac{1}{2}\left(\frac{n}{4}q^2 +  \frac{\quotparam(n-\quotparam)}{n} -  \begin{cases}0 &  \quotparamtwo=0 \\ 1 &  \quotparamtwo = 2 \\ \frac{n}{4} &\quotparamtwo=1,3\end{cases}\right)\,,
\end{equation}
where
\begin{equation}
    \quotparamtwo = \residue{\left(q-\frac{2}{n}\quotparam\right)}{2} + \residue{2\quotparam}{4}
\end{equation}
The other section components vanish to orders
\begin{equation}
\left(\ordvanishtwo{\secx},\ordvanishtwo{\secy},\ordvanishtwo{\secw}\right)= \left(2,3,4\right)\times \ordvanishtwo{\secz} + \begin{cases}(0,0,0) &  \quotparamtwo=0 \\ (1,2,2)&  \quotparamtwo=2 \\ \left(1,\frac{n}{2}, \frac{n}{2}\right)& \quotparamtwo=1,3\end{cases}\,.
\end{equation}
If one knows the orders of vanishing at the codimension-one $\asu(n)$ locus and at the codimension-two antisymmetric locus, the $\au(1)$ charge of the supported matter can still be determined (up to sign) using \cref{eq:summantiordtoq}.

\subsection{$\aso(2n)$}
For models with an $\aso(2n)\oplus\au(1)$ gauge algebra, we label the elements of the center of $\Spin(2n)$ with the parameter $\quotparam$, which can run from 0 to $4$. The section components in a model with the quotient corresponding to $\quotparam$ vanish to the following orders at the codimension-one locus supporting the $\aso(2n)$:
\begin{equation}
\ordvanishone{\secz} = 0\,, \quad \left(\ordvanishone{\secx}, \ordvanishone{\secy},\ordvanishone{\secw}\right) = \begin{cases}(0,0,0)& \quotparam=0\\(1,2,2)& \quotparam=2\\(1,\floor*{\frac{n}{2}},\ceil*{\frac{n}{2}})& \quotparam=1,3 \end{cases}\,.
\end{equation}
The allowed $\au(1)$ charges for the various $\aso(2n)$ representations considered here are summarized in \cref{tab:so2ncharges}.

\paragraph{Vector matter} At a codimension-two locus supporting $\bm{2n}_{q}$ matter, the $\secz$ component should vanish to order
\begin{equation}
\ordvanishtwo{\secz} = \frac{1}{2}\left(q^2 + \begin{cases}0 & \quotparam=0 \\ 1 & \quotparam=2 \\ \frac{n}{4} & \quotparam=1,3\end{cases} -  \begin{cases}0 & \quotparamtwo=0 \\ 1 & \quotparamtwo=2 \\ \frac{n+1}{4} & \quotparamtwo=1,3\end{cases}\right)\,,
\end{equation}
where
\begin{equation}
\quotparamtwo = \residue{(2q-\quotparam)}{4} + \residue{2q}{2}\,.
\end{equation}
The other section components vanish to orders
\begin{equation}
\left(\ordvanishtwo{\secx},\ordvanishtwo{\secy},\ordvanishtwo{\secw}\right) =
 \left(2,3,4\right)\times\ordvanishtwo{\secz} + \begin{cases}(0,0,0)& \quotparamtwo=0\\(1,2,2)& \quotparamtwo=2\\(1,\floor*{\frac{n+1}{2}},\ceil*{\frac{n+1}{2}})& \quotparamtwo=1,3 \end{cases}\,.
\end{equation}
Based on these formulas, if one knows the orders of vanishing of the section components at the codimension-one locus supporting the $\aso(2n)$ algebra and the codimension-two $\bm{2n}$ locus, one can determine the $\au(1)$ charge $q$ of the supported matter (up to sign) with the formula
\begin{equation}
    q^2 = 2 \ordvanishtwo{\secz} - \begin{cases}0 & \ordvanishone{\secw}=0 \\ 1 & \ordvanishone{\secw}=2\\ \frac{n}{4} & \ordvanishone{\secw}=\ceil*{\frac{n}{2}}\end{cases} +  \begin{cases}0 & \ordvanishtwo{\secw}-4\ordvanishtwo{\secz}=0 \\ 1 & \ordvanishtwo{\secw}-4\ordvanishtwo{\secz}=2 \\ \frac{n+1}{4} & \ordvanishtwo{\secw}-4\ordvanishtwo{\secz}=\ceil*{\frac{n+1}{2}}\end{cases}\,.
\end{equation}

\paragraph{Spinor matter for $\aso(8)$} At a codimension-two locus supporting ${\bm{8_\text{s}}}_{,q}$ or ${\bm{8_\text{c}}}_{,q}$ matter, the $\secz$ component should vanish to order
\begin{equation}
\ordvanishtwo{\secz} = \frac{1}{2}\left(q^2 + \begin{cases}0 & \quotparam=0 \\ 1& \quotparam=1,2,3\end{cases} -  \begin{cases}0 & \quotparamtwo=0 \\ 1 & \quotparamtwo=2 \\ \frac{5}{4} & \quotparamtwo=1,3\end{cases}\right)\,,
\end{equation}
where
\begin{equation}
\quotparamtwo = \residue{(2q-2\quotparam)}{4}\,.
\end{equation}
The other section components vanish to orders
\begin{equation}
\left(\ordvanishtwo{\secx},\ordvanishtwo{\secy},\ordvanishtwo{\secw}\right) =
 \left(2,3,4\right)\times\ordvanishtwo{\secz} + \begin{cases}(0,0,0)& \quotparamtwo=0\\(1,2,2)& \quotparamtwo=2\\(1,2,3)& \quotparamtwo=1,3 \end{cases}\,.
\end{equation}
Based on these formulas, if one knows the orders of vanishing of the section components at the codimension-one locus supporting the $\aso(8)$ algebra and the codimension-two $\bm{8_\text{s}}$ or $\bm{8_\text{c}}$ locus, one can determine the $\au(1)$ charge $q$ of the supported matter (up to sign) with the formula
\begin{equation}
q^2 = 2\ordvanishtwo{\secz} -  \begin{cases}0 &  \ordvanishone{\secw} =0 \\ 1 & \ordvanishone{\secw} = 2 \end{cases} +  \begin{cases}0 & \ordvanishtwo{\secw}-4\ordvanishtwo{\secz}=0 \\ 1 & \ordvanishtwo{\secw}-4\ordvanishtwo{\secz}=2 \\ \frac{5}{4} & \ordvanishtwo{\secw}-4\ordvanishtwo{\secz}=3\end{cases}
\end{equation}

\paragraph{Spinor matter for $\aso(10)$}
At a codimension-two locus supporting $\bm{16}_{q}$ matter (along with the conjugate $\overline{\bm{16}}_{-q}$ matter), the $\secz$ component should vanish to order
\begin{equation}
\ordvanishtwo{\secz} = \frac{1}{2}\left(\frac{4}{3}q^2 + \begin{cases}0 & \quotparam=0 \\ 1 & \quotparam=2 \\ \frac{5}{4} & \quotparam=1,3\end{cases} - \frac{2}{3}\quotparamtwo(3-\quotparamtwo)\right)\,,
\end{equation}
where
\begin{equation}
\quotparamtwo = \residue{4q}{3}\,.
\end{equation}
The other section components vanish to orders
\begin{equation}
\left(\ordvanishtwo{\secx},\ordvanishtwo{\secy},\ordvanishtwo{\secw}\right) =\left(2,3,4\right)\times\ordvanishtwo{\secz}+ \left(2,2,3\right)\times\moddist{3}{\quotparamtwo}\,.
\end{equation}
Based on these formulas, if one knows the orders of vanishing of the section components at the codimension-one locus supporting the $\aso(10)$ algebra and the codimension-two $\bm{16}$ locus, one can determine the $\au(1)$ charge $q$ of the supported matter (up to sign) with the formula
\begin{equation}
q^2 = \frac{3}{2}\ordvanishtwo{\secz} - \frac{3}{4}\times\begin{cases}0 & \ordvanishone{\secw}=0 \\ 1 & \ordvanishone{\secw}=2 \\ \frac{5}{4} & \ordvanishone{\secw}=3\end{cases} + \frac{1}{3}\left(\ordvanishtwo{\secw}-4\ordvanishtwo{\secz}\right)\,.
\end{equation}

\paragraph{Spinor matter for $\aso(12)$} At a codimension-two locus supporting $\bm{32}_{q}\oplus\bm{1}_{2q}$ matter or $\bm{32^\prime}_{q}\oplus\bm{1}_{2q}$ matter, the $\secz$ component should vanish to order
\begin{equation}
\ordvanishtwo{\secz} = \frac{1}{2}\left(2q^2 + \begin{cases}0 & \quotparam=0 \\ 1 & \quotparam=2 \\ \frac{3}{2} & \quotparam=1,3\end{cases} - \frac{3}{2}\residue{(2q+\quotparam)}{2}\right)\,,
\end{equation}
while the other components should vanish to orders
\begin{equation}
\left(\ordvanishtwo{\secx},\ordvanishtwo{\secy},\ordvanishtwo{\secw}\right) =\left(2,3,4\right)\times\ordvanishtwo{\secz} + \left(2,3,3\right)\times \residue{(2q+\quotparam)}{2}\,.
\end{equation}
Based on these formulas, if one knows the orders of vanishing of the section components at the codimension-one locus supporting the $\aso(12)$ algebra and the codimension-two spinor locus, one can determine the $\au(1)$ charge $q$ of the supported matter (up to sign) as
\begin{equation}
q^2=\ordvanishtwo{\secz} - \begin{cases}0 & \ordvanishone{\secw}=0 \\ \frac{1}{2} & \ordvanishone{\secw}=2 \\ \frac{3}{4} & \ordvanishone{\secw}=3\end{cases} + \frac{3}{8}\left(\ordvanishtwo{\secx}-2\ordvanishtwo{\secz}\right)\,.
\end{equation}

\paragraph{Spinor matter for $\aso(14)$} At a codimension-two locus supporting $\bm{64}_{q}\oplus\bm{14}_{2q}$ matter (and matter in the conjugate representation), the $\secz$ component should vanish to order
\begin{equation}
\ordvanishtwo{\secz} = \frac{1}{2}\left(4 q^2 + \begin{cases}0 & \quotparam=0 \\1 & \quotparam=2 \\ \frac{7}{4} & \quotparam=1,3\end{cases}\right)\,,
\end{equation}
and the other section components should vanish to orders
\begin{equation}
\left(\ordvanishtwo{\secx},\ordvanishtwo{\secy},\ordvanishtwo{\secw}\right) =\left(2,3,4\right)\times\ordvanishtwo{\secz} \,.
\end{equation}
Based on these formulas, if one knows the orders of vanishing of the section components at the codimension-one locus supporting the $\aso(14)$ algebra and the codimension-two spinor locus, one can determine the $\au(1)$ charge $q$ of the supported matter (up to sign) as
\begin{equation}
q^2 = \frac{1}{2}\ordvanishtwo{\secz}  -  \frac{1}{4}\times \begin{cases}0 & \ordvanishone{\secw}=0 \\ 1 & \ordvanishone{\secw}=2 \\ \frac{7}{4} & \ordvanishone{\secw}=4\end{cases}\,.
\end{equation}

\subsection{$\aE_6$}
For models with an $\aE_6\oplus\au(1)$ gauge algebra, we label the elements of the center of $\gE_6$ with the parameter $\quotparam$, which can run from $0$ to $2$. The section components in a model with the quotient corresponding to $\quotparam$ vanish to the following orders at the codimension-one locus supporting the $\aE_6$ algebra:
\begin{equation}
\ordvanishone{\secz} = 0\,, \quad \left(\ordvanishone{\secx},\ordvanishone{\secy},\ordvanishone{\secw}\right) = \left(2,2,3\right)\times \moddist{3}{\quotparam}\,.
\end{equation}

\paragraph{$\bm{27}$ matter} An $\aE_6\oplus\au(1)$ model can have $\bm{27}$ matter with charge $q=\frac{\quotparam}{3}+k$ for integer $k$. In other words, $\bm{27}_q$ matter can occur when $\quotparam$ is $\residue{3q}{3}$. At the codimension-two locus supporting that $\bm{27}_{q}$ matter, the $\secz$ component vanishes to order
\begin{equation}
\ordvanishtwo{\secz} = \frac{1}{2}\left(\frac{3}{2}q^2+ \frac{2}{3}\quotparam(3-\quotparam) - \frac{3}{2} \residue{3q}{2}\right)\,,
\end{equation}
The other section components vanish to orders
\begin{align}
\left(\ordvanishtwo{\secx},\ordvanishtwo{\secy},\ordvanishtwo{\secw}\right) = \left(2,3,4\right)\times\ordvanishtwo{\secz} + \left(2,3,3\right)\times \residue{3q}{2}\,.
\end{align}
Based on these formulas, if one knows the orders of vanishing of the section components at the codimension-one locus supporting the $\aE_6$ algebra and the codimension-two $\bm{27}$ locus, one can determine the $\au(1)$ charge $q$ of the supported matter (up to sign) as
\begin{equation}
q^2 = {\frac{4}{3}\left(\ordvanishtwo{\secz} - \frac{\tilde{\quotparam}(3-\tilde{\quotparam})}{3} +  \frac{\tilde{\quotparamtwo}}{4} \right)}\,,
\end{equation}
where
\begin{align}
\begin{gathered}
\tilde{\quotparam} = \ordvanishone{\secx} = \ordvanishone{\secy}\,, \\
\tilde{\quotparamtwo} = \ordvanishtwo{\secy} - 3 \ordvanishtwo{\secz} = \ordvanishtwo{\secw} - 4 \ordvanishtwo{\secz}\,.
\end{gathered}
\end{align}

\subsection{$\aE_7$}
For models with an $\aE_7\oplus\au(1)$ gauge algebra, we label the elements of the center of $\gE_7$ with the parameter $\quotparam$, which can be either $0$ or $1$. The section components in a model with the quotient corresponding to $\quotparam$ vanish to the following orders at the codimension-one locus supporting the $\aE_7$ algebra:
\begin{equation}
\ordvanishone{\secz} = 0\,, \quad \left(\ordvanishone{\secx},\ordvanishone{\secy},\ordvanishone{\secw}\right) = \left(2,3,3\right)\times \quotparam\,.
\end{equation}

\paragraph{$\bm{56}$ matter} An $\aE_7\oplus\au(1)$ model can have $\bm{56}$ matter with $\au(1)$ charge $q=\frac{\quotparam}{2}+k$ for integer $k$. In other words, $\bm{56}_{q}$ matter can occur when $\quotparam$ is $\residue{2q}{2}$. At the codimension-two locus supporting this $\bm{56}_{q}$ matter, the $\secz$ component should vanish to order
\begin{equation}
\ordvanishtwo{\secz} = q^2 + \frac{3}{4}\quotparam = q^2 +\frac{3}{4}\times\residue{2q}{2}\,,
\end{equation}
while the other section components vanish to orders
\begin{equation}
\left(\ordvanishtwo{\secx},\ordvanishtwo{\secy},\ordvanishtwo{\secw}\right) = \left(2,3,4\right)\times\ordvanishtwo{\secz}\,.
\end{equation}
Based on these formulas, if one knows the orders of vanishing of the section components at the codimension-one locus supporting the $\aE_7$ algebra and the codimension-two $\bm{56}$ locus, one can determine the $\au(1)$ charge $q$ of the supported matter (up to sign) as
\begin{equation}
q^2 = {\ordvanishtwo{\secz} - \frac{1}{4}\ordvanishone{\secy}} = {\ordvanishtwo{\secz} - \frac{1}{4}\ordvanishone{\secw}}\,.
\end{equation}

\section{The $\scorr$ correction term}
\label{app:scorr}
The results in~\cite{StangeEllTrouble} often use the sequence $\scorr_n$ of numbers in $\Z\cup\{\infty\}$, which is defined as
\begin{equation}
\scorr_n\left(p, b, d, h, s, w\right) =
\begin{cases} b^j s + \frac{b^{j}-1}{b-1} h + d\left(\valuation{p}{n}-j\right)+w & \valuation{p}{n}>j \\
              b^{\valuation{p}{n}}s + \frac{b^{\valuation{p}{n}}-1}{b-1} h &  \valuation{p}{n}\le j \end{cases}\,. \label{eq:scorrdef}
\end{equation}
Here, $v_{p}$ is a $p$-adic valuation on $\Q$ associated with $p$, while $b$, $d$, $h$, $s$, and $w$ are non-negative integers in particular sets:
\begin{align}
b \in& p\Z^{>0}\cup\{1\} & d &\in \Z^{>0} &  h \in& \Z^{\ge0} & s\in& \Z^{>0}\cup\{\infty\} & w&\in \Z^{\ge 0}\cup\{\infty\}\,.
\end{align}
The integer $j$ is $0$ if $b=1$; if $b$ is a multiple of $p$, then $j$ is the smallest non-negative integer such that
\begin{equation}
d \le b^{j}\left(\left(b-1\right)s + h\right)\,.
\end{equation}

For the examples encountered here, $d$ is $\valuation{p}{p}$, which is $1$. This implies that $j$ is $0$ regardless of the value of $b$, and the definition of $\scorr_n$ simplifies to
\begin{equation}
\scorr_n\left(p, b, \valuation{p}{p}, h, s, w\right) =
s + \valuation{p}{n}+\begin{cases}w & \valuation{p}{n}>0 \\
              0 &  \valuation{p}{n}\le 0 \end{cases}\,.
\end{equation}
The variable $w$ is allowed to be 0; in fact, the ideas discussed in Section 5 of~\cite{StangeEllTrouble} suggest that, for all of the cases considered here, $w$ is expected to be 0. When this is the case, we are left with
\begin{equation}
\scorr_n\left(p, b, \valuation{p}{p}, h, s, 0\right) =s + \valuation{p}{n}
\end{equation}
If we are interested in using these formulas to explain the order of vanishing patterns, the $\valuation{p}{n}$ term may seem somewhat strange. While the valuation of an integer is a meaningful quantity, one might expect that the order of vanishing of an integer at some locus should be thought of as 0. It therefore seems that, for the purposes of understanding the order-of-vanishing patterns, the $\valuation{p}{n}$ in $\scorr_{n}$ should be dropped.

To see if this is a reasonable prescription, it is worth understanding why this $\valuation{p}{n}$ occurs. Recall that $\scorr_{n}$ appears in formulas describing the valuations of elliptic divisibility sequences. Therefore, consider an example EDS with the initial values
\begin{equation}
W_{1} = 1\,, \quad W_{2} = 2 p^3\,, \quad W_{3} = 8 p^8-4 p^4-1\,, \quad W_{4} = 4 p^3\left(8 p^{12}-6 p^4 -1\right)\,,
\end{equation}
where $p$ is some prime integer. The $\eds_{n}$ with even $n$ are proportional to $p$, implying that $n_{P}=2$ for this situation. One can then calculate that
\begin{equation}
W_{6} = -2 p^3 \left(8 p^8-4 p^4-1\right) \left(512 p^{24}-768 p^{20}+96 p^{12}+120 p^8+36 p^4+3\right)\,.
\end{equation}
For most choices of $p$, $W_{6}$ is proportional to $p^3$, and $\valuation{p}{W_{6}}$ is 3. If, however, we choose $p$ to be 3, the last factor in the above expression is a factor of 3, and $\valuation{p}{W_{6}}$ enhances to $4$. Similar sorts of $\valuation{p}{W_{n}}$ enhancements may occur in other elliptic divisibility sequences, and the $\valuation{p}{n}$ term accounts for this fact. Based on \cref{eq:scorrdef}, this effect only occurs for particular choices of $n$ and $p$. However, this term is not particularly relevant for the analogous situations in elliptic fibrations where valuations are replaced with orders of vanishing; an integer $n$ would not vanish at a particular locus in the base. Additionally, note that the examples considered here are formed by starting with an elliptic fibration, setting the parameters involved at a particular locus to some arbitrary prime $p$, setting the other parameters to some integers, and analyzing an EDS corresponding to the resulting elliptic curve. We are interested in the behavior for arbitrary $p$, but for a given value of $n$, the ``accidental'' increases in $\valuation{p}{W_n}$ only occur for particular choices of $p$. This effect should therefore be ignored.

Thus, while the $\valuation{p}{n}$ is important for characterizing $p$-adic valuations of elliptic divisibility sequences, it seems that it can be dropped for the purposes of establishing analogies to the orders of vanishing. In other words, if we are looking to justify the expressions we obtain for the order of vanishing, one should be able to replace $\scorr_{n}$ with $s$, a positive constant. However, it would be ideal to have a more formal understanding of how to adapt the formulas in~\cite{StangeEllTrouble}, particularly the $S$ correction term, to scenarios involving the orders of vanishing of section components in elliptic fibrations.

\section{Additional calculations of EDS valuations}
In this appendix, we calculate various EDS valuations corresponding to various singularity types using the formulas in~\cite{StangeEllTrouble}.

\subsection{$\singtype{I}^*_{2k-4}$ singularities}
\label{app:sonedseven}
\subsubsection{$\quotparam=1,3$ or $\quotparamtwo=1,3$}
\label{app:sonedsnu1}
For models with an $\singtype{I}^*_{2k-4}$ locus where $\quotparam$ or $\quotparamtwo$ for the generating section equals 1 or 3, we are interested in the EDS valuations when the components $(\secx,\secy,\secz,\secw)$ vanish to order $(1,k,0,k)$. As an example, we can consider the Weierstrass model in \cref{eq:sunevenantiseedmodelb}, the $\asu(2k)\oplus\au(1)$ seed model with $\quotparam=k+1$ used to find order of vanishing data for the antisymmetric representation. The $\singtype{I}^*_{2k-4}$ singularities are supported at $\locus{\sigma=b_1=0}$, so we construct an elliptic curve $E$ by setting $\sigma$ and $b_1$ to a prime $p$ and setting the remaining parameters to arbitrary numbers. As an example, $E$ can take the form
\begin{equation}
    \begin{aligned}
        y^2 &= x^3 - \frac{p^2}{48}\left[\left(-12+p-4p^{k-2}\right)^2-48p^{k-2}\left(1+p\right)\right]x \\
        &\qquad\quad + \frac{p^3}{864}\left[(p-12)^3-12(p-12)p^{k-2}(7p-6)+48p^{2k-4}(13p-6)\right]\,,
    \end{aligned}
\end{equation}
in which case it admits the rational point $P$ given by
\begin{equation}
    P\colon (x,y) = \left(\frac{p}{12}(p-12), \frac{5}{6}p^k\right)\,.
\end{equation}
Since this elliptic curve has additive reduction modulo $p$, we extend the field by the element $p^\prime = (p^{1/2})$, let $x\rightarrow {p^\prime}^2 x^\prime, y\rightarrow {p^\prime}^3 y^\prime$, and divide through by ${p^\prime}^6$. This gives us an elliptic curve $E^\prime$ of the form
\begin{equation}
    \begin{aligned}
        {y^\prime}^2 &= {x^\prime}^3 +\left[-\frac{1}{48} \left({p^\prime}^2-12-4 {p^\prime}^{2 k-4}\right)^2+{p^\prime}^{2 k-4} \left(1+{p^\prime}^2\right)\right] {x^\prime} \\
        &\qquad\quad +\Big[\frac{1}{864} \left({p^\prime}^2-12-4 {p^\prime}^{2 k-4}\right)^3-\frac{1}{12} \left({p^\prime}^2-12-4 {p^\prime}^{2 k-4}\right) {p^\prime}^{2 k-4} \left(1+{p^\prime}^2\right) \\
        &\qquad\qquad\qquad +\frac{1}{3} {p^\prime}^{4 k-6}+\frac{2}{27} {p^\prime}^{6 k-12} \Big]\,,
    \end{aligned}
\end{equation}
which admits a rational point
\begin{equation}
    P^\prime\colon (x^\prime,y^\prime) = \left(\frac{{p^\prime}^2}{12}-1, \frac{5}{6}{p^\prime}^{2k-3}\right)\,.
\end{equation}
with $3 {x^\prime}^2 + f^\prime$ proportional to ${p^\prime}^{2k-4}$.
The new elliptic curve $E^\prime$ has multiplicative reduction modulo $p^\prime$ with $\valuation{p^\prime}{\Delta^\prime} = 4k-8$. Meanwhile, the point $P^\prime$ has singular reduction, implying that the EDS valuations are given by \cref{eq:valWnpotmultsing} with $a=2k-4$.\footnote{The quantity $w^\prime=3{x^\prime}^2 + f^\prime$ is proportional to ${p^\prime}^{2k-4}$, whereas $y^{\prime}$ is proportional to ${p^\prime}^{2k-3}$. The integer $a$ should equal the smaller of these two exponents, which is $2k-4$.} Even multiples of $P^\prime$ reduce to the identity point with the $(x^\prime, y^\prime)$ coordinates of $2P^\prime$ being proportional to $({p^\prime}^{-2},{p^\prime}^{-3})$. Therefore,  $n_{P^\prime}$ equals $2$, and the $\scorr$ term, which is essentially a constant according to \cref{app:scorr}, should be set to $1$. Combining these observations together, we have
\begin{equation}
    \begin{aligned}
    \valuation{}{\eds_{m}} &= \frac{m^2-1}{2} + \frac{1}{2}\left(\elltrouble{m}{2k-4}{4k-8} +\begin{cases}1 &  2 \mid m\\ 0 &  2 \nmid m \end{cases}\right) \\
    &= \frac{n}{8}m^2 - \frac{1}{2}\begin{cases}0 &  2 \mid m\\ \frac{n}{4} &  2 \nmid m \end{cases}\,.
    \end{aligned}
\label{eq:soevenvaluationsmu1}
\end{equation}

\subsubsection{$\quotparam=2$ or $\quotparamtwo=2$}
\label{app:sonedsnu2}
For models with an $\singtype{I}^*_{2k-4}$ locus where the generating section has $\quotparam=2$ or $\quotparamtwo=2$, we are interested in EDS valuations when the generating section components $(\secx,\secy,\secz,\secw)$ vanish to orders $(1,2,0,2)$. As an example, we can consider the $\asu(2k)\oplus\au(1)$ seed model with $\quotparam=1$, which is used to find order of vanishing data for the fundamental and antisymmetric representations of $\asu(2k)$. The Weierstrass model is given by \cref{eq:suoddseedmodelfund} with the substitutions described around \cref{eq:sunfundevensubs}. The section components are given by \cref{eq:suoddseedfundseccomp} with the same substitutions performed. The $\singtype{I}^*_{2k-4}$ singularities are supported at $\locus{\sigma=b_{1,0}=0}$, so we construct an elliptic curve $E$ by setting $\sigma$ and  $b_{1,0}$ to a prime $p$ and the remaining parameters to some arbitrary numbers. For example, we can consider an elliptic curve $E$ taking the form
\begin{equation}
    \begin{aligned}
        y^2 &= x^3 + p^2 \left(-3 (1+3 p)^2+32 p^{k-1}+p^{2 k-4} (p-1)\right) x \\
        &\quad + p^3 \left(2 (1+3 p)^3-32 (1+3 p) p^{k-1}-(p-1) (7 p-2) p^{2 k-4}+(19-3 p) p^{2k-3}\right)\,,
    \end{aligned}
\end{equation}
which admits a rational point
\begin{equation}
    P\colon (x,y) = \left(2p(5p-1),4p^2(3-7p-p^{k-2})\right)\,.
\end{equation}
This elliptic curve has additive reduction modulo $p$. We therefore extend the field by an element $p^\prime = (p)^{1/2}$, let $x\to{p^\prime}^2 x^\prime, y\to{p^\prime}^3 y^\prime$, and divide through by ${p^\prime}^{6}$. The resulting elliptic curve $E^\prime$ is given by
\begin{equation}
    \begin{aligned}
        {y^\prime}^2 &= {x^\prime}^3 + \left[-3 \left(1+3 {p^\prime}^2\right)^2-{p^\prime}^{4 k-8}+{p^\prime}^{4k-6}+32 {p^\prime}^{2 k-2}\right]{x^\prime}\\
        &\qquad +\left[2 \left(\left(1+3 {p^\prime}^2\right)^3-16 {p^\prime}^{2 k-2} \left(1+3{p^\prime}^2\right)+{p^\prime}^{4 k-8} \left(-1+14 {p^\prime}^2-5{p^\prime}^4\right)\right)\right]\,,
    \end{aligned}
\end{equation}
and it admits a point $P^\prime$ given by
\begin{equation}
    P^\prime\colon (x^\prime,y^\prime) = \left(2\left(5{p^\prime}^2-1\right), 4p^{\prime}\left(3-7{p^\prime}^2-{p^\prime}^{2k-4}\right)\right)\,.
\end{equation}

We can now calculate the EDS valuations according to the formulas in~\cite{StangeEllTrouble}. The $k=2$ case has potential good reduction, and the EDS valuation calculation uses  \cref{eq:valWnpotgood,eq:valWngood}. Even multiples of $P^\prime$ reduce to the identity point, and the $(x^\prime,y^\prime)$ coordinates of $2P^\prime$ are respectively proportional to $\left(p^\prime(7{p^\prime}^2 - 2)\right)^{-2}$ and $\left(p^\prime(7{p^\prime}^2 - 2)\right)^{-3}$. Therefore $n_{P^\prime}=2$ and the $\scorr$ term, which is essentially a constant according to \cref{app:scorr}, should be replaced with 1.

The $k>2$ cases, like those encountered for the antisymmetrics of $\asu(2k+1)$, have potential multiplicative reduction. However, the point $P^\prime$ does not reduce to the singular point, and the EDS valuation calculation therefore involves \cref{eq:valWnpotmult,eq:valWngood}. Even multiples of $2P^\prime$ reduce to the identity point, and the $(x^\prime,y^\prime)$ coordinates of $2P^\prime$ are proportional to $\left(p^\prime \left(-3+7{p^\prime}^2+{p^\prime}^{2k-4}\right)^2\right)^{-2}$ and $\left(p^\prime \left(-3+7{p^\prime}^2+{p^\prime}^{2k-4}\right)^2\right)^{-3}$. Therefore, $n_{P^\prime}=2$, and the $\scorr$ term should be replaced with 1.

Putting everything together, we find that the EDS valuations for $k\ge2$ are given by
\begin{equation}
    \valuation{}{\eds_{m}} = \frac{m^2}{2} -\frac{1}{2}\begin{cases}0 & 2\mid m \\ 1 & 2\nmid m\end{cases}\,.\label{eq:soevenvaluationsmu2}
\end{equation}

\subsection{$\singtype{I}^*_{2k-3}$ singularities with $\quotparam=0$ or $\quotparamtwo=0$}
\label{app:sonoddedsnu0}

For $\singtype{I}^*_{2k-3}$ singular fibers with $\quotparam=0$ or $\quotparamtwo=0$, we are primarily interested in two scenarios: one where the generating section components $(\secx,\secy,\secz,\secw)$ vanish to orders $(0,0,0,0)$ and one where they vanish to orders $(2,3,1,4)$. When the section components vanish to orders $(0,0,0,0)$, the associated EDS valuations are $0$, as can be verified using the formulas in~\cite{StangeEllTrouble}. The $(2,3,1,4)$ situation, however, has nontrivial EDS valuations.

To analyze this situation, we first use the seed model in \cref{eq:so4mvectormodelb} to construct an elliptic curve $E$. Specifically, the $\singtype{I}^*_{2k-3}$ singularities occur at $\locus{v=b_{0,0}=0}$, so we set $v$ and $b_{0,0}$ to a prime $p$ and set the other parameters to some arbitrary numbers. However, the $\secz$ component of the generating section would then be equal to $p$, whereas $\secz$ was not proportional to $p$ in the other situations we have encountered so far. To account for this, we absorb the $\secz^4$ and $\secz^6$ factors in the Weierstrass model into the $f$ and $g$ of $E$. For example, we can take $E$ to be given by the equation
\begin{equation}
    \begin{aligned}
        y^2 &= x^3 -p^6 \left(3-2 p^{k-1}+p^{2 k-2}+p^{2 k-1}\right)x \\
        &\qquad\quad -p^9 \left(-2+2 p^{k-1}-p^{2 k-3}+p^{2 k-2}+p^{2 k-1}\right)\,.
    \end{aligned}
\end{equation}
This elliptic curve admits the rational point
\begin{equation}
    P\colon (x,y) = \left(-p^2(2p-1), -p^3 \left(1-3 p+p^{k+1}\right)\right)\,.
\end{equation}

This elliptic curve is non-minimal and has additive reduction modulo $p$. We therefore extend the field by the element $p^\prime = p^{1/2}$, let $x\to x^\prime {p^\prime}^6, y\to y^\prime {p^\prime}^9$, and divide both sides of the elliptic curve's equation by ${p^\prime}^{18}$. This procedure give us a new elliptic curve $E^\prime$ taking the form
\begin{equation}
    \begin{aligned}
        {y^\prime}^2 &= {x^\prime}^3 + \left(-3+2 {p^\prime}^{2 k-2}-{p^\prime}^{4 k-4}-{p^\prime}^{4 k-2}\right)x \\
        &\qquad\quad + \left(2-2 {p^\prime}^{2 k-2}+{p^\prime}^{4 k-6}-{p^\prime}^{4 k-4}-{p^\prime}^{4 k-2}\right)\,,
    \end{aligned}
\end{equation}
which admits the rational point
\begin{equation}
    P^\prime\colon (x^\prime, y^\prime) = \left(\frac{1-2{p^\prime}^2}{{p^\prime}^2}, -\frac{1-3 {p^\prime}^2+{p^\prime}^{2 k+2}}{{p^\prime}^3}\right)\,.
\end{equation}
The elliptic curve $E^\prime$ has multiplicative reduction modulo $p^\prime$, and hence $E$ has potential multiplicative reduction modulo $p$. The rational point $P^\prime$, meanwhile, reduces to the identity modulo $p^\prime$. Therefore, the EDS valuations should be described by \cref{eq:valWngood,eq:valWnpotmult} with $\valuation{}{f}=6$, $n_{P^\prime}=1$ and $\valuation{1}{x^\prime_{P^\prime}} = -2$. Putting everything together, we have
\begin{equation}
    \valuation{}{\eds_{m}} = \frac{3}{2}\left(m^2-1\right) + \frac{1}{2}\left(-m^2 + \scorr_{m/n_{{P^\prime}}}\left(p^\prime, b, \valuation{1}{p^\prime}, h, s, w\right)\right)\,.
\end{equation}
The $\scorr$ term should be a constant according to \cref{app:scorr}. Since the $\secz$ component of the generating section vanishes to order 1, $\scorr$ should be set to $3$ such that $\valuation{}{\eds_{1}} = 1$. This leaves us with
\begin{equation}
    \valuation{}{\eds_{m}} = m^2\,. \label{eq:sooddedsnu0}
\end{equation}

\subsection{$\singtype{IV}^*$ singularities}
\label{app:e6eds}
For $\singtype{IV}^{*}$ singular fibers, we are primarily interested in the analogous EDS valuations  when the generating section is described by $\quotparam=1,2$ or $\quotparamtwo=1,2$. We first find an elliptic curve $E$ by taking the construction in \cref{eq:so10spinorseed}, setting $\sigma$ and $b_{2,1}$ to a prime integer $p$, and replacing the other parameters with numbers such that $f$ and $g$ are integers. For instance, we can consider an elliptic curve $E$ over $\Q_p$ of the form
\begin{equation}
    y^2 = x^3 + p^3 x + p^4(1-p-p^2)\,,
\end{equation}
which has additive reduction modulo $p$. After making the same substitutions into the generating section components in \cref{eq:so10spinorseedgensec}, we can see that $E$ admits a rational point $P$ with coordinates
\begin{equation}
    P\colon (x,y) = (p^2, p^2)\,.
\end{equation}

We now extend $\Q_p$ by $p^\prime = p^{1/3}$, let $x={p^\prime}^{4}x^\prime$ and $y={p^\prime}^6 y^\prime$, and remove an overall factor of ${p^\prime}^{12}$ from the expression for $E$. This procedure gives us a new elliptic curve $E^\prime$ of the form
\begin{equation}
    {y^\prime}^2 = {x^\prime}^3 + p^\prime x + (1-{p^\prime}^3 -{p^\prime}^6)\,,
\end{equation}
which admits a rational point $P^\prime$ with coordinates
\begin{equation}
    P^\prime\colon (x^\prime, y^\prime) = ({p^\prime}^2,1)\,.
\end{equation}
The elliptic curve $E^\prime$ has good reduction modulo $p^\prime$, and the original elliptic curve $E$ therefore has potential good reduction modulo $p$. In turn, \cref{eq:valWnpotgood} (from Theorem~19 in~\cite{StangeEllTrouble}) suggests that the EDS valuations corresponding to $E$ and $P$ are given by
\begin{equation}
    \valuation{}{W_m} = \frac{2}{3}\left(m^2-1\right) + \frac{1}{3} \valuation{1}{W_m^\prime}\,.
\end{equation}
Here, $\valuation{1}{W_m^\prime}$ represents the EDS valuations corresponding to $E^\prime$ and $P^\prime$ with respect to $p^{\prime}$, which should be given by \cref{eq:valWngood}. If one calculates multiples of $P^\prime$, one finds that $3P^\prime$ reduces to the identity point modulo $p^\prime$ and that $\valuation{1}{W_3}$ is $2$. Therefore, $n_{P}$ in \cref{eq:valWngood} is $3$, and the $\scorr$ term, which should be a constant according to \cref{app:scorr}, can be replaced with $2$. Combining all of this information leads to the formula
\begin{equation}
    \valuation{}{W_m} = \frac{2}{3}\left(m^2-1\right) + \begin{cases} \frac{2}{3} & 3\mid m \\  0 & 3\nmid m\end{cases} = \frac{2}{3}m^2  - \begin{cases} 0 & 3\mid m \\  \frac{2}{3} & 3\nmid m\end{cases}\,. \label{eq:edsvale6}
\end{equation}

\subsection{$\singtype{III}^*$ singularities}
\label{app:e7eds}
\subsubsection{$\quotparam=0$ or $\quotparamtwo=0$}
\label{app:e7edsnu0}
For $\singtype{III}^*$ singularities with either $\quotparam$ or $\quotparamtwo$ equal to 0, we are mostly interested in the EDS valuations corresponding to generating sections whose $(\secx,\secy,\secz,\secw)$ components vanish to orders $(2,3,1,4)$ at a particular locus. As an example, we can consider the Weierstrass model in \cref{eq:so12spinormodel} and the locus $\locus{\sigma=b_{0,0}=0}$. The $\secz$ component in this example, which is equivalent to $b_{0,0}$, vanishes to order one at this locus. To account for this fact, we construct the elliptic curve analogue $E$ of  the fibration given by
\begin{equation}
    y^2 = x^3 + f b_{0,0}^4 + g b_{0,0}^6\,,
\end{equation}
where $f$ and $g$ are taken from \cref{eq:so12spinormodel}. At the end of the calculation, we will address the fact that these factors of $b_{0,0}$ come from $\secz$. As with the other examples, the parameters should be set to appropriate numbers. We set $\sigma$ and $b_{0,0}$ to a prime $p$ and let the other parameters be numbers such that $f$ and $g$ are integers. For a particular choice of these numbers, we obtain an elliptic curve $E$ of the form
\begin{equation}
    y^2 = x^3 -p^7 \left(12 p^3+p^2+p-6\right)x -p^{11} \left(16 p^4+p^3-20 p^2-12 p+11\right)\,,
\end{equation}
which admits a rational point
\begin{equation}
    P\colon (x,y) = \left(-p^2 \left(2 p^3+2 p^2-1\right), -p^3 \left(p^4-3 p^2+1\right)\right)\,.
\end{equation}
The discriminant is proportional to $p^{21}$, and $E$ has additive reduction modulo $p$. We then extend the field with the element $p^\prime = p^{1/4}$ of ramification degree $d=4$, let $x={p^\prime}^{14}{x^\prime}$ and $y={p^\prime}^{21}{y^\prime}$, and divide both sides of the equation for the elliptic curve by ${p^\prime}^{42}$. This procedure leads to a new elliptic curve $E^\prime$ given by
\begin{equation}
    {y^\prime}^2 = {x^\prime}^3 + \left(6-{p^\prime}^4-{p^\prime}^8 -12 {p^\prime}^{12}\right){x^\prime} - {p^\prime}^2 \left(11 - 12 {p^\prime}^4 - 20 {p^\prime}^8 + {p^\prime}^{12} + 16 {p^\prime}^{16}\right)\,.
\end{equation}
$E^\prime$ has good reduction modulo $p^\prime$, and $E$ therefore has potential good reduction. Additionally, $E^\prime$ admits a rational point
\begin{equation}
    P^\prime\colon (x^\prime, y^\prime) = \left({p^\prime}^{-6}\left(1 - 2 {p^\prime}^8 - 2 {p^\prime}^{12}\right), {p^\prime}^{-9}\left(-1 + 3 {p^\prime}^8 -{p^\prime}^{16}\right)\right)
\end{equation}
that reduces to the identity point on $E^\prime$ modulo $p^\prime$. According to \cref{eq:valWnpotgood}, which comes from Theorem~19 of~\cite{StangeEllTrouble}, the EDS valuations associated with $(E,P)$ should be given by the formula
\begin{equation}
    \valuation{}{\eds_m} = \frac{21}{12}(m^2-1) + \frac{1}{4}\valuation{1}{\eds_m^\prime}\,,
\end{equation}
where $\valuation{1}{\eds_m^\prime}$ represents the EDS valuations for $(E^\prime, P^\prime)$. Since $P^\prime$ reduces to the identity point, Theorem~14 of~\cite{StangeEllTrouble} states that
\begin{equation}
	\begin{aligned}
    \valuation{1}{\eds_m^\prime} &= \frac{\valuation{1}{x^\prime_{P^\prime}}}{2}m^2 + \scorr_{m}\left(p^\prime, p^\prime, \valuation{1}{p^\prime}, 0, s, w\right) \\
    &= -3 m^2 + \scorr_{m}\left(p^\prime, p^\prime, \valuation{1}{p^\prime}, 0, s, w\right)\,.
    \end{aligned}
\end{equation}
The discussion in \cref{app:scorr} suggests that the $\scorr$ term can be replaced with a constant $s$. If we put these equations together, we find that
\begin{equation}
    \valuation{}{\eds_m} = \frac{21}{12}(m^2-1) + \frac{1}{4}\left(-3m^2 + s_P\right) = m^2 + \frac{1}{4}(s-7)\,.
\end{equation}
Now we can address the fact that we included extra factors of $b_{0,0}$ in $f$ and $g$. Because $\secz$ for the generating section actually vanishes to order 1 at the locus in question, we set $s$ to 7 such that $\valuation{}{\eds_1}$ is 1. In the end, we have
\begin{equation}
    \valuation{}{\eds_m} = m^2\,.
\end{equation}

\subsubsection{$\quotparam=1$ or $\quotparamtwo=1$}
\label{app:e7edsnu1}
For the EDS valuations when the generating section components vanish to orders $(2,2,0,3)$, we can consider the Weierstrass model in \cref{eq:e6weiermodel} and section components in \cref{eq:e6gensec}. We then obtain an elliptic curve $E$ and a point $P$ by setting $\sigma$ and $y_2$ to a prime $p$ and letting the other parameters be integers. As an example, we could take $(E,P)$ to be
\begin{align}
    E:&\, y^2 = x^3 + p^{3} x - p^{5}, & P:&\,  (x,y) = (p^2,p^3)\,.
\end{align}
This elliptic curve has additive reduction modulo $p$. We can obtain a new elliptic curve by extending the field by $p^\prime = p^{1/4}$, defining $x = (p^{\prime})^6 x^{\prime}$ and $y = (p^{\prime})^9 y^{\prime}$, and removing factors of $(p^{\prime})^{18}$:
\begin{equation}
    {y^\prime}^2 = {x^\prime}^3 +  x^{\prime} - (p^{\prime})^2\,.
\end{equation}
This new curve has good reduction, and the original elliptic curve $E$ therefore has potential good reduction. The point $P$ on $E$ corresponds to a point $P^\prime$ on $E^\prime$ with coordinates
\begin{equation}
    (x^\prime, y^\prime) = \left((p^{\prime})^2,(p^{\prime})^3 \right)
\end{equation}
According to \cref{eq:valWnpotgood,eq:valWngood}, the valuations of an EDS corresponding to a point on the original curve follow the formula
\begin{equation}
    \valuation{p}{\eds_m} = \frac{3}{4}\left(m^2 - 1\right) + \frac{1}{4}\begin{cases}  \scorr_{m/n_{P^\prime}}(p^{\prime}, b, \valuation{p^\prime}{p^\prime}, h, s, w) & n_{P^\prime}\mid m\\ 0 & n_{P^\prime}\nmid m \end{cases}\,. \label{eq:e7curvevals}
\end{equation}
One can show that $2P^\prime$ reduces to the identity point modulo $p^{1/4}$, implying that $n_{P^\prime}$ is $2$. Additionally, the coordinates of $2P^\prime$ show that $\valuation{}{W_2^\prime}$ is 3. The $\scorr$ term, which should essentially be a constant according to \cref{app:scorr}, should therefore be replaced with 3, leading to the formula
\begin{equation}
    \label{eq:e7edsnu1final}
    \valuation{p}{\eds_m} =  \frac{3}{4}\left(m^2 - 1\right) + \begin{cases} \frac{3}{4} & 2\mid m\\ 0 & 2\nmid m \end{cases} = \frac{3}{4}m^2 -\begin{cases} 0 & 2\mid m\\ \frac{3}{4} & 2\nmid m \end{cases}\,.
\end{equation}

\subsection{$\singtype{II}^*$ singularities}
\label{app:e8eds}
For the $\singtype{II}^*$ singularity type, we are primarily interested in situations where the generating section components $(\secx,\secy,\secz,\secw)$ vanish to orders $(2,3,1,4)$ at a particular locus. As an example, we can consider the Weierstrass model of \cref{eq:so14spinorseed} and the locus $\locus{\sigma=b_{0,0}=0}$. We must account for the fact that the $\secz$ component of the generating section, which is given by $b_{0,0}$, vanishes to order 1 at the $\singtype{II}^*$ locus. We therefore construct the $f$ and $g$ of the $p$-adic elliptic curve by plugging in appropriate values of the parameters into $f b_{0,0}^4$ and $g b_{0,0}^6$ from \cref{eq:so14spinorseed};  at the end of the calculation, we will account for the fact that these extra factors of $b_{0,0}$ actually come from $\secz$.  We set both $\sigma$ and $b_{0,0}$ to a prime integer $p$ and set the other parameters to numbers such that $f$ and $g$ are integers. Performing this procedure with a particular choice of numbers leads to an elliptic curve $E$ of the form
\begin{equation}
    \begin{aligned}
        y^2 &= x^3  -p^8 \left(1-12p+p^2+24 p^3 + 12 p^6\right) x \\
        &\qquad\quad + p^{11}\left(-6+p+24 p^2-p^3-26 p^4+24 p^{5}-2 p^{6}-48 p^{7}-16 p^{10}\right)\,.
    \end{aligned}
\end{equation}
The discriminant $\Delta_{E}$ for this elliptic curve is proportional to $p^{22}$. After plugging the same numbers into the $[\secx:\secy:\secz]$ section components, we see that $E$ admits an integral point $P$ with coordinates
\begin{equation}
    P\colon (x,y) = \left(p^2(1-2 p^2-2 p^5),p^3(-1+3 p^2 -p^4)\right)\,.
\end{equation}
The elliptic curve $E$ has additive reduction modulo $p$. We therefore extend the field with $p^\prime = p^{1/6}$, let $x=x^\prime {p^\prime}^{22}$ and $y=y^\prime {p^\prime}^{33}$, and divide both sides of the equation for the elliptic curve by ${p^\prime}^{66}$. This gives a new elliptic curve $E^\prime$ given by the equation
\begin{equation}
    \begin{aligned}
        {y^\prime}^2 &= {x^\prime}^3 - {p^\prime}^4  \left(1-12{p^\prime}^6+{p^\prime}^{12}+24 {p^\prime}^{18} + 12 {p^\prime}^{36}\right){x^\prime} \\
        &\qquad + \left(-6+{p^\prime}^{6}+24{p^\prime}^{12}-{p^\prime}^{18}-26 {p^\prime}^{24}+24 {p^\prime}^{30}-2 {p^\prime}^{36}-48 {p^\prime}^{42}-16 {p^\prime}^{60}\right)\,,
    \end{aligned}
\end{equation}
which admits a rational point $P^\prime$ with coordinates
\begin{equation}
    P^\prime\colon (x^\prime, y^\prime) =  \left({p^\prime}^{-10}(1-2 {p^\prime}^{12}-2 {p^\prime}^{30}), {p^\prime}^{-15}(-1+3 {p^\prime}^{12} -{p^\prime}^{24})\right)\,.
\end{equation}
The elliptic curve $E^\prime$ has good reduction modulo $p^\prime$, and the original elliptic curve $E$ has potential good reduction modulo $p$. Since $\valuation{}{\Delta_{E}}$ is $22$ and $p^\prime$ has ramification degree $d=6$, \cref{eq:valWnpotgood}, which comes from Theorem~19 of~\cite{StangeEllTrouble}, states that the EDS valuations associated with $(E,P)$ should satisfy the formula
\begin{equation}
    \valuation{}{\eds_m} = \frac{22}{12}(m^2-1) + \frac{1}{6}\valuation{1}{\eds_m^\prime}\,,
\end{equation}
where $\valuation{1}{\eds_m^\prime}$ represents the EDS valuations for $E^\prime$ and $P^\prime$. Note that $P^\prime$ reduces to the identity point on $E^\prime$ modulo $p^\prime$, so according to \cref{eq:valWngood} and Theorem~14 of~\cite{StangeEllTrouble},
\begin{equation}
    \begin{aligned}
        \valuation{1}{\eds_m^\prime} &= \frac{\valuation{1}{x^\prime_{P^\prime}}}{2}m^2 + \scorr_{m}\left(p^\prime, p^\prime, \valuation{1}{p^\prime}, 0, s, w\right) \\
        &= -5 m^2 + \scorr_{m}\left(p^\prime, p^\prime, \valuation{1}{p^\prime}, 0, s, w\right)\,.
    \end{aligned}
\end{equation}
According to \cref{app:scorr}, the $\scorr$ term can be replaced with a constant $s_{P}$. Combining everything together, we find that
\begin{equation}
    \valuation{}{\eds_m} = \frac{22}{12}(m^2-1) + \frac{1}{6}\left(-5 m^2 + s_P\right) = m^2 +\frac{1}{6}\left(s_P - 11\right)\,.
\end{equation}
Since the $\secz$ component of the generating section vanishes to order 1 at the matter locus, $\valuation{}{\eds_m}$ should morally be 1 for our purposes. We therefore set $s_P$ to 11, giving us
\begin{equation}
    \valuation{}{\eds_m} = m^2\,. \label{eq:edsvale8}
\end{equation}

\bibliographystyle{JHEP}
\bibliography{references}

\end{document}